\pdfoutput=1
\documentclass[11pt,twoside,a4paper,cmspaper,final,collab]{cms-tdr}
\def\svnVersion{2ad4bfe}\def\svnDate{2025/04/09}\def\cmsCernNoTag{CERN-EP-2024-005}\def\cmsCernDate{\today}\def\cmsMessage{Published in Physics Reports as \href{http://dx.doi.org/10.1016/j.physrep.2024.12.002}{\doi{10.1016/j.physrep.2024.12.002}.}}
\begin{document}\cmsNoteHeader{TOP-23-003}

\newlength\cmsTabSkip\setlength{\cmsTabSkip}{1ex}
\providecommand{\cmsTableShrink}[1]{\resizebox{\textwidth}{!}{#1}}
\ifthenelse{\boolean{cms@external}}{\providecommand{\cmsTable}[1]{#1}}{\providecommand{\cmsTable}[1]{\cmsTableShrink{#1}}}
\ifthenelse{\boolean{cms@external}}{\newcommand\cmsParagraph[1]{\paragraph{#1}}}{\newcommand\cmsParagraph[1]{\textbf{#1:}}}

\newcommand{\asmz}{\ensuremath{\alpS(m_{\PZ})}\xspace}
\newcommand{\asISR}{\ensuremath{\alpS^{\text{ISR}}}\xspace}
\newcommand{\asFSR}{\ensuremath{\alpS^{\text{FSR}}}\xspace}

\newcommand{\mur}{\ensuremath{\mu_{\mathrm{r}}}\xspace}
\newcommand{\muf}{\ensuremath{\mu_{\mathrm{f}}}\xspace}
\newcommand{\muk}{\ensuremath{\mu_{\mathrm{k}}}\xspace}
\newcommand{\muref}{\ensuremath{\mu_{\text{ref}}}\xspace}
\newcommand{\mumass}{\ensuremath{\mu_{\mathrm{m}}}\xspace}
\newcommand{\murISR}{\ensuremath{\mur^{\text{ISR}}}\xspace}
\newcommand{\murFSR}{\ensuremath{\mur^{\text{FSR}}}\xspace}
\newcommand{\murmuf}{\ensuremath{\mur\oplus\muf}\xspace}
\newcommand{\fmur}{\ensuremath{f_{\mur}}\xspace}

\newcommand{\stt}{\ensuremath{\sigma_{\ttbar}}\xspace}
\newcommand{\sttk}{\ensuremath{\stt^k}\xspace}
\newcommand{\sttvis}{\ensuremath{\stt^{\text{vis}}}\xspace}
\newcommand{\sttjet}{\ensuremath{\sigma_{\ttbar\text{+jet}}}\xspace}
\newcommand{\stth}{\ensuremath{\stt^{\text{(h)}}}\xspace}
\newcommand{\sttl}{\ensuremath{\stt^{\text{(l)}}}\xspace}

\newcommand{\mt}{\ensuremath{m_{\PQt}}\xspace}
\newcommand{\mtp}{\ensuremath{\mt^{\text{pole}}}\xspace}
\newcommand{\mtfit}{\ensuremath{\mt^\text{fit}}\xspace}
\newcommand{\mtfiti}{\ensuremath{m_{\PQt,i}^\text{fit}}\xspace}
\newcommand{\mtgen}{\ensuremath{\mt^\text{gen}}\xspace}
\newcommand{\mtreco}{\ensuremath{\mt^\text{reco}}\xspace}
\newcommand{\mtrec}{\ensuremath{\mt^\text{rec}}\xspace}
\newcommand{\mtmc}{\ensuremath{\mt^{\text{MC}}}\xspace}
\newcommand{\mtmsr}{\ensuremath{\mt^{\text{MSR}}}\xspace}
\newcommand{\mtbar}{\ensuremath{m_{\PAQt}}\xspace}
\newcommand{\mtmt}{\ensuremath{\mt(\mt)}\xspace}
\newcommand{\mtmu}{\ensuremath{\mt(\mu)}\xspace}
\newcommand{\mtmum}{\ensuremath{\mt(\mumass)}\xspace}
\newcommand{\mtmuref}{\ensuremath{\mt(\muref)}\xspace}
\newcommand{\delmt}{\ensuremath{\Delta\mt}\xspace}
\newcommand{\delmtp}{\ensuremath{\Delta\mtp}\xspace}
\newcommand{\delmtpm}{\ensuremath{\Delta_{\mt\pm}}\xspace}
\newcommand{\barmt}{\ensuremath{\overline{\mt}}\xspace}

\newcommand{\mW}{\ensuremath{m_{\PW}}\xspace}
\newcommand{\mWreco}{\ensuremath{\mW^\text{reco}}\xspace}
\newcommand{\mWrecoi}{\ensuremath{m_{\PW,i}^\text{reco}}\xspace}
\newcommand{\mZ}{\ensuremath{m_{\PZ}}\xspace}

\newcommand{\mtt}{\ensuremath{m_{\ttbar}}\xspace}
\newcommand{\mttreco}{\ensuremath{m_{\ttbar}^{\text{reco}}}\xspace}
\newcommand{\mttrec}{\ensuremath{m_{\ttbar}^{\text{rec}}}\xspace}
\newcommand{\mttgen}{\ensuremath{m_{\ttbar}^{\text{gen}}}\xspace}
\newcommand{\mttbarjet}{\ensuremath{m_{\ttbarjet}}\xspace}

\newcommand{\mlb}{\ensuremath{m_{\Pell\PQb}}\xspace}
\newcommand{\mlbreco}{\ensuremath{\mlb^\text{reco}}\xspace}
\newcommand{\mlbmin}{\ensuremath{\mlb^\text{min}}\xspace}
\newcommand{\mlbrecomtfit}{\ensuremath{\mlbreco/\mtfit}\xspace}
\newcommand{\Mbl}{\ensuremath{M_{\PQb\Pell}}\xspace}
\newcommand{\mbl}{\ensuremath{m_{\PQb\Pell}}\xspace}

\newcommand{\mjet}{\ensuremath{m_{\text{jet}}}\xspace}
\newcommand{\mjetrec}{\ensuremath{\mjet^{\text{rec}}}\xspace}
\newcommand{\mjetgen}{\ensuremath{\mjet^{\text{gen}}}\xspace}

\newcommand{\ytt}{\ensuremath{y_{\ttbar}}\xspace}

\newcommand{\pTtt}{\ensuremath{p_{\mathrm{T},\ttbar}}\xspace}
\newcommand{\pTnu}{\ensuremath{p_{\mathrm{T},\PGn}}\xspace}
\newcommand{\pZnu}{\ensuremath{p_{z,\PGn}}\xspace}
\newcommand{\pTell}{\ensuremath{p_{\mathrm{T},\Pell}}\xspace}
\newcommand{\pZell}{\ensuremath{p_{z,\Pell}}\xspace}
\newcommand{\pTVECell}{\ensuremath{{\vec p}_{\mathrm{T},\Pell}}\xspace}

\newcommand{\pTsum}{\ensuremath{\pt^{\text{sum}}}\xspace}
\newcommand{\sumpT}{\ensuremath{\sum\pt}\xspace}

\newcommand{\Eell}{\ensuremath{E_{\Pell}}\xspace}
\newcommand{\Eellp}{\ensuremath{E_{\Pellp}}\xspace}
\newcommand{\Eellm}{\ensuremath{E_{\Pellm}}\xspace}
\newcommand{\Enu}{\ensuremath{E_{\PGn}}\xspace}
\newcommand{\Enubar}{\ensuremath{E_{\PAGn}}\xspace}
\newcommand{\Enunu}{\ensuremath{E_{\nunu}}\xspace}
\newcommand{\Ellbar}{\ensuremath{E_{\llbar}}\xspace}
\newcommand{\Ebhad}{\ensuremath{E_{\text{\PQb hadron}}}\xspace}

\newcommand{\PQj}{{\HepParticle{j}{}{}}\xspace}
\newcommand{\PAQqpr}{{\HepAntiParticle{\PQq}{}{\prime}}\xspace}
\newcommand{\PGnell}{{\HepParticle{\PGn}{\!\ell}{}}\xspace}

\newcommand{\tW}{\ensuremath{\PQt\PW}\xspace}
\newcommand{\nunu}{\ensuremath{\PGn\PAGn}\xspace}
\newcommand{\llbar}{\ensuremath{\Pell\PAell}\xspace}
\newcommand{\llnn}{\ensuremath{\llbar\nunu}\xspace}
\newcommand{\qqpr}{\ensuremath{\PQq\PAQqpr}\xspace}
\newcommand{\qqb}{\ensuremath{\qqpr\PQb}\xspace}
\newcommand{\qqbbar}{\ensuremath{\qqpr\PAQb}\xspace}
\newcommand{\lpnb}{\ensuremath{\Pellp\PGn\PQb}\xspace}
\newcommand{\lmnb}{\ensuremath{\Pellm\PAGn\PAQb}\xspace}

\newcommand{\ttobqq}{\ensuremath{\PQt\to\PQb\qqpr}\xspace}
\newcommand{\ttobW}{\ensuremath{\PQt\to\PQb\PW}\xspace}
\newcommand{\ttbartoWbWb}{\ensuremath{\ttbar\to\PWp\PQb\PWm\PAQb}\xspace}
\newcommand{\ZtoLL}{\ensuremath{\PZ\to\Pell\Pell}\xspace}

\newcommand{\sqrts}{\ensuremath{\sqrt{s}}\xspace}
\newcommand{\sqrtseq}[1]{\ensuremath{\sqrts=#1\TeV}\xspace}

\renewcommand{\EE}{\ensuremath{\Pep\Pem}\xspace}
\renewcommand{\MM}{\ensuremath{\PGmp\PGmm}\xspace}
\newcommand{\EM}{\ensuremath{\Pepm\PGmmp}\xspace}

\newcommand{\pp}{\ensuremath{\Pp\Pp}\xspace}
\newcommand{\ppbar}{\ensuremath{\Pp\PAp}\xspace}
\newcommand{\pbarp}{\ensuremath{\PAp\Pp}\xspace}
\newcommand{\ep}{\ensuremath{\Pe\Pp}\xspace}
\newcommand{\GG}{\ensuremath{\Pg\Pg}\xspace}

\newcommand{\xfitter}{\ensuremath{\textsc{xFitter}}\xspace}
\newcommand{\rivet}{\ensuremath{\textsc{rivet}}\xspace}
\newcommand{\PYTHIAEight}{\ensuremath{\PYTHIA8}\xspace}
\newcommand{\PYTHIASix}{\ensuremath{\PYTHIA6}\xspace}
\newcommand{\HitFit}{\ensuremath{\textsc{HitFit}}\xspace}
\newcommand{\KinFitter}{\ensuremath{\textsc{KinFitter}}\xspace}
\newcommand{\Professor}{\ensuremath{\textsc{professor}}\xspace}
\newcommand{\MADGRAPHFive}{\ensuremath{\MADGRAPH5}\xspace}
\newcommand{\POWHEGv}[1]{\ensuremath{\POWHEG\text{~v#1}}\xspace}
\newcommand{\POWHEGPYTHIA}{\ensuremath{\POWHEG\text{+}\PYTHIA}\xspace}
\newcommand{\POWHEGHERWIG}{\ensuremath{\POWHEG\text{+}\HERWIG}\xspace}
\newcommand{\POWHEGPYTHIAEight}{\ensuremath{\POWHEGPYTHIA8}\xspace}
\newcommand{\POWHEGHERWIGSeven}{\ensuremath{\POWHEGHERWIG7}\xspace}
\newcommand{\POWHEGHERWIGpp}{\ensuremath{\POWHEG\text{+}\HERWIGpp}\xspace}

\newcommand{\msbar}{\ensuremath{\mathrm{\overline{MS}}}\xspace}
\newcommand{\CP}{\ensuremath{CP}\xspace}
\newcommand{\CPT}{\ensuremath{CPT}\xspace}
\newcommand{\Run}[1]{\mbox{Run #1}\xspace}
\newcommand{\JSF}{\ensuremath{\text{JSF}}\xspace}
\newcommand{\etaphi}{\ensuremath{\eta\text{-}\phi}\xspace}
\newcommand{\NNPDF}[1]{\mbox{NNPDF#1}\xspace}

\newcommand{\hdamp}{\ensuremath{h_{\text{damp}}}\xspace}
\newcommand{\MTtwobb}{\ensuremath{M_{\mathrm{T}2}^{\PQb\PQb}}\xspace}

\newcommand{\Pgof}{\ensuremath{P_{\text{gof}}}\xspace}
\newcommand{\Pgofi}{\ensuremath{P_{\text{gof},i}}\xspace}
\newcommand{\chisq}{\ensuremath{\chi^2}\xspace}

\newcommand{\gammajet}{\ensuremath{\PGg\text{+jet}}\xspace}
\newcommand{\Zjet}{\ensuremath{\PZ\text{+jet}}\xspace}
\newcommand{\Zjets}{\ensuremath{\PZ\text{+jets}}\xspace}
\newcommand{\Wjets}{\ensuremath{\PW\text{+jets}}\xspace}
\newcommand{\Vjets}{\ensuremath{\PV\text{+jets}}\xspace}
\newcommand{\ttbarjet}{\ensuremath{\ttbar\text{+jet}}\xspace}
\newcommand{\ttbarjets}{\ensuremath{\ttbar\text{+jets}}\xspace}
\newcommand{\mjets}{\ensuremath{\PGm\text{+jets}}\xspace}
\newcommand{\ejets}{\ensuremath{\Pe\text{+jets}}\xspace}
\newcommand{\ttbarnojet}{\ensuremath{\ttbar\text{+0~jet}}\xspace}

\newcommand{\scale}{\ensuremath{\,\text{(scale)}}\xspace}
\newcommand{\fit}{\ensuremath{\,\text{(fit)}}\xspace}
\newcommand{\model}{\ensuremath{\,\text{(model)}}\xspace}
\newcommand{\param}{\ensuremath{\,\text{(param)}}\xspace}
\newcommand{\statJSF}{\ensuremath{\,\text{(stat+JSF)}}\xspace}
\newcommand{\fitPDFas}{\ensuremath{\,\text{(fit+PDF+\alpS)}}\xspace}

\newcommand{\nj}{\ensuremath{N_{\text{jet}}}\xspace}
\newcommand{\njets}{\ensuremath{N_{\text{Jets}}}\xspace}
\newcommand{\njobs}{\ensuremath{N_j^{\text{obs}}}\xspace}
\newcommand{\njexp}{\ensuremath{N_j^{\text{exp}}}\xspace}
\newcommand{\njmttytttwo}{\ensuremath{[\nj^{0,1+},\mtt,\ytt]}\xspace}
\newcommand{\njmttyttthree}{\ensuremath{[\nj^{0,1,2+},\mtt,\ytt]}\xspace}
\newcommand{\nbj}{\ensuremath{N_{\PQb\text{\,jet}}}\xspace}

\newcommand{\Rb}{\ensuremath{R_{\PQb\PQq}^\text{reco}}\xspace}
\newcommand{\vectheta}{\ensuremath{\vec{\theta}}\xspace}
\newcommand{\dnnresponse}{\ensuremath{R_{\text{NN}}}\xspace}
\newcommand{\LQCD}{\ensuremath{\Lambda_{\text{QCD}}}\xspace}
\newcommand{\rbfrag}{\ensuremath{r_{\PQb}}\xspace}
\newcommand{\ebfrag}{\ensuremath{\varepsilon_{\PQb}}\xspace}
\newcommand{\abseta}{\ensuremath{\abs{\eta}}\xspace}

\newcommand{\likelihoodsample}{\ensuremath{\mathcal{L}(\text{sample}|\mt,\JSF)}\xspace}
\newcommand{\likelihoodexp}{\ensuremath{L_{\text{exp}}}\xspace}
\newcommand{\likelihoodpred}{\ensuremath{L_{\text{pred}}}\xspace}
\newcommand{\nuisancevec}{\ensuremath{\vec{\lambda}}\xspace}

\newcommand{\wevt}{\ensuremath{w_{\text{evt}}}\xspace}
\newcommand{\whyb}{\ensuremath{w_{\text{hyb}}}\xspace}
\newcommand{\PJSF}{\ensuremath{P(\JSF)}\xspace}

\newcommand{\zetadefinition}{\ensuremath{\zeta=\ln(\mt/1\GeV)}\xspace}

\newcommand{\dstt}{\ensuremath{\rd\stt/\rd\mtt}\xspace}
\newcommand{\dNcheta}{\ensuremath{\rd N_{\text{ch}}/\rd\eta}\xspace}

\newcommand{\rhotrue}{\ensuremath{\rho_{\text{gen}}}\xspace}
\newcommand{\rhoreco}{\ensuremath{\rho_{\text{reco}}}\xspace}

\newcommand{\DRij}{\ensuremath{\DR_{ij}}\xspace}
\newcommand{\DRqq}{\ensuremath{\DR_{\qqpr}}\xspace}
\newcommand{\DRg}{\ensuremath{\DR_{\Pg}}\xspace}

\newcommand{\distgen}{\ensuremath{\mathbf{g}}\xspace}
\newcommand{\distdet}{\ensuremath{\mathbf{d}}\xspace}
\newcommand{\distgenpr}{\ensuremath{\mathbf{g^\prime}}\xspace}
\newcommand{\distdetpr}{\ensuremath{\mathbf{d^\prime}}\xspace}
\newcommand{\distb}{\ensuremath{\mathbf{b}}\xspace}

\newcommand{\topwidth}{\ensuremath{\Gamma_{\PQt}}\xspace}
\newcommand{\orderas}[1]{\ensuremath{O(\alpS^{#1})}\xspace}
\newcommand{\sensitivity}{\ensuremath{\mathcal{S}}\xspace}
\newcommand{\rhodiffxsec}{\ensuremath{\mathcal{R}}\xspace}

\cmsNoteHeader{TOP-23-003}
\title{Review of top quark mass measurements in CMS}
\date{\today}

\abstract{The top quark mass is one of the most intriguing parameters of the standard model (SM). Its value indicates a Yukawa coupling close to unity, and the resulting strong ties to Higgs physics make the top quark mass a crucial ingredient for understanding essential aspects of the electroweak sector of the SM.
This review offers the first comprehensive overview of the top quark mass measurements performed by the CMS Collaboration using the data collected at centre-of-mass energies of 7, 8, and 13\TeV. Moreover, a detailed description of the top quark event reconstruction is provided and dedicated studies of the dominant uncertainties in the modelling of the signal processes are discussed.
The interpretation of the experimental results on the top quark mass in terms of the SM Lagrangian parameter is challenging and is a focus of an ongoing discussion in the theory community.
The CMS Collaboration has performed two main types of top quark mass measurements, addressing this challenge from different perspectives: highly precise `direct' measurements, based on reconstructed top quark decay products and relying exclusively on Monte-Carlo simulations, as well as `indirect' measurements, where the simulations are employed to determine parton-level cross sections that are compared to fixed-order perturbative calculations. Recent mass extractions using Lorentz-boosted top quarks open a new avenue of measurements based on top quark decay products contained in a single particle jet, with promising prospects for accurate theoretical interpretations.
}

\hypersetup{%
pdfauthor={CMS Collaboration},
pdftitle={Review of top quark mass measurements in CMS},%
pdfsubject={CMS},%
pdfkeywords={CMS, top quark mass}}

\maketitle
\tableofcontents

\section{Introduction}
\label{sec:intro}
In the exploration of the fundamental building blocks of the universe, the study of the top quark, the most massive elementary particle yet known, has emerged as a key area of research at the Large Hadron Collider (LHC) at CERN.
At the Compact Muon Solenoid (CMS) experiment, the properties of this particle have been studied in great detail.

With a multitude of unique features that set it apart from other elementary particles, the top quark plays a crucial role in the standard model (SM) of particle physics. In the SM, the large mass of the top quark (\mt) results in its Higgs Yukawa coupling being close to unity. This leads to a particular significance of the top quark in the context of vacuum stability and cosmology, as well as in alternative models of spontaneous electroweak (EW) symmetry breaking.

The top quark has an extremely short lifetime of approximately $5\times10^{-25}\unit{s}$~\cite{ParticleDataGroup:2022pth}. Therefore it decays through the weak interaction before it would undergo hadronisation (happening at the time scale of
${\sim}10^{-23}\unit{s}$), and well before the strong interaction could affect its spin properties at the spin decorrelation time scale of $\mt /\LQCD^2 \approx 10^{-21}\unit{s}$~\cite{Mahlon:2010gw}, where \LQCD is the Landau pole of quantum chromodynamics (QCD).
Therefore, spin information of the top quark is transmitted to the particles that result from its decay~\cite{Mahlon:2010gw}. This distinct property entails that the top quark exhibits features of a quasi-free observable particle with a Breit--Wigner distributed invariant mass and grants a direct access to its fundamental properties, enabling precise measurements of its mass and polarisation.

This picture of the top quark is the basis of state-of-the-art experimental measurements. It implies approximations such as factorisation of on-shell top quark production and decay, used in most theoretical predictions and implemented in the simulations currently used in the experimental analyses. Furthermore, it neglects subtle quantum and interference effects arising from the top quark colour and electroweak charges. The analogous concept does not apply to any other quark, as the spin and mass are always masked by colour confinement.
However, with
growing precision in the measured top quark properties, in particular \mt, these effects eventually need to be accounted for as well.
The limitations of the picture of the top quark as a free particle
lead to ambiguities in the theoretical interpretation~\cite{Azzi:2019yne,Hoang:2020iah}.

\subsection{Early top quark studies}
In 1972, Kobayashi and Maskawa put forward the existence of a third generation of fermions in the SM~\cite{Kobayashi:1973fv} as an explanation for the violation of the combination of charge conjugation and parity symmetries (\CP), and more precise measurements of this effect paired with progress in the understanding of flavour physics pointed towards a large value of the mass of the hypothetical top quark already in the mid 1980s~\cite{Ginsparg:1983zc, Buras:1983ap}.
Experimental hints to the existence of the top quark emerged in measurements of the \PQb quark isospin from the forward-backward asymmetry in $\EE\to\bbbar$ processes at the DESY PETRA collider~\cite{JADE:1984bmx}, and in the suppression of flavour-changing neutral current decays of \PB mesons through the Glashow--Iliopoulos--Maiani (GIM) mechanism~\cite{Glashow:1970gm}.
The absence of a narrow top quark-antiquark resonance in direct searches at the \EE colliders PETRA~\cite{Adeva:1982bs} and KEK TRISTAN~\cite{TOPAZ:1987pqb} meant that \mt had to be substantially higher than that of the other quarks, setting a lower limit at 23.3 and 30.2\GeV, respectively.
The hadron collider experiments UA1 and UA2 at the S\ppbar S at CERN did not find evidence of the top quark in \PW boson decays $\pbarp\to\PW\to\PQt\PAQb$, excluding $\mt<60$~\cite{UA1:1990rck} and 69\GeV~\cite{UA2:1989tae} at 95\% confidence level (\CL), respectively.
More evidence for a very massive top quark accumulated from measurements of \PBz--\PABz mixing by the ARGUS~\cite{ARGUS:1987xtv} and CLEO~\cite{CLEO:1989fuk} Collaborations, where lower bounds on \mt between 45 and 90\GeV were obtained by exploiting the features of the GIM mechanism~\cite{Altarelli:1987zf}.
In the early 1990s, when the CERN LEP and SLC colliders started operating at the energy of the \PZ resonance, no evidence was found for the decay $\PZ\to\ttbar$, excluding $\mt<45.8\GeV$~\cite{ALEPH:1989tsb, OPAL:1989mxj}. Precise measurements of the \PZ boson mass, partial decay widths, and forward-backward asymmetries were made at the LEP and SLAC SLC colliders.
Since the relation between these quantities and the weak mixing angle is affected by the value of \mt via radiative EW corrections, these measurements at the \PZ pole could be used to indirectly constrain the value of \mt.
Initial constraints indicated \mt to be in the range of 64--169\GeV at 68\% \CL~\cite{LEP:1991hsu}.
With more data, the range narrowed down to
158--199\GeV at 68\% \CL~\cite{ALEPH:1995ac} in the year of the discovery of the top quark, where the extent of this range came mainly from the unknown Higgs boson (\PH) mass.
At the same time, the experimental determinations of the Cabibbo--Kobayashi--Maskawa (CKM) matrix elements had been considerably improved and progress had been made in calculating \PB meson form factors, such that more reliable bounds from \CP violation in \PBz--\PABz and \PKz--\PAKz systems could be calculated~\cite{Lusignoli:1991bm, Buras:1993wr}, resulting in lower limits on \mt of about 100\GeV.
Finally, in 1995, Fermilab experiments CDF and \DZERO, operating at Tevatron proton-antiproton (\ppbar) collider, announced the discovery of the top quark at $\mt=175\pm8\GeV$~\cite{CDF:1995wbb,D0:1995jca}.
In the following years, the properties of the top quark were measured with ever-increasing accuracy by the CDF and \DZERO Collaborations.
While most measurements were done with \ttbar pairs, which are copiously produced by the strong interaction, the production of single top quarks through the EW interaction was also observed for the first time during the Tevatron \mbox{Run II}~\cite{D0:2009isq,CDF:2009itk}.
Combining all \mt measurements performed at the Tevatron, a final result of $\mt=174.30\pm0.65\GeV$ was obtained~\cite{CDF:2016vzt}. A more detailed discussion can be found in Ref.~\cite{Campagnari:1996ai} and references therein.

When the Tevatron shut down in 2011, the CERN LHC became the only collider facility in the world capable of producing top quarks in large quantities.
The LHC increased the number of produced top quarks by orders of magnitude as compared to the Tevatron.

\subsection{Role of the top quark mass in the standard model and beyond}

While the specific values for the elementary couplings or fermion masses, including \mt, are not predicted in the SM, the model provides relationships between \mt and other fundamental parameters.
The value of \mt needs to be determined experimentally by measuring \mt-sensitive observables and comparing those with the theory predictions. These comparisons can be performed either using detector level distributions or measured cross sections.

The value of \mt influences the top quark decay modes and production rates, which are essential for understanding top quark properties and dynamics. Apart from being a reflection of our ability to describe the dynamics of the strong and EW interactions using quantum-field theoretical methods, accurate measurements of \mt provide critical tests of the SM and its extensions. In this context, it needs to be recalled that the quantum aspects of the top quark associated with its colour charge and its finite lifetime imply that \mt is not a directly measurable physical parameter like the masses of hadrons. The value of \mt can only be inferred indirectly through observables that depend on it. Since quantum effects affect this dependence, \mt measurements are only possible on the basis of theoretical predictions of these observables. In these theoretical predictions, it is mandatory to account for the fact that \mt is not a unique physical parameter, but needs to be defined through a certain renormalisation scheme within quantum field theory. Defined this way, \mt  plays a role of a SM coupling and is (like all other quark masses and SM couplings) a renormalisation-scheme dependent quantity, as discussed in Section~\ref{sec:introschemes}.

The top quark appears in quantum loop corrections to various processes, and depending on its mass, it can have a substantial impact on the behaviour of other particles, particularly in rare production processes and precision EW measurements. One example is the \PBz--\PABz mixing mentioned earlier. Another example is the ratio of direct to indirect \CP violation size in kaon decays~\cite{Flynn:1989iu,Buchalla:1989we}.

Further, \mt enters into loop corrections that contribute to the masses of the \PW and \PZ bosons, and therefore indirectly affects the weak mixing angle. Since the sensitivity of EW precision observables to \mt arises through radiative corrections, the choice of the renormalisation scheme for \mt is essential for the precise theoretical description of the EW observables~\cite{Erler:2019hds}. The uncertainty in \mt is among the leading uncertainties in the predictions of the \PW and \PH boson masses~\cite{Haller:2018nnx}, which are crucial for testing the internal consistency of the SM.

The SM Higgs mechanism endows fermions, including the top quark, with mass through their interaction with the Higgs field. The mass of a fermion, $m_{\Pf}$, emerges from a Yukawa interaction with coupling strength $Y_{\Pf}=\sqrt{2}(m_{\Pf}/v)$, where $v=246.22\GeV$~\cite{ParticleDataGroup:2022pth} is the vacuum expectation value of the Higgs field. The top quark has the largest Yukawa coupling in the SM, with a value close to unity.
This can be compared to a direct measurement of the Yukawa coupling strength from the production cross section of final states involving top quarks and the Higgs boson, mostly from $\ttbar\PH$ production,
with further contributing processes, $Y_{\PQt}=0.95\,^{+0.07}_{-0.08}$~\cite{ATLAS:2022vkf, CMS:2022dwd}.
Kinematic distributions in \ttbar production can also be used to probe the top-quark Yukawa coupling through loop-induced corrections from the Higgs field. The most precise such measurement was performed by the CMS experiment, resulting in $Y_{\PQt}=1.16\,^{+0.24}_{-0.35}$~\cite{CMS:2020djy}, consistent with the value obtained from \mt and the direct measurement.
The top quark Yukawa coupling significantly affects the shape of the Higgs potential. The value of \mt is linked to the Higgs boson mass through quantum loop corrections and enhances the quantum contributions to the Higgs potential. Therefore, the value of \mt has a direct impact on the stability of the EW vacuum~\cite{Degrassi:2012ry,Alekhin:2012py}.
In particular, if the potential energy of the Higgs field is too shallow, it could lead to vacuum instability. In such a scenario, the EW vacuum may not be the true minimum of the potential, and the Higgs field could eventually undergo a phase transition to a deeper minimum at very high energies.
This transition would have profound consequences, leading to the collapse of the vacuum and changing the fundamental properties of all particles, which could drastically affect the structure of the universe. Since this sensitivity is generated through quantum effects, accurate control of the renormalisation scheme of \mt is essential.

A deviation of the measured \mt from the prediction using a SM fit when all other free parameters are constrained to their measured values could indicate the presence of new physics beyond the SM (BSM), such as supersymmetry~\cite{Heinemeyer:2013dia}
or the existence of additional Higgs bosons. Further, \mt is related to the evolution of the early universe, and its precise value has implications for cosmology~\cite{Bezrukov:2012sa} and our understanding of dark matter~\cite{Dunsky:2020yhv}.

With the data provided by the LHC so far, there has been no observation of BSM effects in direct searches for new resonant states, which could either point to new physics processes coupling very weakly to the SM sector, or appearing only at energy scales higher than what experiments can probe to date. In the latter case, the BSM contributions can be described by \eg an effective field theory (EFT). In the EFT-extended SM (SMEFT), BSM contributions are parameterised in a model-independent way through higher-dimensional operators~\cite{Buchmuller:1985jz,Giudice:2007fh,Grzadkowski:2010es}. These operators involve the known SM particle fields, while their Wilson coefficients, playing a role of couplings, encode the effects of potential BSM particles and interactions. The value of \mt plays a crucial role in SMEFT interpretations, since it affects the behaviour of higher-dimensional operators and their interplay with known SM interactions. An illustrative example given in Ref.~\cite{Gao:2022srd} is the invariant mass of the \ttbar pair, \mtt, being sensitive to the effective couplings $c_{\mathrm{tG}}$ and $c^8_{\mathrm{tq}}$, which depend on the value of \mt.
In addition, precise knowledge of \mt is essential for reducing uncertainties in theoretical calculations of \PB meson decays~\cite{Buras:2012ru, Misiak:2015xwa, Czakon:2015exa}.

\subsection{Scope of the review}
\label{sec:scope}

The focus of this review is on the measurements of \mt carried out by the CMS Collaboration, based on data collected during the LHC \Run1 at $\sqrts=7$ and 8\TeV in 2010--2012, and \Run2 at \sqrtseq{13} in 2015--2018. Since the initial top quark mass analyses performed at the Tevatron, experimental methods, theoretical calculations, and Monte Carlo (MC) models have evolved in sophistication and accuracy. Modern detector technologies, increased computing power, optimised reconstruction algorithms, and above all the higher centre-of-mass energies and integrated luminosities delivered by the LHC have allowed for the development  of an array of novel top quark mass analyses, exploring new aspects of top quark phenomenology and reaching unprecedented levels of detail and precision.

While all the results included in this review have been published before, it is the first time that a comprehensive overview is presented by the CMS Collaboration, detailing and contrasting the leading approaches and discussing aspects of the theoretical interpretation of the results. To illustrate the broadness of the top mass measurement program of CMS, the summary of the relevant publications to date is given in Table~\ref{tab:all_measurements}, with the details to be discussed in the course of the review.
These investigations have been traditionally classified as either \textit{direct} or \textit{indirect} top quark mass extractions. For consistency with original works, the same classification is adopted in this review.

\begin{table}[!p]
\centering
\topcaption{%
    List of all CMS \mt measurements by using different analysis methods in chronological order of publication. The summary of these measurements is also depicted in Fig.~\ref{conclusions:MtopResults}. The analyses are categorised as direct mass measurements (a), indirect extraction of the Lagrangian mass (b), or boosted measurements (c), as explained in the text.
    The analysis methods of the publications marked with a star (*) are covered in the following sections of this review. All acronyms are defined in Appendix~\ref{sec:glossary}.
}
\label{tab:all_measurements}
\newcommand{\typeA}{\textsuperscript{a}\xspace}
\newcommand{\typeB}{\textsuperscript{b}\xspace}
\newcommand{\typeC}{\textsuperscript{c}\xspace}
\renewcommand{\arraystretch}{1.3}
\cmsTable{\begin{tabular}{rlclcccl}
    Year & Channel & \sqrts & Analysis method & \mt & $\delta\mt^{\text{stat}}$ & $\delta\mt^{\text{syst}}$ & Ref.\\[-2pt]
    &  & [{\TeVns}] &  & [{\GeVns}] & [{\GeVns}] & [{\GeVns}] & \\
    \hline
    2011  & Dilepton & 7 & \typeA KINb and AMWT & 175.5 & 4.6 & 4.6 & \cite{CMS:2011acs}\\
    2012  & Lepton+jets & 7 & \typeA 2D ideogram & 173.49 & 0.43 & 0.98 & \cite{CMS:2012sas}*\\
    2012  & Dilepton & 7 & \typeA AMWT & 172.5 & 0.4 & 1.5 & \cite{CMS:2012tdr}\\
    2013  & Dilepton & 7 & \typeA Kinematic endpoints & 173.9 & 0.9 & $_{-2.1}^{+1.7}$ & \cite{CMS:2013wbt}\\
    2013  & All-jets & 7 & \typeA 2D ideogram & 173.54 & 0.33 & 0.96 & \cite{CMS:2013lqq}*\\
    2014  & Dilepton & 7 & \typeB Cross section & 177.0 & \NA & $_{-3.3}^{+3.6}$ & \cite{CMS:2014rml}*\\
    2015  & Lepton+jets & 8 & \typeA Hybrid ideogram & 172.35 & 0.16 & 0.48 & \cite{CMS:2015lbj}*\\
    & All-jets & 8 & \typeA Hybrid ideogram & 172.32 & 0.25 & 0.59 & \cite{CMS:2015lbj}*\\
    & Dilepton & 8 & \typeA AMWT & 172.82 & 0.19 & 1.22 & \cite{CMS:2015lbj}\\
    & Combination & 7, 8 & \typeA CMS 7 inputs & 172.44 & 0.13 & 0.47 & \cite{CMS:2015lbj}\\
    2016  & Dilepton & 7, 8 & \typeB Cross section & 174.3 & \NA & $_{-2.2}^{+2.1}$ & \cite{CMS:2016yys}*\\
    2016  & 1+2 leptons & 8 & \typeA Lepton + secondary vertex & 173.68 & 0.20 & $_{-0.97}^{+1.58}$ & \cite{CMS:2016iru}\\
    2016  & 1+2 leptons & 8 & \typeA Lepton + \PJGy meson & 173.5 & 3.0 & 0.9 & \cite{CMS:2016ixg}\\
    2017  & Lepton+jets & 13 & \typeB Cross section & 170.6 & \NA & 2.7 & \cite{CMS:2017xrt}\\
    2017  & Single top quark & 8 & \typeA Template fit & 172.95 & 0.77 & $_{-0.93}^{+0.97}$ & \cite{CMS:2017mpr}*\\
    2017  & Boosted & 8 & \typeC CA jet mass unfolded & 170.9 & 6.0 & 6.7 & \cite{CMS:2017pcy}*\\
    2017  & Dilepton & 8 & \typeA \Mbl{}+\MTtwobb hybrid
    fit & 172.22 & 0.18 & $_{-0.93}^{+0.89}$ & \cite{CMS:2017znf}\\
    2018  & Lepton+jets & 13 &  \typeA Hybrid ideogram & 172.25 & 0.08 & 0.62 & \cite{CMS:2018quc}*\\
    2018  & All-jets & 13 & \typeA Hybrid ideogram & 172.34 & 0.20 & 0.70 & \cite{CMS:2018tye}*\\
    & Combination & 13 & \typeA Combined likelihood & 172.26 & 0.07 & 0.61 & \cite{CMS:2018tye}\\
    2018  & Dilepton & 13 & \typeA\ \mbl fit & 172.33 & 0.14 & $_{-0.72}^{+0.66}$ & \cite{CMS:2018fks}*\\
    & Dilepton & 13 & \typeB Cross section & 173.7 & \NA & $_{-2.3}^{+2.1}$ & \cite{CMS:2018fks}*\\
    2019  & Dilepton & 13 & \typeB Multi-differential cross section & 170.5 & \NA & 0.8 & \cite{CMS:2019esx}*\\
    2019  & Dilepton & 13 & \typeB Running mass & \NA & \NA & \NA & \cite{CMS:2019jul}*\\
    2019  & Boosted & 13 & \typeC XCone jet mass unfolded & 172.6 & 0.4 & 2.4 & \cite{CMS:2019fak}*\\
    2021  &  Single top quark & 13 & \typeA $\ln(\mt/1\GeV)$ fit & 172.13 & 0.32 & $_{-0.71}^{+0.69}$ & \cite{CMS:2021jnp}*\\
    2022  & Dilepton & 7, 8 & \typeB ATLAS+CMS cross section & 173.4 & \NA & $_{-2.0}^{+1.8}$ & \cite{ATLAS:2022aof}\\
    2022 & Dilepton & 13 & \typeB\ \ttbarjet differential cross section & 172.13 &  & 1.43 & \cite{CMS:2022emx}*\\
    2022  & Boosted & 13 & \typeC XCone jet mass unfolded & 173.06 & 0.24 & 0.80 & \cite{CMS:2022kqg}*\\
    2023  & Lepton+jets & 13 & \typeA Profile likelihood & 171.77 & 0.04 & 0.37 & \cite{CMS:2023ebf}*\\
    2024  & Combination & 7, 8 &\typeA  CMS 9 inputs & 172.52 & 0.14 & 0.39 & \cite{CMS:2023wnd}\\
    & Combination & 7, 8 &\typeA  ATLAS+CMS 15 inputs & 172.52 & 0.14 & 0.30 & \cite{CMS:2023wnd}\\
\end{tabular}}
\end{table}

The direct measurements are based on the picture of the top quark as a free and asymptotic particle, which implies that the invariant mass of the top quark decay products is directly related to the pole mass of the original top quark particle.
In this picture, the main challenges are to identify the top quark decay products and reconstruct their invariant mass with the best possible experimental resolution. Furthermore, the final-state particles not originating from the top quark decay have to be accounted for, and the uncertainties related to theoretical limitations of the MC simulations need to be properly estimated, including off-shell and colour-neutralisation corrections.
The direct measurements rely on MC simulations for the precise modelling of the event decay topologies and experimental effects, but also for the calibration of the analysis in terms of a built-in \mt parameter that is extracted from the simulation. Therefore the value of \mt obtained in such a way corresponds to the respective top quark mass parameter, \mtmc, used in the particular MC simulation.
In the simulated top quark samples used by CMS in this report, top quarks are generated with a mass following a relativistic Breit--Wigner distribution centered around \mtmc, with a width \topwidth equal to the SM prediction. Each top quark then decays independently.
The direct measurements have the smallest experimental uncertainties since these are based on the most \mt-sensitive observables. However, due to limitations of the current theoretical knowledge implemented in the MC simulations, an additional conceptual uncertainty has to be accounted for when the result is interpreted in terms of \mt defined in the field theory of QCD.
The quantification of this conceptual uncertainty has not been part of the direct \mt measurements so far and requires separate investigations.
The measurements in this category typically employ a full reconstruction of the top quark and are performed by analysing top quark-antiquark pair (\ttbar) events in multiple decay channels.
In the dilepton channel, a full kinematic analysis (KINb)~\cite{CMS:2011acs}, the analytical matrix weighting technique (AMWT)~\cite{CMS:2011acs,CMS:2012tdr,CMS:2015lbj}, an \Mbl{}+\MTtwobb hybrid fit, taking into account external constraints on the jet energy scale (hybrid)~\cite{CMS:2017znf}, as well as an \mbl fit~\cite{CMS:2018fks} have been employed.
In the lepton+jets and all-jets channels the techniques have evolved from a simultaneous fit of \mt and the jet energy scale (2D ideogram)~\cite{CMS:2012sas,CMS:2013lqq} to the hybrid ideogram method~\cite{CMS:2015lbj,CMS:2018quc,CMS:2018tye} and, in the most recent measurement~\cite{CMS:2023ebf}, to a 5D profile likelihood fit.
Template fits were used to extract \mt in single top quark~\cite{CMS:2017mpr,CMS:2021jnp} events.
While the single top quark analyses currently have relatively large uncertainties compared to the analyses using \ttbar events, they offer complementary information and have an excellent potential for improvement with the large data sets expected in future LHC runs.

The indirect extraction of \mt is realised by comparison of the measured inclusive or differential parton-level cross sections of the on-shell \ttbar production to the corresponding fixed-order calculations at next-to-leading order (NLO) or next-to-next-to-leading order (NNLO) accuracy in QCD, which operate with \mt being a parameter of the QCD Lagrangian defined in a certain renormalisation scheme. Therefore, the results of the indirect \mt extractions are traditionally also referred to as the Lagrangian \mt obtained in the renormalisation scheme of the fixed-order prediction used.
The measurement of the parton-level cross sections for the indirect \mt extraction is also based on MC simulations, as detailed in Sections~\ref{sec:unfolding} and~\ref{sec:particlePartonLevel}. As for the case of the direct measurements, there are conceptual uncertainties in the determination of the parton-level cross sections due to the limited theoretical knowledge implemented in the MC simulations. The estimate of the uncertainty associated with the assumptions in the simulations, including the definition of the \mt parameter in the MC, used to connect the particle and the parton levels, is an integral part of the cross section measurement. The results of the indirect \mt extractions currently have larger uncertainties than those of the direct ones. The reasons are additional uncertainties in the calculations of the \ttbar production cross sections and a reduced sensitivity to \mt, since more inclusive observables like production rates are used.
The indirect \mt extractions were performed using the measured \ttbar inclusive~\cite{CMS:2014rml,CMS:2016yys,CMS:2017xrt,CMS:2018fks,ATLAS:2022aof} and differential~\cite{CMS:2019esx}, as well as \ttbarjet differential~\cite{CMS:2022emx} production cross sections.

Recently, \mt extractions were also carried out using events where the top quarks are produced with a high Lorentz boost~\cite{CMS:2017pcy,CMS:2019fak,CMS:2022kqg}.
These boosted top quark events are characterised by the top quark decay products being collimated within a single jet.
In contrast to the direct measurements at lower top quark boosts, where individual top quark decay objects are reconstructed, in this case the \mt sensitive observable is the invariant mass of a single top quark jet. This novel approach ultimately aims to use the measured differential jet mass cross section as a basis to obtain \mt, similar to
indirect \mt extractions. However, in the absence of corresponding fixed-order calculations, currently the MC simulation is used to obtain \mt, as in direct measurements.
The boosted topology, which allows for a clearer separation of the decay products from top quarks and antiquarks, combines a kinematic \mt sensitivity and the ability to make systematic theoretical predictions at the experimentally observable level, namely quantum-field theoretical predictions of the invariant mass of top quark jets consisting of stable particles. This fact may be used to eventually establish a clear relation between the direct and indirect \mt measurements in the future~\cite{ATLAS:2021urs}. In this review, we therefore discuss these \mt measurements~\cite{CMS:2017pcy,CMS:2019fak,CMS:2022kqg} separately from the already established direct and indirect approaches.

Finally, CMS conducted an extensive program of measurements using alternative methods.
These are conceptually close to the direct measurements but were designed aiming at reduced or orthogonal systematic uncertainties. The \mt measurements from kinematic endpoints~\cite{CMS:2013wbt} and from \PQb hadron decay products~\cite{CMS:2016ixg,CMS:2016iru} are considered the most promising. The first two employ the lepton+jets channel, while the latter combines the lepton+jets and dilepton channels.
The \PJGy method~\cite{CMS:2016ixg} had been proposed already in the CMS technical design report~\cite{CMS:2006myw} as a particularly clean method, relying only on the reconstruction of three leptons in the final state: one lepton from the \PW boson decay, and two from the decay of a \PJGy  produced in the decay of the \PQb-flavoured hadron in the \PQb jet. The results have demonstrated the viability of the method, however its full potential can only be reached with the much larger data sets expected at the High-Luminosity LHC (HL-LHC) (as discussed in Section~\ref{sec:ProspectsHiLumi}). The secondary vertex method~\cite{CMS:2016iru} uses a similar approach, but replaces the leptonic decay of the \PJGy particle by the secondary vertex of the decay of the \PQb hadron in the \PQb jet, thus obtaining a much larger selection of events, and still only using tracking information, however sacrificing the much cleaner experimental signature of the leptonic \PJGy meson decay.

Measurements performed using alternative methods or in single top quark enriched topologies, despite reaching lower precision compared to standard measurement with the current data sets, can already have a beneficial effect in \mt combinations. These measurements, in fact, have different sensitivity to systematic uncertainties both from the experimental and modelling points of view, and therefore provide independent information. For example, measurements based on the reconstruction of \PQb-hadron decay products do not rely on the precise calibration of the \PQb jet energy, at the cost of a stronger dependence on the modelling of the \PQb quark fragmentation. This can be seen explicitly in the updated CMS \Run1 combination presented in Ref.~\cite{CMS:2023wnd} and resulting in a value of $\mt = 172.52 \pm 0.42\GeV$.
By performing the combination of CMS inputs excluding the single top quark and alternative measurements of Refs.~\cite{CMS:2017mpr,CMS:2016iru,CMS:2016ixg}, a total uncertainty of 0.44\GeV is obtained, which corresponds to adding in quadrature an extra uncertainty of about 0.15\GeV. This is equivalent to more than half the size of the leading systematic uncertainty in the combination, \ie the jet energy response of \PQb quark jets. The work of Ref.~\cite{CMS:2023wnd} also provides the combination of ATLAS and CMS measurements in \Run1, resulting in a value of $\mt = 172.52 \pm 0.33\GeV$, with a precision demonstrating the importance of combination of results obtained at different experiments.

The focus of this review is the development of analysis strategies in CMS leading to the high-precision \mt results in direct determination, indirect extraction of \mt, and measurements in boosted topologies. Before highlighting recent examples of the major approaches to measure \mt in Sections~\ref{sec:direct}--\ref{sec:boosted}, the general aspects in common between the different analyses are discussed in Section~\ref{sec:extraction}. The measurements are summarised and the future perspectives are given in Section~\ref{sec:conclusions}.

\section{Conceptual and experimental aspects of top quark mass measurements}
\label{sec:extraction}

Measurements of the top quark mass rely on the detection and accurate reconstruction of events containing a \ttbar pair or a single top quark. Depending on the final state formed in the top quark decay, as described in Section~\ref{sec:production}, the details of the event reconstruction may differ. Sophisticated algorithms have been developed to identify final-state particles and their momenta with optimal efficiency and resolution, as described in Section~\ref{sec:reconstruction}. In many of the analyses discussed in this review, it is advantageous to use a kinematic reconstruction of the full event, using the laws of energy and momentum conservation to improve the knowledge of the final state objects beyond the detector resolution.
This is particularly important for final states that contain an energetic neutrino.
The approaches used in the reconstruction of \ttbar events are described in Section~\ref{sec:kinreco}. The full event reconstruction also aims to resolve ambiguities in the assignment of final-state objects as decay products of a given top quark. This task becomes more challenging in the presence of energetic gluon radiation creating additional jets, and the presence of remnants of the colliding protons (underlying event, UE) as well as multiple simultaneous proton collisions (pileup, PU). In order to account for these effects, all analyses rely on MC simulation programs, tuned to describe the event properties as accurately as possible, as reported in Section~\ref{sec:mcsetup}. Besides uncertainties in the MC models, the analyses are also affected by experimental uncertainties, briefly summarised in Section~\ref{sec:experimentalunc}. Finally, to perform a measurement of \mt, the features of the events observed in data are compared with the theoretical predictions or MC simulations, for a range of hypothetical \mt values, and a fit is performed to extract the best fit \mt, and uncertainties are evaluated.
This procedure can be based on distributions reconstructed at the detector level (via a so-called `template fit') or by comparing theoretical predictions to the distributions corrected for experimental effects using unfolding techniques as discussed in Section~\ref{sec:unfolding}.
The unfolding procedure can rely on the MC generator to correct back to a hypothetical picture of on-shell top quarks (`parton level') or to reproduce the event distributions at the level of stable particles in the final state (`particle level'). The latter approach is particularly useful to provide experimental distributions that can be compared to new MC generator predictions for the purpose of MC tuning, as discussed in Section~\ref{sec:particlePartonLevel}. To interpret the measured \mt as a parameter of the SM, quantum aspects related to the short lifetime and colour charge of the top quark must be considered, as outlined in Section~\ref{sec:introschemes}.

\subsection{Top quark production and decay}
\label{sec:production}

At the LHC, top quarks can either be produced in \ttbar pairs, via the strong interaction, or as single top quarks through the EW interaction. Enhanced by the strong coupling, the rate of \ttbar production is about four times larger than that of the single top quark process.

\begin{figure}[!htb]
\centering
\includegraphics[width=0.8\textwidth]{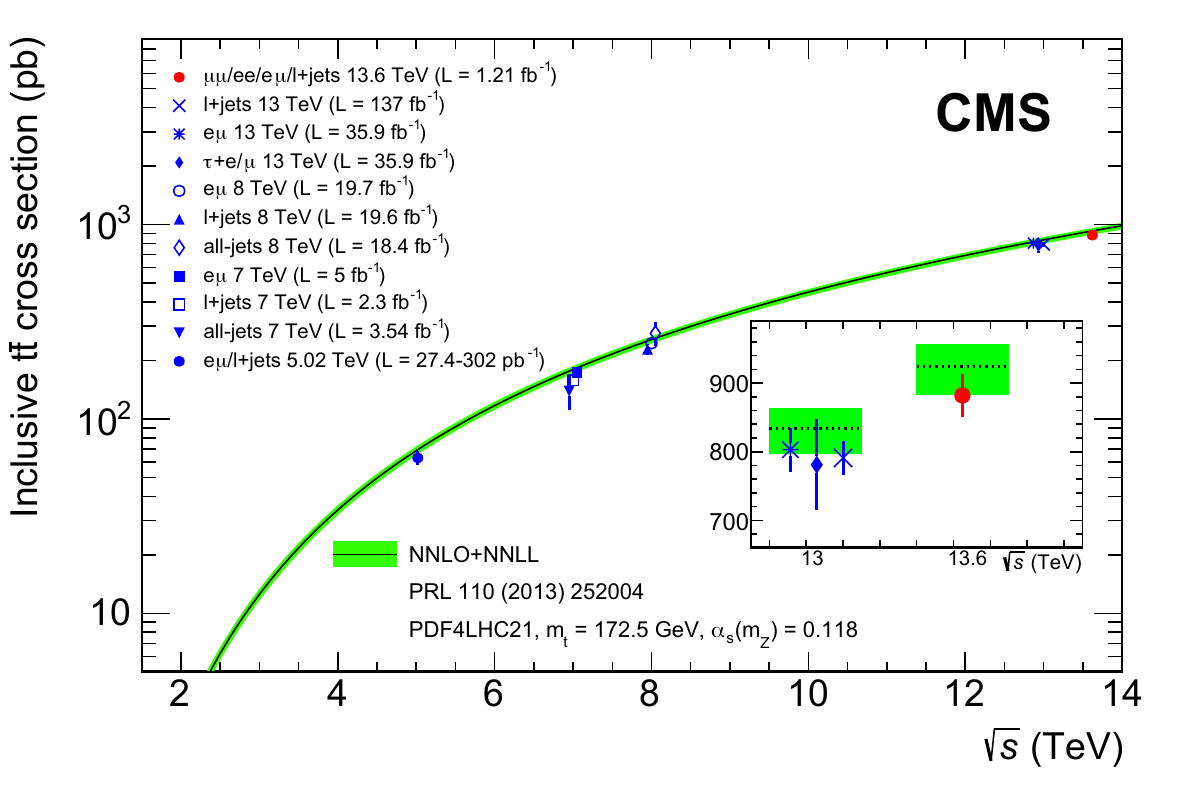}
\caption{%
    Summary of CMS measurements of the \ttbar production cross section as a function of \sqrts compared to the NNLO QCD calculation complemented with NNLL resummation (\TOPpp v2.0~\cite{Czakon:2011xx}). The theory band represents uncertainties due to
    the renormalisation and factorisation scales, parton distribution functions, and the strong coupling. The measurements and the theoretical calculation are quoted at $\mt=172.5\GeV$. Measurements made at the same \sqrts are slightly offset for clarity. An enlarged inset is included to highlight the difference between 13 and 13.6\TeV predictions and results.
    Figure taken from Ref.~\cite{CMS:2023qyl}.
}
\label{fig:tt_xsection_vs_s}
\end{figure}

In leading order (LO) in QCD, hadronic collisions at higher energies produce \ttbar pairs through quark-antiquark (\qqbar) annihilation or gluon-gluon (\GG) fusion. In contrast to \ppbar collisions at the Tevatron, where \ttbar production is dominated by \qqbar annihilation, in proton-proton (\pp) collisions at the LHC, the \GG fusion process is dominant~\cite{Nason:1987xz,Beenakker:1988bq,Bernreuther:2004jv}. The QCD predictions for \ttbar production have reached next-to-next-to leading order (NNLO) precision and include next-to-next-to-leading-logarithmic (NNLL) soft-gluon resummation~\cite{Czakon:2011xx, Kidonakis:2014pja, Czakon:2015owf, Czakon:2018nun, Catani:2019hip,Kidonakis:2019yji}, and electroweak corrections~\cite{Czakon:2017wor,Czakon:2020qbd}. The cross section of \ttbar production has been studied by the experiments at the Tevatron and the LHC
at different centre-of-mass energies and is found to be well described by the QCD predictions, as shown in Fig.~\ref{fig:tt_xsection_vs_s}.

Single top quark production is mediated by virtual \PW bosons in $s$- and $t$-channels, with the latter being kinematically enhanced and resulting in a sizeable cross section both at the Tevatron and the LHC~\cite{Cortese:1991fw, PhysRevD.34.155}. The cross sections for single top quark production in $s$- and $t$-channels are calculated at NNLO~\cite{Campbell:2020fhf,Brucherseifer:2014ama,Berger:2017zof, Kidonakis:2010tc}. In \ppbar collisions at the Tevatron, the \PQt and \PAQt quarks are produced with identical cross sections in each channel. In contrast, in \pp collisions at the LHC these differ because of the charge-asymmetric initial state. Furthermore, at the LHC, the \PW-associated production (\tW) becomes relevant. From the theory perspective, \tW production is well defined only at leading order. At NLO, the presence of real corrections, \ie the emission of an additional particle, such as $\GG\to\PQt\PW\PQb$ leads to overlap between \tW and $\GG\to\ttbar$. The contribution of the latter, with a cross section about an order of magnitude larger than that of \tW production, needs to be subtracted~\cite{Frixione:2008yi, White:2009yt}. This is only possible with approximations and leads to ambiguities that must be carefully estimated~\cite{Bevilacqua:2010qb, Denner:2010jp,Denner:2012yc}.

\begin{figure}[!ht]
\centering
\includegraphics[width=0.8\textwidth]{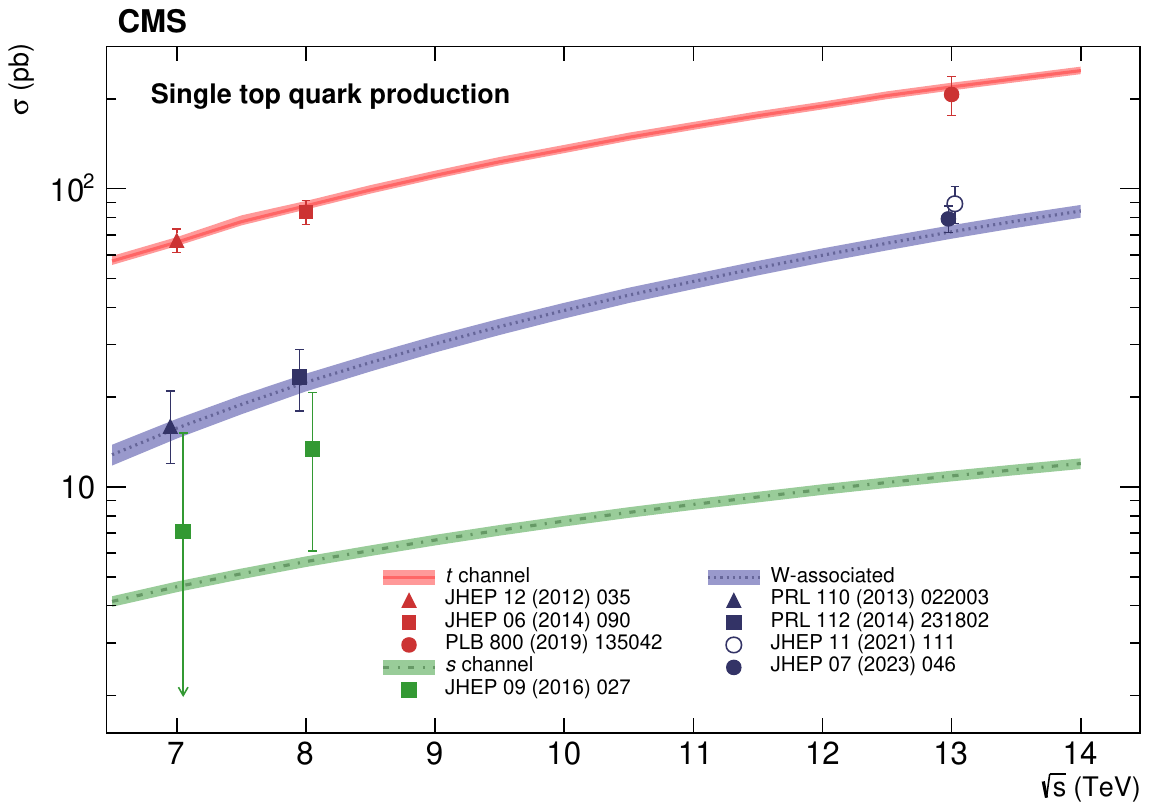}
\caption{%
    Summary of single top quark production cross section measurements by CMS. Theoretical calculations for $t$-channel, $s$-channel, and \PW-associated production are courtesy of N.~Kidonakis~\cite{Kidonakis:2010tc,Kidonakis:2010ux}.
}
\label{fig:st_xsection_vs_s}
\end{figure}

In Fig.~\ref{fig:st_xsection_vs_s}, the CMS measurements of single top quark production cross sections in different channels are presented as functions of the centre-of-mass energy in comparison to the theoretical predictions.

The decay width of the top quark is predicted at NNLO~\cite{Czarnecki:1998qc, Chetyrkin:1999ju,Blokland:2004ye,Blokland:2005vq}, with the most precise analytic result being 1.331\GeV with an uncertainty of less than 1\%~\cite{Chen:2022wit} for $\mt=172.69\GeV$, increasing with increasing \mt. With the correspondingly short lifetime of about $5\times10^{-25}\unit{s}$, the top quark decays before forming top-flavoured hadrons or \ttbar quarkonium-bound states~\cite{Bigi:1986jk}. Instead, the top quark decays weakly into a \PW boson and a down-type quark, most probably a \PQb quark. The branching fraction is given by $\mathcal{B}_{\PQb\PW}=\abs{V_{\PQt\PQb}}^2/(\abs{V_{\PQt\PQb}}^2 + \abs{V_{\PQt\PQs}}^2  + \abs{V_{\PQt\PQd}}^2)$, with $V_{\PQt\PQq}$ ($\PQq=\PQd,\PQs,\PQb$) denoting the elements of the CKM matrix, in particular $V_{\PQt\PQb}=0.998$~\cite{ParticleDataGroup:2022pth}.
In the SM, the denominator of $\mathcal{B}_{\PQb\PW}$ is unity.

Events with \ttbar production are categorised by the final states of the \PW bosons emitted in the decays of \PQt and \PAQt quarks. In the \textit{dilepton} channel, both \PW bosons decay leptonically, \ie into a charged lepton and neutrino; in the \textit{lepton+jets}
channel one \PW boson decays leptonically while the other one decays to a \qqbar
pair; in the \textit{all-jets} channel, both \PW bosons decay into \qqbar, forming hadronic jets in the final state:
\begin{itemize}
\item dilepton (10.5\%), $\ttbartoWbWb\to\lpnb\lmnb$,
\item lepton+jets (43.8\%), $\ttbartoWbWb\to\qqb\lmnb$ or $\lpnb\qqbbar$ ,
\item all-jets (45.7\%), $\ttbartoWbWb\to\qqb\qqbbar$.
\end{itemize}
For each channel, the relative contributions are indicated in parentheses and include hadronic corrections and assume lepton universality~\cite{ParticleDataGroup:2022pth}.
The charged leptons \Pell denote electrons \Pe, muons \PGm, or tau leptons \PGt. Since \PGt leptons are more difficult to reconstruct experimentally compared to \Pe or \PGm, these are implicitly included in the experimental measurements via their leptonic decays. Excluding hadronic decays of \PGt leptons, the branching ratios for the dileptonic and lepton+jets channels decrease to 6.5\% and 34.4\%, respectively. Further in this review, the notation `lepton' refers to \Pe and \PGm if not specified otherwise.

Despite the lowest relative contribution, top quark dilepton decays are widely used in physics analyses
since they can be experimentally identified with the highest purity. While the
all-jets channel accounts for almost half of the \ttbar decays, it is difficult to
distinguish those from QCD multijet production. The lepton+jets channel has intermediate properties, with moderate background contamination and large relative contribution.

In addition to the quarks resulting from the top quark decays, extra QCD radiation can lead to additional jets. Although the neutrinos remain undetected, their transverse momenta \pt are obtained from the imbalance in the transverse momentum measured in each event.

\subsection{Reconstruction of physics objects in CMS}
\label{sec:reconstruction}

All top quark measurements rely on the efficient reconstruction of its decay products from electrical signals in the detector.
A detailed description of the CMS detector, together with a definition of its coordinate system, can be found in Ref.~\cite{Chatrchyan:2008zzk}.
Particles are reconstructed using the particle-flow (PF) algorithm~\cite{CMS:2017yfk}, which follows the trajectory of particles through the various detector systems of the CMS experiment and combines the measurements in the tracking system, calorimeters, and muon system in order to achieve an optimised reconstruction.
For each event, the PF algorithm returns a list of PF candidates that are categorised either as electron, muon, photon, neutral hadron, or charged hadron, depending on their signature in the detector systems.
Electrons are identified by combining hits in the silicon tracker, the energy measured in a corresponding cluster in the electromagnetic calorimeter (ECAL), and the sum of all bremsstrahlung photons compatible with the electron trajectory.
Muons are reconstructed from hits in the tracker and muon system.
Charged hadrons are measured by a combination of tracker and the connected energy clusters in the ECAL and hadronic calorimeter (HCAL).
Photons and neutral hadrons are reconstructed from energy clusters in the ECAL and a combination of ECAL and HCAL, respectively.

The primary \pp interaction vertex is taken to be the vertex corresponding to the hardest scattering in the event, evaluated using tracking information alone, as described in Section 9.4.1 of Ref.~\cite{CMS-TDR-15-02}.
In order to reduce effects from additional \pp collisions in each event, we use pileup mitigation tools that act on the list and remove PF candidates that can be associated with a pileup vertex.
The CMS Collaboration uses two algorithms for pileup mitigation.
The charge-hadron subtraction (CHS)~\cite{CMS-PAS-JME-14-001} technique
removes charged hadrons that are associated with a pileup vertex by calculating the distance
of closest approach of each track to the reconstructed primary vertices.
The PU-per-particle identification (PUPPI)~\cite{,CMS:2020ebo,Bertolini:2014bba} algorithm goes one step further and also acts on neutral PF candidates.
Each PF candidate is assigned a weight between 0 and 1 that scales the four-momentum according
to the probability of the particle to originate from a pileup interaction.
The weight is calculated as a function of a variable defined by the energy deposits
in the vicinity of the PF candidate.
The PUPPI algorithm makes the additional pileup corrections to jets unnecessary,
and has improved the performance and pileup stability of jet substructure tagging.

The modified list of PF candidates is subsequently used as input for jet clustering algorithms, such that ha\-dronic decay products of the top quark can be identified with jets.
In CMS, the anti-\kt~\cite{Cacciari:2008gp} jet clustering algorithm is commonly used, as implemented in the \FASTJET software package~\cite{Cacciari:2011ma} using a distance parameter of $R=0.4$.
The missing transverse momentum vector \ptvecmiss is computed as the negative vector sum of the transverse momenta of all the PF candidates in an event, and its magnitude is denoted as \ptmiss~\cite{CMS:2019ctu}.
The jet energy scale (JES)~\cite{CMS:2016lmd} is corrected for pileup effects, detector effects, and residual differences between data and simulation.
The jet energy resolution (JER)~\cite{CMS:2016lmd} is smeared in simulated events in order to match the resolution observed in data.
Both corrections are propagated to \ptmiss in each event.

Jets originating from \PQb quarks are identified (tagged) with multivariate approaches that make use of global event, secondary vertex, displaced track, and jet constituent information~\cite{CMS:2017wtu}.

\subsection{Kinematic reconstruction of the \texorpdfstring{\ttbar}{tt} system}
\label{sec:kinreco}

The top quarks are investigated experimentally by measuring their decay products and their
kinematic properties. In the all-jets decay channel, all decay products are reconstructed.
In the dilepton channel, however, the two neutrinos from the \PW boson decay are not measured,
thus leading to ambiguities in the reconstruction of neutrino momenta.
The lepton+jets channel exhibits intermediate properties with only one neutrino in the final state, leading to fewer ambiguities.
Several methods of kinematic reconstruction of \ttbar pairs have been developed, which are described in the following.

\subsubsection{Reconstruction in the lepton+jets and all-jets channels}

In the lepton+jets and all-jets channels, kinematic fits~\cite{Abbott:1998dc,CMS-NOTE-2006-023} are employed to check the compatibility of an event with the \ttbar hypothesis and to improve the resolution of the reconstructed quantities.
The fit parameters are the three-vectors of the momenta of the six decay products resulting in 18 unknowns.
The following constraints are applied in the fit: the invariant masses of the top quark and antiquark candidates should be the same and the invariant masses of both \PW boson candidates should be 80.4\GeV~\cite{ParticleDataGroup:2022pth}. The intrinsic width effects are negligible with respect to the experimental resolution, with the latter taken into account in the kinematic fit.

In the lepton+jets channel, the four-momenta of the lepton and the four highest-\pt (leading) jets,  and  \ptvecmiss are the inputs that are fed together with their resolutions to the fit algorithm~\cite{Abbott:1998dc}. With these input values, the fit has two degrees of freedom.
In the all-jets channel, the momenta and resolutions of the six leading jets are the inputs to the fitter~\cite{CMS-NOTE-2006-023} resulting in a fit with three degrees of freedom.
The kinematic fit then minimises $\chisq\equiv(\mathbf{x}-\mathbf{x}^m)^{\mathrm{T}}G(\mathbf{x}-\mathbf{x}^m)$,
where $\mathbf{x}^m$ and  $\mathbf{x}$ are the vectors of the measured and fitted momenta, respectively, and $G$ is the inverse covariance matrix, which is constructed from the uncertainties in the measured momenta. The above-mentioned constraints are added to the minimisation procedure with Lagrange multipliers.

The fit is performed for all possible assignments of the jets to the decay products.
To reduce combinatorics, exactly two of the selected leading jets are required to be identified as originating from a \PQb quark (\PQb tagged).
In the lepton+jets channel, the two \PQb-tagged jets are candidates for the \PQb quarks in the \ttbar hypothesis, while the two jets that are not \PQb tagged serve as candidates for the light quarks from the hadronically decaying \PW boson.
In addition, there are  two solutions for the start value of the longitudinal component of the neutrino momentum per parton-jet assignment.
Hence, the fit is performed for four different permutations per event.
In the all-jets channel, the two \PQb-tagged jets are the candidates for the \PQb quarks and the four jets that are not \PQb tagged serve as candidates for the light quarks from the hadronically decaying \PW bosons. Hence, the fit is performed for six different permutations.

The \chisq probability \Pgof of the kinematic fits is used to rank the permutations, since the permutations with wrongly assigned jets typically have very low \Pgof values. For simulated \ttbar events, the parton-jet assignments can be classified as correct, wrong, and unmatched permutations. In the first case, all quarks from the \ttbar decay are matched within a distance of $\DR=\sqrt{\smash[b]{(\Delta\eta)^2+(\Delta\phi)^2}}<0.3$, where $\phi$ is the
azimuthal angle and $\eta$ is the pseudorapidity, to a selected jet and assigned with the correct flavour assumption to the correct top quark. If all quarks are matched to a selected jet, but the wrong permutation is chosen, it is labelled `wrong', while `unmatched' indicates that not all quarks are matched unambiguously to a selected jet.

\begin{figure}[!tp]
\centering
\includegraphics[width=0.48\textwidth]{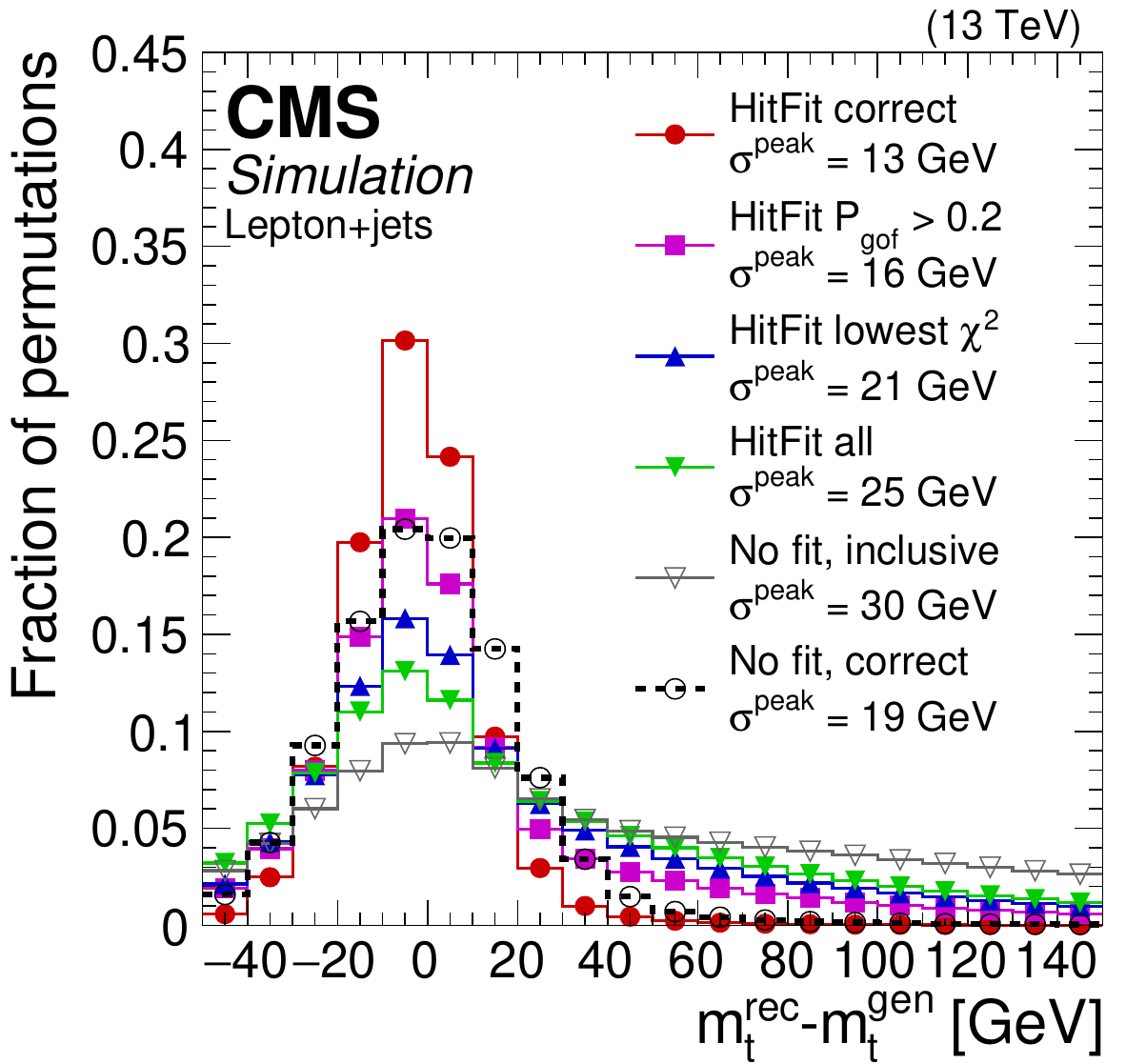}%
\hfill%
\includegraphics[width=0.48\textwidth]{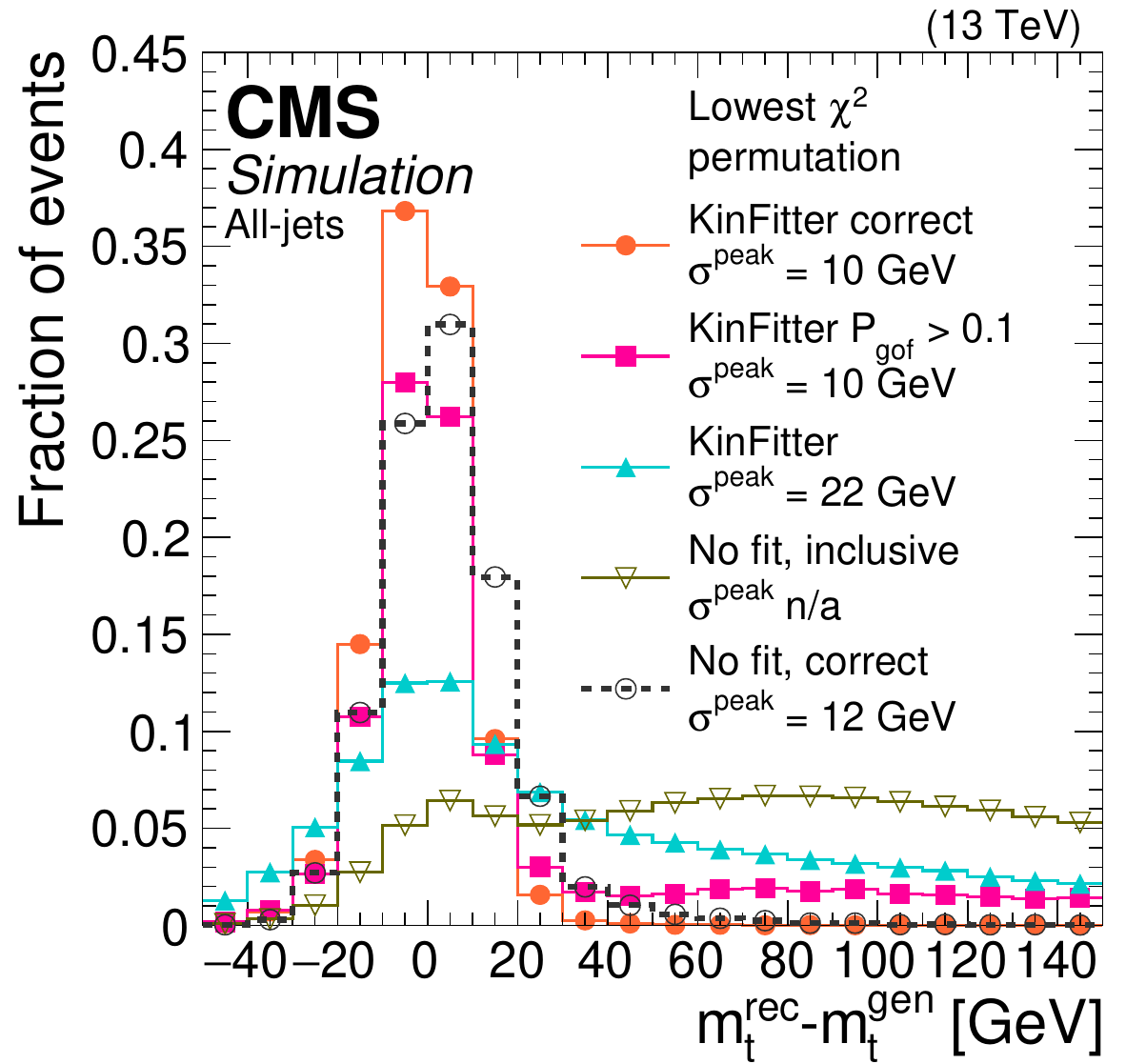}
\caption{%
    Reconstructed top quark mass resolution with and without the \HitFit/\KinFitter kinematic reconstruction in the lepton+jets (left) and all-jets (right) channels.
    Multiple reconstruction options with and without kinematic fit are represented by lines of different colour, and ``correct'' denotes the correct parton-jet assignments as discussed in the text.
    The \HitFit/\KinFitter reconstruction with a cutoff on \Pgof is used for measuring the top quark mass~\cite{CMS:2018quc,CMS:2018tye}.
}
\label{fig:hitfit_mt}
\end{figure}

\begin{figure}[!tp]
\centering
\includegraphics[width=0.48\textwidth]{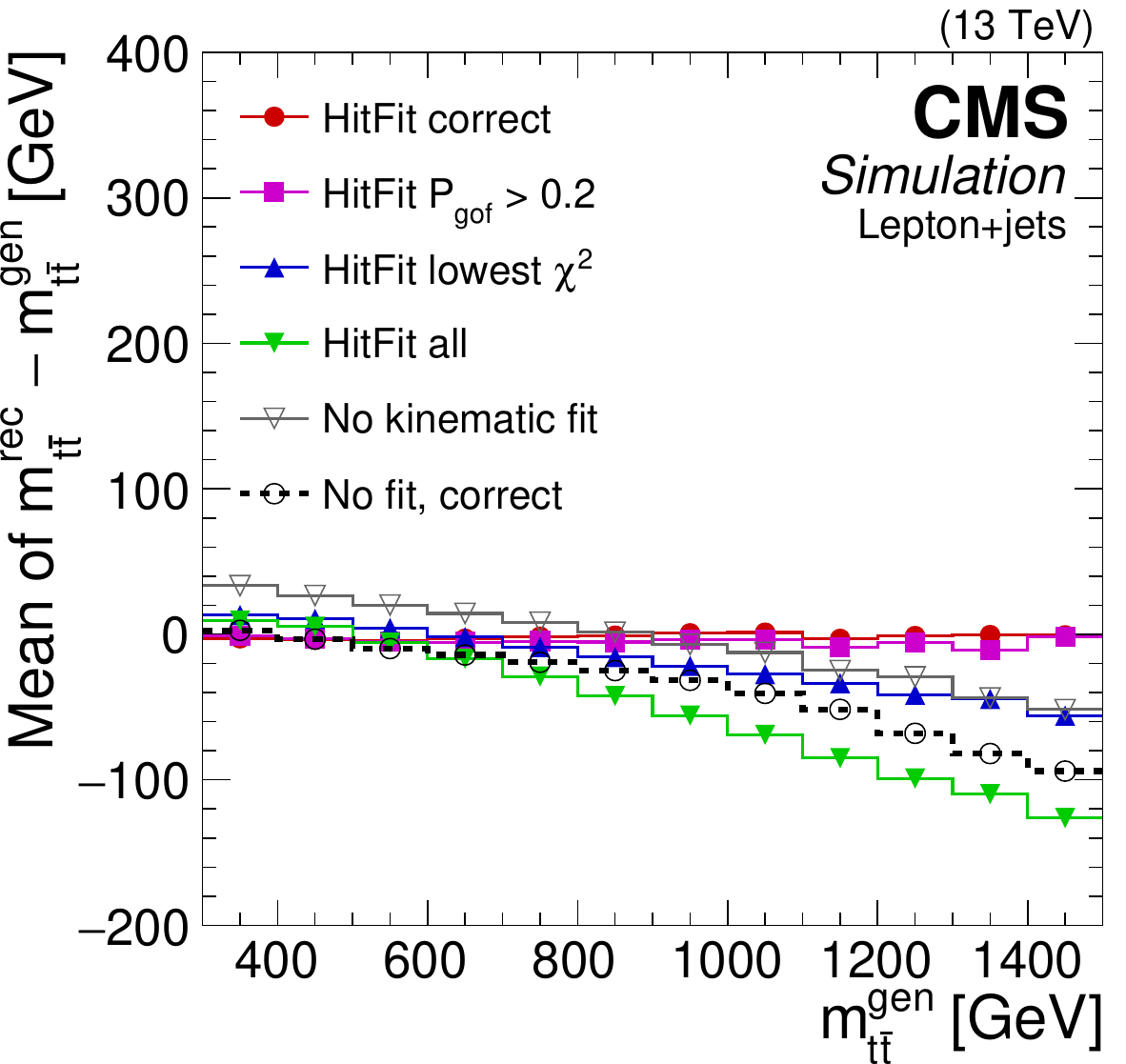}%
\hfill%
\includegraphics[width=0.48\textwidth]{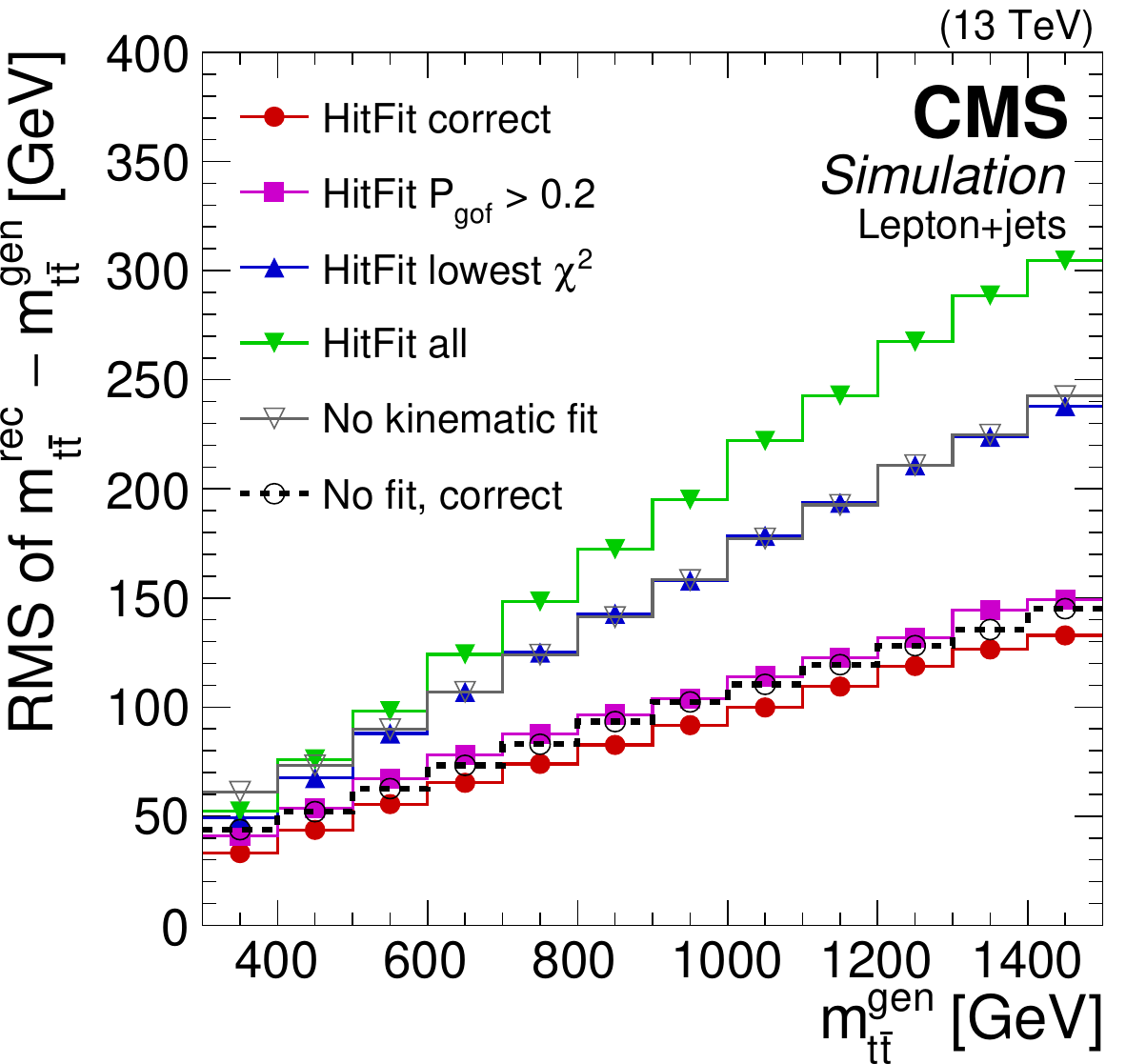}
\caption{%
    Reconstructed \ttbar mass bias (left) and resolution (right) with and without the \HitFit kinematic reconstruction in the lepton+jets channel, as functions of the \ttbar invariant mass at generator level. Multiple reconstruction options with and without kinematic fit are represented by lines of different colour, and ``correct'' denotes the correct parton-jet assignments as discussed in the text.
    The \HitFit reconstruction with a cutoff on \Pgof is used for measuring the top quark mass~\cite{CMS:2018quc}.
}
\label{fig:hitfit_mtt}
\end{figure}

\begin{figure}[!tp]
\centering
\includegraphics[width=0.48\textwidth]{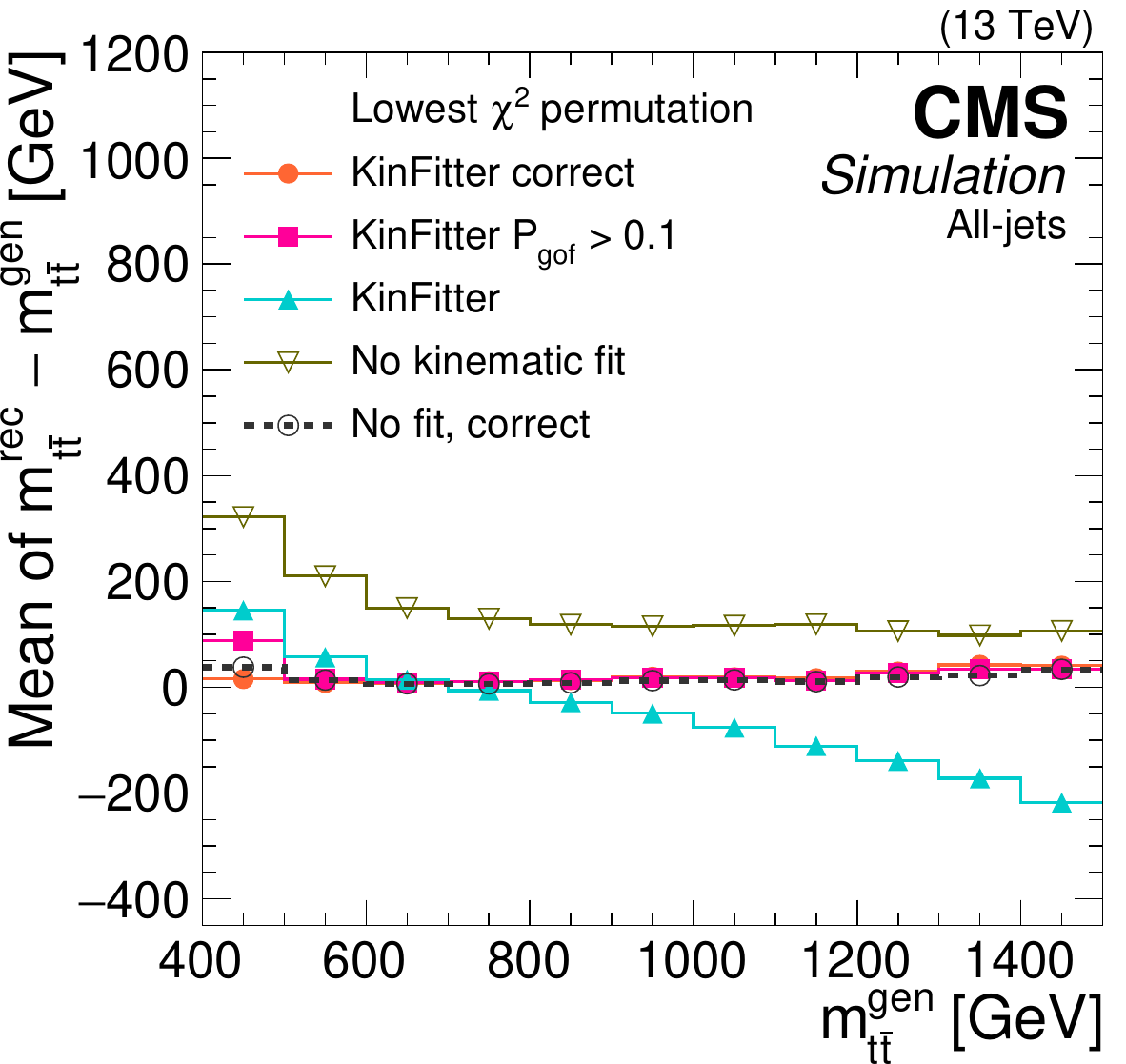}%
\hfill%
\includegraphics[width=0.48\textwidth]{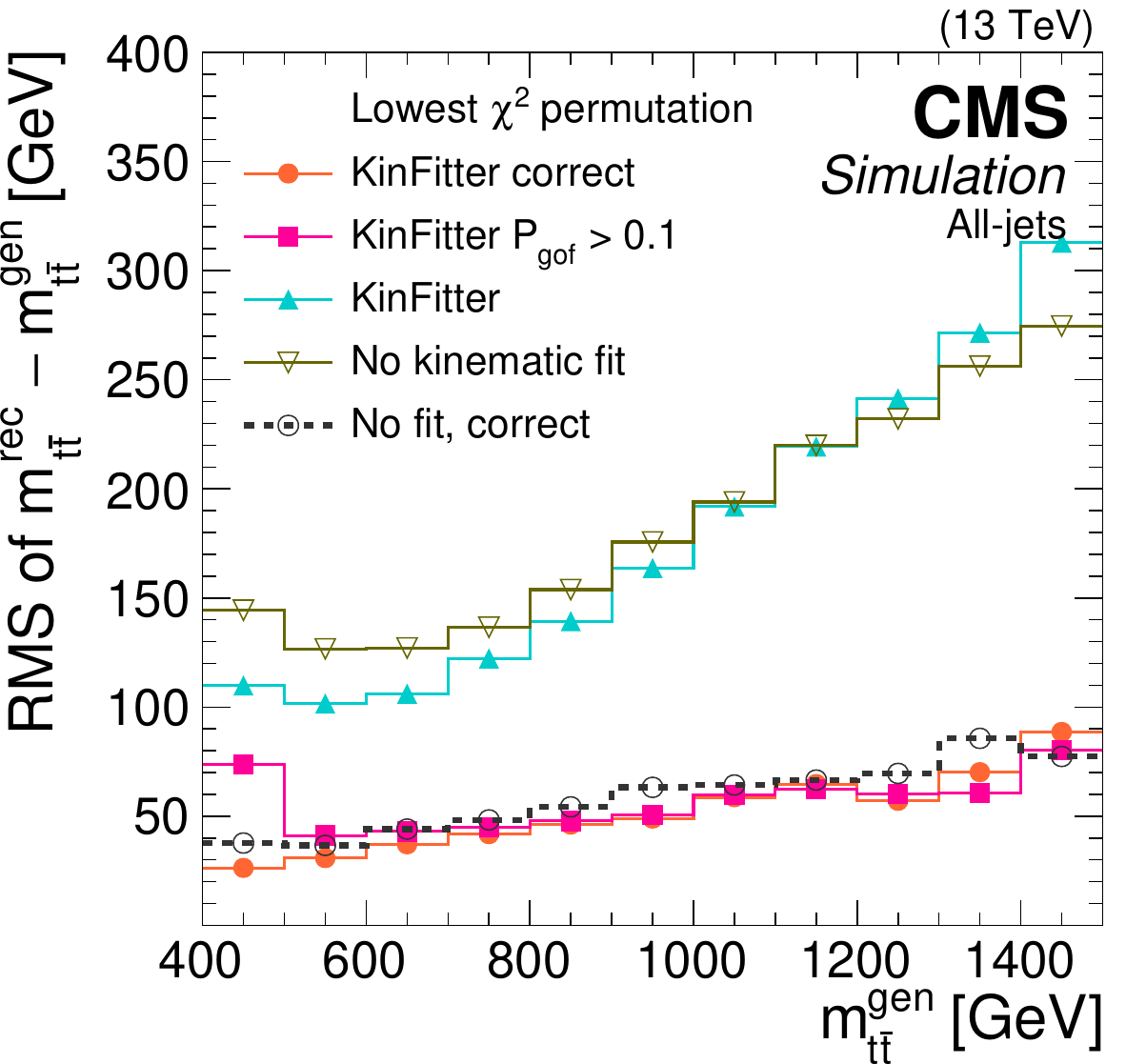}
\caption{%
    Reconstructed \ttbar mass bias (left) and resolution (right) with and without the \KinFitter kinematic reconstruction in the all-jet channel, as functions of the \ttbar invariant mass at generator level.
    Multiple reconstruction options with and without kinematic fit are represented by lines of different colour, and ``correct'' denotes the correct parton-jet assignments as discussed in the text.
    The \KinFitter reconstruction with a cutoff on \Pgof is used for measuring the top quark mass~\cite{CMS:2018tye}.
}
\label{fig:kinfitter_mtt}
\end{figure}

Due to the constraints, the kinematic fits improve the resolution of the reconstructed mass of the top quark candidates. The resolution of the reconstructed mass of the top quark with and without applying the kinematic fit is presented in Fig.~\ref{fig:hitfit_mt} for the lepton+jets (multiple permutations) and all-jets channels (permutation with lowest \chisq).
In the all-jets channel, only the permutation with the lowest \chisq in each event is considered for further analysis.
The resolution $\sigma^{\text{peak}}$ is extracted by fitting a Gaussian distribution within the range $-40<\mtrec-\mtgen<+40\GeV$.
Without a kinematic fit, the resolution of the reconstructed top quark mass is relatively poor in the case of the lepton+jets channel, while the peak is hardly discernible at all in the all-jets channel.
In both \ttbar decay channels, the kinematic fit improves the resolution using either all jet-parton permutations or the one with the lowest \chisq.
Finally, a cut on $\Pgof>0.2$ (0.1) is used in the lepton+jets (all-jets) channel, which matches the resolution of the case where only correct permutations are considered with their pre-fit momenta.
The selection efficiency of the \Pgof cut is 27.4 (5.3)\% in the lepton+jets (all-jets) channel.
Besides the mass, the kinematic fits can also improve the reconstruction of other kinematic variables of the \ttbar system, such as its invariant mass \mtt.
The bias and resolution of the reconstructed \mttrec with regard to the generated \mttgen is shown for the lepton+jets channel in Fig.~\ref{fig:hitfit_mtt} and for the all-jets channel in Fig.~\ref{fig:kinfitter_mtt}.
The resolution is defined as the root-mean-square (RMS) of the difference between the reconstructed and the generated parton-level quantity, and the bias as its mean.
The kinematic fit with a \Pgof cutoff improves the resolution and is almost free of bias over the examined range in \mttgen.

\subsubsection{Reconstruction in the dilepton channel}
\label{sec:kinrecodillep}

In contrast to the lepton+jets channel, direct measurements of \mt in the dilepton channel are challenging because of the ambiguity due to the two neutrinos in the final state, reconstructed as \ptvecmiss.
Therefore, the dilepton \ttbar events are mostly used for extraction of \mt through comparisons of the measurements of inclusive or differential \ttbar cross sections~\cite{CMS:2019esx,CMS:2019jul,CMS:2022emx} to the theoretical predictions, as explained in Section~\ref{sec:indirect}. In this case, the reconstruction method aims to obtain good resolution of the observable of interest and a high reconstruction efficiency.

For the \ttbar reconstruction in the dilepton channel, several methods have been developed, with the primary task of obtaining solutions for the two unknown neutrino momenta.
Depending on the observable of interest, either the individual top quark and antiquark, \eg in the measurement of single-particle kinematics, or only the \ttbar system, 
are reconstructed.

The \textbf{full kinematic reconstruction} (FKR) of the \ttbar pair is based on the algebraic approach suggested in Ref.~\cite{Sonnenschein:2006ud}. A system of kinematic equations describing the \ttbar system is solved using the four-momenta of the six final-state particles, \ie two leptons, two \PQb jets, and the two neutrinos.
It is assumed that the total measured missing transverse momentum is due to the two neutrinos and can be decomposed as follows:
\begin{equation}
\label{EQ:kinrecodilep1}
    p_x^{\text{miss}}=p_{x,\PGn}+p_{x,\PAGn},\qquad
    p_y^{\text{miss}}=p_{y,\PGn}+p_{y,\PAGn}.
\end{equation}
The invariant mass of the lepton and the neutrino from the same top quark should correspond to the mass of the \PW boson, resulting in the following equations:
\begin{align}
    m_{\PWp}^2&=(\Eellp+\Enu)^2-(p_{x,\Pellp}+p_{x,\PGn})^2-(p_{y,\Pellp}+p_{y,\PGn})^2-(p_{z,\Pellp}+p_{z,\PGn})^2, \\
    m_{\PWm}^2&=(\Eellm+\Enubar)^2-(p_{x,\Pellm}+p_{x,\PAGn})^2-(p_{y,\Pellm}+p_{y,\PAGn})^2-(p_{z,\Pellm}+p_{z,\PAGn})^2.
\end{align}
Finally, the masses of the top quark and antiquark are given, respectively, by:
\begin{align}
    &\begin{aligned}
        \mt^2=(\Eellp+\Enu+E_{\PQb})^2 - (p_{x,\Pellp}+p_{x,\PGn}+p_{x,\PQb})^2 - (p_{y,\Pellp}+p_{y,\PGn}+p_{y,\PQb})^2& \\ - (p_{z,\Pellp}+p_{z,\PGn}+p_{z,\PQb})^2&,
    \end{aligned} \\
    \label{EQ:kinrecodilep5}
    &\begin{aligned}
        \mtbar^2=(\Eellm+\Enubar+E_{\PAQb})^2 - (p_{x,\Pellm}+p_{x,{\PAGn}}+p_{x,\PAQb})^2 - (p_{y,\Pellm}+p_{y,{\PAGn}}+p_{y,\PAQb})^2& \\ - (p_{z,\Pellm}+p_{z,{\PAGn}}+p_{z,\PAQb})^2&.
    \end{aligned}
\end{align}
The masses of the \PQb quarks are set to the values used in the simulation, while lepton masses are assumed to be negligible. The masses of the top quark and of the \PW boson need to be fixed in order to solve the system of equations~\eqref{EQ:kinrecodilep1}--\eqref{EQ:kinrecodilep5}. For analyses where the choice does not directly affect the result of the measurement, they are typically fixed to the default values of $\mt=172.5\GeV$ and $\mW=80.4\GeV$. The equation system can then be solved analytically with a maximum four-fold ambiguity. Selected is the solution which yields the minimum invariant mass of the \ttbar system, as it was shown that this choice provides the best solution in most cases.
In analyses that target direct reconstruction of \mt in the dilepton channel, a dedicated method~\cite{CMS:2011acs,CMS:2012tdr,CMS:2015lbj} is used that tests different \mt hypotheses. In contrast, in differential measurements of the \ttbar cross section, the dependence on the choice of \mt in the reconstruction is usually estimated by varying the top quark mass assumption in the MC simulation.

To capture the effects of the finite detector resolution, the kinematic reconstruction is repeated 100 times, each time randomly smearing the measured energies and directions of the reconstructed leptons and jets within their resolutions. This smearing procedure recovers events that initially yielded no solution because of limited experimental resolution.
Further, in the same smearing procedure, the mass of the \PW boson is varied according to a relativistic Breit--Wigner function, estimated using the generator-level \PW boson mass distribution.
For each solution, a weight is calculated based on the expected true spectrum of the invariant mass of a lepton and a \PQb jet (\mlb) stemming from the decay of a top quark and taking
the product of the two weights for the top quark and antiquark decay chains:
$w=w_{m_{\PAell\PQb}} w_{m_{\Pell\PAQb}}$.
The final three-momenta of the top quarks $j$ and $k$ are then determined as a weighted average over all smeared solutions summing over all 100 kinematic reconstructions:
\begin{equation}
	\langle\vec{p}_{\PQt}^{\,k,j}\rangle = \frac{1}{w_s}\sum_{i=1}^{100}w_i\vec{p}_{\PQt,i}^{\,k,j},
	\quad\text{with}\quad
    w_s=\sum_{i=1}^{100}w_i.
\end{equation}
All possible lepton-jet combinations in the event that satisfy the requirement for the invariant mass of the lepton and jet $\mlb<180\GeV$ are considered.
Combinations are ranked, based on the presence of \PQb-tagged jets in the assignments, \ie a combination with
both leptons assigned to \PQb-tagged jets is preferred over those with one or zero \PQb-tagged jet.
Among assignments with an equal number of \PQb-tagged jets, the one with the highest sum of weights is chosen.
Events with no solution after smearing are discarded.
The four-momentum vector of the top quark is determined by its energy, which is calculated from $\langle\vec{p}_{\PQt}\rangle$, and the top quark mass of 172.5\GeV. The kinematic properties of the top antiquark are determined analogously.
The efficiency of the kinematic reconstruction, defined as the number of events where a solution
is found divided by the total number of selected \ttbar events, is studied in data and simulation, and consistent results of about 90\% are found in analyses at \sqrtseq{13}.

The value of the invariant mass \mtt of the \ttbar pair obtained using FKR is highly sensitive to the predefined value of the top quark mass used as a kinematic constraint. However, the objective of the analyses described in this paper is the extraction of \mt, in some cases exploiting the \mtt distribution or related observables.
For such cases, the \textbf{loose kinematic reconstruction} (LKR) was developed~\cite{CMS:2019esx}, where the value of the top quark mass is not constrained.
In this algorithm, the \nunu system is reconstructed, rather than the individual \PGn and \PAGn.
As a consequence, only the \ttbar system can be reconstructed in LKR, but not the individual top quark and antiquark.
As in FKR, all possible lepton-jet combinations in the event that satisfy the requirement
for the invariant mass of the lepton and jet $\mlb<180\GeV$ are considered.
Combinations are ranked, based on the presence of \PQb-tagged jets in the assignments, but from all the combinations with an equal number of the \PQb-tagged jets, the ones with the highest \pt jets are chosen.
The kinematic variables of the \nunu system are derived as follows:
\begin{enumerate}
\item the transverse momentum \ptvec of the \nunu system is set equal to \ptvecmiss;
\item the \nunu longitudinal momentum $p_{z,\nunu}$ is set to that of the lepton pair, $p_{z,\nunu}=p_{z,\llbar}$, for $p_{\mathrm{T},\nunu}<\Ellbar$, and to zero otherwise;
\item the energy of the \nunu system \Enunu is defined as $\Enunu=\Ellbar$ for $p_{\nunu}<\Ellbar$, and $\Enunu=p_{\llbar}$ otherwise, ensuring that $m_{\nunu}\geq0$;
\item the four-momentum sum of \llnn is calculated;
\item for $m_{\llnn}<2\mW=2\times80.4\GeV$, the mass component of the four-momentum of \llnn is set to $2\mW$, ensuring that $m_{\PWp\PWm}\geq2\mW$;
\item the four-momentum of the \ttbar system is calculated by using the four-momenta of the \llnn system and of the two \PQb jets as \llnn{}+\bbbar.
\end{enumerate}
The additional constraints that are applied on the invariant mass of the neutrino pair, \mbox{$m_{\nunu}\geq0$} (item 3)
and on the invariant mass of the \PW bosons, $m_{\PWp\PWm}\geq2\mW$ (item 5) have only minor effects on the performance of the reconstruction.
The method yields similar \ttbar kinematic resolutions and reconstruction efficiency as for the FKR method.
In the CMS analysis~\cite{CMS:2019esx}, the LKR was exclusively used to measure triple-differential \ttbar cross sections as functions of the invariant mass and rapidity of the \ttbar system, and the additional-jet multiplicity.

\begin{figure}[!tp]
\centering
\includegraphics[width=0.48\textwidth]{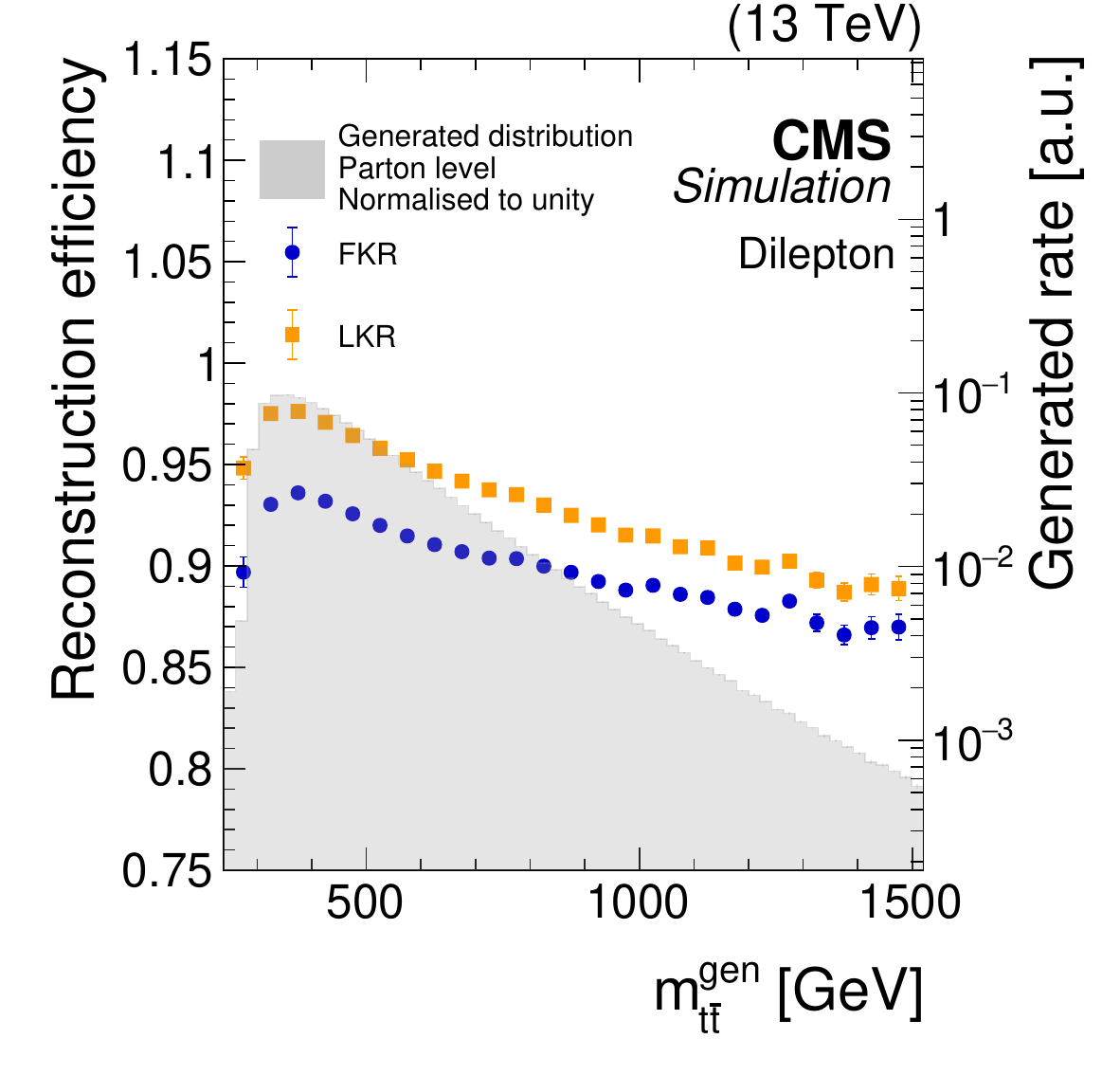}%
\hfill%
\includegraphics[width=0.48\textwidth]{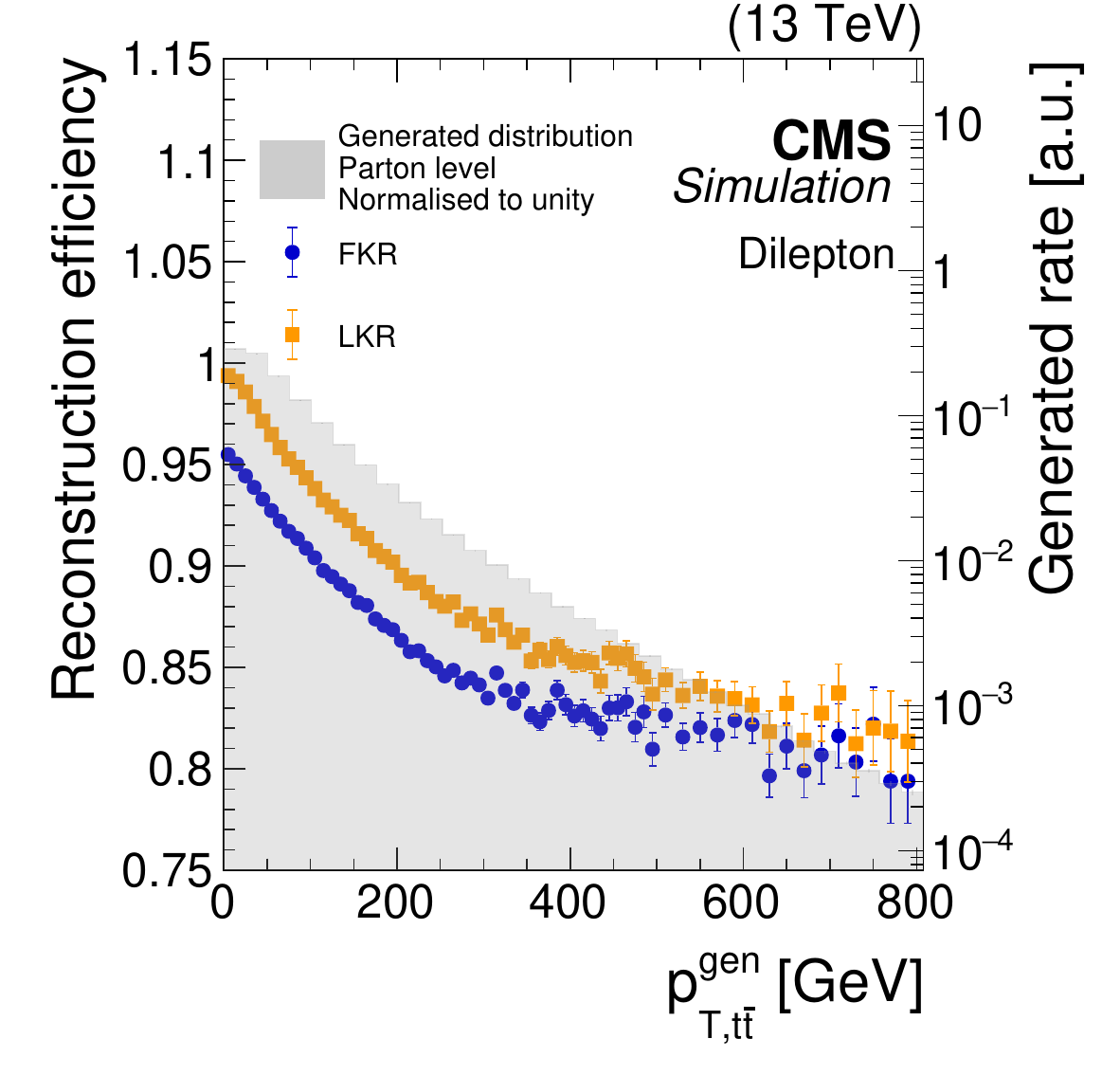}\\
\includegraphics[width=0.48\textwidth]{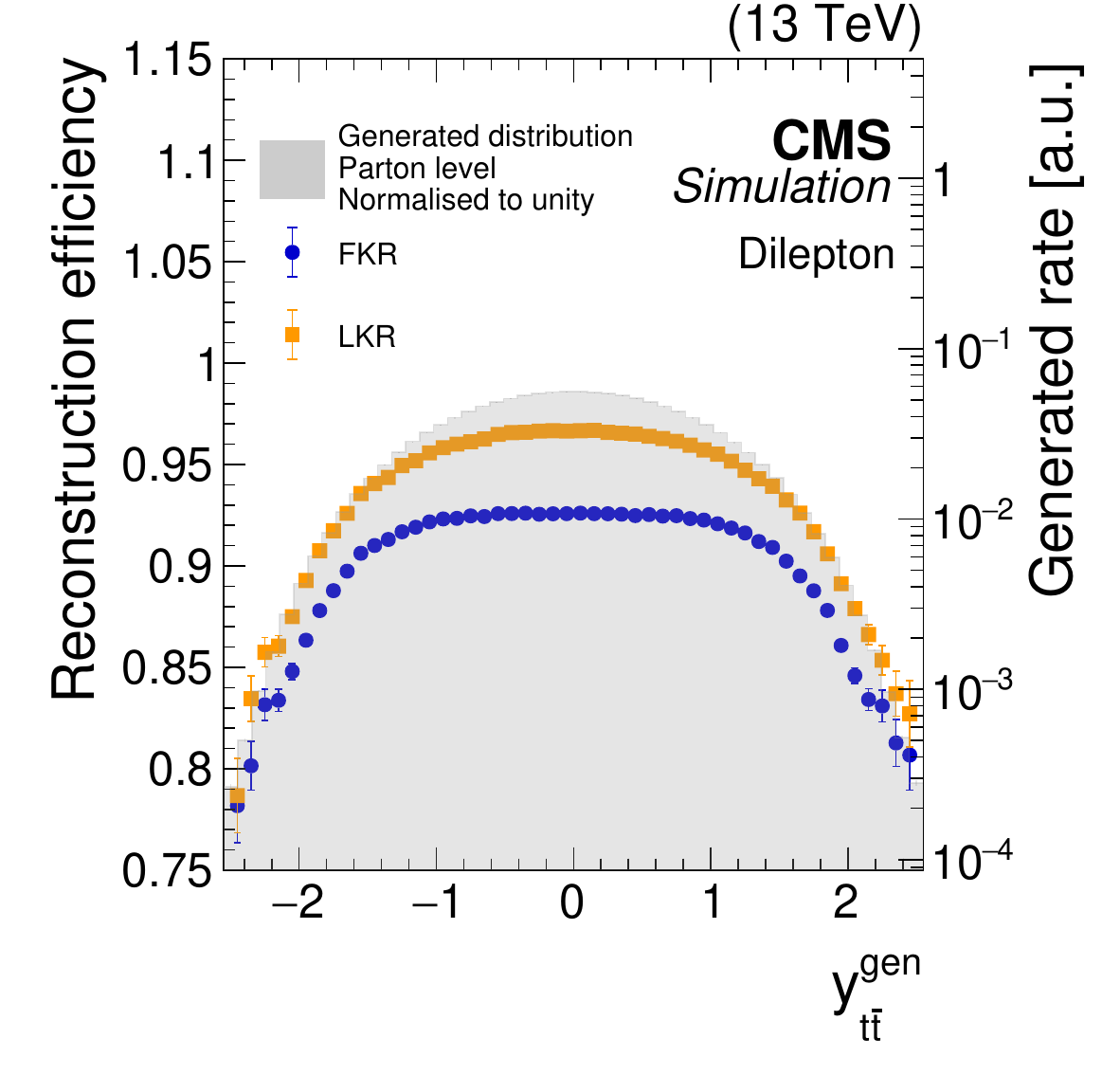}
\caption{%
    The reconstruction efficiencies for the full kinematic reconstruction (FKR, blue circles) and loose kinematic reconstruction (LKR, orange squares) are shown as functions of the invariant mass, transverse momentum, and rapidity of the reconstructed \ttbar system. The averaged efficiencies are 92 (96)\% for the FKR (LKR). The corresponding parton-generator-level distributions, normalised to unit area, for \ttbar production are represented by the grey shaded areas, shown on the logarithmic scale (right $y$ axis). The \POWHEGPYTHIAEight\ \ttbar simulated samples are used.
}
\label{fig:tt_kinRecoEff_comp}
\end{figure}

For the presented performance studies, the \POWHEGPYTHIAEight~\cite{Frixione:2007nw,Nason:2004rx,Alioli:2010xd,Frixione:2007vw,Sjostrand:2014zea} \ttbar simulated samples are used, which are explained in detail in Section~\ref{sec:mcsetup}.
The reconstruction efficiency for both methods is shown in Fig.~\ref{fig:tt_kinRecoEff_comp} as a function of the reconstructed \ttbar kinematic variables \mtt, \pTtt, and \ytt. An event is considered as reconstructed if the reconstruction method yields at least one solution as described above.
The overall efficiency for the LKR is about 4\% higher than for the FKR, and shows the same kinematic properties. The maximum efficiency is achieved for low \mtt, central \ytt, and low \pTtt. The efficiency drops rapidly with increasing \pTtt as the leptons and jets become less separated. For Lorentz-boosted configurations with $\pTtt>700\GeV$, the reconstruction fails in 20\% of the cases.

\begin{figure}[!tp]
\centering
\includegraphics[width=0.48\textwidth]{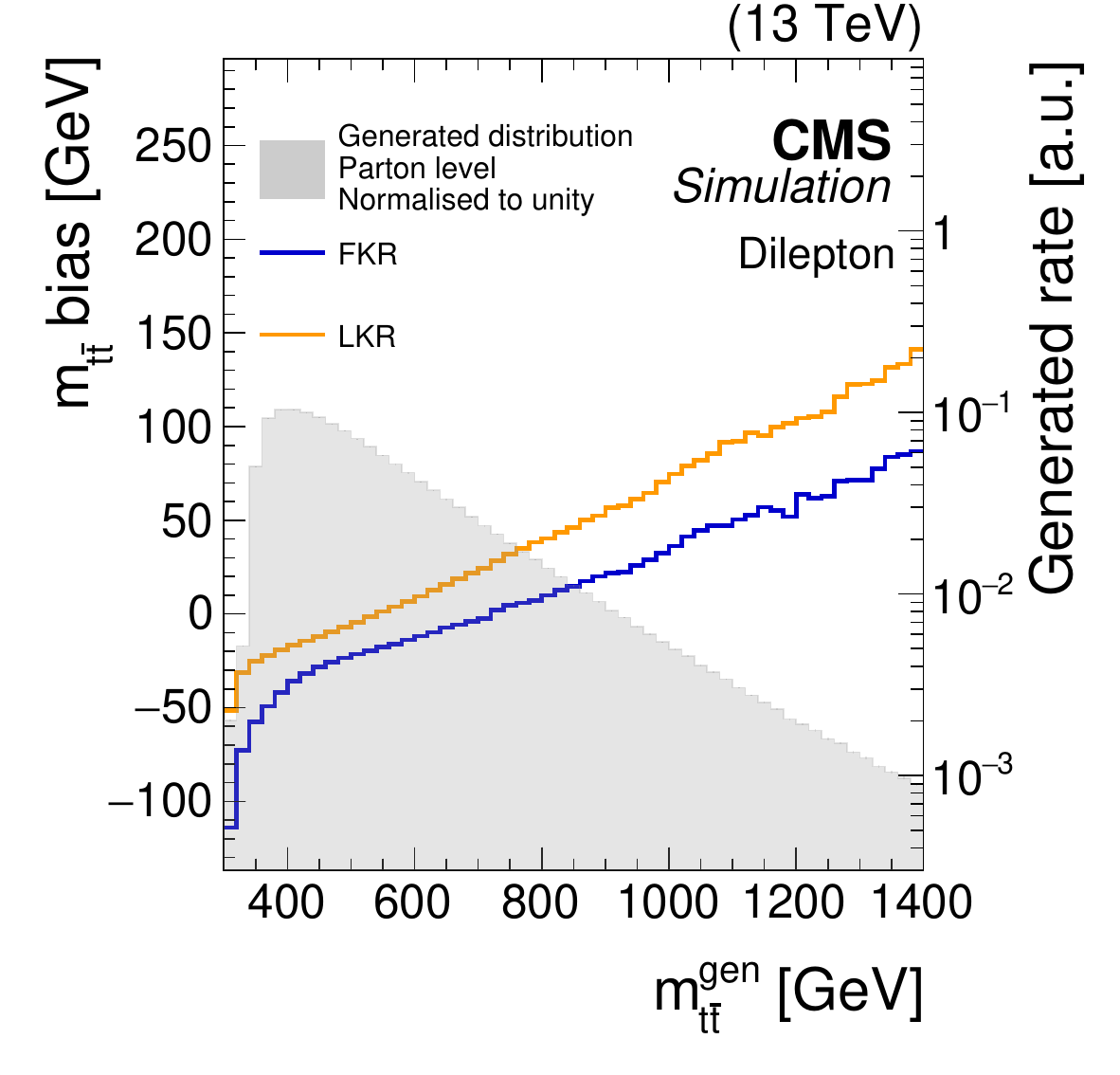}%
\hfill%
\includegraphics[width=0.48\textwidth]{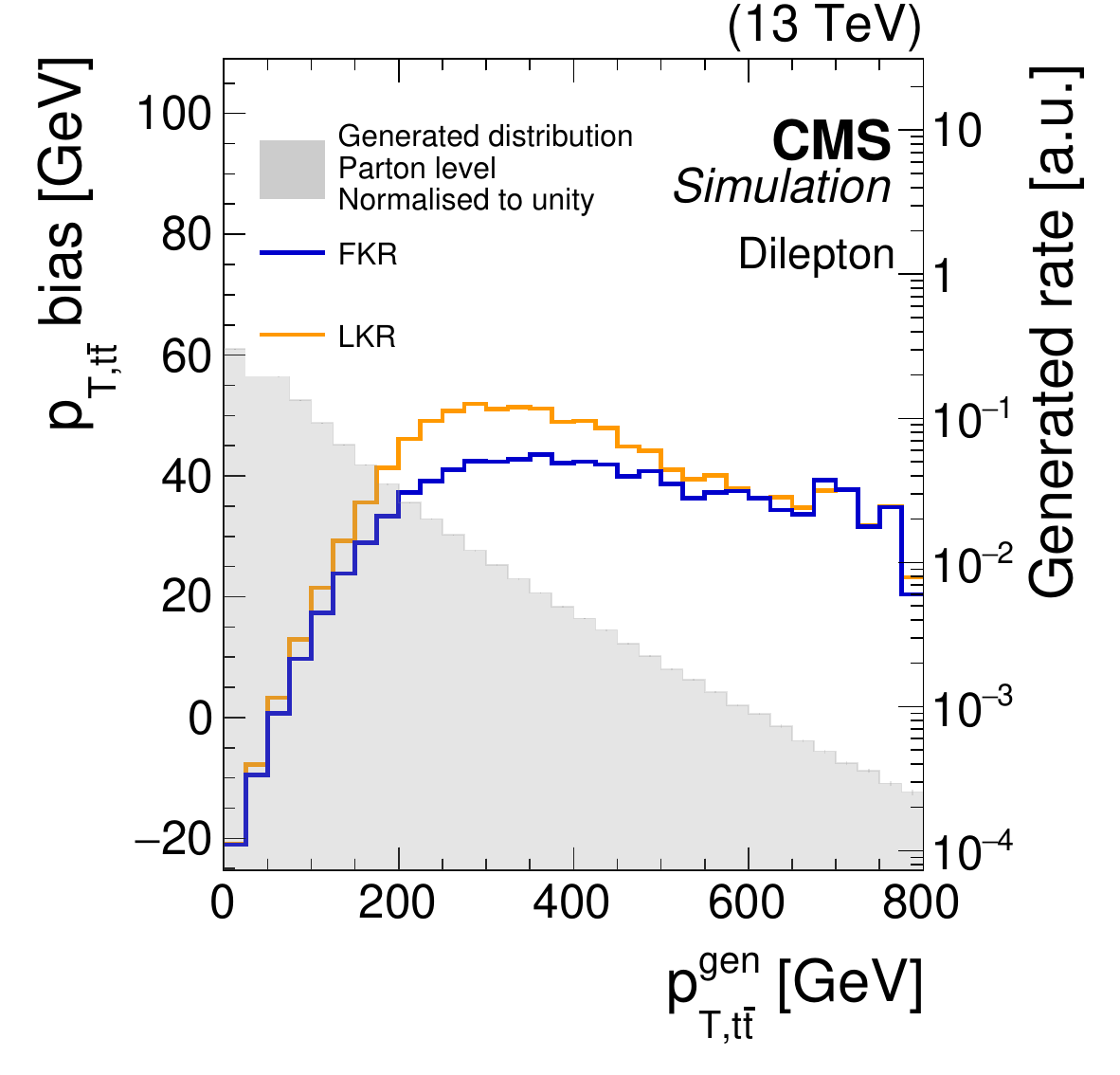}\\
\includegraphics[width=0.48\textwidth]{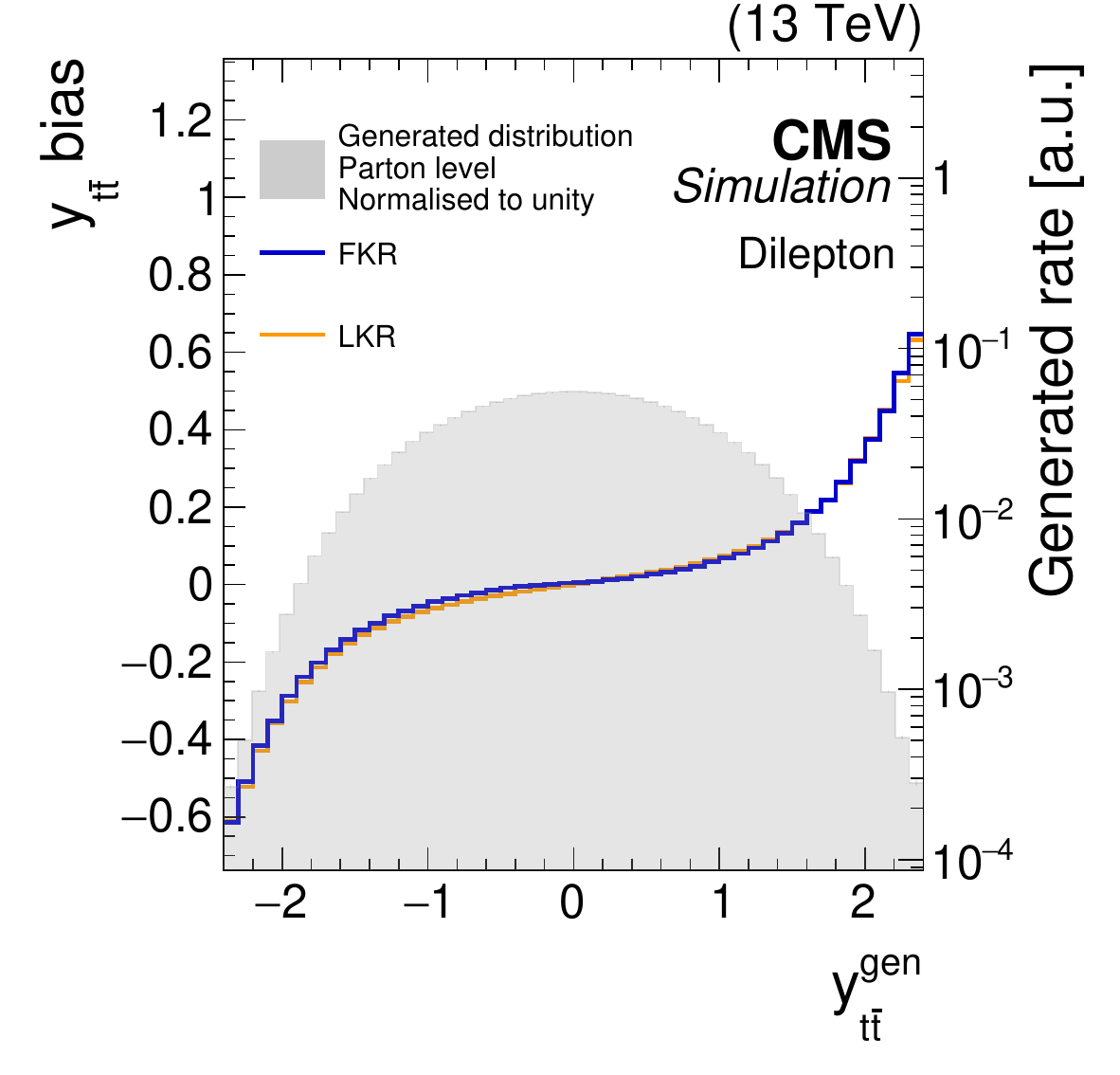}
\caption{%
    The biases (solid lines), as defined in the text, for the full kinematic reconstruction (FKR, blue) and loose kinematic reconstruction (LKR, orange) are shown for the invariant mass, transverse momentum, and rapidity of the \ttbar system, as a function of the same variables at the generator level. The corresponding parton-generator-level distributions, normalised to unit area, for \ttbar production are represented by the grey shaded areas, shown on the logarithmic scale (right $y$ axis). The \POWHEGPYTHIAEight\ \ttbar simulated samples are used.
}
\label{fig:tt_kinRecoResolution_comp_bias}
\end{figure}

\begin{figure}[!tp]
\centering
\includegraphics[width=0.48\textwidth]{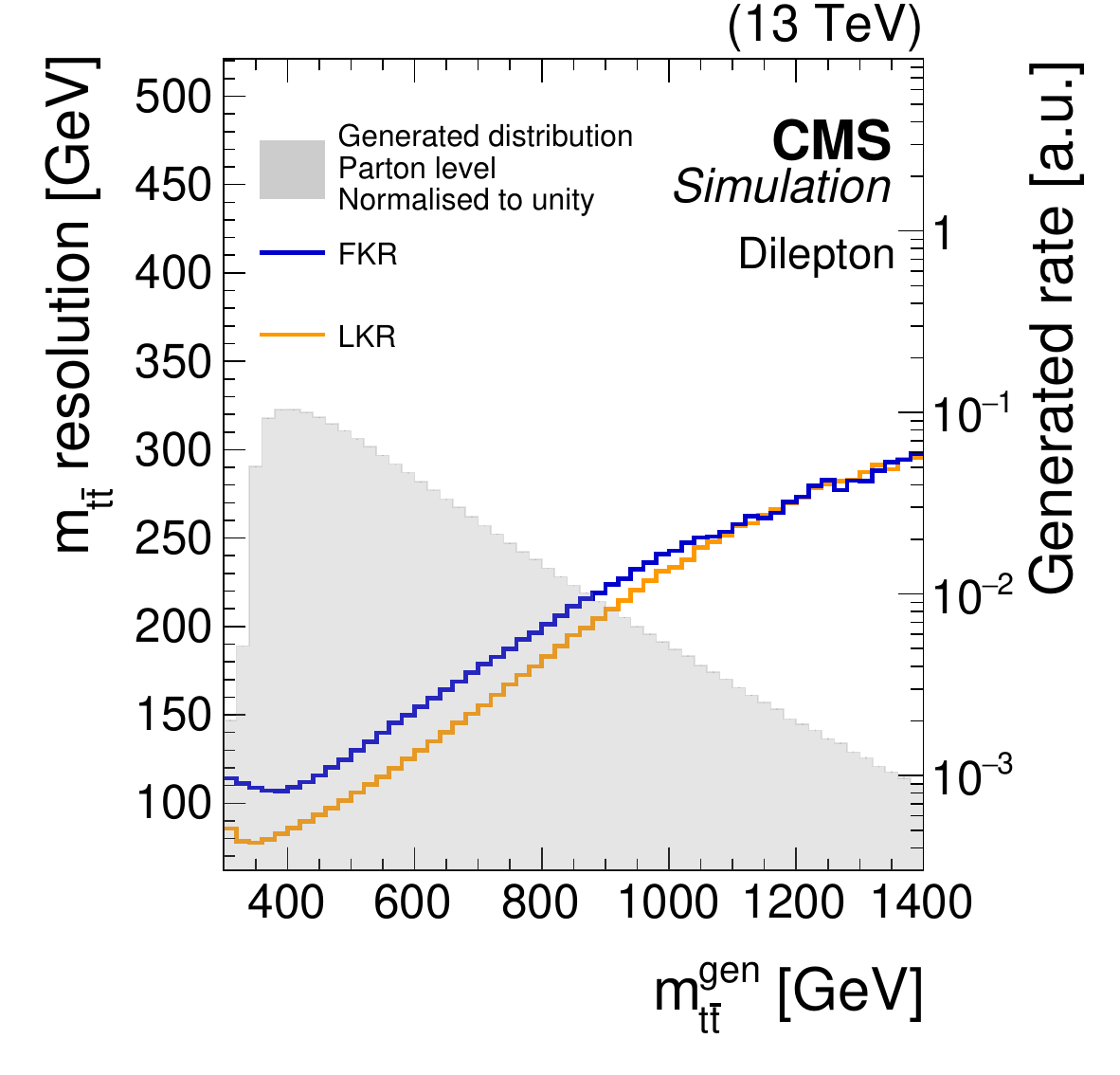}%
\hfill%
\includegraphics[width=0.48\textwidth]{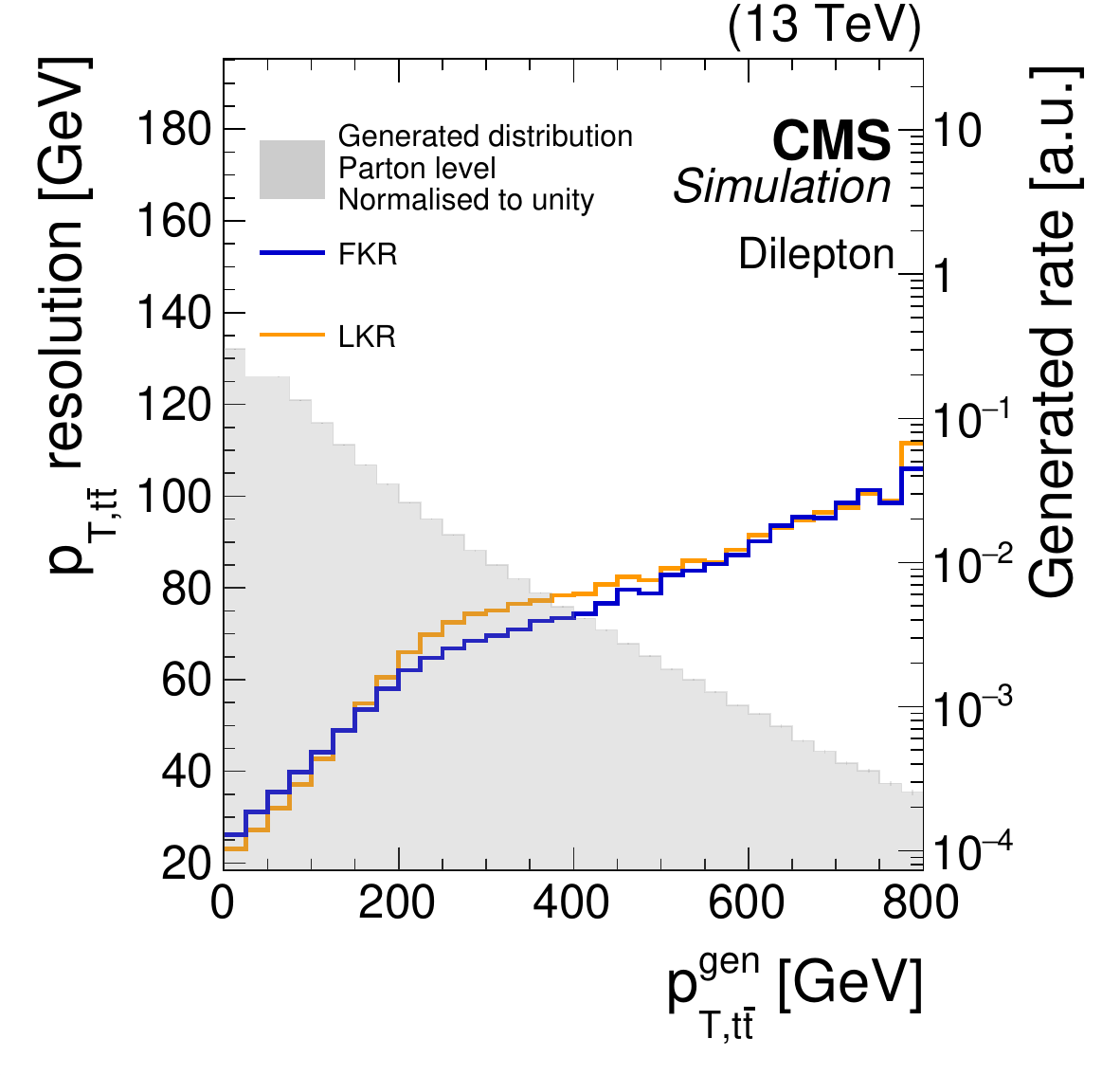}\\
\includegraphics[width=0.48\textwidth]{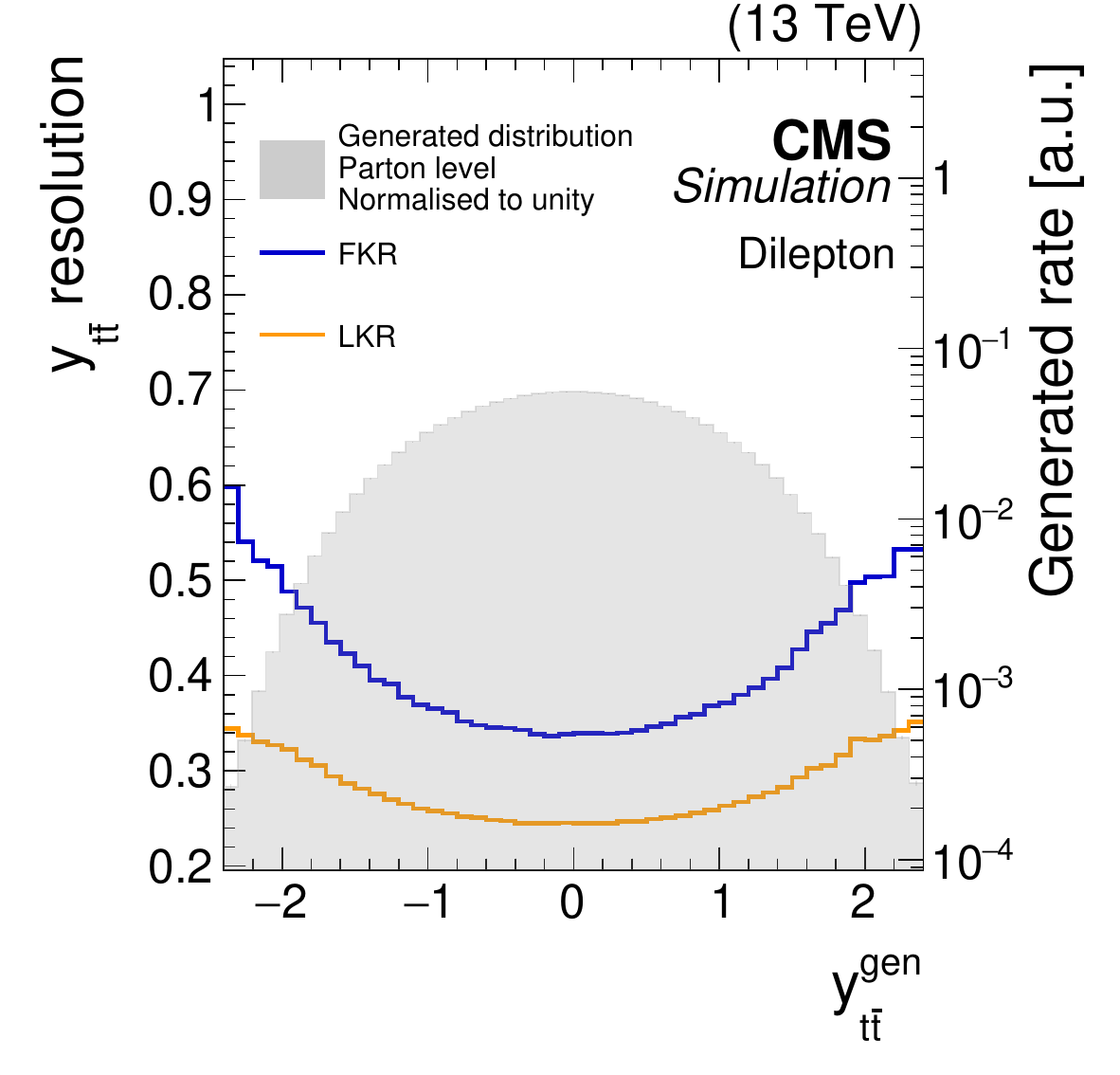}
\caption{%
    The resolutions (solid lines), as defined in the text, for the full kinematic reconstruction (FKR, blue) and loose kinematic reconstruction (LKR, orange) are shown as functions of the invariant mass, transverse momentum, and rapidity of the \ttbar system at the generator level. The corresponding parton-generator-level distributions, normalised to unit area, for \ttbar production are represented by the grey shaded areas, shown on the logarithmic scale (right $y$ axis). The \POWHEGPYTHIAEight\ \ttbar simulated samples are used.
}
\label{fig:tt_kinRecoResolution_comp_res}
\end{figure}

The resolution and bias for both algorithms are shown in Figs.~\ref{fig:tt_kinRecoResolution_comp_bias} and \ref{fig:tt_kinRecoResolution_comp_res}, respectively, as functions of the same three observables at the generator level.
As described above for the lepton+jets decay channel, the resolution is defined as the RMS of the difference between the reconstructed and the parton-level quantity, and the bias as its mean.
As in the case of the efficiencies, the LKR shows better performance. Its bias is often closer to zero in the low-\mtt regime, but becomes larger than in the case of the FKR for very large values of \mtt. The LKR shows better resolution over the whole spectra, but it should be noted that the resolution definition is sensitive to outliers, \eg in the tails of the distribution, affecting the performance of the FKR, \eg in the low-\mtt region.
For probing \mt in the dilepton channel, the resolution at low \mtt, close to the production threshold, is of key importance. The resolution is about 100--150\GeV, which defines the minimal bin width in the differential \mtt measurement.

Since the FKR and LKR methods are developed to be agnostic to additional radiation for \ttbar production, a multivariate method was developed in CMS~\cite{CMS:2022emx} to optimise the resolution for an observable related to the invariant mass of the \ttbarjet system, denoted as $\rho$, which is defined for \ttbar events with at least one additional jet:
\begin{equation}
    \rho=\frac{340\GeV}{\mttbarjet}.
\end{equation}
In the definition of $\rho$, the leading jet is considered and \mttbarjet is the invariant mass of the \ttbarjet system.
This observable shows a large \mt sensitivity and is measured in a CMS analysis~\cite{CMS:2022emx} described in Section~\ref{sec:ttjetanalysis} to extract \mtp. The result of the measurement is independent of the choice of the scaling constant in the numerator, which is introduced to define $\rho$ dimensionless, and is on the order of two times \mt.
Set up as a regressional neural network (NN), a fully connected feed-forward NN is trained.
The benefit of using a regression NN is the maximised reconstruction efficiency, increasing the acceptance of the measurement, as it yields a solution for every event.
The NN uses a set of low-level inputs, \eg particle four-momenta, and high-level input variables, such as geometric and kinematic properties of the systems of the final-state objects. Starting from a set of 100 variables, the ten variables with the highest impact on the output of the NN are selected. These also include solutions of the LKR and FKR algorithms.
Simulated events are used for the training of the regression NN if they contain at least three reconstructed jets with $\pt>30\GeV$ and $\abseta<2.4$.
The ten input variables, ordered by their impact, used for the regression NN are:
\begin{itemize}
    \item the calculation for $\rho$ using the LKR;
    \item the calculation for $\rho$ using the FKR;
    \item the invariant mass of the dilepton and subleading jet system;
    \item the invariant mass of the leading lepton and subleading jet system;
    \item the \pt of the subleading lepton;
    \item the invariant mass of the dilepton system;
    \item the invariant mass of the subleading lepton and subleading jet system;
    \item the invariant mass of the subleading lepton and leading jet system;
    \item the invariant mass of the dilepton and leading jet system;
    \item \ptmiss.
\end{itemize}
The training is performed using an independent data set, which is produced with the \MGvATNLO~\cite{Alwall:2011uj} event generator at NLO accuracy, interfaced with \PYTHIAEight~\cite{Sjostrand:2014zea}. Afterwards, the resulting performance is also evaluated using the \POWHEGPYTHIAEight simulation, and is checked for possible overtraining.

\begin{figure}[!ht]
\centering
\includegraphics[width=0.48\textwidth]{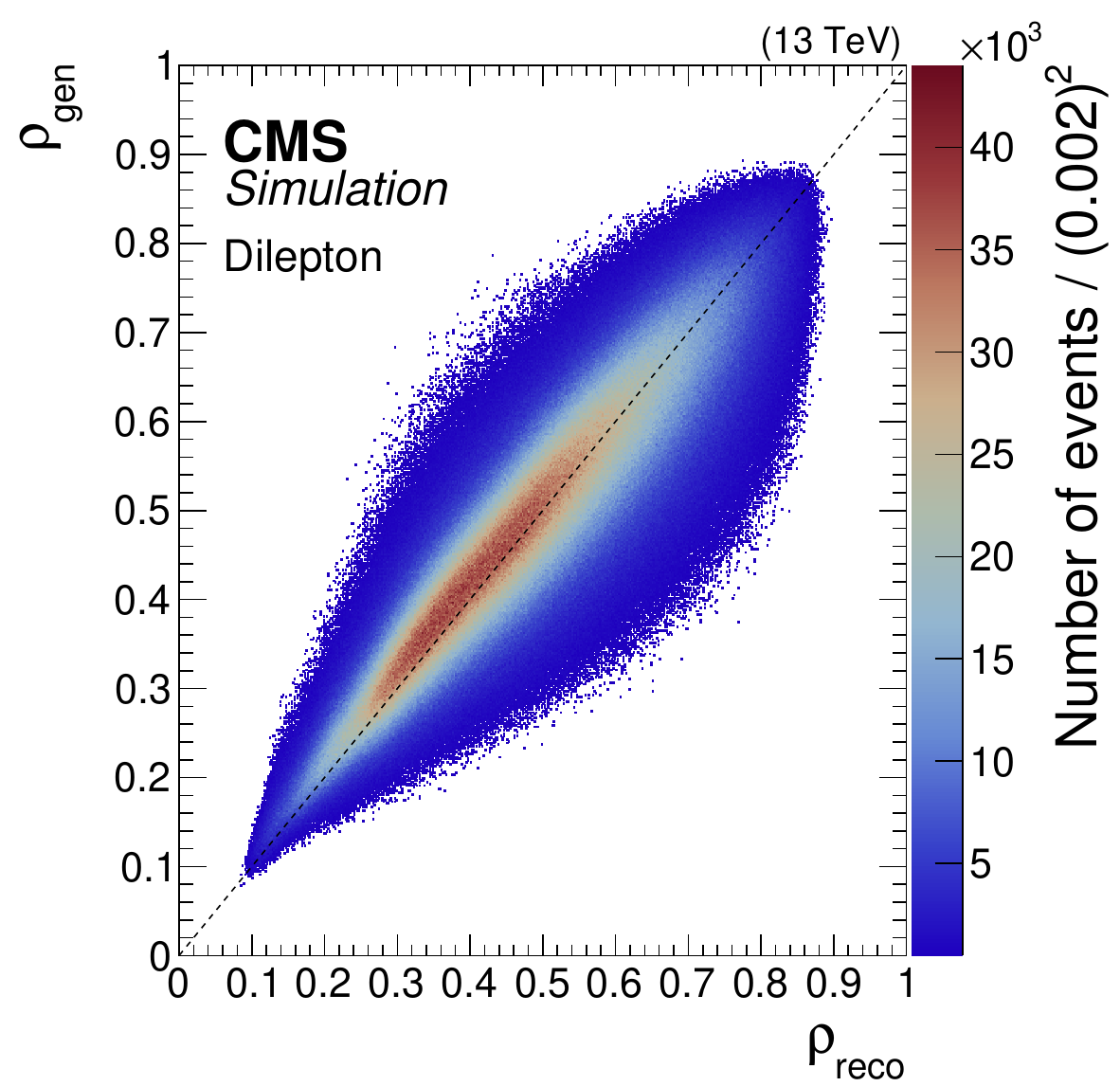}%
\hfill%
\includegraphics[width=0.48\textwidth]{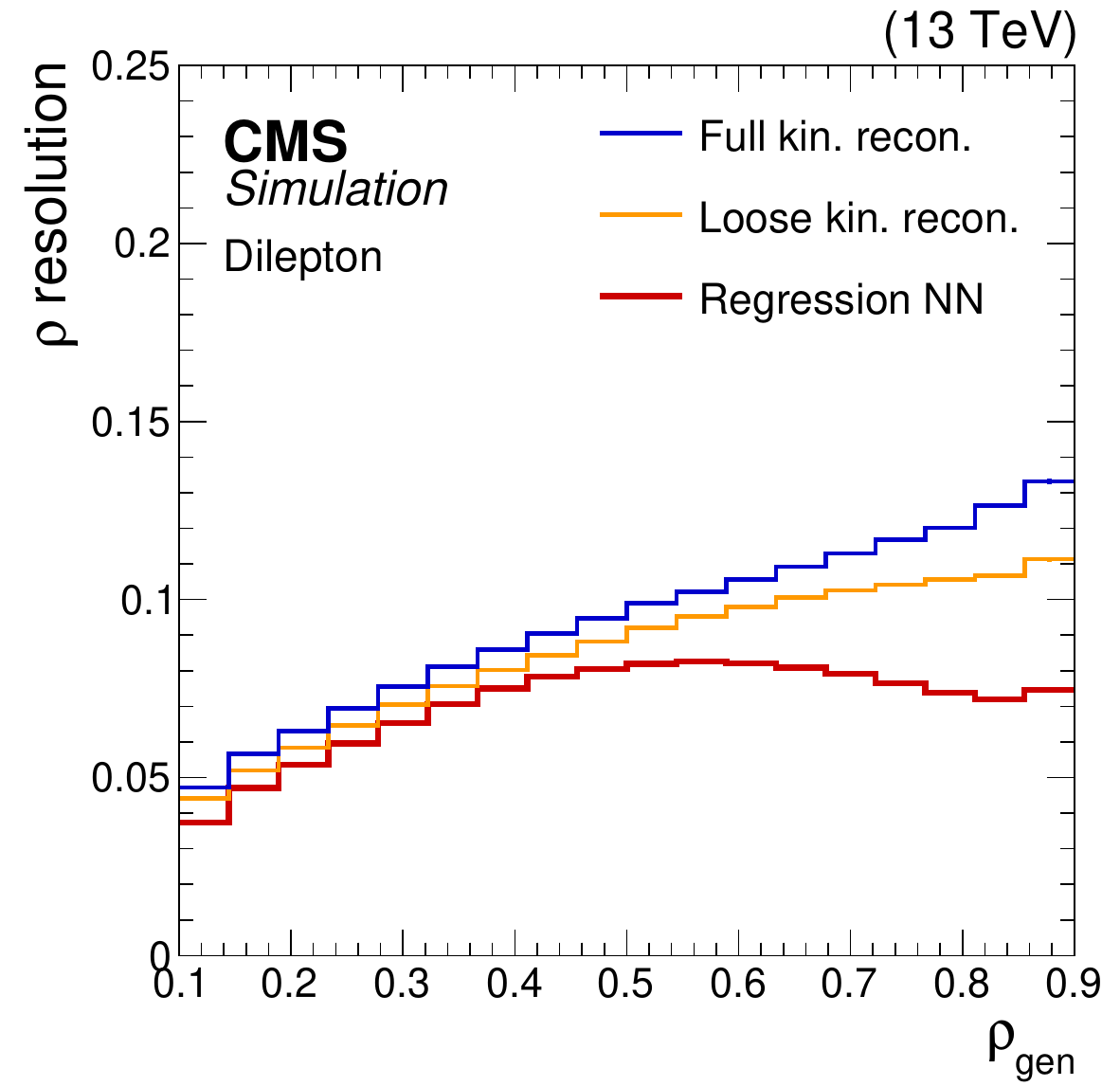}
\caption{%
    The correlation between \rhotrue and \rhoreco is shown for the regression NN reconstruction method (left). The \rhoreco resolution, defined in the text, as a function of \rhotrue (right) for the full (blue line) and loose (orange line) kinematic reconstructions and the regression NN (red line) methods. The number of events per bin in the left plot is shown by the colour scale.
    Figure taken from Ref.~\cite{CMS:2022emx}.
}
\label{fig:ttj_rho_performance_comp}
\end{figure}

The performance of the NN regression is shown in Fig.~\ref{fig:ttj_rho_performance_comp}. The left plot shows the correlation between the parton-level value (\rhotrue) and the reconstructed value (\rhoreco). The correlation coefficient for the regression is 0.87, compared to 0.78 (0.84) for the loose (full) kinematic reconstruction.

The resolution of the regression NN is compared to that of the FKR and LKR  in Fig.~\ref{fig:ttj_rho_performance_comp} (right).
The resolution is defined as the RMS of the difference between the true value \rhotrue at parton level and the reconstructed value \rhoreco of the regression NN in a given \rhotrue bin, divided by $1+\langle\rhotrue-\rhoreco\rangle$ to account for the bias in the reconstruction and to evaluate the response corrected resolution.
The advantage of the multivariate method is the final resolution ranging between 0.05 and 0.08 in the full spectrum, which is an improvement by as much as a factor of two with respect to earlier approaches. The most significant improvement is achieved for the values of \rhotrue close to unity. Since this kinematic regime corresponds to small values for the invariant mass of the \ttbarjet system, it is the most sensitive region for the \mt measurement.
An additional advantage is the 10--15\% higher reconstruction efficiency since the described method is 100\% efficient.

\subsection{Monte Carlo simulations and modelling uncertainties}
\label{sec:mcsetup}

\cmsParagraph{Physics generator configurations for top quark mass measurements}
Proton-proton collisions are modelled and studied using MC event generators, which split the prediction into several steps, each tackled with different techniques, depending on the typical energies involved: the hard scattering, computed with a pure perturbative approach; the parton shower (PS), evolving the partons emerging from the hard scattering down to energies where the perturbative approach is no longer viable; the hadronisation, which is based on phenomenological models; UE, and the decays of unstable hadrons.
The calculation of the cross section is factorised as a convolution of a perturbatively calculable parton scattering process with the structure of the colliding protons, described by the parton distribution functions (PDFs), at a certain factorisation scale \muf, commonly chosen as a typical energy scale of the process $Q^2$.
The proton PDFs are functions of the fraction $x$ of the proton momentum carried by the parton participating in the interaction, and the scale $Q^2$.
While the $x$ dependence of the PDFs cannot yet be calculated from first principles and has to be determined empirically by using dedicated measurements, their $Q^2$ dependence is encoded in the pQCD DGLAP evolution equations~\cite{Gribov:1972ri,Gribov:1972rt, Lipatov:1974qm,Dokshitzer:1977sg,Altarelli:1977zs}.
The hard scattering is generated by using matrix element (ME) codes such as \MADGRAPH~\cite{Alwall:2011uj}, and initial-state radiation (ISR) and final-state radiation (FSR) are simulated with parton shower algorithms with general-purpose MC codes such as \PYTHIAEight.
The UE is composed of additional interactions of the initial beam hadrons besides the hard scattering, the particles from multiple-parton interactions (MPI), and their radiation. Hadronisation, underlying event, colour reconnection (CR), and MPI can only be calculated nonperturbatively, and require tuning of the involved phenomenological parameters to describe the data reliably. Another nonperturbative ingredient to event generators is given by the parton distribution functions (PDFs) used in the hard partonic ME calculation, the PS simulation, and the MPI model.
Typically, the generated events are processed with the CMS detector simulation based on \GEANTfour~\cite{Agostinelli:2002hh} using the conditions appropriate for each period of data taking.
As a convention among the Tevatron and LHC experiments and the theory community, from the beginning of the LHC running, the reference value for the top quark mass in the MC simulations is set to $\mtmc=172.5\GeV$~\cite{ParticleDataGroup:2007}.

In the LHC \Run1, \ttbar signal samples were generated at LO in QCD with up to three additional partons using the \MADGRAPHFive.1 ME generator~\cite{Alwall:2011uj}.
The top quark decays were treated without spin correlations in the samples produced for the analysis of the 7\TeV dataset. The 8\TeV CMS samples employed \textsc{madspin}~\cite{Artoisenet:2012st} to improve the description of angular correlations between the top quark decay products.
For parton showering, hadronisation, and underlying event simulation, \PYTHIASix.4~\cite{Sjostrand:2006za} was used with the Z2~\cite{CMS:2011qzf} and Z2* tunes~\cite{CMS:2015wcf} at 7 and 8\TeV, respectively. The tune Z2* is a result of retuning a subset of the parameters of the Z2 tune using the automated \Professor tuning package~\cite{Buckley:2009bj}.

The top quark MC samples produced for the analyses of LHC \Run2 data, in particular those used in the analyses of data taken at 13\TeV and collected during the years 2015 and 2016, were generated with the \POWHEGv2~\cite{Frixione:2007nw,Nason:2004rx,Alioli:2010xd,Frixione:2007vw} NLO generator interfaced with \PYTHIAEight.2~\cite{Sjostrand:2014zea} using the CUETP8M2T4 tune~\cite{CMS:2016kle}. This tune included a fit to CMS \ttbarjet data taken at \sqrtseq{8} to obtain an improved description of ISR in \ttbar events.

Later \Run2 samples (so-called ``legacy'' samples, referring to the updated data reconstruction and calibrations) were produced with the CP5 tune~\cite{CMS:2019csb}, which for the first time incorporated fits to data taken at 13\TeV and employed an identical NNLO PDF set and the corresponding value of the strong coupling \alpS at NNLO for both the \POWHEG ME generator and the \PYTHIAEight components, \ie ISR, FSR, and MPI.

In the measurements of the top quark mass, the uncertainties related to simulations need to be considered. Ideally, different MC generators and implied setups should provide an adequate description of the observables of interest. In practice, the default MC setups were validated most extensively in CMS analyses.
The modelling uncertainties are factorised into individual components associated with the aforementioned setups, as summarised in Table~\ref{tab:MC}, and are discussed in more detail in the following.

\begin{table}[!ht]
\centering
\topcaption{%
    Overview of CMS MC setups for \ttbar production used in analyses of \Run1 and \Run2 data, and their associated modelling uncertainties.
    Variations marked with a dagger ($\dagger$) are evaluated via event weights, which mitigates the uncertainty associated with the size of MC samples without the need for additional simulations.
}
\label{tab:MC}
\cmsTable{\renewcommand{\arraystretch}{1.1}\begin{tabular}{lccc}
& \Run1 & Early \Run2 & \Run2 legacy \\
\hline
Default setup & & & \\
ME generator & \MADGRAPHFive & \POWHEGv2 & \POWHEGv2 \\
& \ttbar + $\leq$3 jets @ LO & \ttbar @ NLO & \ttbar @ NLO \\
PDF & CT10 NLO & \NNPDF{3.0} NLO & \NNPDF{3.1} NNLO \\
PS/UE generator & \PYTHIASix.4 & \PYTHIAEight.2 & \PYTHIAEight.2 \\
PS/UE tune   & Z2(*) & CUETP8M2T4 & CP5 \\ [\cmsTabSkip]
Uncertainties \\
PDF & CT10 eigenvectors, & NNPDF replicas $\dagger$ & NNPDF eigenvectors, \\
& MSTW08, \NNPDF{2.3} $\dagger$ & & CT14, MMHT14 $\dagger$ \\
ME scales & \murmuf up/down & \murmuf 7-point $\dagger$ & \murmuf 7-point $\dagger$ \\
ME-PS matching & threshold up/down & \hdamp up/down & \hdamp up/down \\
Alternative ME & \POWHEGv1 & \MGvATNLO & \MGvATNLO \\
Alternative PS & \multicolumn{3}{c}{\PYTHIASix.4 / \HERWIGpp2.3 jet flavour uncertainty} \\
Top quark \pt & ratio to 7/8\TeV data & ratio to 13\TeV data & ratio to 13\TeV data \\
ISR & \murISR up/down & \murISR up/down & \murISR up/down $\dagger$ \\
& (correlated with ME) & & \\
FSR & \NA  & \murFSR up/down & \murFSR up/down $\dagger$ \\
UE   & P11, P11 mpiHi/TeV & CUETP8M2T4 up/down & CP5 up/down \\
CR & P11, P11noCR     & ERD on/off, CR1 (ERD on), & ERD on/off, \\
&                    &  CR2 (ERD off) & CR1, CR2 (both ERD off) \\
\PQb fragmentation    & \rbfrag up/down $\dagger$  & \rbfrag up/down, & \rbfrag up/down, un/tuned, \\
&  & Peterson $\dagger$ & Peterson $\dagger$ \\
\end{tabular}}
\end{table}

\cmsParagraph{PDF uncertainties}
PDF uncertainties are evaluated through reweighting, without the need of generating additional MC samples.
The \MADGRAPHFive LO samples used in analyses of \Run1 data were reweighted a posteriori using LHAPDF5.6~\cite{Giele:2002hx,Whalley:2005nh,Bourilkov:2006cj} following the formula
\begin{equation}
    w^{\text{new}}=\frac{f_1^{\,\text{new}}(x_{1};Q^2)\,f_2^{\,\text{new}}(x_2;Q^2)}{f_1^{\,\text{ref}}(x_1;Q^2)\,f_2^{\,\text{ref}}(x_2;Q^2)}.
\end{equation}
Here, $f_i$ refers to the distribution of the interacting parton $i$ in each of the two colliding protons and is a function of the fraction $x_i$ of the proton momentum carried by that parton, and of the factorisation scale denoted here as $Q$.
The PDF uncertainty was evaluated as an envelope of the individual uncertainties encoded in Hessian CT10 NLO~\cite{Lai:2010vv} and MSWT2008~\cite{Martin:2009iq} eigenvectors, and in \NNPDF{2.3} NLO~\cite{Ball:2012cx} replicas.

{\tolerance=800
Since \Run2, PDF weights are calculated directly during the \POWHEGv2 NLO event generation and stored in the event.
In particular, in early \Run2 analyses, the PDF uncertainty was evaluated using replicas of the \NNPDF{3.0} NLO PDF set~\cite{NNPDF:2014otw}.
The \Run2 legacy setup includes the Hessian eigenvectors of \NNPDF{3.1} NNLO by default, and, alternatively, of CT14 NNLO~\cite{Dulat:2015mca} and MMHT2014 NNLO~\cite{Harland-Lang:2014zoa}.
\par}

\cmsParagraph{Matrix element scales}
For the \Run1 \MADGRAPHFive predictions, additional samples were generated varying the renormalisation (\mur) and factorisation (\muf) scales in the matrix element by factors of 1/2 and 2, in parallel with the ISR renormalisation scale prefactor and the FSR \LQCD (outside resonance decays) in \PYTHIASix.
The \POWHEGv2 samples in \Run2 include weights for variations of \mur and \muf that allow for independent, simultaneous, or full 7-point scale variations, avoiding the cases in which $\mur/\muf=1/4$ or 4, following Ref.~\cite{Cacciari:2003fi}.

{\tolerance=800
\cmsParagraph{Parton shower matching}
The \Run1 samples were generated with MLM matching~\cite{Alwall:2007fs} to interface the \MADGRAPHFive matrix elements with the \PYTHIASix PS.
The matching threshold was varied from a default of 40\GeV to 30 and 60\GeV, respectively.
For the early \Run2 \POWHEGPYTHIAEight samples, the \POWHEG \hdamp parameter, regulating the high-\pt radiation, and the value of \asISR were tuned to CMS \ttbarjets data in the dilepton channel at 8\TeV~\cite{CMS:2016kle,CMS:2015ilk}, yielding
$\hdamp=1.58\,_{-0.59}^{+0.66}\,\mt$ and $\asISR=0.111\,_{-0.014}^{+0.014}$.
For the \Run2 legacy samples, \asISR was fixed to 0.118 and only the damping parameter was retuned to $\hdamp=1.38\,_{-0.51}^{+0.93}\,\mt$.
\par}

{\tolerance=800
\cmsParagraph{Initial-state radiation}
In \Run1 simulations, the ISR renormalisation scale in \PYTHIASix was varied simultaneously with the matrix-element scales in dedicated samples by factors of 1/2 and 2.
For the early \Run2 analyses, additional samples were produced with the ISR scale in \PYTHIAEight varied by the same factors, to approximate the \asISR variations found in the tuning to \ttbar data.
For production of \Run2 legacy samples and later, ISR scale variations are included as weights~\cite{Mrenna:2016sih}, providing reduced (factor $\fmur=\sqrt{2}$ and $1/\sqrt{2}$), default ($\fmur=2$ and 1/2), and conservative ($\fmur=4$ and 1/4) variations.
In addition, $\fmur=2$ (and 1/2) and nonsingular term variations~\cite{Mrenna:2016sih} are available for each ISR splitting $\Pg\to\GG$, $\Pg\to\qqbar$, $\PQq\to\PQq\Pg$, and $\PQb\to\PQb\Pg$ separately.
The nonsingular terms are ambiguous terms that appear away from the soft collinear singular infrared limits. These terms are sensitive to missing higher-order ME corrections, the effect of which could be ameliorated by NLO scale compensation terms, as discussed in Ref.~\cite{Mrenna:2016sih}.
\par}

\cmsParagraph{Final-state radiation}
Both \PYTHIASix and \PYTHIAEight include NLO matrix-element corrections for the top quark and \PW boson decays so that the leading gluon emission has LO precision.
There was no variation for FSR from the top quark and W boson decay products in the \Run1 samples.
For early \Run2, additional samples were produced with the FSR scale in \PYTHIAEight varied by factors of 1/2 and 2.
The \Run2 legacy samples include weights providing reduced (factor $\fmur=\sqrt{2}$), default ($\fmur=2$), and conservative ($\fmur=4$) variations for FSR.
As for ISR, $\fmur=2$ and nonsingular term variations are available for each FSR splitting $\Pg\to\GG$, $\Pg\to\qqbar$, $\PQq\to\PQq\Pg$, and $\PQb\to\PQb\Pg$ separately.
In particular, this allows for a decorrelation of radiation from the quarks within the \PW boson decay (which typically is constrained by the reconstructed \PW boson mass) and the radiation from \PQb quarks in the top quark decay.

\begin{figure}[!ht]
\centering
\includegraphics[width=0.48\textwidth]{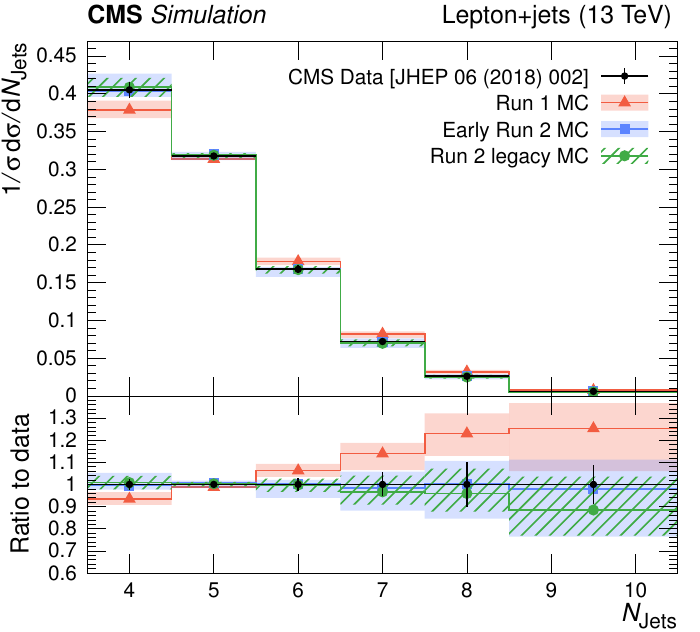}%
\hfill%
\includegraphics[width=0.48\textwidth]{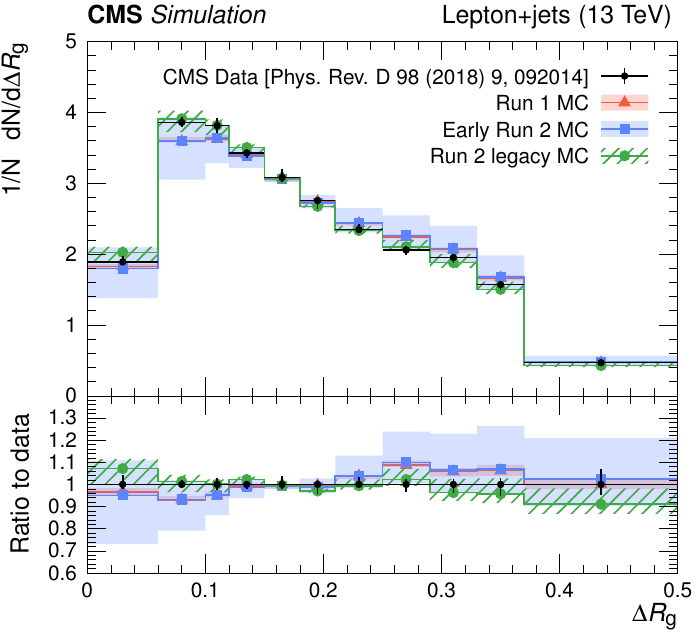}
\caption{%
    Distributions of the jet multiplicity \njets~\cite{CMS:2018tdx} (left) and the jet substructure observable \DRg, the angle between the groomed subjets, normalised to the number of jets~\cite{CMS:2018ypj} (right) in \ttbar events at 13\TeV (black symbols).
    The data are compared to the MC simulation setups used in \Run1, early \Run2, and \Run2 legacy analyses, presented by bands of different style and colour.
    The uncertainty bands include ME scale, ME-PS matching, ISR, and FSR uncertainties.
}
\label{fig:Run1Run2Rad}
\end{figure}

Figure~\ref{fig:Run1Run2Rad} shows the evolution of central prediction and radiation uncertainties from \Run1 to \Run2 compared to measurements at 13\TeV.
The jet multiplicity \njets~\cite{CMS:2018tdx} is sensitive to ME scale, ME-PS matching, and ISR uncertainties, while the angle between groomed subjets \DRg~\cite{CMS:2018ypj} strongly depends on the FSR and its uncertainties.
The FSR uncertainty in the \Run2 legacy sample is significantly reduced due to an NLO scale compensation term~\cite{Mrenna:2016sih}.

{\tolerance=800
\cmsParagraph{Alternative ME generators}
Alternative MC samples were generated using the \POWHEGv1 NLO generator in \Run1, and their difference was included as a systematic uncertainty.
In \Run2, alternative samples have been produced with \MGvATNLO and FxFx merging~\cite{Frederix:2012ps}, including up to three additional partons at NLO.
As these samples were missing matrix-element corrections to the top quark decays, they were not suitable for the top quark mass measurements and were not included in systematic uncertainty estimation.
Note that neither a \MGvATNLO sample with decay ME corrections~\cite{Frixione:2023hwz} nor a \POWHEG \texttt{bb4l}~\cite{Jezo:2016ujg,Jezo:2023rht} sample was available in a form to be considered for the measurements presented in this review.}
\par

\cmsParagraph{Alternative parton shower models}
The comparison of different PS algorithms is an important validation of the factorised uncertainty model.
Usually, this was done by comparing samples generated with \PYTHIA6/8 or \HERWIG6/++/7, using the generator versions available at the time.
However, it should be noted that these general-purpose shower MC generators also differ in ME-PS matching, UE, CR, fragmentation functions, and accuracy of decay ME corrections, such that a comparison between them contains ambiguities that can result in underestimating or overestimating the impact on a given \mt-sensitive observable.
In particular, \HERWIGpp2.7 tune EE5C shows an excess of additional jets in the direction of the top quark with respect to the CMS 13\TeV data~\cite{CMS:2018htd} that was deemed concerning for measurements of \mt.
Therefore, the difference in jet energy response between the two generators is isolated and taken into account as a systematic uncertainty.
This has been determined by comparing the jet energy response in \PYTHIASix.4 tune Z2* and \HERWIGpp~\cite{Bahr:2008pv} tune EE3C~\cite{Gieseke:2012ft} with regard to the quark-dominated \Zjet reference sample~\cite{CMS:2016lmd}.
For the early \Run1 measurements at 7\TeV,
the full impact on a QCD sample with events containing mostly gluon-initiated jets was propagated as an uncertainty into the \PQb jet energy response.
Since the \Run1 measurements at 8\TeV, this uncertainty is evaluated for each jet flavour separately but applied to the measurement of \mt in a correlated way, mimicking the effect of switching the simulation used for jet energy calibration and analysis from \PYTHIA to \HERWIG.
The latest versions of modern shower MC generators such as \PYTHIAEight and \HERWIG7~\cite{Bellm:2015jjp} also provide alternative PS algorithms, which should enable an improved comparison of PS algorithms in the near future.

\begin{figure}[!ht]
\centering
\includegraphics[width=0.32\textwidth]{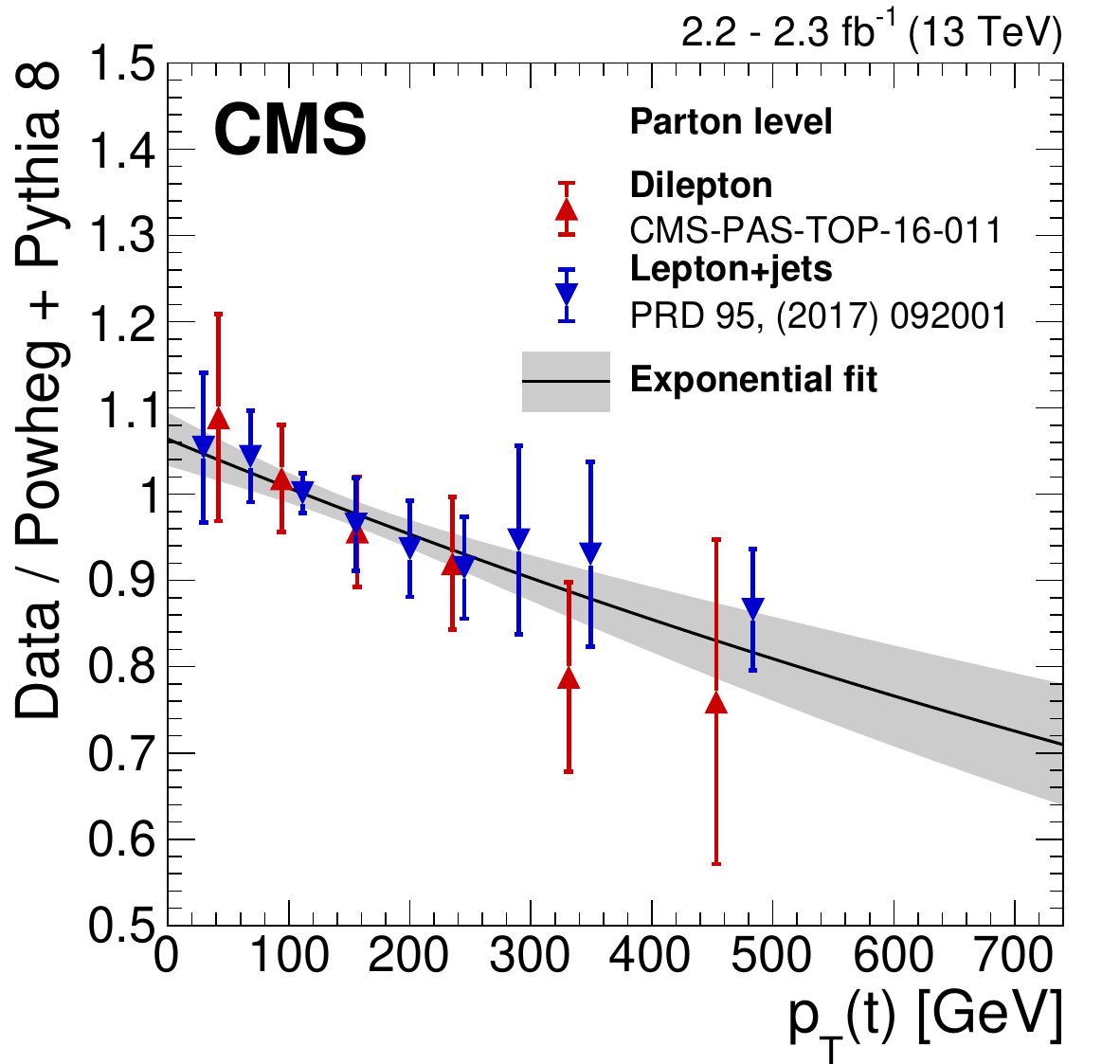}%
\hfill%
\includegraphics[width=0.32\textwidth]{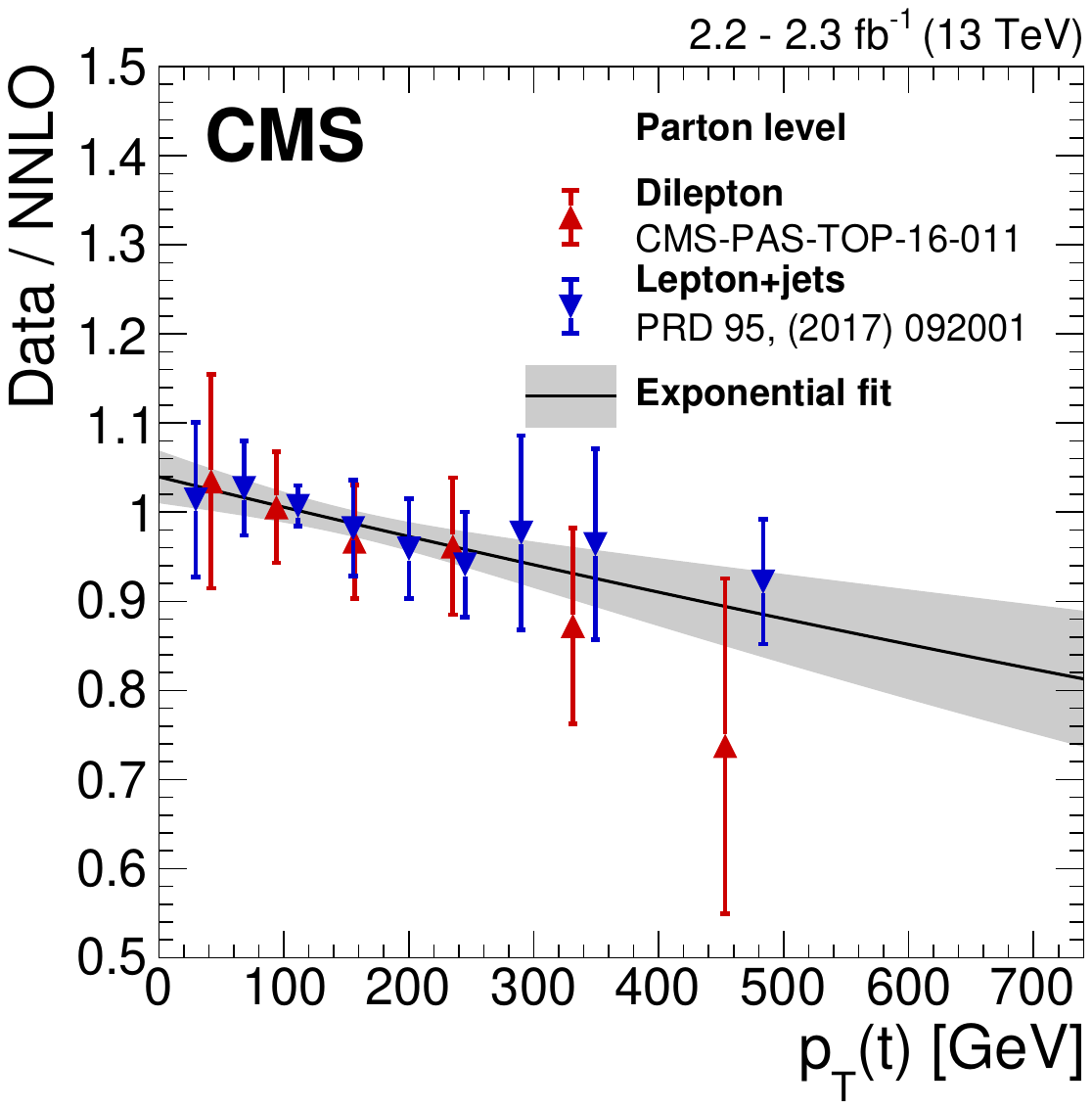}%
\hfill%
\includegraphics[width=0.34\textwidth]{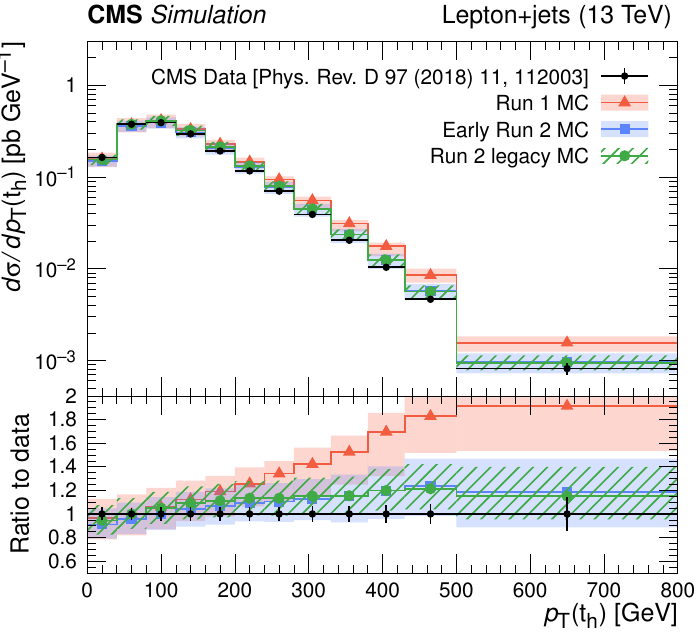}
\caption{%
    Left: Ratio of data to \POWHEGPYTHIAEight (early \Run2) predictions for top quark \pt at the parton level in the dilepton (red symbols) and lepton+jets (blue symbols) channels along with an exponential fit (solid line).
    Centre: same for the ratio of data to the NNLO QCD+NLO EW prediction~\cite{Czakon:2017wor}.
    Right: Distribution of the transverse momentum of the hadronically decaying top quark as measured by CMS~\cite{CMS:2018htd} at the particle level (black symbols) compared to MC simulations for the generator setups used in \Run1, early \Run2, and \Run2 legacy  analyses, presented by bands of different styles. The uncertainty bands include ME scale, ME-PS matching, ISR, and FSR uncertainties.
}
\label{fig:Run1Run2TopPt}
\end{figure}

\cmsParagraph{Top quark \pt}
In the context of \Run1 analyses, it was observed that the \pt spectra of top quarks in data are considerably softer than predicted by the then available NLO MC generators.
While the central MC prediction was not altered, an additional uncertainty was introduced to cover this difference, derived from the ratio of data to NLO MC prediction.
In Fig.~\ref{fig:Run1Run2TopPt}, this ratio is shown for 2015 data and \POWHEGPYTHIAEight simulation used in early \Run2, in dilepton
and lepton+jets~\cite{CMS:2016oae} events.
To evaluate the systematic uncertainty, the fitted exponential function $\exp (0.0615-0.0005\pt)$ is applied to \pt of each top quark at the parton level.
The NNLO QCD + NLO EW prediction~\cite{Czakon:2017wor} agrees better with the measured data, as exemplified by the flatter fitted ratio of $\exp (0.0389-0.0003\pt)$.
Also shown is the top quark \pt measurement at the particle level using 13\TeV data recorded in 2016~\cite{CMS:2018htd}, compared to the predictions of the generator setups used in \Run1, early \Run2, and \Run2 legacy samples with ME scale, ME-PS matching, and ISR/FSR uncertainties.
The \Run2 \POWHEG simulation shows an improved agreement with the data.

{\tolerance=800
\cmsParagraph{Underlying event}
For the simulations used in CMS \Run1 measurements, the \PYTHIASix Z2 tune~\cite{CMS:2011qzf} was employed.
This tune is obtained by fitting 900\GeV and 7\TeV CMS UE data and is based on the CTEQ6L PDF set and uses \pt-ordered showers.
The variations for the Z2 tune have not been provided, therefore corresponding UE uncertainties are estimated by comparing the Perugia 2011 (P11) tune to the P11 mpiHi, and P11 Tevatron tunes~\cite{Skands:2010ak}. The Perugia Tevatron tunes family is derived using hadronic \PZ boson decays at LEP, Tevatron minimum bias (MB) data taken at \sqrtseq{0.63}, Tevatron MB and Drell--Yan data at 1.8\TeV and 1.96\TeV, and SPS MB data taken at 0.2, 0.546, and 0.9\TeV. As in the Z2 tune, it is based on \pt-ordered showers. The Perugia tunes and their corresponding variations were updated in 2011~\cite{Skands:2010ak} to use the same value of \LQCD for both ISR and FSR in the shower and to take into account the early 0.9 and 7\TeV LHC MB and UE data. With this update, a variant, called P11 mpiHi, with MPI that also uses the same \LQCD used for ISR and FSR is also provided.
\par}

In the \mt analyses in \Run2 the differences between the nominal tunes and their corresponding variations, obtained by their eigentunes, are considered as the UE uncertainty.
In early \Run2 top quark analyses, the simulations employ the CUETP8M2T4 tune~\cite{CMS:2016kle}, which is derived using $\asISR(\mZ)$ constrained by the \ttbar kinematic properties of the jet (also using the ISR rapidity ordering~\cite{CMS:2019csb} to cure the overestimation of high jet multiplicities).
In legacy \Run2 analyses, the \PYTHIAEight UE tune CP5~\cite{CMS:2019csb} is used. This tune is based on an NNLO version of the \NNPDF{3.1} set (NNPDF31\_nnlo\_as\_0118)~\cite{NNPDF:2017mvq}, and the strong coupling evolution at NLO. The CP5 consistently uses the same value of $\asmz=0.118$  in various components of the parton shower: initial and final state radiation, and MPI. The tune uses the MPI-based CR model.
The CMS UE tunes are detailed in Table~\ref{tab:MC}.

\begin{figure}[!tp]
\centering
\includegraphics[width=0.48\textwidth]{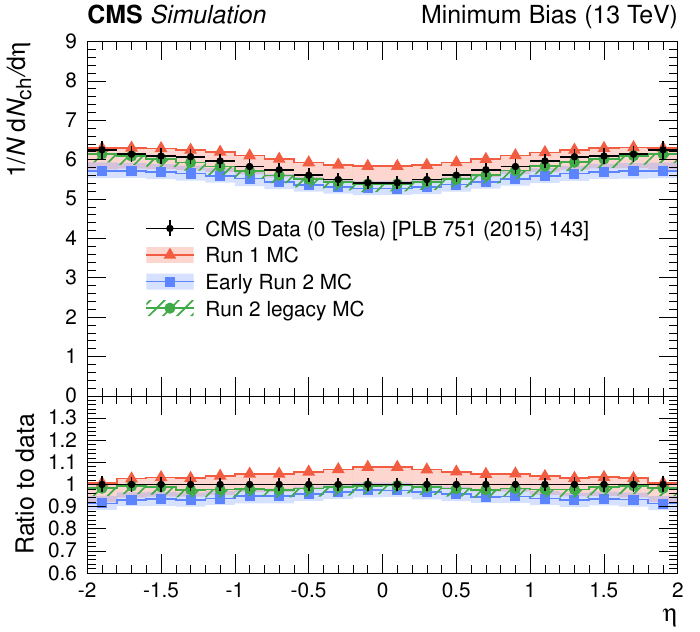}%
\hfill%
\includegraphics[width=0.48\textwidth]{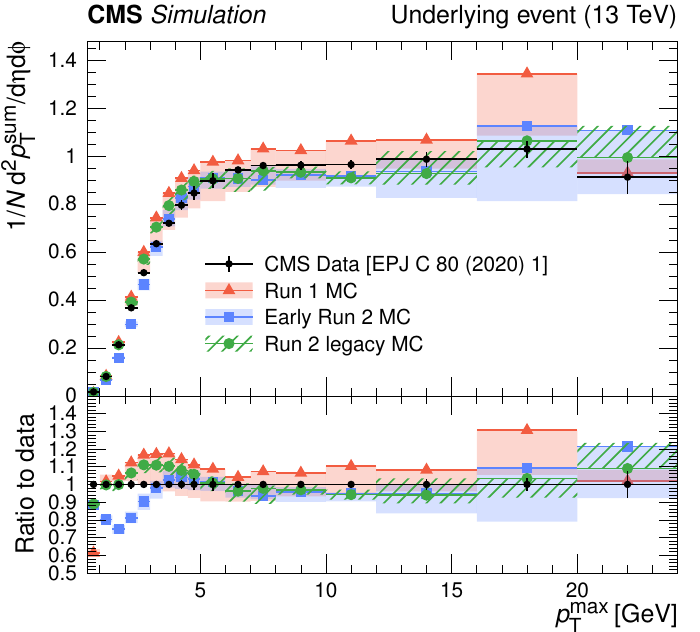}
\caption{%
    Left: The pseudorapidity density of charged hadrons, \dNcheta, using data from about 170\,000 MB events from inelastic \pp collisions using both hit pairs and reconstructed tracks by the CMS experiment~\cite{CMS:2015zrm} at \sqrtseq{13}. Right: The charged-particle \pTsum density in the azimuthal region transverse to the direction of the leading charged particle as a function of the \pt of the leading charged particle, $\pt^{\text{max}}$, measured by the CMS experiment~\cite{CMS:2015zev} at \sqrtseq{13}. The predictions of the CMS UE tunes from \Run1 to \Run2 legacy evaluated at 13\TeV are compared with data. The coloured bands represent the variations of the tunes, and error bars on the data points represent the total experimental uncertainty in the data including the model uncertainty. Both distributions are normalised to the total number of events.
}
\label{extraction:UEtunes_mb_transmin}
\end{figure}

In Fig.~\ref{extraction:UEtunes_mb_transmin}, a minimum bias observable is displayed, the pseudorapidity density of charged hadrons (\dNcheta) from inelastic \pp collisions, within $\abseta=2$ using both hit pairs and reconstructed tracks by the CMS experiment at \sqrtseq{13}~\cite{CMS:2015zrm} operated at zero magnetic field (left diagram). Also an UE observable is shown, the density of the scalar sum of \pt of charged particles (\pTsum density)  in the azimuthal region transverse to the direction of the leading charged particle as a function of the \pt of the same particle, $\pt^{\text{max}}$, measured by the CMS experiment at \sqrtseq{13}~\cite{CMS:2015zev} compared with different UE predictions simulated by \PYTHIAEight. The leading charged particle is required to be produced in the central region $\abseta<2$ with transverse momentum $\pt>0.5\GeV$. The coloured band in these plots represents the variations of the tunes.
For the \Run1 predictions, uncertainties are estimated from the envelope of the three tunes Z2*, P11, and P11 mpiHi, since Z2* eigentune variations were not available. This causes the one-sided variation in the \Run1 sample in the left diagram of Fig.~\ref{extraction:UEtunes_mb_transmin}. For the early \Run2 and \Run2 legacy predictions, the uncertainties are estimated from the eigentune variations provided by the \Professor tuning package.
For practical purposes, the eigentune variations are condensed in two effective variations: ``up'' and ``down''.
The ``up'' (``down'') variation is calculated using the positive (negative) differences in each bin between each eigentune and the central prediction of the nominal tune for the distributions used in the tuning procedure, added in quadrature.
The resulting ``up'' and ``down'' variations are fit using the same fitting procedure that is used to obtain the nominal tune to obtain parameter sets for ``up'' and ``down'' that can be used in the uncertainty estimation in the nominal tune.

\begin{figure}[!ht]
\centering
\includegraphics[width=0.41\textwidth]{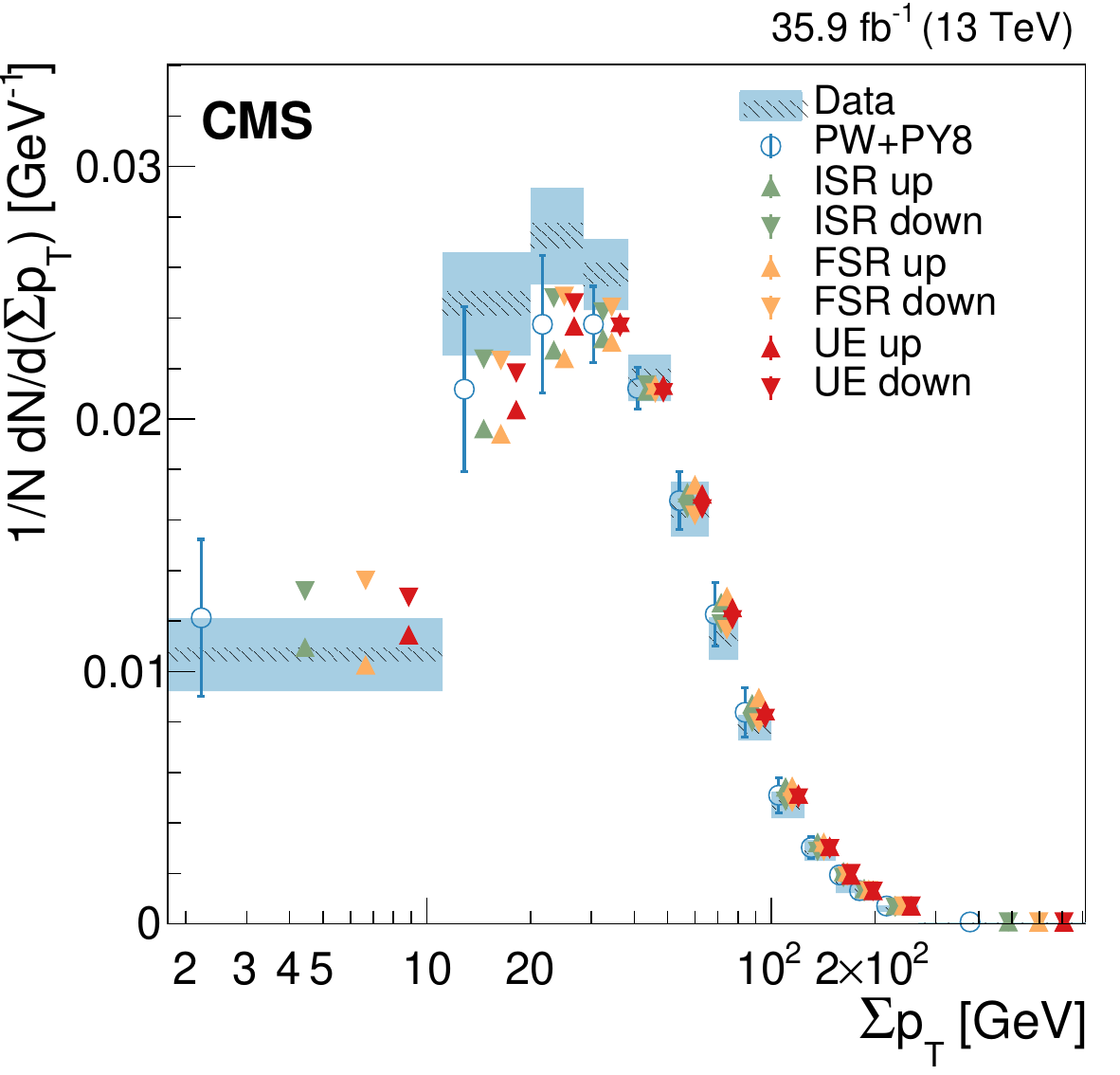}%
\hfill%
\includegraphics[width=0.58\textwidth]{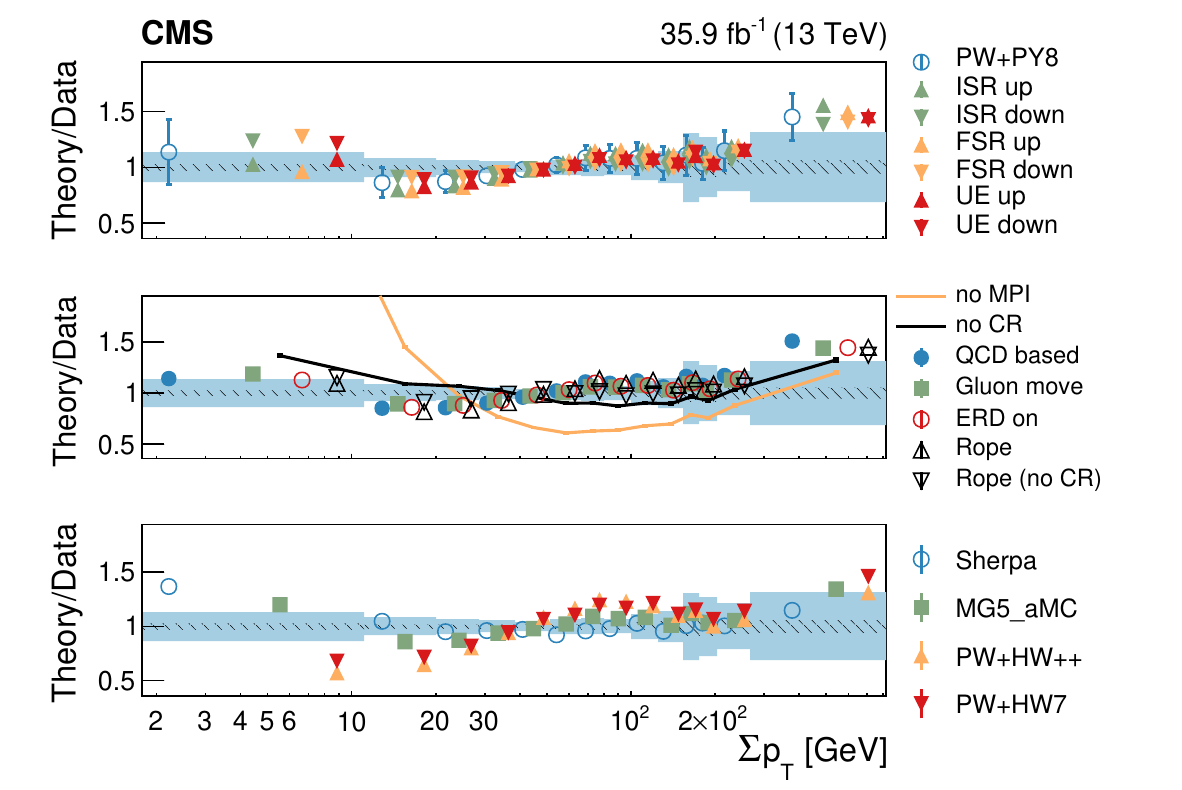}
\caption{%
    Left: Normalised differential cross section as a function of \sumpT of charged particles in the UE in \ttbar events, compared to the predictions of different models. The data (coloured boxes) are compared to the nominal \POWHEGPYTHIAEight predictions and to the expectations obtained from varied $\asISR(\mZ)$ or $\asFSR(\mZ)$ \POWHEGPYTHIAEight setups (markers).
    In the case of the \POWHEGPYTHIAEight setup, the error bar represents the envelope obtained by varying the main parameters of the CEUP8M2T4 tune, according to their uncertainties. This envelope includes the variation of the CR model, $\asISR(\mZ)$, $\asFSR(\mZ)$, the \hdamp parameter, and the \mur/\muf scales at the ME level.
    Right: The different panels show the ratio between each model tested and the data. The shaded (hatched) band represents the total (statistical) uncertainty of the data, while the error bars represent either the total uncertainty of the \POWHEGPYTHIAEight setup, or the statistical uncertainty of the other MC simulation setups.
    Figures taken from Ref.~\cite{CMS:2018mdd}.
}
\label{extraction:UEinttbar}
\end{figure}

The underlying event, together with CR, has been one of the dominant systematic uncertainties for the most precise CMS top quark measurements.
Therefore, more dedicated studies have been performed.
UE activity in \ttbar dilepton events is measured, for the first time, by CMS at \sqrtseq{13}~\cite{CMS:2018mdd}.
This is achieved by removing charged particles associated with the decay products of the \ttbar event candidates as well as with removing pileup interactions for each event.
Normalised differential cross sections in bins of the multiplicity and kinematic variables of charged-particle tracks from the UE in \ttbar events are studied.
The observables and categories chosen for the measurements enhance the sensitivity to \ttbar modelling, MPI, CR, and \asmz in \PYTHIAEight.
The normalised differential cross section measured as a function of \sumpT in the UE of \ttbar-dilepton events is shown in Fig.~\ref{extraction:UEinttbar} (left).
The distribution is obtained after unfolding the background-subtracted data and normalising the result to unity. The ratio between different predictions and the data is shown in Fig.~\ref{extraction:UEinttbar} (right).
The comparisons indicate a fair agreement between the data and \POWHEG~\cite{Nason:2004rx,Frixione:2007vw,Alioli:2010xd} matched with \PYTHIAEight using the CUETP8M2T4 tune, but disfavour the setups in which MPI and CR is switched off or
the default configurations of \POWHEGHERWIGpp with the EE5C UE tune~\cite{Seymour:2013qka} and the
CTEQ6 (L1)~\cite{Pumplin:2002vw} PDF set, \POWHEGHERWIGSeven~\cite{Bahr:2008pv,Bellm:2015jjp} with its default tune and the MMHT2014 (LO)~\cite{Harland-Lang:2014zoa} PDF set and \SHERPA2.2.4~\cite{Gleisberg:2008ta} + \textsc{openloops} (v1.3.1)~\cite{Cascioli:2011va} with a PS-based on the Catani--Seymour dipole subtraction scheme~\cite{Schumann:2007mg}.
It has been furthermore verified that, as expected, the choice of the NLO ME generator does not impact significantly the expected characteristics of the UE by comparing predictions from \POWHEG and \MGvATNLO, both interfaced with \PYTHIAEight.
The UE measurements in \ttbar events test the hypothesis of universality of UE at an energy scale of two times \mt, considerably higher than the ones at which UE models have been studied in detail.
The results also show that a value of $\asmz^{\text{FSR}}=0.120 \pm 0.006$
is consistent with these data.

\cmsParagraph{Colour reconnection}
In the limit of large number of colours $N_{\text{c}}$, quarks and gluons are assigned unique colour charges during the parton shower stage, and Lund string hadronisation describes the formation of hadrons from the colour string formed between each colour and anti-colour pair. Colour reconnection (CR) is a reconfiguration of the colour assignments, finding states with lower potential energy and allowing interactions between the partons from the hard collision and the UE, independent of their history of production.
The CR uncertainty in the \Run1 (2009--2013) analyses at $\sqrts=7$ and 8\TeV was calculated comparing two values of \mt, using predictions with the same UE tune with and without CR effects using the P11 tune~\cite{CMS:2015lbj}. However, the data completely disfavours the setups in which CR is switched off (as discussed, \eg in Ref.~\cite{CMS:2018mdd}). Because of this, comparing setups with CR switched on and off may be nonoptimal for uncertainty calculations. Instead, a more realistic estimation of the CR uncertainty may be obtained by comparing different CR models that describe the data.
In order to do this, we compare MPI-based, QCD-inspired, and gluon-move models in \PYTHIAEight for which the details, and further references, can be found in Ref.~\cite{CMS:2022awf}.
In addition, the early resonance decay (ERD)~\cite{Argyropoulos:2014zoa}, which allows top quark decay products to take part in CR, was investigated.
This was first done in Ref.~\cite{CMS:2018quc} for \mt measurements with \ttbar events, and in Refs.~\cite{CMS:2017mpr,CMS:2021jnp} with single top quark events, using the CUETP8M2T4 tune and the QCD-inspired and gluon-move CR models compared to the default CR model.
New sets of tunes for two of the CR models implemented in \PYTHIAEight, QCD-inspired (CR1) and gluon-move (CR2), have been derived by CMS~\cite{CMS:2022awf}.
The new CMS CR tunes are based on \sqrtseq{1.96} CDF, and 7 and
13\TeV CMS data.
They are obtained by changing the CR model in the default CMS CP5 tune and retuning. These new CR tunes are tested against a wide range of measurements from LEP, CDF, and CMS.
The new CMS CR tunes for MB and UE describe the data significantly better than the ones with the default parameters.

\begin{figure}[!b]
\centering
\includegraphics[width=0.48\textwidth]{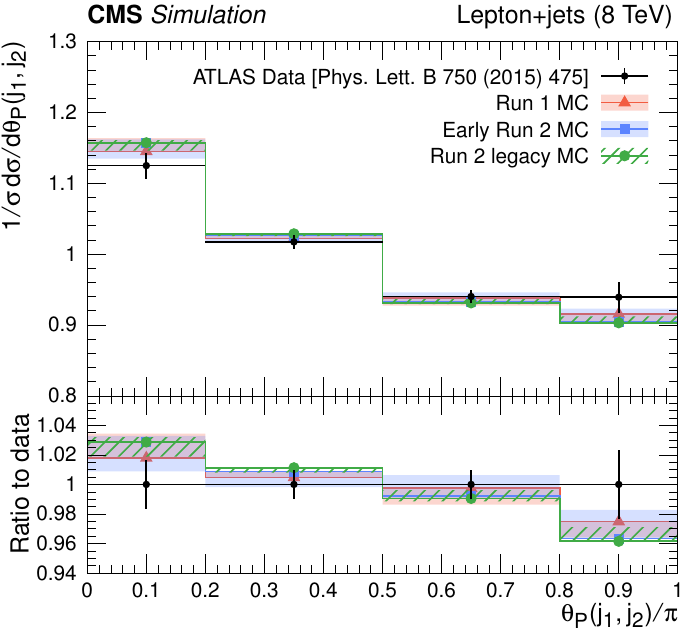}
\caption{%
    Measured distribution of the pull angle in \ttbar events taken at 8\TeV recorded by ATLAS~\cite{ATLAS:2015ytt} (points with vertical error bars) compared to MC simulations for the generator setups used in \Run1, early \Run2, and \Run2 legacy analyses, presented by bands of different styles.
    The uncertainty bands illustrate the uncertainties resulting from colour reconnection effects, as estimated by variations described in the main text. The same variations are applied in CMS top quark mass measurements.
}
\label{fig:Run1Run2CR}
\end{figure}

\begin{figure}[!t]
\centering
\includegraphics[width=0.48\textwidth]{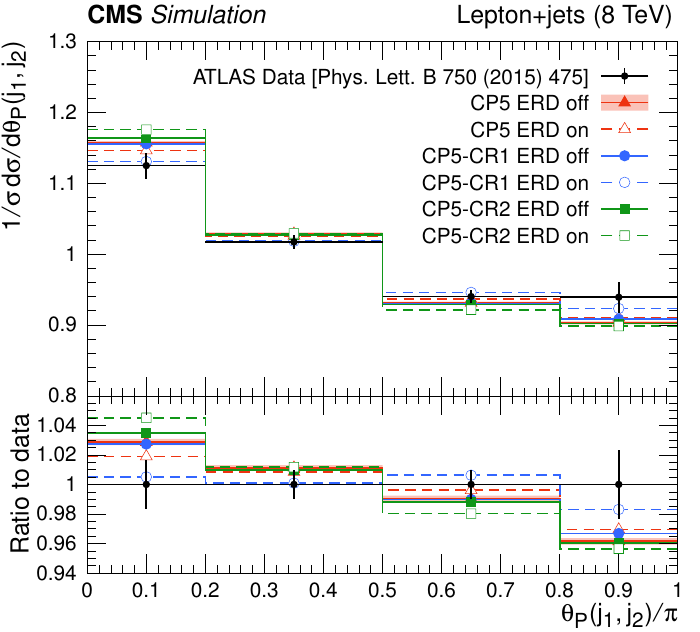}%
\hfill%
\includegraphics[width=0.48\textwidth]{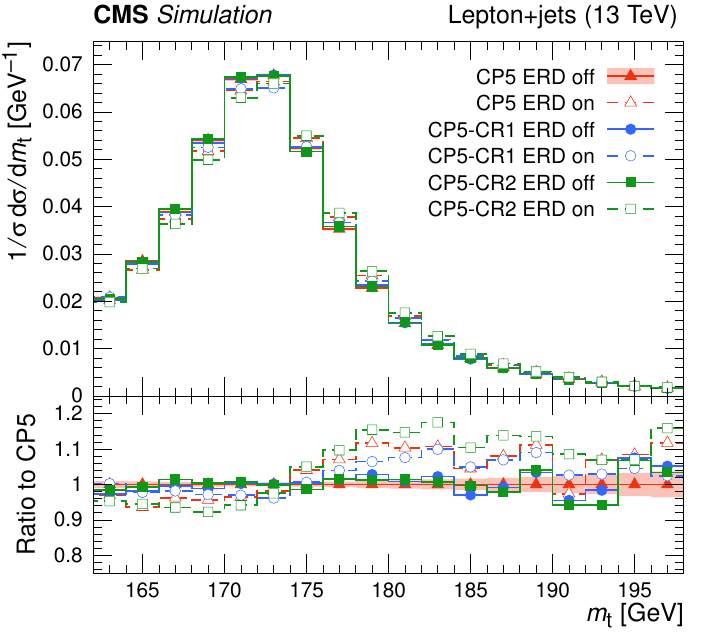}
\caption{%
    Normalised \ttbar differential cross section for the pull angle between jets from the \PW boson in hadronic top quark decays, calculated from the charged constituents of the jets, measured by the ATLAS experiment using \sqrtseq{8} data~\cite{ATLAS:2015ytt} to investigate colour flow (left). The predictions from \POWHEGPYTHIAEight using different tune configurations are compared with data. The statistical uncertainties in the predictions are represented by the coloured band and the vertical bars. The coloured band and error bars on the data points represent the total experimental uncertainty in the data. The invariant mass reconstructed from the hadronically decaying top quark candidates at the generator level (right). The coloured band and the vertical bars represent the statistical uncertainty in the predictions.
    Figures adapted from Ref.~\cite{CMS:2022awf}.
}
\label{extraction:CRtunes}
\end{figure}

Figure~\ref{fig:Run1Run2CR} shows the evolution of colour reconnection uncertainties from \Run1 to \Run2 compared to the ATLAS measurement of the colour flow in \ttbar events at 8\TeV~\cite{ATLAS:2015ytt}. Colour flow is measured using the jet pull angle, $\theta_{\text{p}}(\PQj_1,\PQj_2)/\pi$ where the jets $\PQj_1$ and $\PQj_2$ originate from the \PW boson decays and reconstructed using only charged constituents. Figure~\ref{extraction:CRtunes} (left) displays the colour flow in \ttbar events measured in data,  compared to \POWHEGPYTHIAEight predictions using different tune configurations: CP5, CP5-CR1, CP5-CR2, and these three tunes with the ERD option.
Colour flow exhibits a high degree of sensitivity to the ERD option. Without ERD, \PW boson decay products are not colour reconnected, therefore the predictions of the tunes are closer to each other compared to the tunes with ERD for which CR modifies the angle between the two jets visibly in Fig.~\ref{extraction:CRtunes}.
It can also be observed from this figure that CP5-CR1 (QCD-inspired) tune with ERD provides the best description of colour flow, and CP5-CR2 (gluon-move) tune with ERD  displays the largest deviation from the data.

Figure~\ref{extraction:CRtunes} (right) displays the invariant mass of the hadronically decaying top quark constructed at the particle level, comparing theoretical predictions with different tunes. Although CR is one of the  dominant uncertainties in top quark mass measurements, it is difficult to demonstrate its direct effect on the measurements.  Therefore, here, we show comparisons at the particle level for which the differences are not diluted by detector and reconstruction effects. As for colour flow, the largest deviation from the prediction of the default CP5 tune is by the CP5-CR2 (gluon-move) tune with ERD. The deviation visible here is consistent with what is found in the top quark mass measurement at \sqrtseq{13}~\cite{CMS:2018mdd} using the CUETP8M2T4 tune.

\begin{figure}[!ht]
\centering
\includegraphics[width=0.495\textwidth]{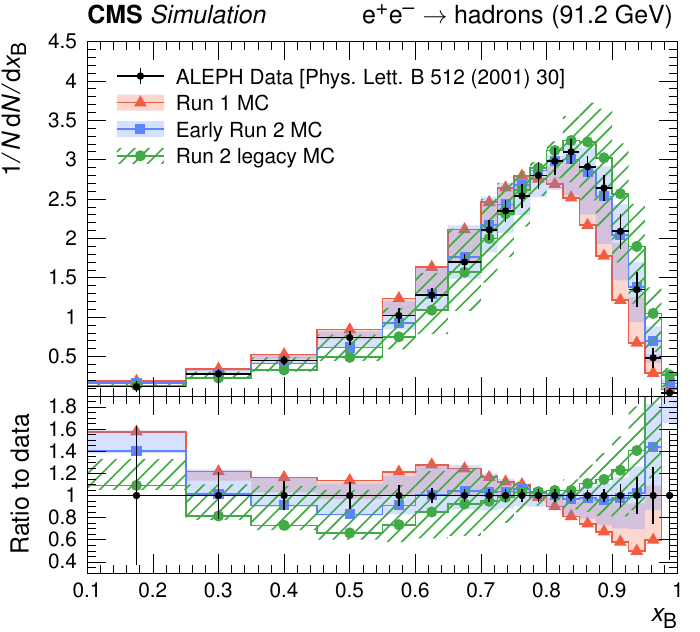}%
\hfill%
\includegraphics[width=0.465\textwidth]{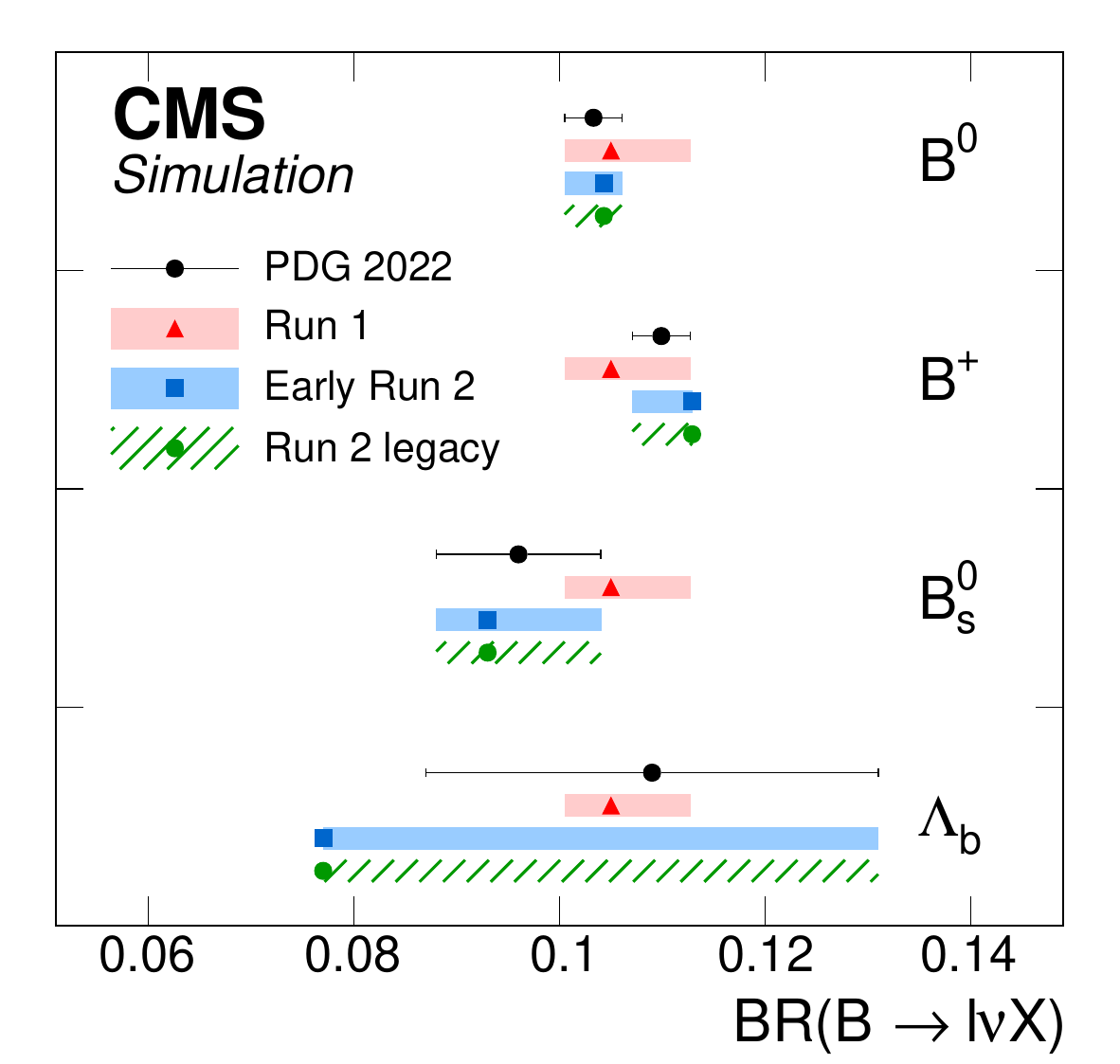}
\caption{%
    Distribution of the \PQb quark fragmentation function normalised to the number of \PQb hadrons measured by ALEPH in \EE collisions at $\sqrts=91.2\GeV$~\cite{Heister:2001jg} (black symbols with vertical error bars showing the total measurement uncertainties) compared to \EE MC simulations for the generator setups used in \Run1, early \Run2, and \Run2 legacy analyses, presented by bands of different styles (left).
    The uncertainty bands are constructed around the default prediction and illustrate the \PQb quark fragmentation uncertainties.
    The measured semileptonic branching ratios of \PQb  hadrons~\cite{ParticleDataGroup:2022pth} (black symbols) compared to the values in the generator setups (coloured symbols) and their uncertainties, illustrated by shaded bands (right).
}
\label{fig:Run1Run2bfrag}
\end{figure}

\cmsParagraph{\PQb quark fragmentation and semileptonic \PQb hadron decays}
In the Bowler--Lund fragmentation function~\cite{Bowler:1981sb} used in \PYTHIA,
\begin{equation}
    f(z)\propto\frac{1}{z^{1+\rbfrag b\mT^2}}(1-z)^a\exp\left(\frac{-b \mT^2}{z}\right),
\end{equation}
the parameter \rbfrag steers the distribution of the momentum fraction $z$ carried by the \PQb quark containing hadron (\PQb hadron), defined as $z=\Ebhad/E_{\text{quark}}$.
The parameter \rbfrag is tuned to the distribution of $x_{\PQb}=\Ebhad/E_{\text{beam}}$ measured in $\PZ\to\bbbar$ events at the LEP and SLC colliders~\cite{Heister:2001jg,DELPHI:2011aa,OPAL:2002plk,SLD:2002poq} as a proxy for $z$.
The parameter \mT is the transverse mass defined by $\mT=\sqrt{\smash[b]{m^2+\pt^2}}$, where $m$ is the mass and \pt is the transverse momentum of the \PQb hadron.
The resulting modelling of the \PQb quark fragmentation is compared to ALEPH data~\cite{Heister:2001jg} in Fig.~\ref{fig:Run1Run2bfrag} (left) and described in more detail in the following paragraphs.

For the \PYTHIASix setup used in \Run1, the default value $\rbfrag=1.0$ leads to \PQb quark fragmentation which appeared too soft, and was subsequently tuned to the $x_{\PQb}$ data provided by the ALEPH and DELPHI experiments.
While the central Z2* prediction was left unchanged, the difference to the tuned $\rbfrag=0.591\,^{+0.216}_{-0.275}$ was taken as the systematic uncertainty, as it was larger than the uncertainties in the retuning.

In early \Run2, the \PYTHIAEight fragmentation function was pre-tuned by the \PYTHIA authors to $\rbfrag=0.855$, and only a minor change in the central value was found by tuning to ALEPH, DELPHI, OPAL, and SLD data: $\rbfrag=0.895\,^{+0.184}_{-0.197}$.
In addition to the uncertainties in \rbfrag, the Peterson fragmentation function~\cite{Peterson:1982ak}
\begin{equation}
    f(z)\propto\frac{1}{z}\left(1-\frac{1}{z}-\frac{\ebfrag}{1-z}\right)^{-2},
\end{equation}
with the tuned $\ebfrag=3.27\,^{+3.98}_{-2.06}\times10^{-3}$,
was considered as an alternative parameterisation of the \PQb quark fragmentation.

The CP5 tune used for \Run2 legacy samples featured a lower value of \alpS for FSR which resulted in the prediction of a harder \PQb quark fragmentation compared to the $x_\PQb$ data when using the default value of $\rbfrag=0.855$.
While the central prediction was again left unchanged, the difference between the default value and the newly tuned $\rbfrag=1.056\,^{+0.193}_{-0.196}$ is considered as an uncertainty in addition to the variations of \rbfrag and of the tuned parameter of Peterson fragmentation ($\ebfrag=6.038\,^{+4.382}_{-2.466}\times10^{-3}$), thus covering the data as well.

\cmsParagraph{Semileptonic \PQb hadron decays}
These constitute a source of unobservable neutrinos inside \PQb jets, lowering the jet response with respect to the original \PQb quark.
For the \Run1 \PYTHIASix samples, a common semileptonic branching fraction was used for multiple \PQb hadron species.
The uncertainty in this was estimated from the envelope of the measured values and uncertainties for charged and neutral \PB mesons (\PBpm and \PBz) reported by the PDG~\cite{ParticleDataGroup:2022pth}, and propagated to all \PQb hadron species.
For \Run2, \PYTHIAEight includes decay tables specific to \PBz, \PBpm, \PBzs, and \PGLb.
These are simultaneously reweighted within their respective PDG uncertainties. By construction, the uncertainty bands become highly asymmetric in cases where the generator value is outside the PDG value with its uncertainty range.
The values and uncertainties used for semileptonic branching fractions are shown in Fig.~\ref{fig:Run1Run2bfrag} (right).

\subsection{Experimental uncertainties}
\label{sec:experimentalunc}

The observables used in top quark mass measurements are sensitive to systematic effects related to the uncertainties in the calibration of the final-state objects used in the physics analyses. These include for example the calibration of the JES and JER, the measurement of the missing transverse momentum in the event, the efficiency in reconstructing and identifying leptons and jets originating from \PQb quarks, the integrated luminosity of the considered data set (mostly relevant in absolute cross section measurements), and the average number of PU interactions. Correction factors are obtained by comparing data with simulation, and are used to correct the relevant quantities in simulated events.

The JES and JER corrections are derived as functions of the jet transverse momentum and pseudorapidity~\cite{CMS:2016lmd}. The measurements are obtained by exploiting momentum balance in dijet, \gammajet, \Zjet, and multijet events, and take into account systematic dependencies related to uniformity of the detector response, the number of pileup interactions, and residual differences between data and simulation.
The absolute JES calibration is determined with the highest precision in \Zjet events at $\pt=200\GeV$, where approximately 20\% of the jets stem from gluons, 70\% from light (\PQu, \PQd, \PQs) quarks, and 10\% from heavy (\PQc and \PQb) quarks.
In order to extrapolate to different flavour compositions, notably pure \PQb jets, the \PYTHIASix and \HERWIGpp parton-shower generators are used with their respective hadronisation models, resulting in additional flavour-dependent jet energy uncertainties.
The energy scale of central-rapidity jets with $\pt>30\GeV$, which are the most relevant in the context of \mt measurements, is measured with a precision better than 1\%, excluding the flavour-dependent components,
while the total uncertainty varies between 1 and 3.5\%, depending on the jet kinematics~\cite{CMS:2016lmd}. The energy resolution of particles that are not clustered in jets is also taken into account in the estimate of the missing transverse momentum in the event~\cite{Khachatryan:2014gga}.

The efficiencies of electron and muon identification algorithms are corrected as functions of the lepton's (\Pell) kinematic quantities, making use of \ZtoLL events. This is commonly achieved by means of the so-called `tag-and-probe' method, where one of the leptons is used to tag the \ZtoLL event, while the other is used as a probe to estimate the efficiency. In order to achieve a pure sample of neutral Drell--Yan events, the invariant mass of the lepton pair is required to be compatible with that of the \PZ boson. The corresponding uncertainties lie in the range 0.5--1.5\% for muons and 2--5\% for electrons~\cite{CMS:2020uim,CMS:2018rym}. The energy scale of the leptons is also calibrated using \ZtoLL events and the corresponding uncertainty is propagated to the analyses. Typical values of the lepton scale uncertainties are 0.1 (0.3)\% for electrons and 0.2 (0.3)\% for muons in the barrel (endcap)~\cite{CMS:2020uim,CMS:2018rym}. Leptons are also reconstructed at the trigger level and are used to pre-select events during data taking~\cite{CMS:2016ngn}. The trigger efficiencies are often estimated by each individual analysis, and are derived as functions of the lepton kinematics making use of an orthogonal data set. The corresponding uncertainty is then propagated to the final result, and is often dominated by the statistical uncertainty of the utilised data set.

To select \PQb jets, three working points are defined based on fixed light-quark jet misidentification probabilities of 10, 1, and 0.1\%. Correction factors for the \PQb tagging efficiencies and light jet misidentification probabilities are derived as functions of the jet kinematic properties and the generator-level flavour of the jet. Different calibration methods make use of independent \PQb jet and light jet enriched regions, \eg in muon-enriched inclusive jet production or \ttbar phase spaces. The resulting corrections have uncertainties of 1--5\% and 5--10\% for \PQb jets and light jets, respectively~\cite{CMS:2017wtu}.

The PU in an event can also affect the calibration of the final-state objects. Simulated PU events are weighted according to Ref.~\cite{CMS:2020ebo} in order to match the PU distribution observed in data.
For the reweighting procedure, PU-sensitive distributions, such as the number of vertices ($N_{\text{vtx}}$) are used to determine an effective value for the inelastic cross section.
The remaining disagreement between data and MC simulation in the PU-sensitive observables is accounted for by an uncertainty, determined by varying the average number of PU interactions.

In the measurements, the uncertainty due to the integrated luminosity is also taken into account.
The expected signal and background yields in simulation are normalised to the measured integrated luminosity and the related uncertainty is accounted for.  For this purpose, the simulated distributions are obtained by varying the yields within the uncertainty in the integrated luminosity, which in \Run1 ranges between 2.2 and 2.6\%~\cite{CMS-PAS-SMP-12-008,CMS-PAS-LUM-13-001}, and in \Run2 ranges between 1.2 and 2.5\%, depending on the year of data taking~\cite{CMS:2021xjt,CMS-PAS-LUM-17-004,CMS-PAS-LUM-18-002}.  The uncertainty in the integrated luminosity is particularly relevant in the context of indirect \mt extraction based on the measurements of the absolute \ttbar cross sections.

\subsection{General aspects of unfolding}
\label{sec:unfolding}

The MC simulations described in Section~\ref{sec:mcsetup} are generally processed through the CMS detector simulation based on \GEANTfour~\cite{Agostinelli:2002hh} so that predicted and observed distributions for observables such as the reconstructed top quark mass can be compared at the reconstructed detector level.
In order to compare to theoretical calculations at the parton or particle level (Section~\ref{sec:particlePartonLevel}), an unfolding procedure has to be applied in order to remove experimental effects from the measured detector-level distributions.
This is the case also for the indirect top quark mass extraction, where \mt is obtained by comparing measured (differential) cross sections to standalone calculations.

Depending on the purpose of the measurement and on the details of the theoretical calculation, the unfolding can be performed to the particle or the parton level, discussed in detail in Section~\ref{sec:particlePartonLevel}. Once the generator level  in the simulation is defined, the unfolding procedure to either particle or parton level is identical. However, unfolding to parton level requires a larger degree of extrapolation from the measured distributions, and often comes at the cost of increased dependence on the modelling uncertainties. On the other hand, unfolding to particle level does not allow for a comparison of the obtained results to fixed-order calculations.  In the following, general aspects of the unfolding problem are discussed, while the details of the unfolding methods are presented in the context of each particular analysis in the following sections. In the following, ``generator level'' refers to both parton and particle levels.

The goal of unfolding is the inference of a distribution corrected for experimental effects, such as resolution, misreconstruction, inefficiencies, and detector acceptance. The problem can be formulated as a maximum likelihood estimate. A generator-level distribution \distgen can be mapped to the corresponding detector-level distribution \distdet using the so-called response matrix $R$ as $\distdet=R\distgen$. The elements of the response matrix $R_{ij}$ represent the probabilities to observe in bin $i$ an event generated in bin $j$. The response matrices are typically obtained by using the simulated events and incorporate all experimental effects.

Assuming a Poisson distribution of the observed yields \distdetpr, the likelihood for the unfolding problem can be written as
\begin{equation}
    L=\prod\limits_i\text{Poisson}\bigg(d^\prime_i,\sum\limits_j R_{ij}g^\prime_j\bigg) .
 \label{eq:likelihood_unfolding}
\end{equation}
The maximum likelihood estimate for the unfolded distribution \distgenpr can then be obtained as  $\distgenpr=R^{-1}\distdetpr$. When detector resolution effects are larger or of comparable size to the desired binning in the unfolded distribution, the unfolding problem can become ill-conditioned. This means that small differences in \distdetpr can lead to large effects on the evaluated \distgenpr. In such cases, the statistical fluctuations in \distdetpr can result in extremely large variances in estimates of \distgenpr. However, in cases where $R$ is sufficiently diagonal, this simple approach is the preferred method, as it provides an unbiased estimate of \distgenpr.

When the approach described above is found to be ill-conditioned, the likelihood function in Eq.~\eqref{eq:likelihood_unfolding} can be extended
by adding to $\chisq=-2\ln L$ a so-called regularisation term, such as~\cite{Schmitt:2012kp,Schmitt:2016orm}
\begin{equation}
    \tau^2(\distgenpr-\distb)^{\mathrm{T}}D^{\mathrm{T}}D(\distgenpr-\distb),
\end{equation}
where the quantity \distb is set to the expected \distgen as estimated in the simulation, and the matrix $D$ is the discrete second-order derivative operator. In this way, the regularisation term penalises solutions whose curvatures deviate from the expectation. The regularisation strength is controlled by the parameter $\tau$, which is then optimised, \eg by minimising the average global
correlation coefficient or using the so-called L-curve scan~\cite{Schmitt:2012kp,Schmitt:2016orm}. While such an approach prefers solutions that do not suffer from large oscillations, the obtained solution can be biased towards the simulation. Analyses making use of this approach therefore perform dedicated tests in order to verify that biases from regularisation are covered by the measurement uncertainties.

The unfolding procedure, especially in the presence of large off-diagonal components in the response matrix, can introduce large statistical correlations among the bins of the unfolded distribution. To take this into account, the statistical uncertainties in \distdetpr and the systematic uncertainties in $R$ are propagated to the final result in order to obtain the full covariance matrix of the measured \distgenpr.
Whenever a \chisq is calculated between unfolded distributions and a theoretical prediction, \eg for a fit extracting  \mt,  the full covariance matrix with all bin-to-bin correlations is utilised.

Several unfolding and regularisation procedures were proposed~\cite{Schmitt:2012kp,Schmitt:2016orm,DAGOSTINI1995487,dagostini2010improved,Hocker:1995kb,TIKHONOV,Brenner:2019lmf,Stanley:2021xce}, which are not reviewed in the scope of this work. Different procedures may lead to differences in the unfolded results, and the most appropriate method is chosen in each analysis based on the nature of the unfolding problem to solve.

\subsection{Particle- and parton-level top quark definitions}
\label{sec:particlePartonLevel}

In the simulations at NLO, a finite width of the top quark is assumed. This is important for accurate modelling of the off-shell top quark production and the interference with background processes. However, in such simulations, the concept of a top quark particle is not precisely defined and is model-dependent. An unambiguous object can be constructed only using the kinematic quantities of the final-state particles without extra assumptions. A particle-level top quark (or pseudo-top quark) can be defined using the final-state objects after hadronisation and is less affected by nonperturbative effects or acceptance corrections. Similar phase space definitions at the particle and detector levels lead to mitigation of the model dependence. More details of particle-level top quark definitions, maximising the correlation of reconstructed quantities with the parton-level definition, are discussed in Ref.~\cite{Collaboration:2267573} as a fundamental aspect of top quark measurements. The algorithms implemented in \rivet routines~\cite{Buckley:2010ar} that describe the measurements at particle level allow for testing the quality of top-quark modelling.
The results reported in Ref.~\cite{Collaboration:2267573} suggest that the choice of a particle-level top quark definition is not universal and should be optimised depending on the production mode, the final state, or the variable and the phase space under study. Below, a typical particle-level definition used in the CMS top quark mass measurements is described.

Pseudo-top quarks are reconstructed from a sample of simulated lepton+jets \ttbar events using a \rivet routine. These events fulfil specific criteria for leptons and jets to define top quarks at the particle level, similar to the ones described in Ref.~\cite{Collaboration:2267573} and summarised in Table~\ref{tab:pseudotop}. Each charged lepton is ``dressed'', i.e. its momentum is corrected to take into account any photons within a cone defined in Table~\ref{tab:pseudotop}, and the invariant mass that it forms with a neutrino is required to be within $75.4<m_{\Pell\PGn}<85.4\GeV$.
In the jet clustering process, hadrons stemming from charm and bottom quark  fragmentation, and regardless of the decay channel \PGt leptons are included, with their momenta scaled to a negligible value, such that  the jet kinematic properties are not modified by the presence of these hadrons, which are therefore referred to as ``ghost'' particles.
A jet can encompass one or more ghost particles, which can be utilised for the purpose of flavour assignment and are included in the list of constituents of the jet.
The events are required to include a minimum of four jets, which are defined in Table~\ref{tab:pseudotop}. Among these jets, at least two must be unequivocally associated to the fragmentation of bottom quarks, while the remaining two jets, \ie light-quark jets, must not stem from the bottom quarks.
A leptonically decaying top quark is reconstructed by combining the (dressed) lepton, the neutrino, and one of the jets originating from a bottom quark in the event. A hadronically decaying top quark candidate is reconstructed by combining the other jet originating from a bottom quark with two remaining jets. Typically, it is required that the difference between the invariant masses of top quark reconstructed in the leptonic leg and the hadronic leg in an event must not exceed 20\GeV. Additionally, the invariant mass of the system of the two light-quark jets should fall within a window of 10\GeV, centred at 80.4\GeV. In situations where multiple combinations of jets satisfy these criteria, along with the charged lepton and neutrino, we employ a selection process to choose the most appropriate combination. This selection is based on two factors: the closeness of the invariant masses of the two top quark candidates to each other, and the closeness of the invariant mass of the light-quark jets to the \PW boson mass value of 80.4\GeV.

\begin{table}[!ht]
\centering
\topcaption{%
    Typical object definitions, and configuration parameters used for defining top quarks at the particle level (pseudo-top candidate). The pseudo-top candidate definition is not universal and may be optimised for the production mode, final state, the variable, and the phase space being studied. The details of particle-level top quark definitions adopted in the \rivet~\cite{Buckley:2010ar,Bierlich:2019rhm} framework by CMS codes are described
    in Ref.~\cite{Collaboration:2267573} as a fundamental aspect for current and future measurements of differential
    production cross sections in both \ttbar and single-top quark production.
}
\label{tab:pseudotop}
\renewcommand{\arraystretch}{1.1}
\cmsTable{\begin{tabular}{ll}
    Requirement & Comment \\ \hline
    {\textit{All final-state particles}}\\
    $\abseta<5.0$ & matching the detector coverage \\[1ex]
   {\textit{Charged leptons, neutrinos, photons}} \\
    must be prompt & exclude those stemming from hadron decays \\[1ex]
    {\textit{Leptons}}\\
    $R_{\Pell}=0.1$ & radius in \etaphi, used to dress the leptons \\
    $\pt(\Pell)>15\GeV$, $\abs{\eta(\Pell)}<2.5$ & matching the tracker coverage \\
    & (\Pe/\PGm from $\PGt\to\Pe/\PGm$ are also accepted)\\[1ex]
    {\textit{Jet clustering and selection}}\\
    exclude prompt leptons & include only leptons from hadron decays \\
    $R=0.4$ (0.8) & anti-\kt jet cone parameter for resolved (boosted) jets \\
    $\pt>30$ (400)\GeV, $\abseta<2.4$ (2.4) & selection for resolved (boosted) jets \\
\end{tabular}}
\end{table}

Parton-level object definitions allow for direct comparisons to fixed-order theoretical calculations and extractions of SM parameters.
The kinematic properties of the top quarks and the \ttbar system are defined with respect to the on-shell top quark and antiquark before decay, as given by the MC generator.
The used definitions vary for \Run2 with respect to \Run1 analyses. For \Run1 analyses, top quarks were typically defined at the matrix-element level before radiation was added by the parton-shower algorithms.
For measurements described in this review, the partonic top quark is defined as its last produced instance by the parton shower code, \ie, after ISR emissions and correcting for the intrinsic transverse momentum of the initial-state partons.
As a consequence, the description depends on the generator used and is model-dependent.
For \PYTHIAEight, the top quark parton level corresponds to the status code 62.
Measurements are usually performed in the visible phase space (within acceptance) and are extrapolated to the full (not measured) phase space using the MC simulation. In this procedure, the results are corrected for detector and hadronisation effects.
Unless further specified, all presented parton-level results use the given \Run2 definition.

\subsection{Top quark mass definitions}
\label{sec:introschemes}

Due to the quantum aspects of the top quark related to its colour and electroweak charges, \mt is not a unique physical parameter but needs to be defined through renormalisation schemes within quantum field theory. The top quark mass (and likewise the masses of all other quarks) therefore plays a role similar to the couplings of the SM Lagrangian. There are many possibilities to define \mt, but theoretical control can be maintained only when renormalisation schemes, defined in perturbation theory, are employed such that the values of \mt in different schemes can be related to each other reliably~\cite{Herren:2017osy,Hoang:2021fhn} and \mt-dependent perturbative cross section predictions can be expressed in these schemes. Formally, theoretical predictions for  (differential) cross sections are independent of a choice of renormalisation scheme. However, the fact that these theoretical predictions can be made only at some finite truncation order in perturbation theory entails that for a particular observable only certain scheme choices are adequate, so that the scheme provides an absorption of sizeable quantum corrections in the \mt dependence.
For example, the impact of the choice of renormalisation scheme for \mt is very large in the theoretical predictions for single Higgs boson or Higgs boson pair production~\cite{Mazzitelli:2022scc}, expected to be measured with high precision in the upcoming HL-LHC era.

Top quark mass renormalisation schemes, defined within perturbation theory, include the pole mass scheme, the modified minimal-subtraction (\msbar) scheme, and the low-scale short-distance mass (MSR) scheme~\cite{Hoang:2017suc}. The \msbar and MSR schemes  furthermore depend on the renormalisation scales \mumass and $R$, respectively.

\begin{figure}[!ht]
\centering
\includegraphics[width=0.8\textwidth]{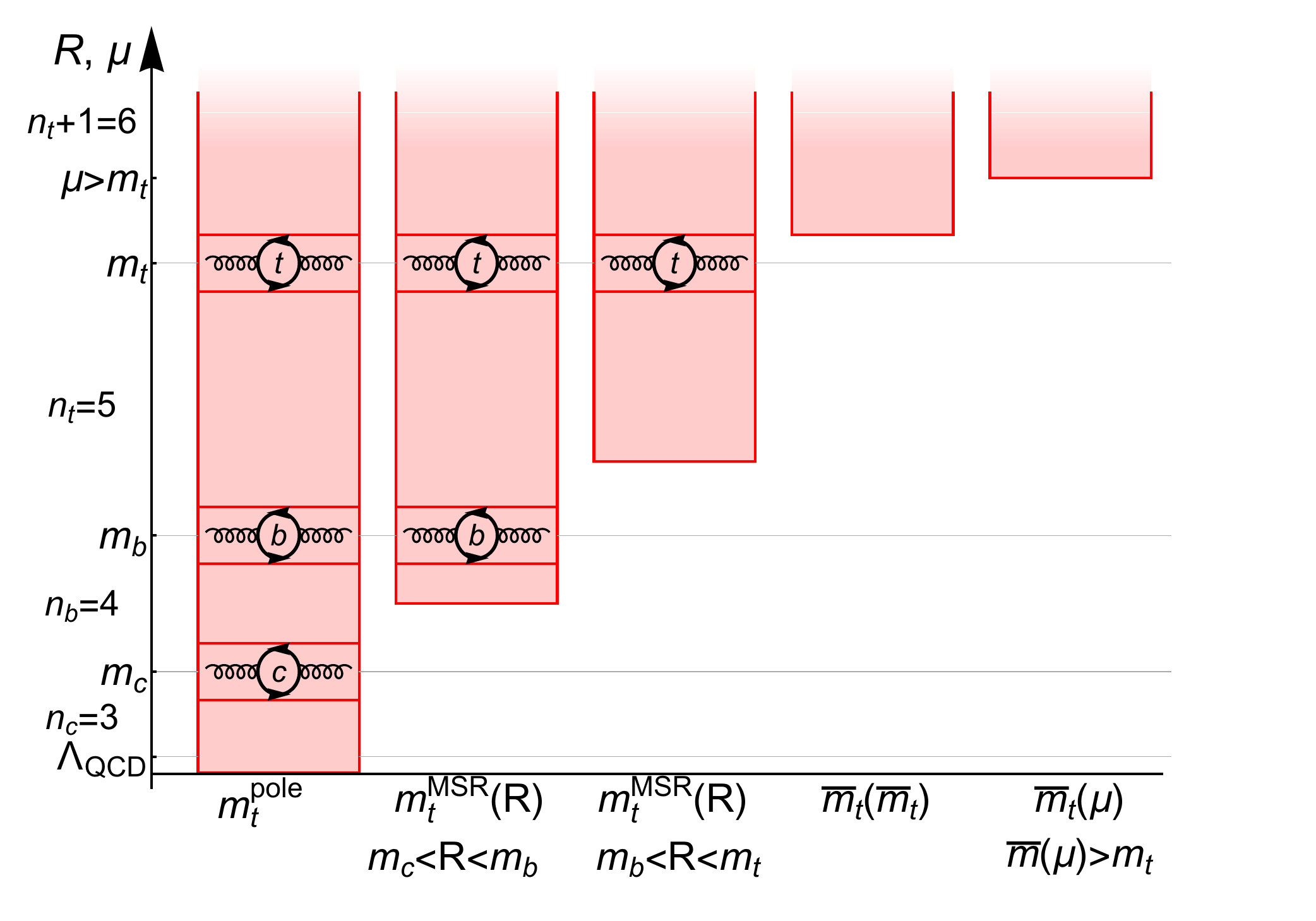}
\caption{%
    Momenta of the self-energy quantum corrections in the top quark rest frame (red segments), absorbed into the top quark mass parameter in the pole (very left), MSR and \msbar schemes for different mass renormalisation scales with respect to the charm and bottom quark masses. The red segments extend to infinite momenta for all top quark mass schemes. The loops inside the red segments illustrate contributions of the virtual top, charm, or bottom quark loops, and $n_{\PQq}$ stands for the number of quarks lighter than quark \PQq, indicating that the MSR and the \msbar masses run with different flavour numbers between flavour thresholds, as does the strong coupling constant \alpS.
    Figure taken from Ref.~\cite{Hoang:2017btd}.
}
\label{fig:massesschematics}
\end{figure}

The pole mass \mtp is defined as the pole of the top quark propagator in the approximation of a free particle. It is used most frequently for theoretical calculations of the top quark production cross sections in fixed-order perturbation theory. Because its colour does not prohibit the definition of the top quark as an asymptotic state within perturbation theory, \mtp can be formally defined at any order~\cite{Tarrach:1980up,Kronfeld:1998di}. However, the concept of an asymptotic ``top particle'' is unphysical because it assumes that the virtual QCD self energy quantum corrections (absorbed into the mass) can be distinguished from the real radiation effects at arbitrarily small scales $\mu$, as shown in the very left part of Fig.~\ref{fig:massesschematics}. This unphysical aspect results in \mtp having an intrinsic renormalon ambiguity of 110--250\MeV~\cite{Beneke:2016cbu,Hoang:2017btd}.

The \msbar scheme implies \mt as a function of the mass-renormalisation scale \mumass, \mtmum, sometimes also denoted as $\barmt(\mumass)$. At the scale of the mass itself, $\barmt(\barmt)$ is also referred to as \mtmt. The MSR scheme interpolates between the pole and the \msbar schemes, as detailed in the following, and operates with $\mtmsr(R)$.
The \msbar and MSR masses do not have the renormalon ambiguity of \mtp. Their scales \mumass and $R$ represent the energy scales, above which the self-energy corrections are absorbed into the mass parameter. Below these scales, the real and virtual quantum corrections are treated as unresolved, as shown by the other parts of Fig.~\ref{fig:massesschematics}. This more physical treatment of QCD self-energy corrections avoids the renormalon ambiguity.

The freedom in the choice of \mumass or $R$ allows to set them equal to the dynamical momentum scale of the \mt dependence of an observable. This dynamical scale is related to the size of the typical momenta involved in the quantum corrections to this \mt dependence. For example, in the case of a reconstructed top quark invariant mass resonance, where the \mt sensitivity arises from the shape and position of the peak, this dynamical scale can be as small as the top quark width \topwidth, depending on the reconstruction procedure. On the other hand, for an inclusive total cross section, the dynamical scale is at least of the order of \mt or the energy of the hard interaction. In general, the more inclusive the observable, the larger the dynamical scale of the \mt dependence. An adequate choice of \mumass or $R$ can reduce the size of higher order perturbative corrections and make the theoretical predictions, which are always based on truncated perturbative expansions, more reliable.
As far as QCD corrections are concerned, \mtp is about 9\GeV larger than the \msbar mass \mtmt, which is a quite sizeable effect. This conversion, however, suffers from the renormalon ambiguity mentioned in the previous paragraph. The renormalon-free mass schemes MSR and \msbar, for any choice of their renormalisation scales, can be related to each other with a precision of about 10--20\MeV~\cite{Hoang:2017suc}. Libraries for numerical conversion of different top quark mass schemes are provided in Refs.~\cite{Herren:2017osy,Hoang:2021fhn}.

While the \msbar mass $\barmt(\mumass)$ is suitable for dynamical scales $\mumass>\mt$, the choice of $\mtmsr(R)$ is preferred for smaller dynamical scales $R<\mt$. For $R=\mtmt$ the MSR mass is approximately equal to \mtmt, and in the limit of vanishing $R$, the MSR mass approaches the pole mass,  $\mtmsr(R)\stackrel{R\to 0}{\longrightarrow}\mtp$. However, this limit is formal since the MSR mass can only be used for $R$ scales that are still in the realm of perturbation theory. For small $R$ values of 1--3\GeV, shown by the second bin in Fig.~\ref{fig:massesschematics}, the MSR mass can serve as a renormalon-free proxy for the pole mass.
A proper choice of the scheme or of the renormalisation scales is straightforward in the context of analytic theoretical predictions, \eg through the analysis of logarithmic terms in the perturbative coefficients and convergence studies (as demonstrated, \eg in Refs.~\cite{Hoang:2017kmk,Bachu:2020nqn}). However, corresponding analyses in the context of purely numerical predictions, which is the case for the calculations for top quark production at the LHC, are more involved and also need to account for correlations with other input quantities and renormalisation scales related, \eg to the strong coupling and PDFs.

The direct \mt measurements and the \mt measurements from boosted top quarks reviewed in Secs.~\ref{sec:direct} and
\ref{sec:boosted}, respectively, rely entirely on MC simulations, operating with the top quark mass parameter \mtmc as a proxy for \mtp.
With increasing measurement precision, subtle effects in the modelling of top quark production and decay become increasingly relevant, and there is ongoing work in the theory community to understand them (see \eg Refs.~\cite{Azzi:2019yne,Hoang:2020iah}).

With the continuously increasing precision of the experimental top mass analyses, the proper interpretation and respective consistency of the results become increasingly relevant.
In the works on \mt determination carried out by the CMS Collaboration so far, measurements of  \mtmc, \mtp, and $\barmt(\mumass)$ have been provided.

\section{Direct measurements from top quark decays}
\label{sec:direct}

The top quark mass can be measured directly using the top quark decay products.
This section focuses mainly on two direct measurements.
One is performed in the lepton+jets channel of \ttbar production using a profile likelihood method and the other analyses single top final states using a template method.

\subsection{Top quark mass measurements in top quark pair events}

In the direct mass measurements, \mt-dependent templates are fit to data to measure \mt directly.
These templates  are derived from simulations of different top quark mass values.
They are described by probability density functions $p(x|\mt,\vectheta)$, where $x$ is an observable and \vectheta a list of possible additional fit parameters.
The considered observable should have a strong dependence on \mt.
In the CMS measurements, this is usually the invariant mass of the top quark decay products in the lepton+jets and all-jets channels and the invariant mass of a lepton and a \PQb-tagged jet in the dilepton channel.

In the lepton+jets channel, a second observable was already used in the measurements at the Tevatron: the invariant mass of the two jets assigned to the decay products of hadronically decaying \PW bosons.
In \ttbar events, the position of the maximum of the invariant mass distribution is expected to be near the precisely known \PW boson mass and depends strongly on the calibration of the reconstructed jets. This allows the introduction of an additional jet energy scale factor (\JSF) in the probability density function to reduce the impact of the uncertainty in the the JES corrections on the measurement.
An ideogram method was utilised in the \Run1 and early \Run2 measurements, while a profile likelihood method was applied in the latest CMS measurement using lepton+jets final states.

\subsubsection{Ideogram method in the lepton+jets channel}

Besides the  JES, the statistical uncertainty was a major uncertainty in the measurement of \mt due to the limited data sample sizes in the measurements at the Tevatron and the early CMS measurements.
Hence, a couple of steps were taken to get the best \mt sensitivity from each \ttbar candidate event, as described in the following.

\begin{figure}[!t]
\centering
\includegraphics[width=0.48\textwidth]{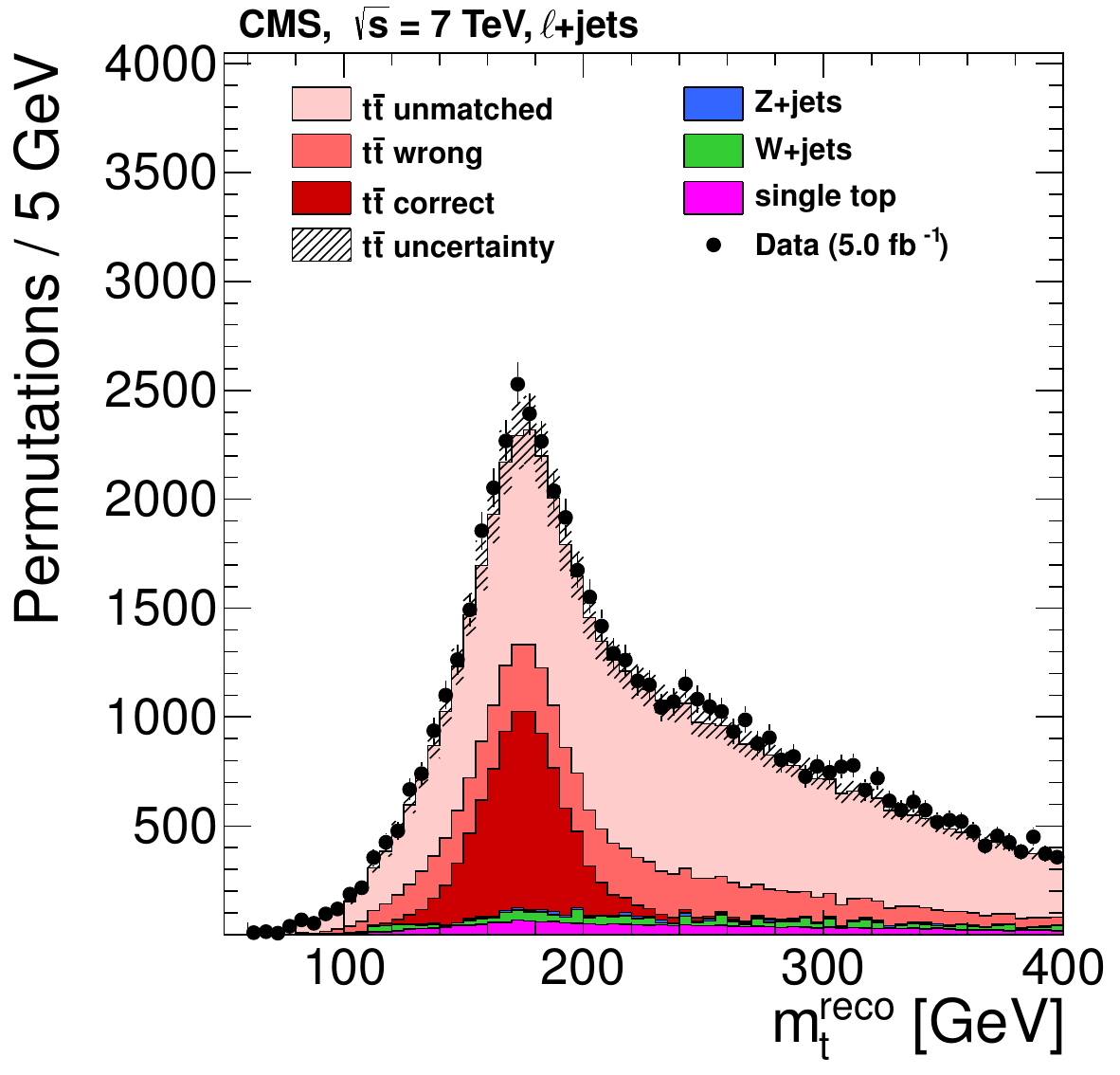}%
\hfill%
\includegraphics[width=0.48\textwidth]{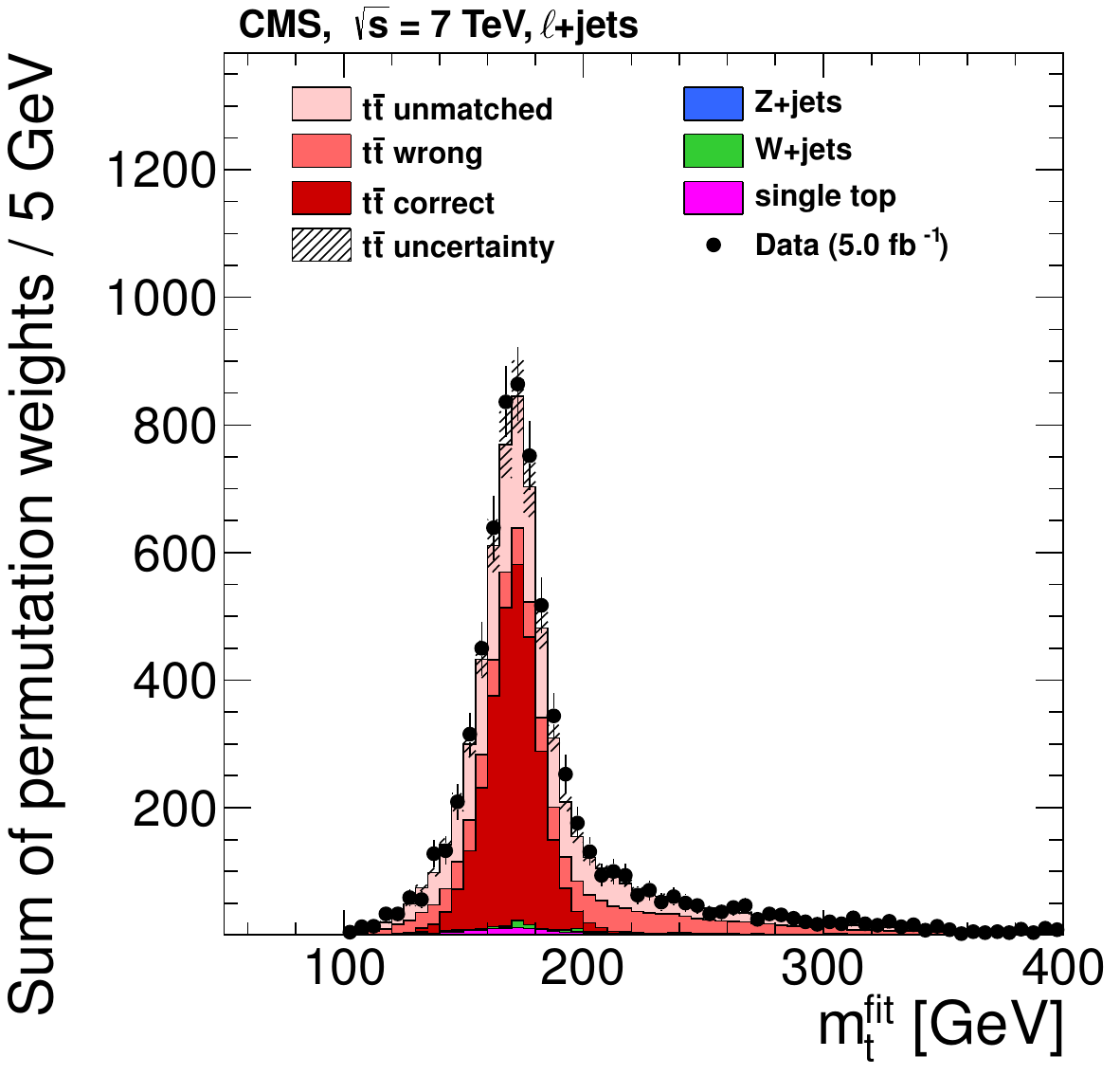}
\caption{%
    Left: The distribution of the reconstructed top quark mass \mtreco using the jet assignment from the kinematic fit, but the reconstructed jet momenta and no addition selection. Right: The distribution of the top quark mass from the kinematic fit \mtfit with the $\Pgof>0.2$ selection.
    Data are shown as points with vertical error bars showing the statistical uncertainties.
    The coloured histograms show the simulated signal and background contributions.
    The simulated signal is decomposed into the contributions from correct, wrong, or unmatched permutations as introduced in Section~\ref{sec:kinreco}.
    The uncertainty in the predicted \ttbar cross section is indicated by the hatched area.
    In the figures, the default value of  $\mtgen=172.5\GeV$ is used.
    The reduction of permutations with wrongly assigned jets and the much narrower peak are clearly visible in the \mtfit measurement.
    Figures taken from Ref.~\cite{CMS:2012sas}.
}
\label{fig:ljets_recotofit}
\end{figure}

At first, the kinematic fit described in the previous chapter is employed.
The \PW boson mass constraint enforced in the fit drastically improves the estimates of the momenta of the two quarks from the \PW boson decay.
In addition, the top quark mass from the kinematic fit, \mtfit, includes information from the lepton+jets decay branch due to the requirement of equal invariant masses for both top quark candidates.
An alternative to the kinematic fit and \mtfit is to compute the invariant mass of the hadronically decaying top quark, \mtreco, from the reconstructed momenta, \ie before the kinematic fit, of the assigned jets.
For correct permutations, where the jets can be matched to corresponding partons, the resolution of \mtfit is 30\% better than the resolution of \mtreco.
For the measurements discussed in this section, only permutations with a \chisq goodness-of-fit probability $\Pgof>0.2$ are used to increase the fraction of well-reconstructed and correctly assigned jets. Figure~\ref{fig:ljets_recotofit} shows the improvements in the mass resolution and the fraction of permutations with correctly assigned jets obtained for the measurement using data collected at \sqrtseq{7} in \Run1~\cite{CMS:2012sas}, corresponding to an integrated luminosity of 5.0\fbinv.

If one assumes that the peak position or the average is used as an estimator of \mt, the statistical uncertainty in the measurement scales with $\sigma/\sqrt{N}$ where $\sigma$ is the standard deviation of the observable and $N$ is the number of events. Hence, an improvement in the resolution by 30\% is equivalent to an increase in the number of events in the peak by a factor of two.
However, this simplistic approach only works if the jets are  correctly assigned to the decay products.
As illustrated in Fig.~\ref{fig:ljets_recotofit} (left), a large fraction of the events are in the unmatched category, \ie at least one of the selected jets cannot be matched to a parton from the top quark decay.
These unmatched permutations dilute the measurement and are the reason for the $\Pgof>0.2$ selection, which helps to effectively suppress their contribution.

The use of the ideogram method~\cite{Abdallah:2008xh, CMS:2012sas} was the second step in order to reduce the statistical uncertainty in the direct \mt measurements.
The details of the procedure outlined below are identical with the approach taken in the \Run1 CMS measurement~\cite{CMS:2015lbj} and the first \Run2 CMS measurement~\cite{CMS:2018quc}.
The observable used to measure \mt is the mass \mtfit evaluated after applying the kinematic fit.
It takes the reconstructed \PW boson mass \mWreco, before it is constrained by the kinematic fit, as an estimator for measuring the additional JES factor to be applied in addition to the standard CMS JES corrections.
An ideogram is the likelihood per event for certain values of \mt and \JSF.
It is the weighted sum of the probabilities of all selected permutations of an event: $\sum_i\Pgofi p(\mtfiti,\mWrecoi|\mt,\JSF)$, where $p(\mtfit,\mWreco|\mt,\JSF)$ is a probability density function obtained from simulation and \Pgofi, \mtfiti, and  \mWrecoi are the values of the respective variable of the $i$-th permutation.
As the momenta of the jets from the \PW boson decay are strongly modified in the kinematic fit by the mass constraint $mW^{\text{fit}}=80.4\GeV$, \mtfit and \mWreco can be assumed as independent random variables and the ansatz  $P(\mtfit,\mWreco|\mt,\JSF)=P(\mtfit|\mt,\JSF) P(\mWreco|\mt,\JSF)$ is used.
The distributions of \mtfit and \mWreco are obtained from simulation for different \mt and \JSF values.
From these distributions, the probability density functions $P(\mtfit|\mt,\JSF)$ and $P(\mWreco|\mt,\JSF)$  are derived separately for the three  permutation cases, \ie correct, wrong, and unmatched.
Analytical functions are used to describe the shape of the distributions.
The parameters of these functions are themselves linear functions of \mt and \JSF and the product of the two.

\begin{figure}[!ht]
\centering
\includegraphics[width=0.48\textwidth]{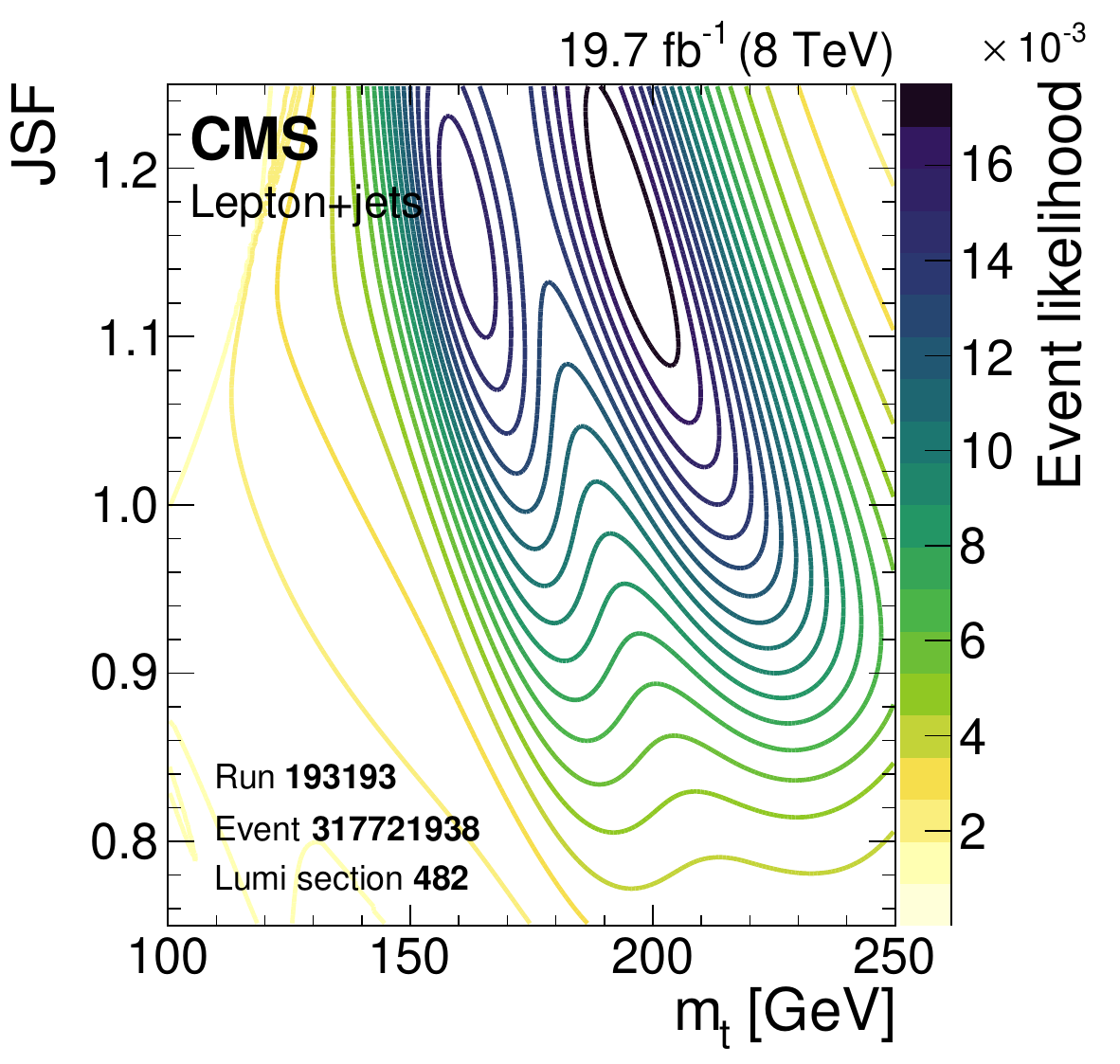}%
\hfill%
\includegraphics[width=0.48\textwidth]{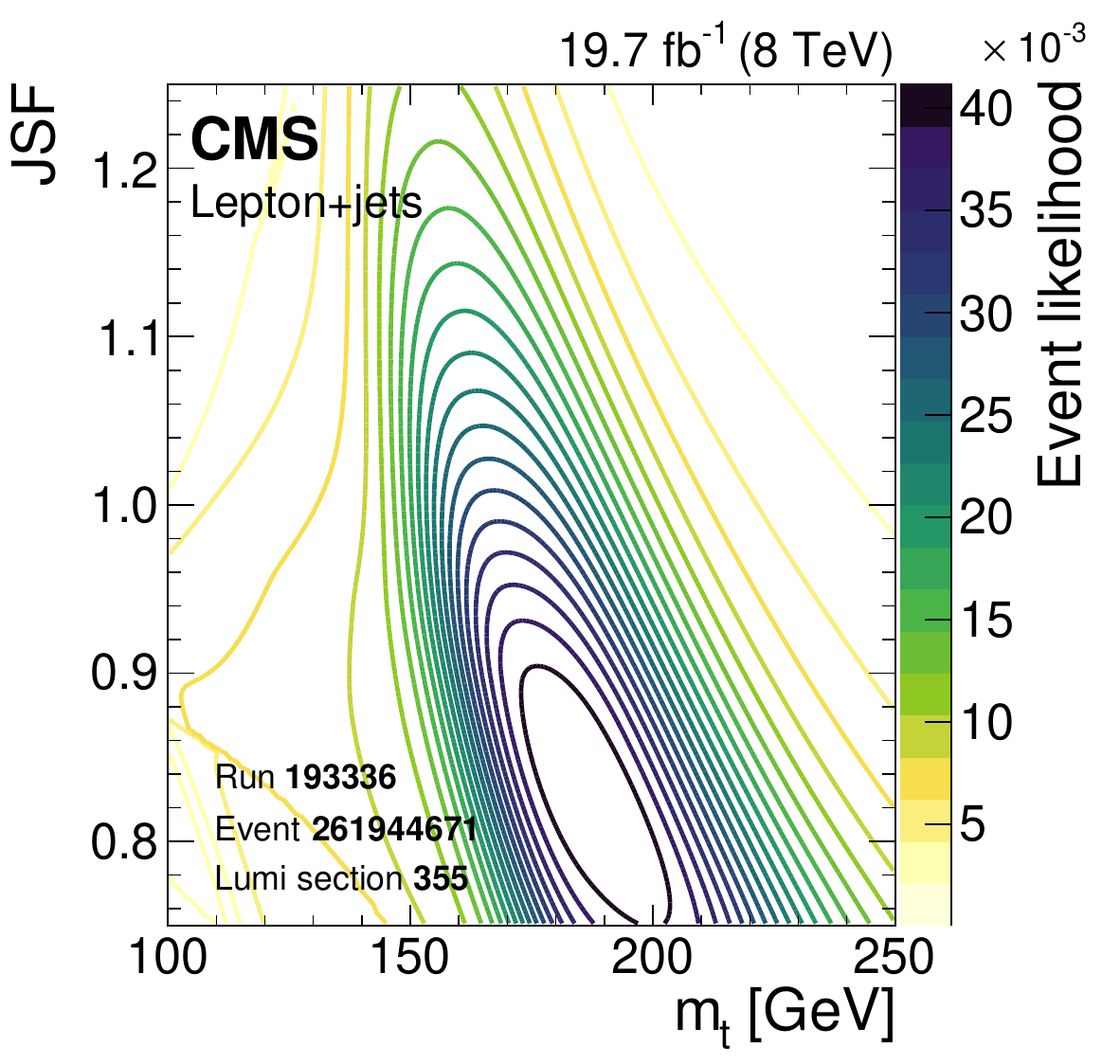}\\
\includegraphics[width=0.48\textwidth]{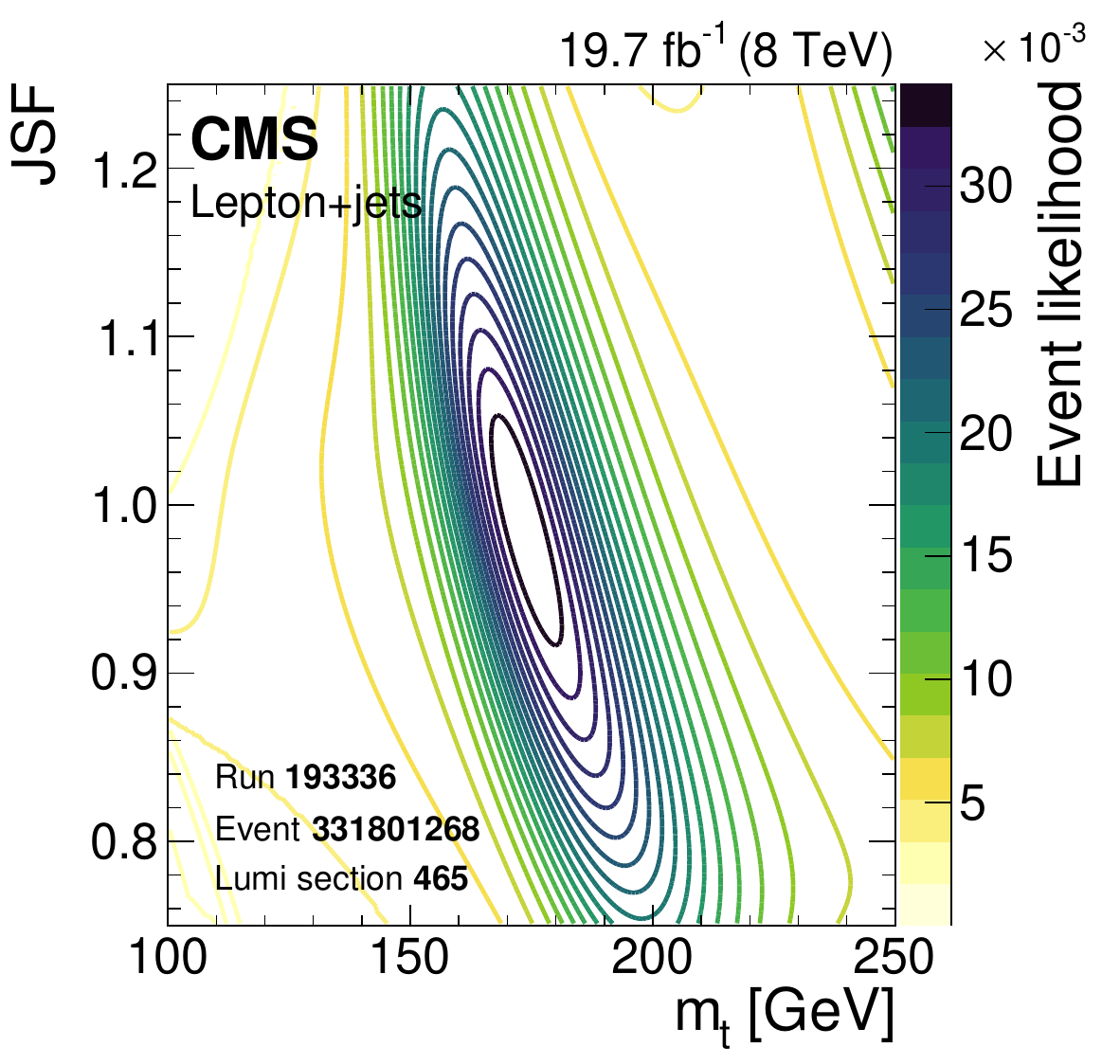}%
\hfill%
\includegraphics[width=0.48\textwidth]{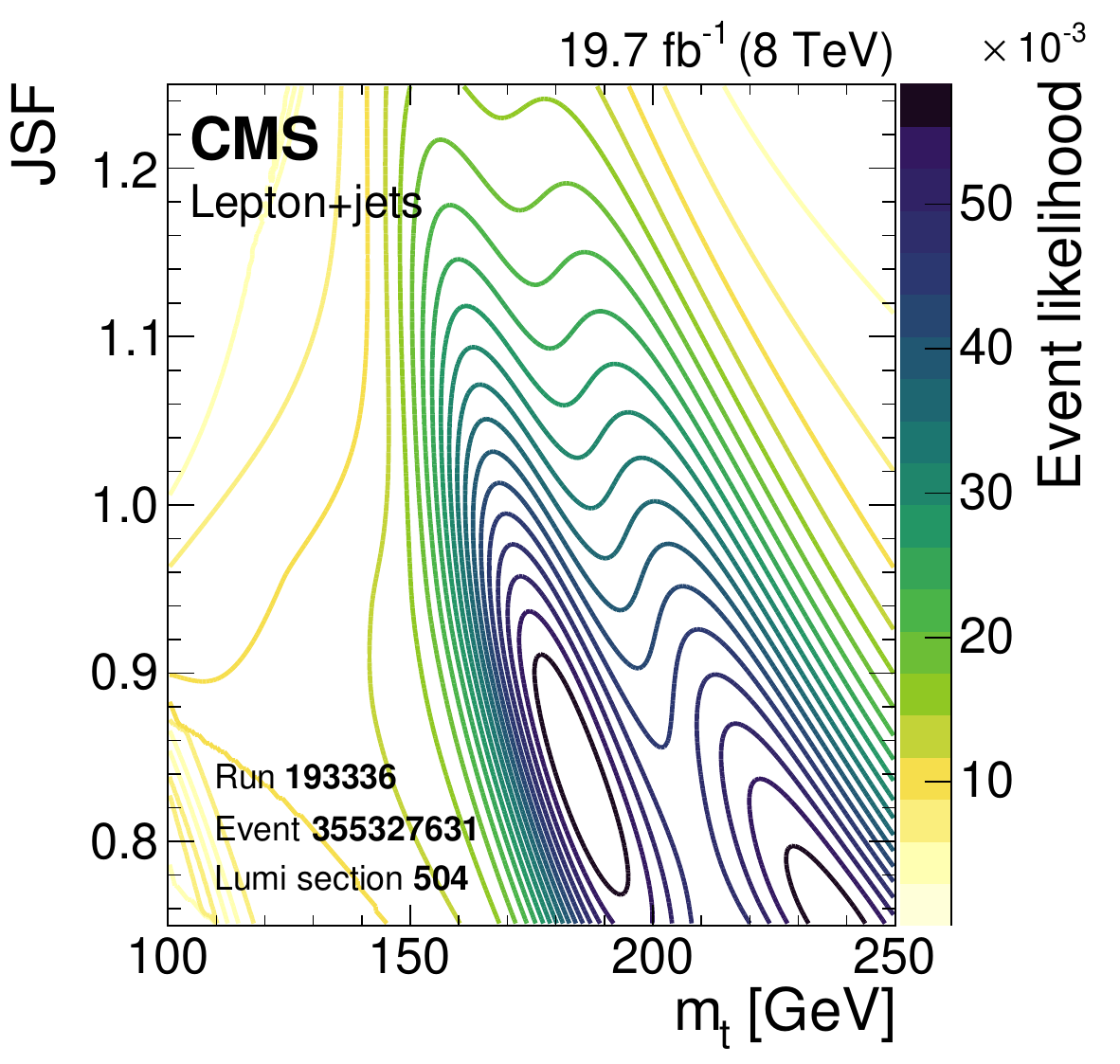}
\caption{%
    Contours of the likelihood of \mt and \JSF values for single events in the \Run1 CMS measurement~\cite{CMS:2015lbj}.
}
\label{fig:ljets_ideos}
\end{figure}

{\tolerance=5500
The most likely \mt and \JSF values are obtained by minimising $-2\ln[\likelihoodsample]$.
With an additional probability density function  \PJSF, the likelihood \likelihoodsample is defined as:
\begin{equation}
    \likelihoodsample=\PJSF\prod_\text{events}\Bigg(\sum_{i=1}^n \Pgof(i)\Big(\sum_jf_jP_j(\mtfiti|\mt,\JSF) P_j(\mWrecoi|\mt,\JSF)\Big)\Bigg)^{\wevt},
\end{equation}
where $n$ denotes the number of the (at most four) permutations in each event, $j$ labels the permutation cases, and $f_j$ represents their relative fractions.
The event weight $\wevt=c\,\sum_{i=1}^n\Pgof(i)$ is introduced to reduce the impact of events without correct permutations, where $c$ normalises the average \wevt to 1.
Examples of ideograms from the \Run1 CMS measurement~\cite{CMS:2015lbj} can be seen in Fig.~\ref{fig:ljets_ideos}.
\par}

As background contributions are neglected in the derivation of the probability density functions, the measurement needs to be calibrated.
This is done with pseudo-experiments where events are drawn from signal samples generated for different top quark mass values, \mtgen,  and background samples according to their expected occurrence in data.
Usually, the corrected bias  amounts to 0.5\GeV for \mt.
Corrections for the statistical uncertainty reported by the method are also derived from pseudo-experiments and have a size of 5\%.

The systematic uncertainties in the final measurement are determined from pseudo-experi\-ments.
Events are drawn from samples where the parameters in the simulation that are related to a systematic uncertainty are changed by $\pm1$ standard deviation.
Then, the pseudo-data is fit with the ideogram method yielding \mt and \JSF values for the up and down varied samples for each systematic uncertainty source.
These values are compared to the values for the default simulation and the absolute value of the largest observed shifts in \mt and \JSF are assigned as systematic uncertainties.
The only exception to this is if the statistical uncertainty in the observed shift is larger than the value of the calculated shift. In this case the statistical uncertainty is taken as the best estimate of the uncertainty in the parameter.
This (over)cautious approach guarantees that systematic effects that are known from particle-level studies to have a sizeable impact on \mt are not underestimated because of finite sample sizes.

Different choices are made for the probability density function \PJSF in the fit.
When the \JSF is fixed to unity, the $P_j(\mWrecoi|\mt,\JSF)$ can be approximated by a constant, as they barely depend on \mt. Hence, only the \mtfit observable is used in the fit, and this approach is called the 1D analysis. The approach with an unconstrained \JSF is called the 2D analysis.
Finally, in the hybrid analysis, the prior \PJSF is a Gaussian centred at one.
Its width depends on the relative weight \whyb that is assigned to the prior knowledge on the \JSF, $\sigma_{\text{prior}}=\delta\JSF^{\text{2D}}_{\text{stat}} \sqrt{\smash[b]{1/\whyb-1}}$, where $\delta\JSF^{\text{2D}}_{\text{stat}}$ is the statistical uncertainty in the 2D result of the \JSF.

The optimal value of \whyb is determined  from pseudo-experiments.
The constraint on the \JSF gets stronger, the lower the experimental uncertainty in the JES is.
However, it is important to note that the introduction of the \JSF reduces not only experimental uncertainties, but also all modelling uncertainties that affect the \mtfit and \mWreco distributions similarly to a JES change.
In other words, the effects of these  uncertainties would shift the position of the \PW boson and top quark peaks in the same direction, and are mitigated by a corresponding change in the \JSF.
Hence, the optimisation of the hybrid approach also results in a strong reduction of most modelling uncertainties.
This approach leads to the most precise single measurement of \mt with \Run1 data of $\mt=172.35\pm0.16\statJSF\pm0.48\syst\GeV$~\cite{CMS:2015lbj}.
Its application to \Run2 data resulted in $\mt=172.25\pm0.08\statJSF\pm0.62\syst\GeV$~\cite{CMS:2018quc} where the larger systematic uncertainty stems from the changes in the evaluation of the modelling uncertainties described in Section~\ref{sec:mcsetup}.

Although the ideogram method has proven itself to be very successful, its implementation has some drawbacks: the neglect of the background in the probability density function and the way the ideograms are constructed require an iterative calibration of estimated mass values with pseudo-experiments.
In addition, the evaluation of the systematic uncertainties via pseudo-experiments is computationally challenging with the growing number of considered sources.
However, the main concern is the choice of the hybrid weight and the fact that the \JSF parameter reduces not just the jet energy correction uncertainties but also many modelling uncertainties in an opaque way.
The large data sample collected during \Run2 makes the use of complicated ideograms that achieve the best statistical precision unnecessary.

\subsubsection{Profile likelihood method}

To overcome the shortcomings of the ideogram method, a profile likelihood method with nuisance parameters was chosen for the latest top quark mass measurement~\cite{CMS:2023ebf}.
The incorporation of all systematic effects via nuisance parameters has multiple advantages.
There is no need anymore to perform dedicated pseudo-experiments for each systematic effect.
All parameters are determined by the fit to give the best agreement with data and precision and, hence, no additional optimisation of an external parameter such as the hybrid weight in the ideogram method is needed.
The nuisance parameter values and uncertainties after the fit show directly how each uncertainty is constrained by the measurement procedure.

However, there are some differences between a direct top quark mass measurement and the application of the profile likelihood method in other analyses.
The main difference is that \mt is estimated from the shape of the data distributions and not from the rate in distinct phase space regions as is done to measure cross sections.
The most characteristic feature of the \mtfit distribution is the position of the top quark mass peak and this is not easily described by changes in the content of coarse bins in \mtfit.
Instead of the (linear) interpolation of bin contents, \ie vertical morphing, used in most implementations of the profile likelihood method, it is desirable to still use analytic functions to describe the \mtfit distribution where one parameter is directly the peak position.
The probability density function for the \mtfit histograms is approximated by the sum of a Voigt profile (the convolution of a relativistic Breit--Wigner distribution and a Gaussian distribution) for the correctly reconstructed \ttbar candidates and Chebyshev polynomials for the remaining event contributions.
Unlike the previous measurements with the ideogram method, this ansatz includes the effect of backgrounds and does not need an iterative calibration of the estimator with pseudo-experiments.
For other distributions, which do not feature a narrow peak, a binned probability density function  is used that returns the relative fraction of events per histogram bin.
Here, eight bins are used for each observable and the widths of the bins are chosen so that each bin has a similar number of selected events for the default simulation ($\mtgen=172.5\GeV$).
The dependence of bin contents of the first seven bins on \mt and the nuisance parameters is implemented with vertical morphing.
The content of the eighth bin is given by the normalisation to data.

\begin{figure}[!ht]
\centering
\includegraphics[width=0.48\textwidth]{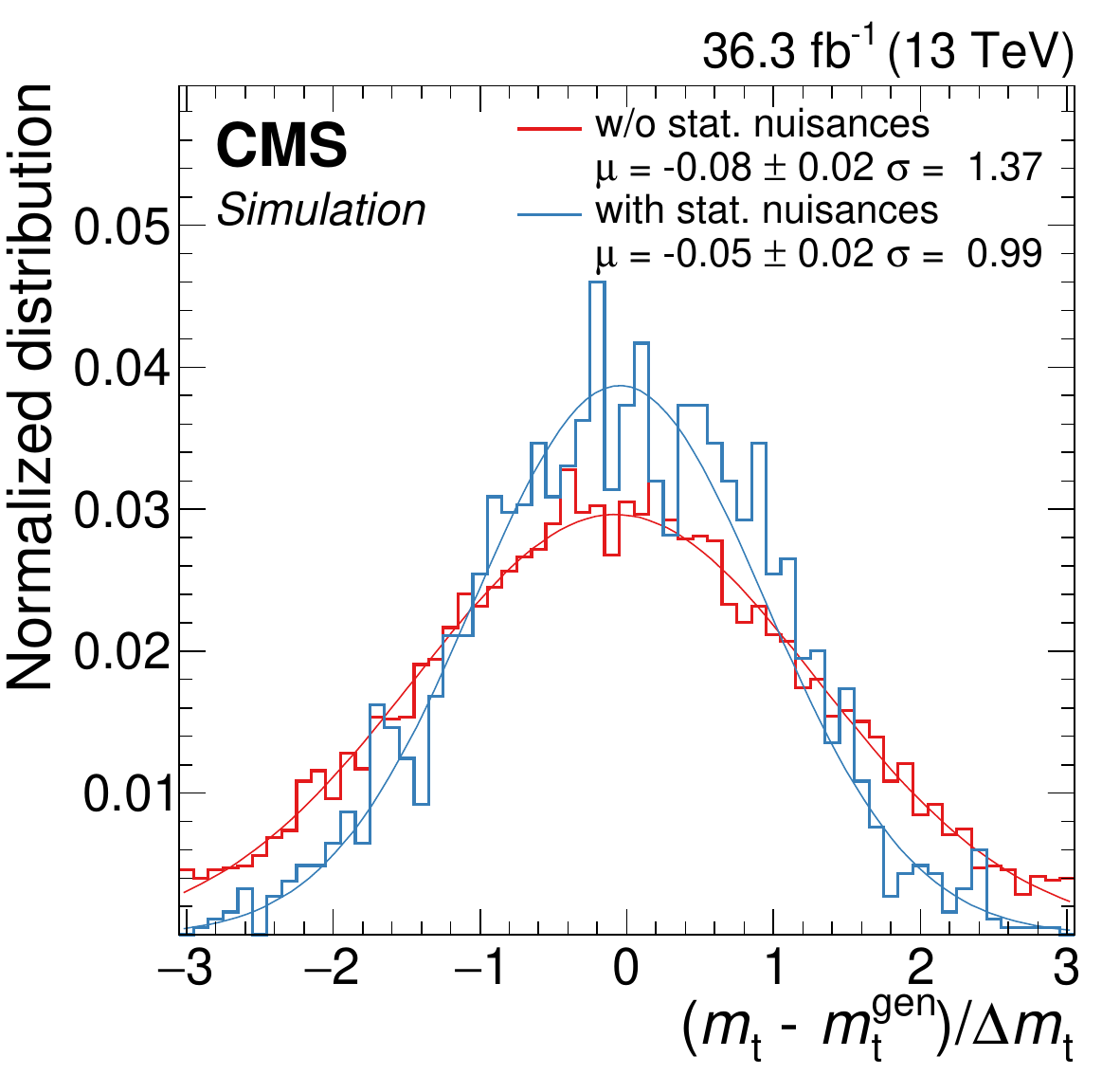}
\caption{%
    The difference between the measured and generated \mt values, divided by the uncertainty reported by the fit from pseudo-experiments without (red) or with (blue) the additional nuisance parameters for the finite sample sizes.
    Also included in the legend are the $\mu$ and $\sigma$ parameters of Gaussian functions (red and blue lines) fit to the histograms.
    Figure taken from Ref.~\cite{CMS:2023ebf}.
}
\label{fig:ljets_pull}
\end{figure}

A custom implementation was also developed for the inclusion of the effects of finite sample sizes~\cite{Barlow:1993dm, Conway:2011in}.
Random fluctuations in the shapes predicted for a systematic variation can cause overly strong constraints on the corresponding nuisance parameter.
This was seen in the first application of a profile likelihood method for a direct \mt measurement in the dilepton channel~\cite{CMS:2018fks}.
Already in the measurements with the ideogram method, the statistical uncertainties in the samples used for estimation of the systematic effects were sizeable, and a special treatment was introduced to include them to avoid a possible underestimation of the systematic uncertainties.
However, the profile likelihood method introduces a clear bias towards too small systematic uncertainties from finite sample sizes.
In the dilepton analysis described in Ref.~\cite{CMS:2018fks}, the size of this effect is estimated by repeating the measurement with alternative simulation templates representing $\pm1$ standard deviation variations of a systematic source that are varied within their Poisson uncertainties.
In the lepton+jets analysis, additional nuisance parameters were introduced directly into the likelihood that account for the statistical uncertainty.
The implementation is different from the approach of Refs.~\cite{Barlow:1993dm, Conway:2011in} and the formulas can be found in Ref.~\cite{CMS:2023ebf}.
This approach is validated with pseudo-experiments.
Here, multiple steps are performed for each pseudo-experiment. At first, new probability density functions that describe how the observables depend on \mt and the nuisance parameters are derived using templates from simulated samples that are varied within their statistical uncertainties.
Then \mt is drawn from a uniform distribution with a mean of 172.5\GeV and a standard deviation of 1\GeV.
The values of the nuisance parameters for systematic effects are drawn from standard normal distributions.
For these parameter values, pseudo-data are generated from the new probability density functions. Then, a fit with the same probability density functions that are applied to the collider data is performed on the pseudo-data.
The fit is performed twice, once with and once without the additional nuisance parameters that account for the finite sample sizes.
Figure~\ref{fig:ljets_pull}  shows the distribution of the differences between the measured and generated \mt values, divided by the uncertainty reported by the fit for both cases.
A nearly 40\% underestimation of the measurement uncertainty can be seen for the case without the additional nuisance parameters, while consistency is observed for the method that is employed on data.
This demonstrates that the limited sample sizes have a big effect on the total uncertainty of the measurement and that the additional nuisance parameters can account for these effects.

\subsubsection{Observables and systematic uncertainties}

In the lepton+jets channel, events are selected with exactly one isolated electron or muon and at least four jets.
Only the four jets with the highest transverse momentum are used in the kinematic fit.
Exactly two \PQb-tagged jets are required among the four selected jets.
In the latest CMS measurement~\cite{CMS:2023ebf} using a data set corresponding to an integrated luminosity
of 36.3\fbinv at \sqrtseq{13}~\cite{CMS:2021xjt}, this  yields 287\,842 (451\,618) candidate events in the electron+jets (muon+jets) decay channel.

The goodness-of-fit probability, \Pgof, computed from the \chisq value of the kinematic fit is used to determine the most likely parton-jet assignment.
For each event, the observables from the permutation with the highest \Pgof value are the input to the \mt measurement.
In addition, the events are categorised as either $\Pgof<0.2$ or $\Pgof>0.2$, matching the value chosen in Ref.~\cite{CMS:2018quc}.
Requiring  $\Pgof>0.2$ yields 87\,265 (140\,362) \ttbar candidate events in the electron+jets (muon+jets) decay channel and has a predicted signal fraction of 95\%.
This selection improves the expected fraction of correctly reconstructed events from 20 to 47\%.

\begin{figure}[!ht]
\centering
\includegraphics[width=0.48\textwidth]{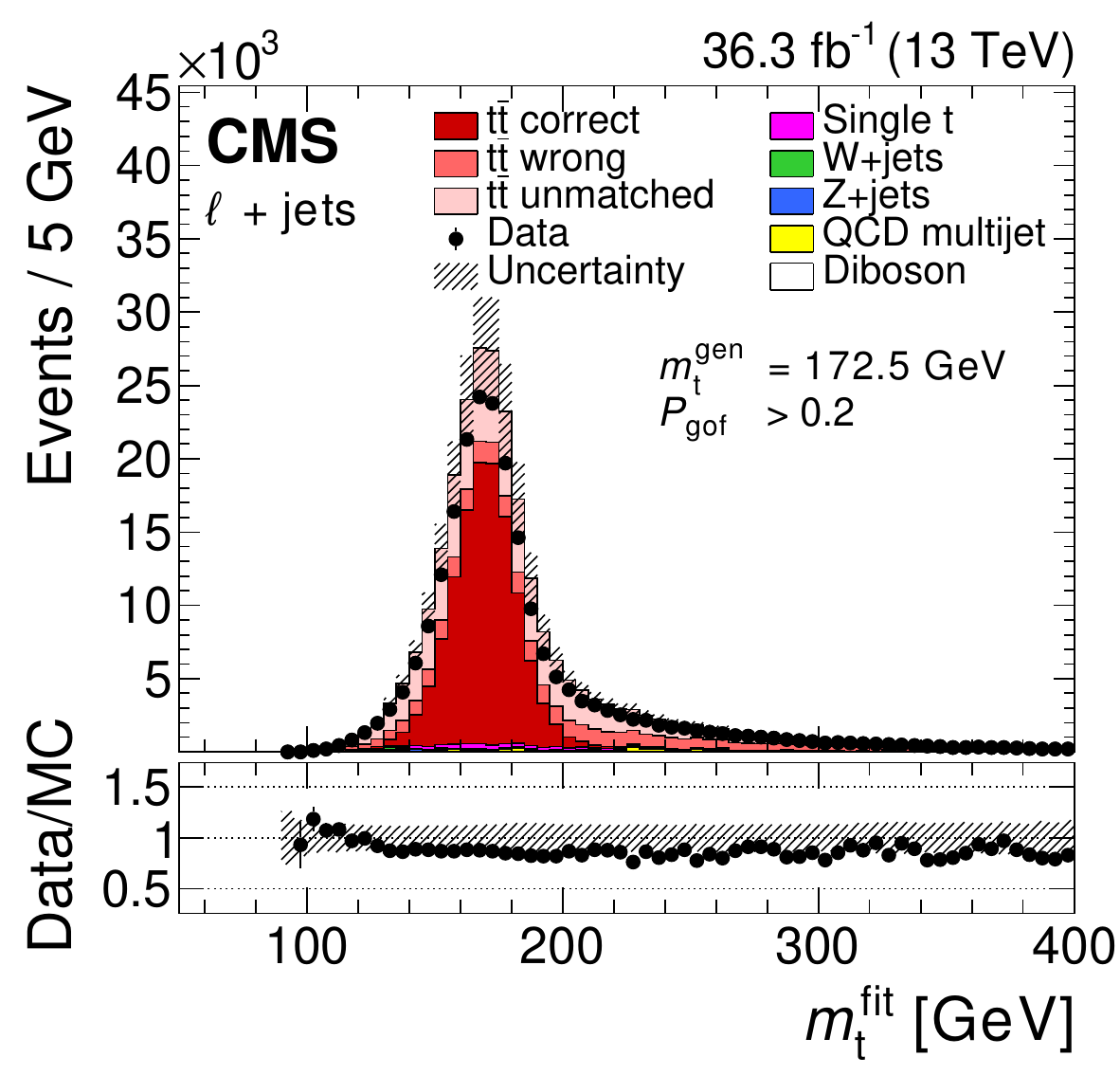}%
\hfill%
\includegraphics[width=0.48\textwidth]{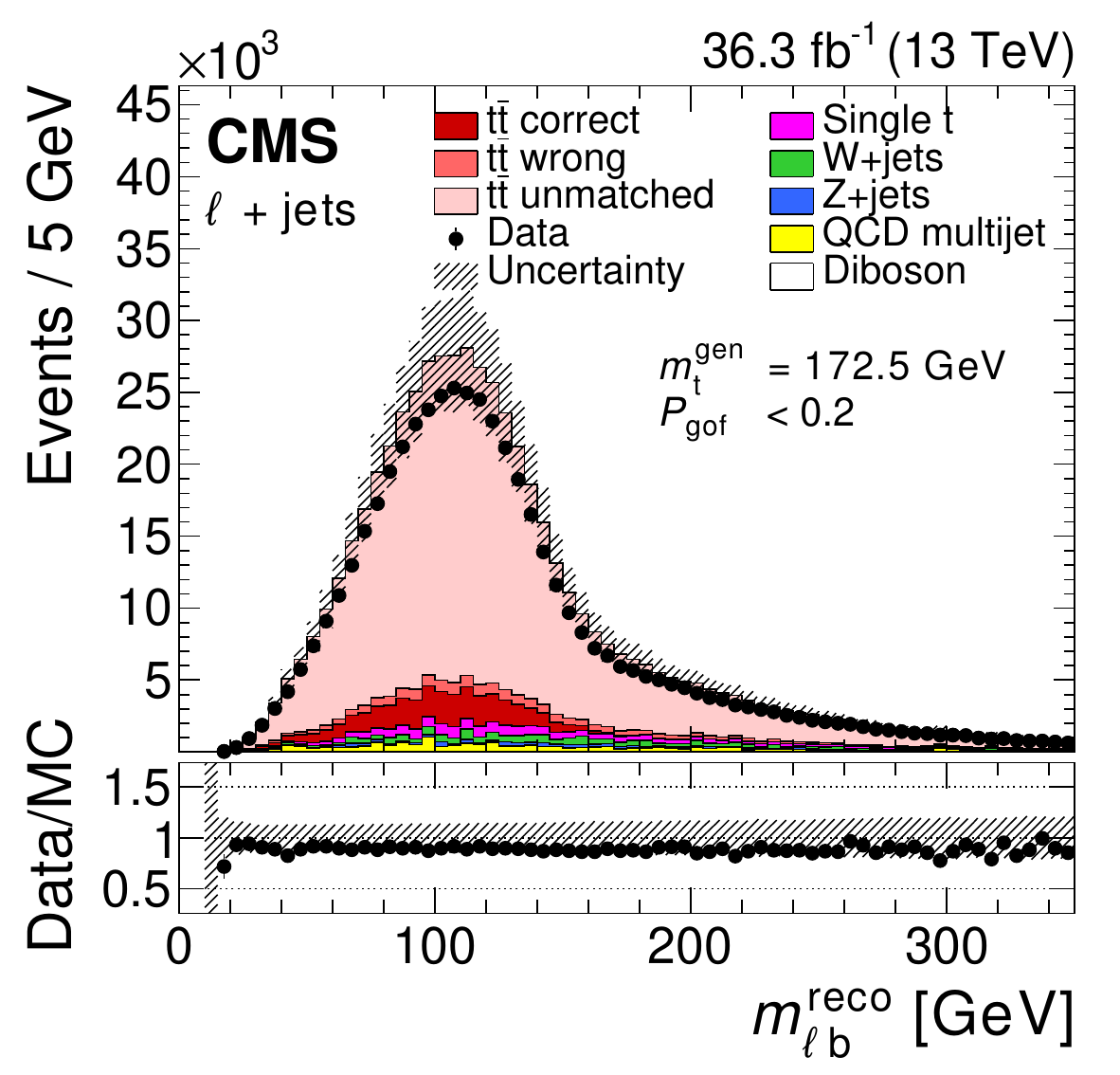}
\caption{%
    The distributions of the top quark mass from the kinematic fit for the $\Pgof>0.2$ category (left) and of the invariant mass of the lepton and the jet assigned to the top quark decaying in the lepton+jets channel for the $\Pgof<0.2$ category (right).
    Data are shown as points with vertical error bars showing the statistical uncertainties.
    The coloured histograms show the simulated signal and background contributions.
    The simulated signal is decomposed into the contributions from correct, wrong, or unmatched permutations, as introduced in Section~\ref{sec:kinreco}.
    The uncertainty bands contain statistical uncertainties in the simulation, normalisation uncertainties due
    to the integrated luminosity and cross section, JES correction, and all uncertainties that are evaluated from event-based weights. A large part of the depicted uncertainties in the expected event yields are correlated. The lower panels show the ratio of data to the prediction. In the figures, the default value of $\mtgen=172.5\GeV$ is used.
    Figures taken from Ref.~\cite{CMS:2023ebf}.
}
\label{fig:ljets_obs}
\end{figure}

The distributions of the two main observables for the \mt measurement in the lepton+jets channel are shown in Fig.~\ref{fig:ljets_obs}.
A large part of the depicted uncertainties in the expected event yields are correlated.
Hence, the overall normalisation of the simulation agrees with the data within the uncertainties, although the simulation predicts 10\% more events in all distributions.
For the final measurement, the simulation is normalised to the number of events observed in data.

For events with $\Pgof>0.2$, the mass of the top quark candidates from the kinematic fit, \mtfit, shows a very strong dependence on \mt and is the main observable in this analysis.
For events with  $\Pgof<0.2$, the invariant mass of the lepton and the \PQb-tagged jet assigned to the top quark, decaying in lepton+jets channel, \mlbreco is used.
For most \ttbar events, a low \Pgof  value is caused by assigning a wrong jet to the \PW boson candidate, while the two \PQb-tagged jets are the correct candidates for the \PQb quarks.
Hence, \mlbreco preserves a good \mt dependence and adds additional sensitivity to the measurement.
While a similar observable has routinely been used in \mt measurements in the dilepton channel~\cite{Aaboud:2016igd,CMS:2018fks}, this is the first application by CMS of this observable in the lepton+jets channel.

\begin{figure}[!htp]
\centering
\includegraphics[width=0.48\textwidth]{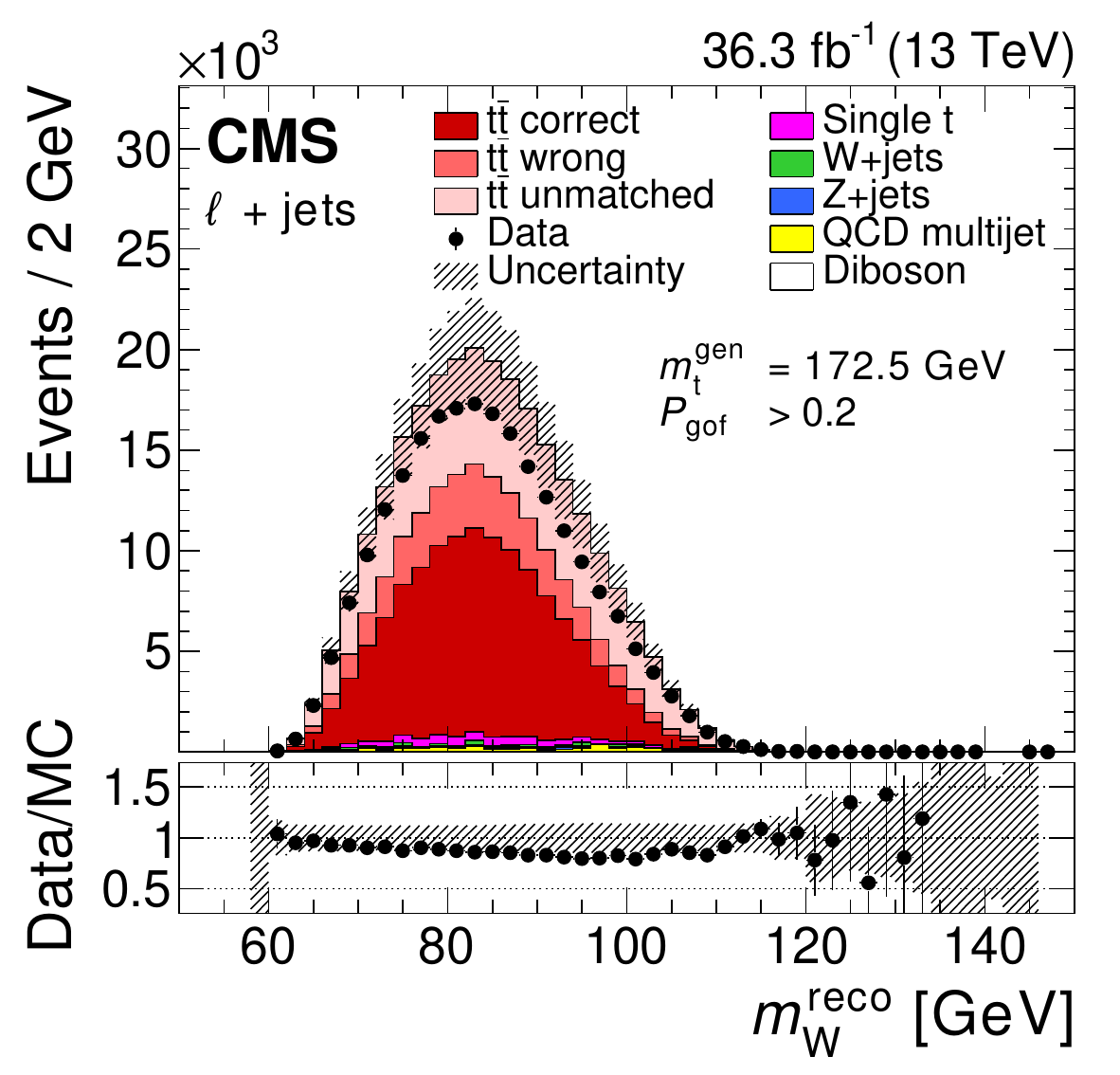}%
\hfill%
\includegraphics[width=0.48\textwidth]{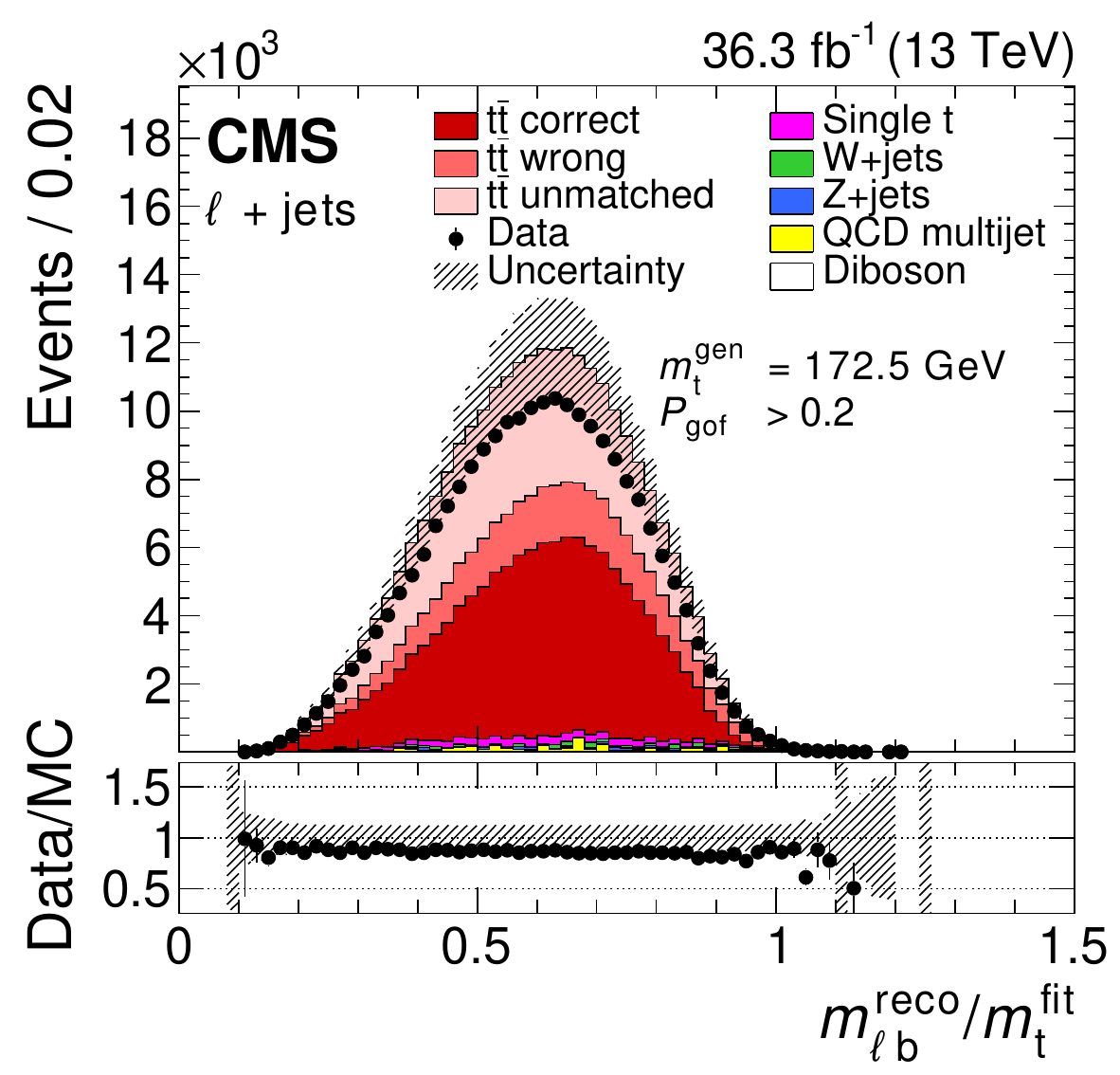}\\
\includegraphics[width=0.48\textwidth]{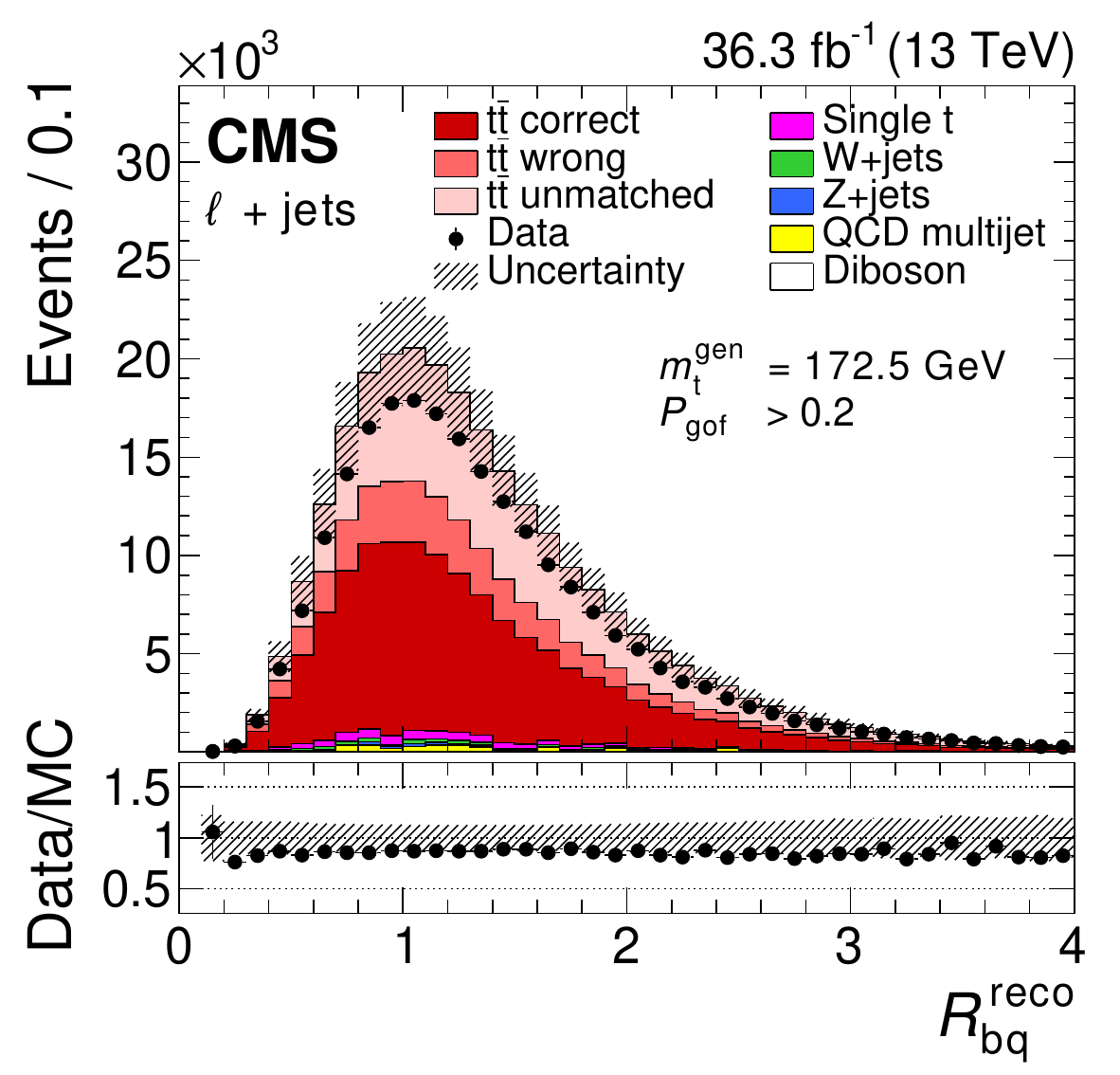}
\caption{%
    The distributions of \mWreco (upper left), \mlbrecomtfit (upper right), and \Rb (lower) for the $\Pgof>0.2$ category.
    Symbols and patterns are the same as in Fig.~\ref{fig:ljets_obs}.
    In the figures, the default value of $\mtgen=172.5\GeV$ is used.
    Figures taken from Ref.~\cite{CMS:2023ebf}.
}
\label{fig:ljets_addobs}
\end{figure}

Additional observables are used in parallel for the mass extraction to constrain systematic uncertainties.
In previous analyses by the CMS Collaboration in the lepton+jets channel~\cite{CMS:2018quc,CMS:2015lbj}, the invariant mass of the two non-\PQb -tagged jets before the kinematic fit, \mWreco, has been used together with \mtfit, mainly to reduce the uncertainty in the JES and the jet modelling.
As \mWreco is only sensitive to the energy scale and modelling of light-flavour jets, two additional observables are employed to improve sensitivity to the scale and modelling of jets originating from \PQb quarks.
These are the ratio \mlbrecomtfit as well as the ratio of the scalar sum of the transverse momenta of the two \PQb-tagged jets ($\PQb1$, $\PQb2$)  and the two non-\PQb-tagged jets ($\PQq1$, $\PQq2$), $\Rb=(\pt^{\PQb1}+\pt^{\PQb2})/(\pt^{\PQq1}+\pt^{\PQq2})$.
The distributions of all three additional observables are shown in Fig.~\ref{fig:ljets_addobs}.
While \mtfit and \mWreco have been used  by the CMS Collaboration in previous analyses in the lepton+jets channel, \mlbrecomtfit, and \Rb are new additions.
However, \Rb has been used in the lepton+jets channel by the ATLAS Collaboration~\cite{Aad:2015nba,Aaboud:2018zbu}.

\begin{table}[!htp]
\centering
\topcaption{%
    The overall list of different input histograms and their inclusion in a certain histogram set. A histogram marked with ``$\times$'' is included in a set (measurement).
}
\label{tab:ljets_obs}
\renewcommand{\arraystretch}{1.1}
\begin{tabular}{llccccc}
    \multicolumn{2}{c}{Histogram} & \multicolumn{5}{c}{Set label}\\
    Observable & Category &  \textit{1D} &  \textit{2D} &  \textit{3D} &  \textit{4D} &  \textit{5D}   \\  \hline
    \rule{0pt}{2.3ex}\mtfit  & $\Pgof>0.2$ & $\times$ &$\times$ &$\times$ & $\times$ & $\times$  \\
    \mWreco &  $\Pgof>0.2$ & & $\times$ & $\times$ & $\times$ & $\times$  \\
    \mlbreco &  $\Pgof<0.2$ &  &   &  $\times$ &  $\times$  &  $\times$  \\
    \mlbrecomtfit  & $\Pgof>0.2$ & & & & $\times$ & $\times$ \\
    \Rb &  $\Pgof>0.2$ & &  &  &   & $\times$ \\
\end{tabular}
\end{table}

The value of \mt is determined  with the profile likelihood fit for different sets of data histograms.
As shown in Table~\ref{tab:ljets_obs}, the \textit{1D} measurement set fits just the \mtfit distribution for events with $\Pgof>0.2$ and the \textit{2D} measurement set simultaneously fits this distribution and the \mWreco for events with $\Pgof>0.2$.
These sets allow the comparison with the analyses using the ideogram method.
The \textit{5D} measurement performs a simultaneous fit of the \mtfit, \mWreco, \mlbrecomtfit, and \Rb distributions for $\Pgof>0.2$ and the \mlbreco distribution for $\Pgof<0.2$.

\begin{figure}[!htp]
\centering
\includegraphics[width=0.48\textwidth]{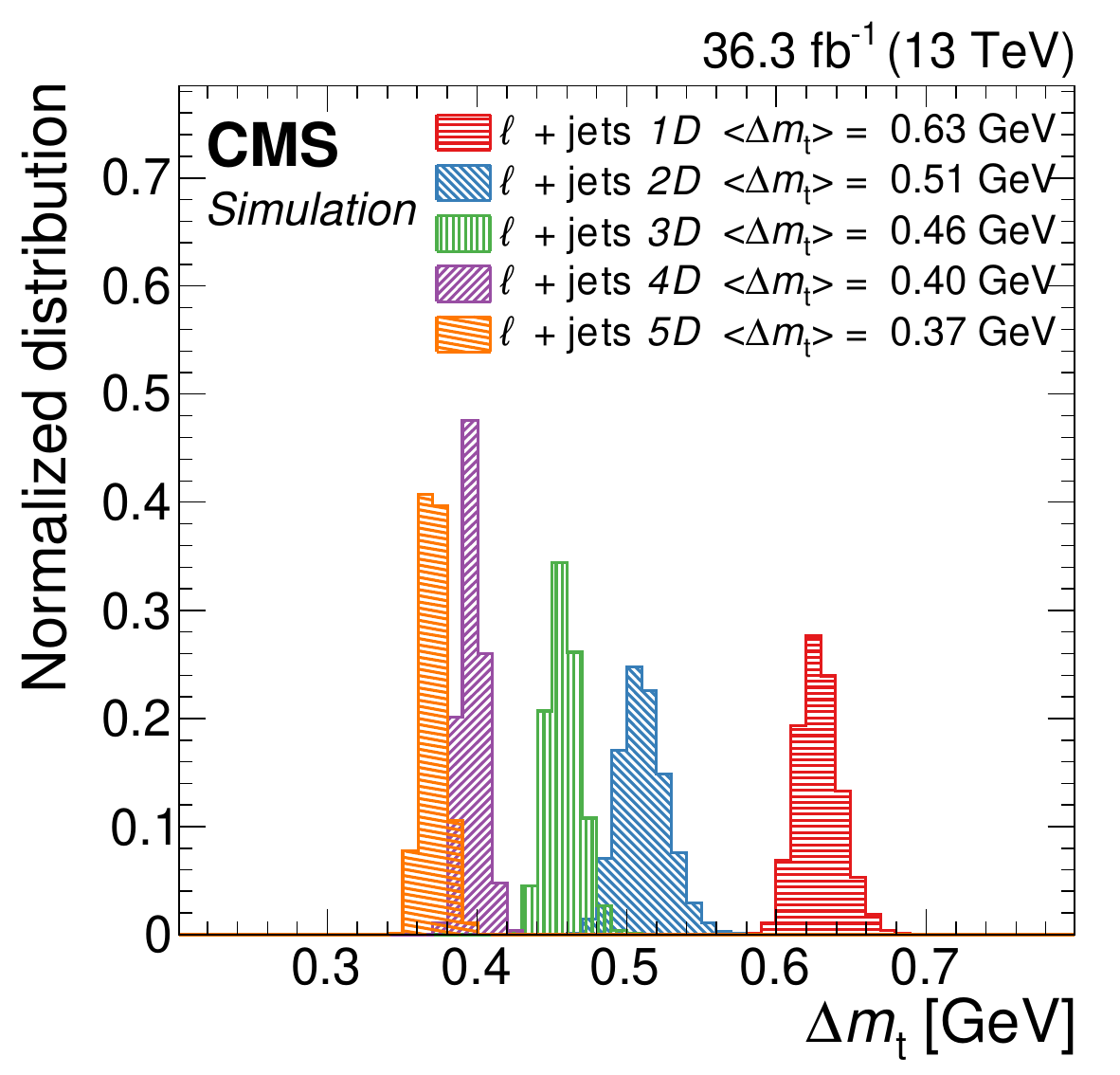}
\caption{%
    Comparison of the expected total uncertainty in \mt in the combined lepton+jets channel and for
    different observable categories defined in Table~\ref{tab:ljets_obs}.
    Figure taken from Ref.~\cite{CMS:2023ebf}.
}
\label{fig:ljets_tots}
\end{figure}

The expected total uncertainty in \mt is evaluated for each set defined in Table~\ref{tab:ljets_obs} with pseudo-experiments using the default simulation.
The results of the pseudo-experiments are shown in Fig.~\ref{fig:ljets_tots}.
The improvements in the data reconstruction and calibration, event selection, simulation, and mass extraction method reduce the uncertainty in  the \textit{1D} measurement from 1.09 to 0.63\GeV, when compared to the previous measurement~\cite{CMS:2018quc}, which used the same data set.
The uncertainty in the \textit{2D} measurement improves from 0.63 to 0.51\GeV.
The additional observables and the split into categories further reduce the expected uncertainty down to 0.37\GeV for the \textit{5D} set.

{\tolerance=800
The statistical uncertainty is obtained from fits that only have \mt as a free parameter.
From studies on simulation, it is expected to be 0.07, 0.06, and 0.04\GeV in the electron+jets, muon+jets, and the combined (lepton+jets) channels, respectively.
\par}

\subsubsection{Mass extraction method and results}

\begin{figure}[!ht]
\centering
\includegraphics[width=\textwidth]{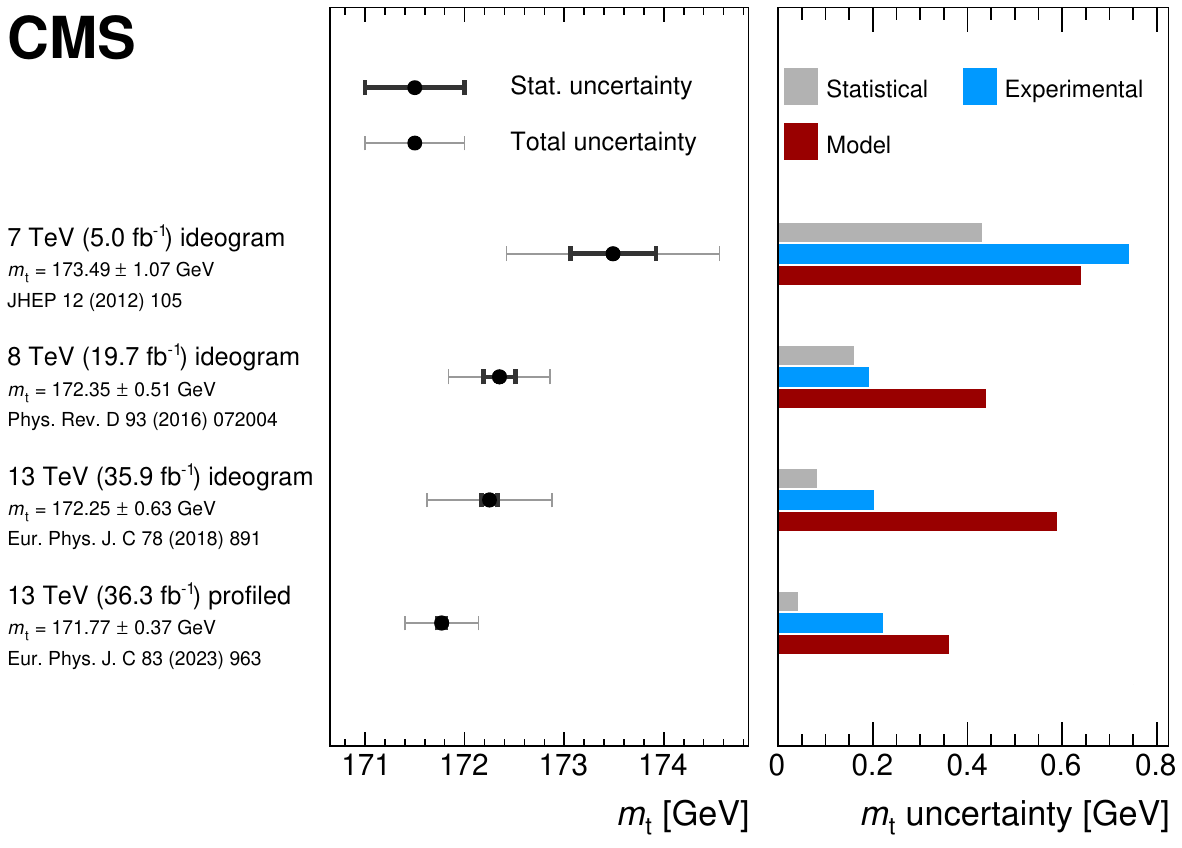}
\caption{%
    Summary of the direct \mt measurements in the lepton+jets channel by the CMS Collaboration. The left panel shows the measured value of \mt (marker) with statistical (black bars) and total (grey bars) uncertainties. The right panel displays a breakdown of contributing uncertainty groups and their impact on the uncertainty in the measurement. The two results at 13\TeV are derived from the same data.
    The figure is compiled from Refs.~\cite{CMS:2012sas, CMS:2015lbj, CMS:2018quc, CMS:2023ebf}.
}
\label{fig:ljets_results}
\end{figure}

The result of the \textit{5D} fit to data~\cite{CMS:2023ebf} and the previous direct \mt measurements in the lepton+jets channel~\cite{CMS:2012sas, CMS:2015lbj, CMS:2018quc} are displayed in Fig.~\ref{fig:ljets_results}.
The uncertainties in the measurements are broken down into statistical, experimental, and modelling uncertainties.

For the statistical uncertainty in the three ideogram measurements, the expected reduction is observed, proportional to the inverse of the square root of the number of selected \ttbar candidates. The increase in the number of candidates stems not only from the increase in the recorded luminosity from 5.0 to 36.3\fbinv, but also in the increased \ttbar production cross section from \sqrtseq{7} to 13\TeV.
While the statistical uncertainty for the three ideogram measurements is obtained from a fit with two free parameters, \mt and \JSF, the statistical uncertainty for the profile likelihood method is derived when only \mt is free in the fit.
This explains a large part of the difference in the statistical uncertainty in the ideogram and the profile likelihood (\textit{5D}) measurements on the same data, but with slightly different reconstruction and calibration.
However, the \mt-only fit with the ideogram method~\cite{CMS:2018quc} yields still a roughly 50\% larger statistical uncertainty of 0.06\GeV compared to 0.04\GeV in the \textit{5D} method.
This remaining reduction stems from the inclusion of previously discarded events that fail the \Pgof criterion via the \mlbreco observable in the \textit{5D} measurement.

The main experimental uncertainties are in the JES and JER.
The energy scale and resolution corrections are mainly derived from QCD dijet events.
Due to the high cross section for these processes for the relatively soft jets ($\pt\lesssim100\GeV$) from top quark decays, the sample size is not limited by the integrated luminosity but by the bandwidth allocated to the dijet triggers.
Hence, one cannot expect an improvement with rising integrated luminosity or centre-of-mass energy.
A lot of time and effort was invested after the end of the \Run1 data taking to reduce the uncertainty in the JES corrections for the legacy \Run1 measurement at \sqrtseq{8}~\cite{CMS:2016lmd}, and, hence, this measurement has the smallest experimental uncertainty.
Similarly, the second measurement using the 13\TeV data with the profile likelihood method~\cite{CMS:2023ebf} should profit from the improved JES corrections that were used in comparison to the ideogram measurement on the same data.
Nevertheless, the ideogram measurement has a slightly smaller experimental uncertainty.
For the profile likelihood measurement, the JES uncertainties are split in many categories and the FSR PS scale is varied independently for different emission processes.
The latter reduces the constraint from the \PW boson peak position on the JES as out-of-cone radiation from the quarks of the \PW boson decay has a stronger impact on the \mWreco distribution than a single JES variation.
In addition, the non-\ttbar background, which is included in the experimental uncertainties, has become more important by the inclusion of events that fail the \Pgof criterion, which have a higher contribution from background processes.

The main modelling uncertainties are related to \PQb jets, FSR, and CR.
The small experimental uncertainties, especially in the JES corrections, in the legacy \Run1 measurement also lead to reduced modelling uncertainties with the hybrid approach.
For the \Run2 measurements, new procedures for the CR and FSR uncertainty  lead to larger modelling uncertainties.
In part, this is just caused by the increase in the number of alternative signal samples for CR/ERD modelling from one to three samples and, hence, more statistical effects on the size of the uncertainty.
In contrast, weights are used to vary parameters of the FSR modelling in the profile likelihood measurement removing the statistical component on the size of the FSR uncertainty.
While this reduces the estimated uncertainty, the introduction of separate scales per splitting type leads to an overall increase in the size of the FSR uncertainty. The introduction of \mlbrecomtfit and \Rb reduces the impact of the \PQb jet modelling on the \mt measurement by 30\% comparing the ideogram and the profile likelihood measurements with 2016 data.

\subsubsection{Other channels and outlook}

\begin{figure}[!tp]
\centering
\includegraphics[width=\textwidth]{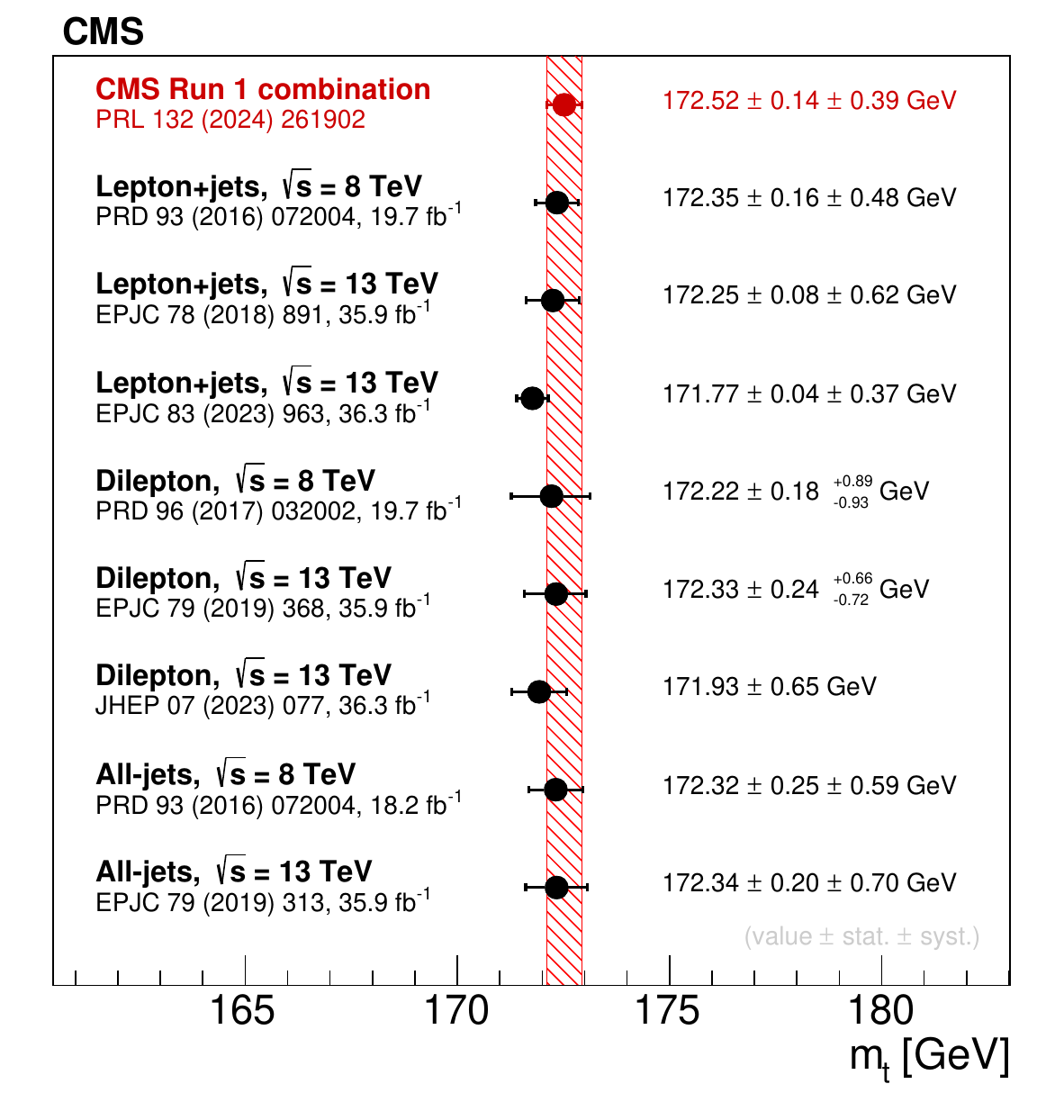}
\caption{%
    Comparison of the CMS direct \mt measurements from the \Run2 data collected in 2016 at \sqrtseq{13} to the best \Run1 measurements at \sqrtseq{8} in each channel.
    The horizontal bars display the total uncertainty in the measurements and the red band shows the uncertainty in the \Run1 combination~\cite{CMS:2023wnd}.
    The figure is compiled from Refs.~\cite{CMS:2015lbj, CMS:2018quc, CMS:2023ebf, CMS:2017znf, CMS:2018fks, CMS:2022emx, CMS:2018tye, CMS:2023wnd}.}
\label{fig:run2direct}
\end{figure}

Besides the lepton+jets channel, also the dilepton and the all-jets channels can be used to measure \mt using its decay products.
Figure~\ref{fig:run2direct} compares the best CMS measurements from \sqrtseq{8} \Run1 data for each channel with the corresponding \sqrtseq{13} \Run2 data collected in 2016.

In contrast to the lepton+jets channel, both \Run2 measurements in the dilepton channel~\cite{CMS:2018fks, CMS:2022emx} utilise a profile likelihood approach and, hence, surpass the \Run1 precision.
The later measurement~\cite{CMS:2022emx} has the same tendency to lower \mt values as the latest measurement in the lepton+jets channel. Both analyses were derived on simulated \Run2 legacy samples described in Section~\ref{sec:mcsetup} and the lower \mt value might be a consequence of the specific parameters used in these simulated samples.

The all-jets channel requires a very pure event selection to suppress QCD multijet background and, hence, suffers from low event count. This is partly compensated by the two fully reconstructed top quark candidates and superior resolution in the predicted top quark mass from the kinematic fit.
The only published analysis with \Run2 data in this channel~\cite{CMS:2018tye} still employed the ideogram method derived on early \Run2 simulation and could not improve on the \Run1 result.

\begin{figure}[!ht]
\centering
\includegraphics[width=0.48\textwidth]{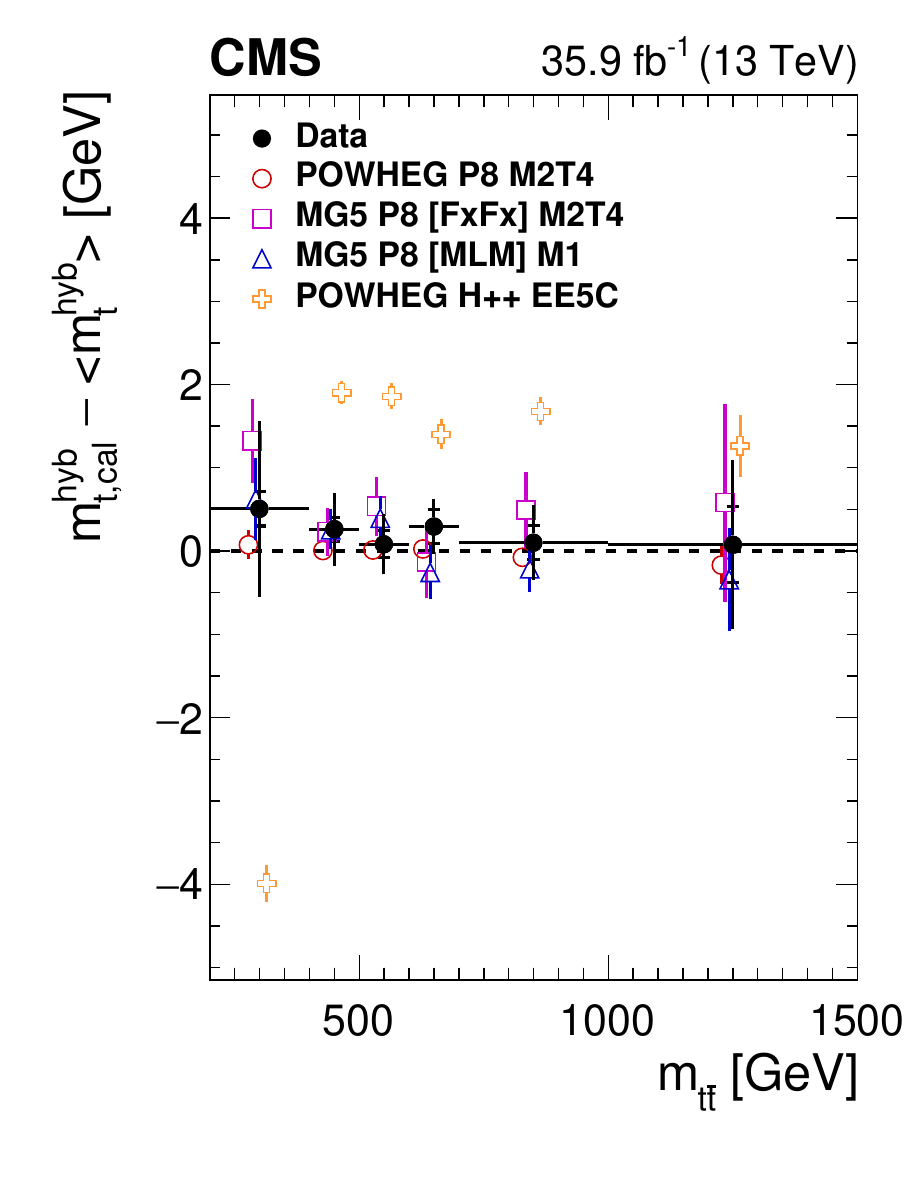}%
\hfill%
\includegraphics[width=0.48\textwidth]{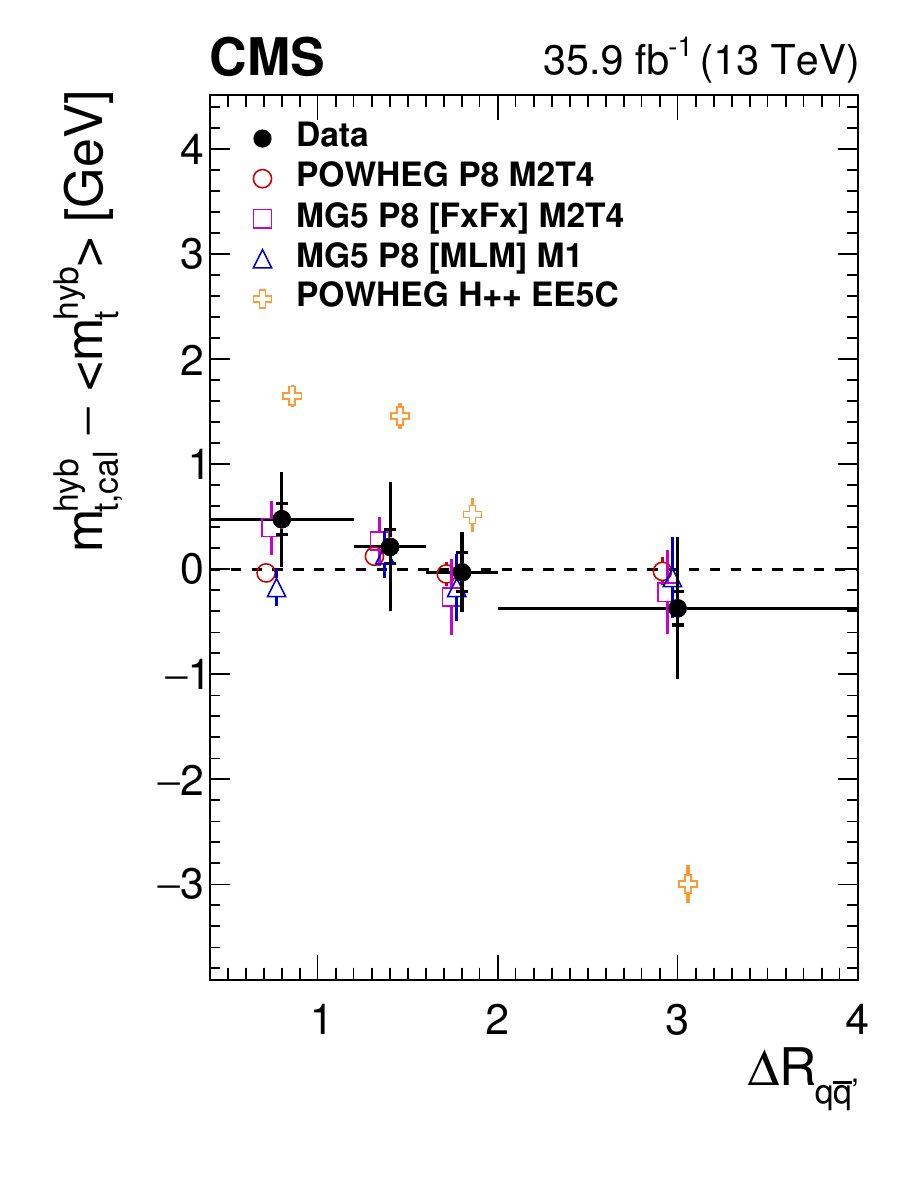}
\caption{%
    Difference of the \mt extracted after calibration in each bin and from the inclusive sample as a function of the invariant mass of the \ttbar system \mtt (left) and the $\DR$ between the light-quark jets \DRqq (right), obtained from the hybrid fit~\cite{CMS:2018quc}, compared to different generator models. The filled circles represent the data, and the other symbols are for the simulations. For reasons of clarity, the horizontal bars indicating the bin widths are shown only for the data points and each of the simulations is shown as a single offset point with a vertical error bar representing its statistical uncertainty. The statistical uncertainty of the data is displayed by the inner error bars. For the outer error bars, the systematic uncertainties are added in quadrature.
    Figures taken from Ref.~\cite{CMS:2018quc}.
}
\label{fig:mtopdiff}
\end{figure}

Measurements of \mt for different phase space regions allow us to experimentally test the universality of the \mt values measured by direct methods and appraise the quality of the modelling by simulation.
The results obtained in Ref.~\cite{CMS:2018quc} and depicted in Fig.~\ref{fig:mtopdiff} show the difference between the measured \mt value in a particular bin and \mt from the inclusive sample in bins of the invariant mass of the \ttbar system, \mtt, and the $\DR$ between the light-quark jets, \DRqq, with comparisons to four generator models. The models use either \POWHEG or \MADGRAPH for the hard interaction interfaced into either \PYTHIAEight or \HERWIGpp. The data and models that use \PYTHIAEight show agreement within 0.5\GeV, while the model using \HERWIGpp shows variations of several \GeVns.

\subsection{Measurement of the top quark mass in single top quark events}

\subsubsection{Motivation}

\begin{figure}[!htb]
\centering
\includegraphics[width=0.35\textwidth]{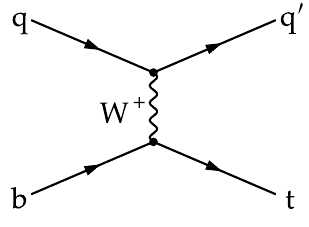}%
\hspace*{0.1\textwidth}%
\includegraphics[width=0.35\textwidth]{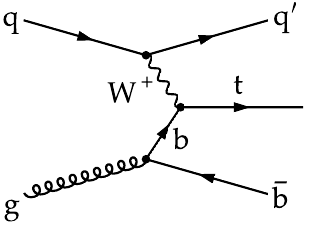}
\caption{%
    Feynman diagrams of the $t$-channel single top quark production at LO corresponding to five- (left) and four-flavour (right) schemes, assuming five (\PQu, \PQd, \PQs, \PQc, \PQb) or four (\PQu, \PQd, \PQs, \PQc) active quark flavours in the proton, respectively. At NLO in perturbative QCD, the right diagram is also part of the five-flavour scheme.
}
\label{fig:SingleTopProcesses}
\end{figure}

At the LHC, single top quark production occurs through charged-current electroweak (EW) interactions. The different production modes can be distinguished at the tree level, depending on the virtuality of the \PW boson: the $t$ channel (spacelike), the \PW-associated or \tW channel (on-shell), and the $s$ channel (timelike). In Fig.~\ref{fig:SingleTopProcesses}, the Feynman diagrams for the $t$ channel, which is the dominant mode for single top quark production in \pp collisions at the LHC, are shown.

The $t$-channel single top quark production offers a partially independent event sample for \mt measurements in a complementary region of phase space as compared to \ttbar events.
Being an EW production process, it provides different sensitivity to systematic and modelling effects, such as PDFs and CR. In fact, in the case of \ttbar, both top quarks, as well as their decay products, are colour connected to the colliding protons, which complicates the modelling of the colour reconnection of final-state particles. On the contrary, in single-top events, the top quark is colour connected only to the parton that participated in the $\PQt\PW\PQb$ vertex.

\begin{figure}[!t]
\centering
\includegraphics[width=0.48\textwidth]{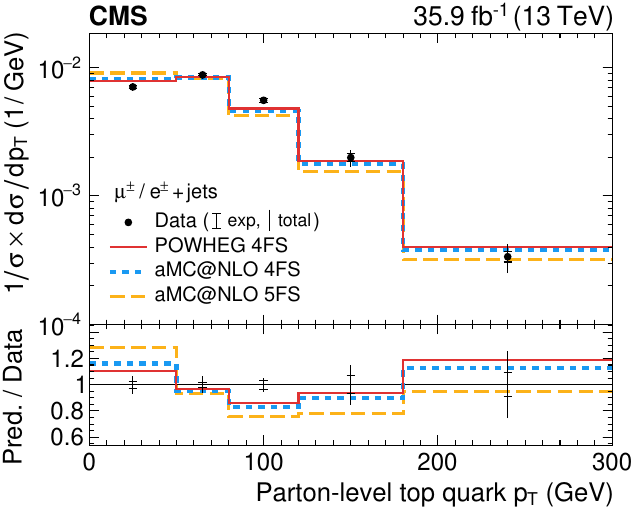}%
\hfill%
\includegraphics[width=0.48\textwidth]{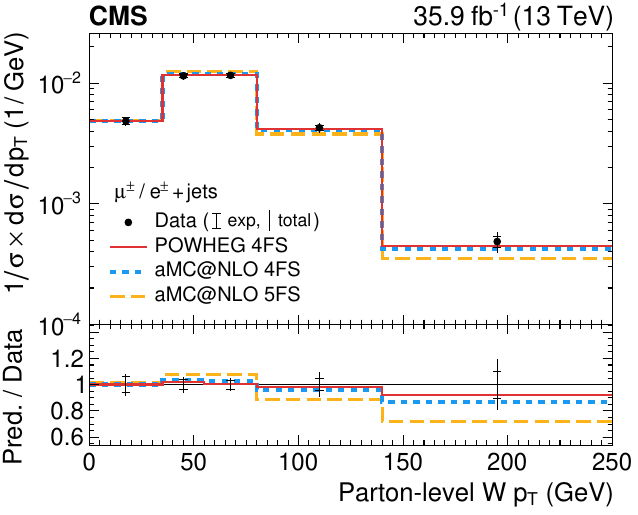}
\caption{%
    Normalised differential cross section of the $t$-channel single top quark production as a function of the \pt of the parton-level top quark (left) and the \PW boson (right).
    Figures taken from Ref.~\cite{CMS:2019jjp}.
}
\label{fig:tCh_diff_xsec}
\end{figure}

{\tolerance=800
The $t$-channel single top quark production is simulated with \POWHEGv2 in the four-flavour number scheme (4FS)~\cite{Alioli:2009je}, where \PQb quarks are produced via gluon splitting, as shown in Fig.~\ref{fig:SingleTopProcesses} (right).
This scheme is expected to yield a better description of the kinematic properties of the top quark and its decay products for the $t$-channel events, as compared to the five-flavour number scheme (5FS)~\cite{Frederix:2012dh,ATLAS:2017rso,CMS:2019jjp} shown in Fig.~\ref{fig:SingleTopProcesses} (left), since it accounts for the mass of the \PQb quark.
This is illustrated in Fig.~\ref{fig:tCh_diff_xsec}, presenting the differential cross section measurements at 13\TeV~\cite{CMS:2019jjp}, together with the 4FS and 5FS predictions.
On the other hand, 5FS calculations offer the advantage of the resummation of potentially large logarithmic corrections, and are therefore often used to predict total rates. For the same reason, the simulated samples are normalised using the total cross section calculated at NLO in the 5FS using the \textsc{hathor} 2.1 package~\cite{Aliev:2010zk,Kant:2014oha}.
\par}

\subsubsection{Event selection and categorisation}

The considered final-state signature of $t$-channel single top quark production used for \mt measurement consists of an isolated high-momentum charged muon or electron, a neutrino from the \PW boson decay, which results in an overall transverse momentum imbalance, a light-quark jet often produced in the forward direction, and another jet arising from the hadronisation of a \PQb quark
from the top quark decay.
The second \PQb jet arising from the initial-state gluon splitting, as shown in Fig.~\ref{fig:SingleTopProcesses} (right), is found to have a softer \pt spectrum and a broader $\eta$ distribution compared to the \PQb jet originating from the top quark. Therefore these jets often escape the final-state object selection or lie outside the detector acceptance.

Based on the above considerations, candidate events are required to contain one isolated electron or muon  with $\pt>20$ or 30\GeV, respectively, and $\abseta<2.4$, exactly two jets with $\pt>40\GeV$, and  $\abseta<4.7$, one of which is \PQb tagged and has $\abseta<2.5$.
The \PQb-tagged jet is required to satisfy a stringent identification criterion corresponding to approximately 0.1\% misidentification probability for light-quark or gluon jets. Additionally, the transverse mass of the charged lepton and neutrino system is required to exceed 50\GeV to further suppress the QCD multijet background.

The selected events are then assigned to two categories (labelled $n$J$m$T), depending on the number of jets ($n$) and number of \PQb-tagged jets ($m$).
The 2J1T category has the largest contribution from $t$-channel single top quark production events and is referred to as the signal category for the measurement.
The contribution from the QCD multijet background is determined from a side-band in data, by inverting the isolation (identification) criteria of the charged muons (electrons)~\cite{CMS:2021jnp}.

\subsubsection{Single top quark reconstruction}
\label{sssec:topreco}

The top quark mass and four-momentum are reconstructed by combining the momenta of its decay products.
The transverse momentum of the neutrino from the \PW boson decay, \pTnu, is inferred from \ptmiss, while the momenta of the lepton and \PQb-tagged jet are measured in the detector.
The longitudinal momentum of the neutrino, \pZnu, can be calculated by imposing energy-momentum conservation at the $\PW\to\Pell\PGn$ vertex while assuming $\mW=80.4\GeV$~\cite{ParticleDataGroup:2022pth}:
\begin{equation}
\label{eq:mW_quad}
    \mW^2=\bigg(\Eell+\sqrt{(\ptmiss)^2+\pZnu^2}\bigg)^2-\big(\pTVECell+\ptvecmiss\big)^2-\big(\pZell+\pZnu\big)^2.
\end{equation}
Here, \pZell is the $z$ component of the charged-lepton momentum and \Eell is its energy.
Two possible solutions for \pZnu can be obtained from Eq.~\eqref{eq:mW_quad}:
\begin{equation}
\label{eq:pz_Nu}
    \pZnu=\frac{\Lambda\pZell}{(\pTell)^2} \pm \frac{1}{(\pTell)^2}\sqrt{\Lambda^2\pZell^2-(\pTell)^2\big[\Eell^2 (\ptmiss)^2-\Lambda^2\big]},
\end{equation}
with $\Lambda=\mW^2/2+\pTVECell\cdot\ptvecmiss$.

The finite resolution of \ptmiss can lead to negative values in the radical of Eq.~\eqref{eq:pz_Nu}, giving rise to complex solutions. In the case of real solutions, the one with the smaller magnitude is retained~\cite{CDF:2009itk, D0:2009isq}. This choice is found to have higher accuracy of the inferred values of \pZnu when compared to the true values in simulated events. If complex solutions are obtained, the radical in Eq.~\eqref{eq:pz_Nu} is set to zero, and the value of \pTnu satisfying Eq.~\eqref{eq:mW_quad} and with the smallest $\abs{\Delta\varphi}$ with respect to \ptmiss is chosen.

This reconstruction method, however, leads to a softer reconstructed \ptmiss distribution compared to the true \ptmiss in the simulation. This results in a bias in the reconstructed \mt distribution, which is one of the reasons that the mass extraction needs to be calibrated a posteriori. The value of the extracted \mt from the final fit, when applied to a sample of simulated $t$-channel single top quark and \ttbar simulations with a given
\mtmc, is plotted for a range of \mtmc values, and fitted with a linear dependence~\cite{CMS:2021jnp}. The uncertainty in the calibration is then propagated to the final result as an additional systematic uncertainty~\cite{CMS:2017mpr,CMS:2021jnp}.

\subsubsection{Top quark mass extraction}
\label{sec:mt_fit_ST}

The primary challenge in measuring \mt in single top quark events lies in controlling the large irreducible \ttbar background. Improved analysis techniques, such as multivariate and likelihood approaches, have contributed to significant reduction of the impact of the \ttbar background and to improvement of precision of single top quark mass measurements~\cite{CMS:2021jnp}. The main changes with respect to the corresponding \Run1 analysis are summarised in Table~\ref{tab:RunI_vs_RunII}. In this section, the main aspects of such improvements are discussed.

\begin{table}[!ht]
\centering
\topcaption{%
    Advancement in analysis strategies between \Run1~\cite{CMS:2017mpr} and \Run2~\cite{CMS:2021jnp} measurements of \mt in single top events.
    Primary improvements that resulted in a higher precision in the \Run2 measurement are indicated by boldface type.
}
\label{tab:RunI_vs_RunII}
\renewcommand{\arraystretch}{1.1}
\cmsTableShrink{\begin{tabular}{lll}
    & \Run1 & \Run2 \\ \hline
    Final state & \mjets  & \mjets and \textbf{\ejets} \\[\cmsTabSkip]
    \multirow{3}{*}{Strategy} & Cutoff-based: & \bf{Multivariate:} \\
    & untagged jet $\abseta>2.5$ & \textbf{Boosted decision tree} (BDT) per lepton flavour \\
    & \PGm charge${}=+1$ & \bf{Any lepton charge} \\
    & & \bf{Optimised thresholds on BDT responses} \\[\cmsTabSkip]
    Fit observable & Reconstructed \mt ($m_{\PGm\PGn\PQb}$) & \zetadefinition \\[\cmsTabSkip]
    Signal and background norm. & No constraints & \bf{Constrained in final fit} \\[\cmsTabSkip]
    \multirow{2}{*}{QCD multijet background} & Absorbed into EW (\Vjets and $\PV\PV$) & Subtracted from data before final fit; \\
    & category during final fit & separate systematic uncertainty for its modelling\\[\cmsTabSkip]
    Fit model validation & Using events with \PGm charge${}=-1$ & Using orthogonal region based on the BDT values \\
\end{tabular}}
\end{table}

\begin{figure}[!ht]
\centering
\includegraphics[width=0.48\textwidth]{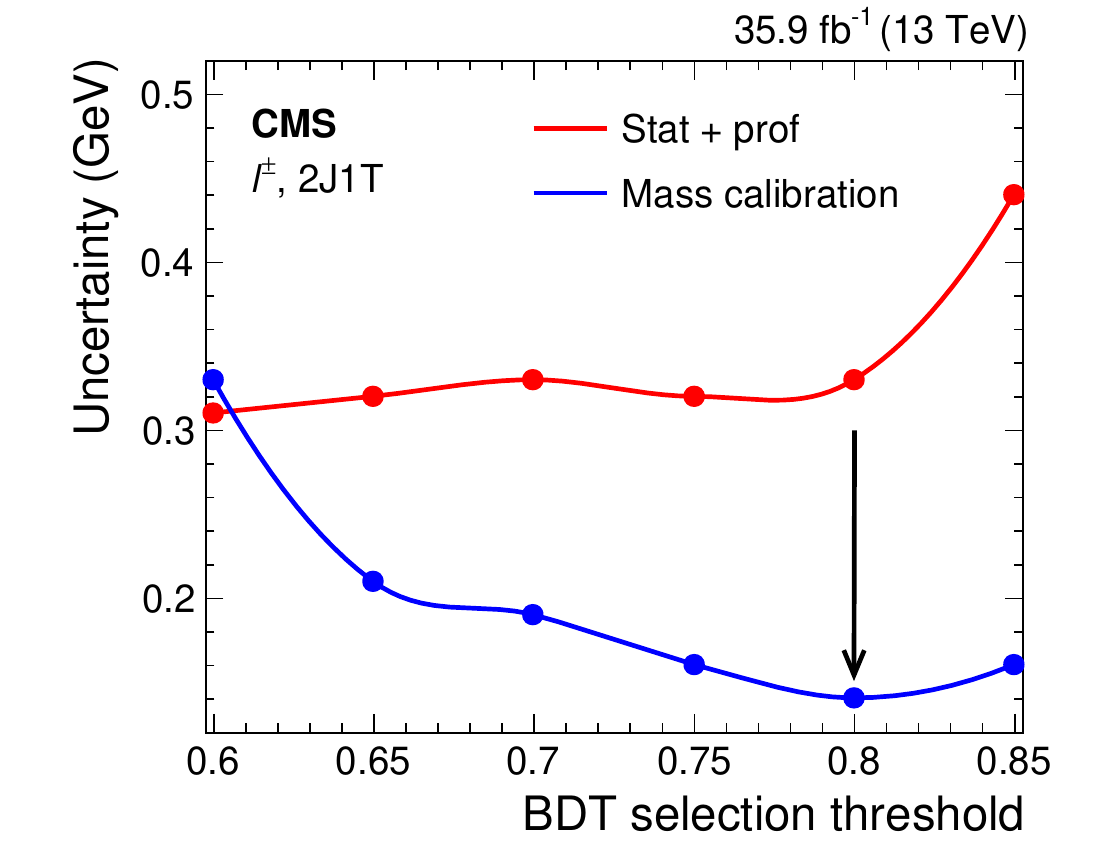}
\caption{%
    The uncertainty in \mt from the statistical and profiled systematic components (red) and uncertainty in the \mt calibration (blue) as a function of the cutoff on the BDT score.
    Figure taken from Ref.~\cite{CMS:2021jnp}.
}
\label{fig:bdtPerf}
\end{figure}

In the analysis of Ref.~\cite{CMS:2021jnp}, a boosted decision tree (BDT) is trained in each lepton flavour in the 2J1T event category in order to separate $t$-channel single top quark production from a combination of other top quark (\ttbar, \tW, and $s$-channel), EW, and QCD multijet processes. A minimal set of observables that provide good discrimination power while being loosely correlated with the reconstructed \mt is used in the BDT training~\cite{CMS:2021jnp}. The correlation between the BDT score and the reconstructed \mt is found to be 13\%, which ensures that the selection based on the BDT score does not significantly affect the reconstructed mass spectrum. The value of the BDT cutoff that minimises the calibration uncertainty mentioned in Section~\ref{sssec:topreco} is used in the analysis (Fig.~\ref{fig:bdtPerf}). This cutoff corresponds to an expected signal purity of 65 (60)\% in the muon (electron) channels.

\begin{figure}[!ht]
\centering
\includegraphics[width=0.48\textwidth]{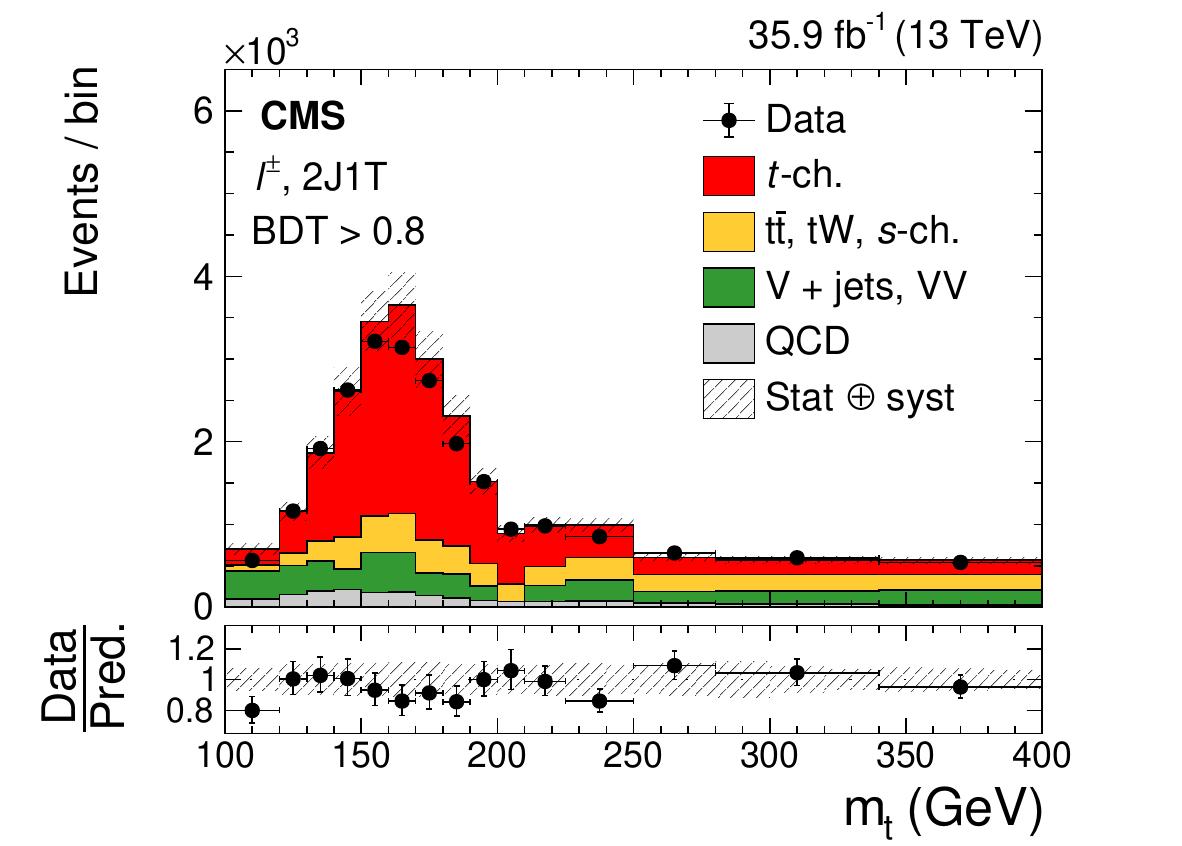}%
\hfill%
\includegraphics[width=0.48\textwidth]{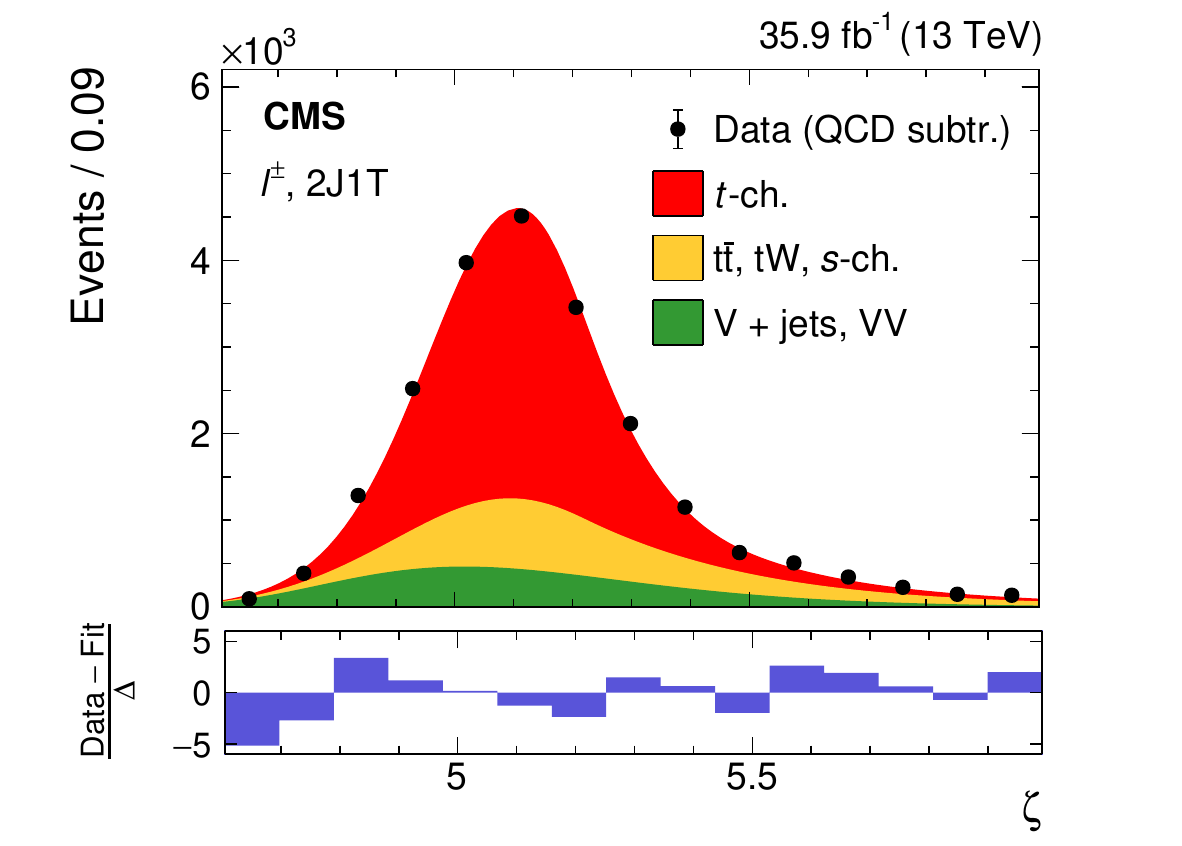}
\caption{%
    Data-to-simulation comparison of the reconstructed top quark mass (left) and postfit \zetadefinition (right) distributions after BDT selection. The lower panel in the left plot shows the data-to-simulation ratios for each bin, while the lower panel in the right plot shows the normalised residuals or pulls, determined using the bin contents of the data  distributions (after background QCD subtraction) and the $F(\zeta)$ values evaluated at the centre of the bins.
    Figures taken from Ref.~\cite{CMS:2021jnp}.
}
\label{fig:recoMass}
\end{figure}

The asymmetric shape of the reconstructed \mt distribution (Fig.~\ref{fig:recoMass}, left) makes it challenging to obtain an accurate analytic description of signal and background shapes, which is desirable when the position of the peak of a distribution has to be determined. This can be solved by introducing the variable \zetadefinition, which exhibits a more symmetric distribution around the peak (Fig.~\ref{fig:recoMass}, right).
A simultaneous maximum likelihood fit is performed with the $\zeta$ distributions obtained from the muon and electron channels.
The fit is carried out separately for a positively charged lepton (\Pellp), negatively charged lepton (\Pellm), as well as inclusive in lepton charge (\Pellpm) in the final state. The dependence of the fitted distributions on \mt is taken into account in the fit for both the signal and the \ttbar background.
The estimated QCD multijet contribution is subtracted from data before the fit in the absence of a reliable analytic shape to model this background.
A separate systematic uncertainty is assigned due to the QCD multijet background, by conservatively varying its per-bin contribution independently by 50\%.
The binned $\zeta$ distribution, obtained after the QCD background subtraction, is parameterised with an analytic function $F_{\Pell}(\zeta)$ for each lepton flavour ($\Pell=\PGm$ or \Pe).
The total likelihood is given by
\begin{equation}
    \mathcal{L}_{\text{tot}}=\prod_{\Pell=\PGm,\Pe}\mathcal{L}_{\Pell}
    \quad\text{with}\quad
    \mathcal{L}_{\Pell}=\prod_{i,j}\mathcal{P}\Big[N^{\text{obs}}_{i,\Pell}|F_{\Pell}(\zeta;\zeta_0,f_j)\Big]\Theta(f_j),
\end{equation}
where $i$ is the bin index, $\zeta_0$ represents the value of $\zeta$ corresponding to the  true value of \mt, $\mathcal{P}$ denotes the Poisson probability of the analytic model, $F_{\Pell}(\zeta;\zeta_0,f_j)$, to describe the observed $\zeta$ distribution, and $\Theta$ represents penalty terms for the normalisation parameters $f_j$.
These parameters are defined for the rates of various processes denoted by $j$, namely $t$-channel signal, \ttbar, and EW backgrounds, as
\begin{equation}
\label{eq:sf}
    f_j=\frac{\njobs}{\njexp},
    \quad
    j\in\big\{\text{$t$-ch.},\,\ttbar,\,\text{EW}\big\},
\end{equation}
where \njobs (\njexp) is the observed (expected) yield for the process $j$.
The function $F_{\Pell}(\zeta;\zeta_0,f_j)$ is then expressed as
\begin{equation}
    F_{\Pell}(\zeta;\zeta_0,f_j)=f_{\text{sig}}F_{\text{sig}}(\zeta;\zeta_0)+f_{\ttbar}F_{\ttbar}(\zeta;\zeta_0)+f_{\text{EW}}F_{\text{EW}}(\zeta),
\end{equation}
where $F_{\text{sig}}$, $F_{\ttbar}$, and $F_{\text{EW}}$ represent the analytic shapes for the signal, \ttbar, and EW background, respectively.

The $F_{\text{sig}}$ shape is described by a sum of an asymmetric Gaussian ($\zeta_0$) function convolved with a Landau distribution to account for asymmetry at higher $\zeta$, while the $F_{\ttbar}$ shape is modelled by a Crystal Ball function~\cite{CrystalBallRef}.
The \tW and $s$-channel single top quark processes are absorbed into the dominant \ttbar component.
The $F_{\text{EW}}$ shape comprises contributions from the \Wjets, \Zjets, and diboson processes and is modelled with a Novosibirsk function~\cite{Ikeda:1999aq}.
The parameter $\zeta_0$ is then treated as a free parameter of the fit, and is used to directly extract the fitted \mt.
Other parameters that alter the analytic shapes of the signal and background models are fixed to their estimated values from simulated events.
around their estimated values and are considered as sources of systematic uncertainties.
The parameters $f_{\text{sig}}$, $f_{\ttbar}$, and $f_{\text{EW}}$ are constrained in the fit
within their corresponding uncertainties of 15, 6, and 10\%, respectively.
The postfit $\zeta$ distributions for the \Pellpm case are shown in Fig.~\ref{fig:recoMass} (right).
The fit model described above is validated in a control sample obtained using an orthogonal cutoff in the BDT score.

\subsubsection{Systematic uncertainties and results}

All relevant sources of systematic uncertainties described in Section~\ref{sec:extraction} are considered. Similarly to the measurements in \ttbar events, the dominant sources of systematic uncertainties are those related to the JES, the signal modelling,  the colour reconnection, and \PQb quark hadronisation model. The largest impact originates from the JES calibration, and can be attributed to the requirement of a jet in the endcap region of the detector. In fact, the jet energy calibrations are known to have large uncertainties in the endcap regions, because of their coarse granularity~\cite{Chatrchyan:2011ds}.

In the \Run2 simulation, the models of CR (Section~\ref{sec:mcsetup}) have evolved in sophistication, as compared to those used in \Run1 analyses, and correspond to larger estimated uncertainties. The uncertainty associated with the \PQb quark hadronisation is also increased with respect to \Run1, since alternative fragmentation functions are considered (Section~\ref{sec:mcsetup}).

Similar to the case of the \ttbar analyses, the impact due to the possible mismodelling of the signal process is determined by considering the variation of parton-shower and matrix element scales, and by varying the PDF within uncertainties, for which \NNPDF{3.0} NLO set~\cite{NNPDF:2014otw} is used. In addition, \NNPDF{3.0} sets with the value of the strong coupling constant changed from the default value 0.118 to 0.117 and 0.119 are evaluated and the observed mass difference is added in quadrature. In the case of single top quark, the matrix-element renormalisation and factorisation scales are set to a nominal value of $\mt=172.5\GeV$, and are varied up and down by a factor of two.

{\tolerance=1200
As a cross check, the value of \mt is also extracted using alternative MC models for the parton shower (\HERWIGpp), the matrix element generator (\MGvATNLO), the flavour scheme, and the underlying event tune. Resulting changes in the value of \mt are found to be covered by the signal modelling uncertainties used in the fit.
\par}

\begin{figure}[!ht]
\centering
\includegraphics[width=\textwidth]{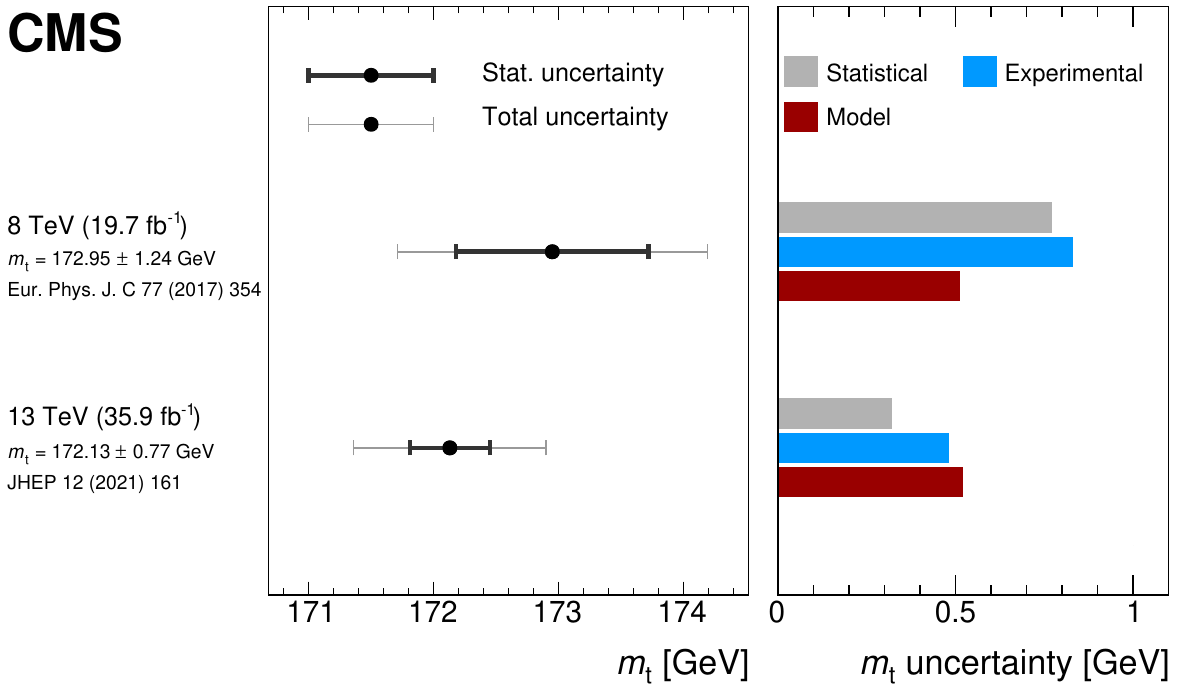}
\caption{%
    Summary of \mt measurements in single top quark events.
    The left panel shows the measured value of \mt (marker) with statistical (thick bars) and total (thin bars) uncertainties. In the case of the 13\TeV measurement~\cite{CMS:2021jnp}, the statistical component of the uncertainty includes contributions from the statistical and profiled systematic uncertainties.
    The right panel displays a breakdown of contributing uncertainty groups and their impact on the uncertainty in the measurement.
    The figure is compiled from Refs.~\cite{CMS:2017mpr,CMS:2021jnp}.
}
\label{fig:mtComp}
\end{figure}

The fit in the \Pellpm inclusive channel yields
\begin{equation}
\label{eq:mass_incl}
    \mt=172.13\,^{+0.76}_{-0.77}\GeV,
\end{equation}
resulting in the first \mt measurement in the $t$-channel with sub-\GeVns precision. The result is consistent with the CMS 8\TeV measurement in single top quark events~\cite{CMS:2017mpr}, as shown in  Fig.~\ref{fig:mtComp}. Thanks to the improvements in the analysis techniques, the larger data set, and the inclusion of the electron channel in the fit, the \Run2 measurement improves the precision by about 30\% compared to the \Run1 result, despite the fact that the impact of the signal modelling uncertainties has remained mostly unchanged. Therefore, this class of measurements can benefit significantly from future advancements in the modelling of the signal process.

\subsection{Top quark-antiquark mass difference and ratio}

In quantum field theory, the equality of the mass of a particle and its antiparticle is a consequence of the \CPT theorem, according to which all Lorentz-invariant local gauge theories are invariant under a \CPT transformation~\cite{Greenberg:2002uu}. Therefore, the validity of the \CPT  theorem can be tested experimentally by measuring the mass of a particle and its antiparticle.

In CMS, the difference in mass between the top quark and the top antiquark has been probed
in measurements using top quarks produced both singly and in pairs.
Using the single top production process,
 the masses of the top quark and antiquark were independently determined by performing the fit described in Section~\ref{sec:mt_fit_ST} in the \Pellp and \Pellm final states, respectively, resulting in
\begin{equation}\begin{aligned}
    \mt&=172.62\,^{+1.04}_{-0.75}\GeV, \\
    \mtbar&=171.79\,^{+1.44}_{-1.51}\GeV,
\end{aligned}\end{equation}
in good agreement with each other and with the result of the combined-channel fit. The uncertainty in \mtbar is found to be larger due to a lower production rate of top antiquarks compared to top quarks in single top quark production in \pp collisions.

\begin{figure}[!ht]
\centering
\includegraphics[width=\textwidth]{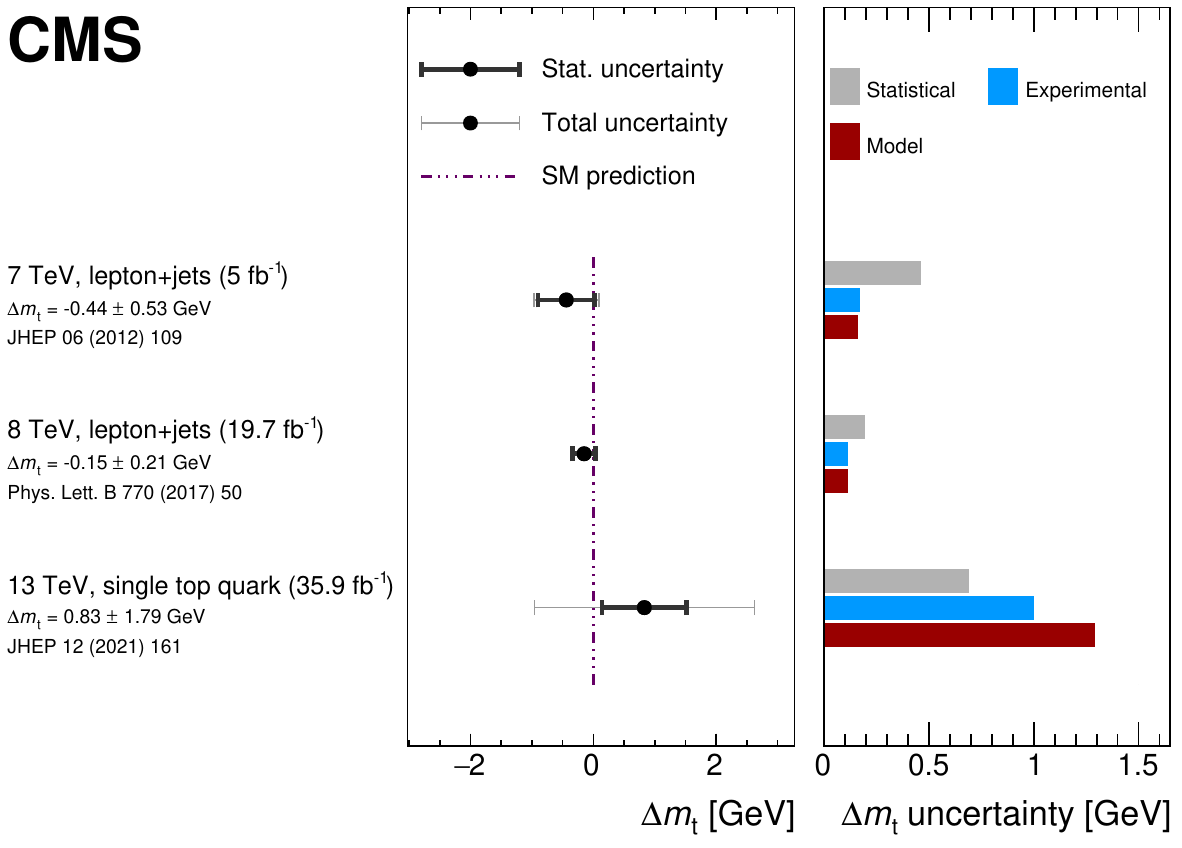}
\caption{%
    Summary of \delmt measurements in \ttbar and single top quark events.
    The left panel shows the measured value of \delmt (marker) with statistical (thick bars) and total (thin bars) uncertainties. In the case of the single top quark measurement~\cite{CMS:2021jnp}, the statistical component of the uncertainty includes contributions from the statistical and profiled systematic uncertainties.
    The right panel displays a breakdown of contributing uncertainty groups and their impact on the uncertainty in the measurement. The figure is compiled from Refs.~\cite{CMS:2012hzy, CMS:2016les, CMS:2021jnp}.
}
\label{fig:delM_Comp}
\end{figure}

The mass ratio and the mass difference are then derived accounting for the correlation between the systematic uncertainties in the two cases, resulting in:
\begin{equation}\begin{aligned}
    R_{\mt}&=\frac{\mtbar}{\mt}=0.9952\,^{+0.0079}_{-0.0104},\\
    \delmt&=\mt-\mtbar=0.83\,^{+1.79}_{-1.35}\GeV.
\end{aligned}\end{equation}
The estimated values of $R_{\mt}$ and \delmt are consistent with unity and zero, respectively, within uncertainties, showing no evidence for violation of \CPT invariance. In Fig.~\ref{fig:delM_Comp}, the result for \delmt is compared to those of previous CMS measurements in \ttbar events~\cite{CMS:2012hzy, CMS:2016les}, which were based on a modified ideogram analysis method in the lepton+jets channel, allowing \mt and \mtbar to have different values, and separating the event samples using the lepton charge. The results in \ttbar events are of better precision compared to single top quark results. All measurements of \delmt are compatible with zero. Currently, the most stringent test of \CPT invariance is obtained from the measurements of the antiproton to proton mass ratio in so-called Penning-trap experiments~\cite{BASE:2022yvh, Cheng:2022omt}. However, the CMS 8\TeV result from \ttbar events~\cite{CMS:2016les} remains the most precise measurement of the
mass ratio for the top quark to antiquark.

\section{Indirect extractions of the top quark mass}
\label{sec:indirect}

An alternative to the direct measurements is the extraction of \mt from the measured parton-level cross section of \ttbar pair production\footnote{Here, \ttbar includes \ttbarjet production.}, \stt.
In this indirect approach, the \mt dependence of \stt is used to determine \mt by comparing the measured parton-level \ttbar cross sections to the corresponding theoretical predictions, using \mt defined in a given top quark mass renormalisation scheme.

The theoretical predictions for \stt, which require NLO or higher precision, describe the production of the on-shell top quark and antiquark and are inclusive with respect to other radiation in the event, therefore an unfolding procedure from the detector to the parton level needs to be employed in the experimental data analysis.
First measurements of this kind were performed at the Tevatron~\cite{D0:2011hwd} using the inclusive \stt. In this approach, \mt can in principle be determined in any renormalisation scheme, but suitable choices of mass schemes are tied to convergence properties of the respective prediction, in close analogy to suitable renormalisation scale choices of the strong coupling \alpS. The values of \mt, obtained by using this approach, are less precise than the direct measurements. This is because the \stt is more sensitive to the hard production mechanism and, in general, less sensitive to the kinematic dependence on \mt than the observables in direct measurements. These indirect \mt extractions are affected by very different systematic uncertainties, and therefore represent important alternatives to direct \mtmc determinations.

The first extraction of \mtp using inclusive \stt in proton-proton collisions at the LHC was performed by the CMS Collaboration at \sqrtseq{7}~\cite{CMS:2014rml}. This analysis identified a general issue of such determinations, that is the further dependence of the \stt prediction on \asmz and the PDFs. Another problem was represented by the remaining dependence of the measured \stt on the value of \mtmc, inherited from the extrapolation of the fiducial measurement to the full phase space, which relies on the simulation of the final state. These problems were addressed by the CMS Collaboration in a series of follow-up studies~\cite{CMS:2018fks,CMS:2022emx, CMS:2019esx,CMS:2021yzl}, where novel experimental analyses techniques have been developed, and specific observables in \ttbar and \ttbarjet production have been measured.

To assure the highest purity of the \ttbar signal, most of the \stt measurements used to extract \mt have been performed in the dilepton channel. The experimental techniques of the cross section measurements have constantly been improved. More recent measurements use template fits to multi-differential distributions in the selected final state, taking into account features of the topology of the \ttbar signal and the background. As a result, the systematic uncertainties were further reduced and correlations between systematic uncertainties were treated consistently, resulting in a significantly improved experimental precision of the cross section measurements~\cite {CMS:2016yys,CMS:2018fks}. Since the first \mt measurement in CMS, also the technique of reconstructing the \ttbar pairs in dilepton final states have experienced significant developments. As detailed in Section~\ref{sec:extraction}, the determination of the momenta of the two neutrinos in the dilepton channel required assumptions on the masses of the \PW boson and the top quark. Releasing these requirements in \mt measurements has triggered methodical improvements, such as the so-called loose kinematic reconstruction and the DNN-based reconstruction of \ttbar pairs, discussed in details in Section~\ref{sec:kinreco}.

Further, novel observables in top quark production and decay have been explored, as suggested by theoretical investigations. The inverse of the invariant mass of the \ttbarjet system, $\rho$, in events where the \ttbar pair is produced with an associated energetic jet, and the invariant mass of the \PQb quark and the lepton from the \PW boson decay, \mlb~\cite{Biswas:2010sa}, exhibit a strong dependence on \mtp. In particular, by considering the \mlbmin distribution in the \stt measurement, its dependence on \mtmc is used for the simultaneous extraction of \stt and \mtmc. This way, the remaining dependence of \stt on \mtmc is mitigated and one of the major problems of \mt extractions via inclusive or differential \stt measurements is resolved. This approach made it possible to extract \mtp and \mtmt without an additional uncertainty related to the prior assumption of \mtmc in inclusive and differential measurements, leading to the first experimental confirmation of the running of the scale dependent \msbar top quark mass~\cite{CMS:2019jul}.

The 3-fold correlations of \mtp, PDFs, and \alpS in the QCD prediction of \stt was further investigated by the CMS Collaboration~\cite{CMS:2019esx} using multi-differential \stt measurements. In particular, by including the measurements of \mtt and \ytt in a comprehensive QCD analysis at NLO, the PDFs, \asmz, and \mtp could be extracted simultaneously and their correlations were demonstrated to significantly reduce. This analysis resulted in the most precise value of \mtp at NLO to that date, with simultaneously reduced uncertainty in the gluon PDF. At the same time, a low value of \asmz was obtained, in tension with the results of other measurements at the LHC. In a follow-up analysis~\cite{CMS:2021yzl}, this issue was resolved by including the CMS jet production measurements which have additional strong sensitivity to PDFs and \asmz.

In the following, the aforementioned analyses are discussed in more details, with the emphasis on the progress of analysis strategies with respect to the state-of-the-art at the time of the measurements.
The aspects of the parton-level cross section measurements, including unfolding and related uncertainties, can be found in the original publications. In each of the mentioned analyses, the extraction of \mt is performed under the assumption that the measured \ttbar cross sections are not affected by physics phenomena beyond the SM.

\subsection{Extractions of \texorpdfstring{\mtp}{mt-pole} from inclusive \texorpdfstring{\ttbar}{ttbar} cross sections}
\label{sec:inclusive}

\begin{figure}[!ht]
\centering
\includegraphics[width=0.6\textwidth]{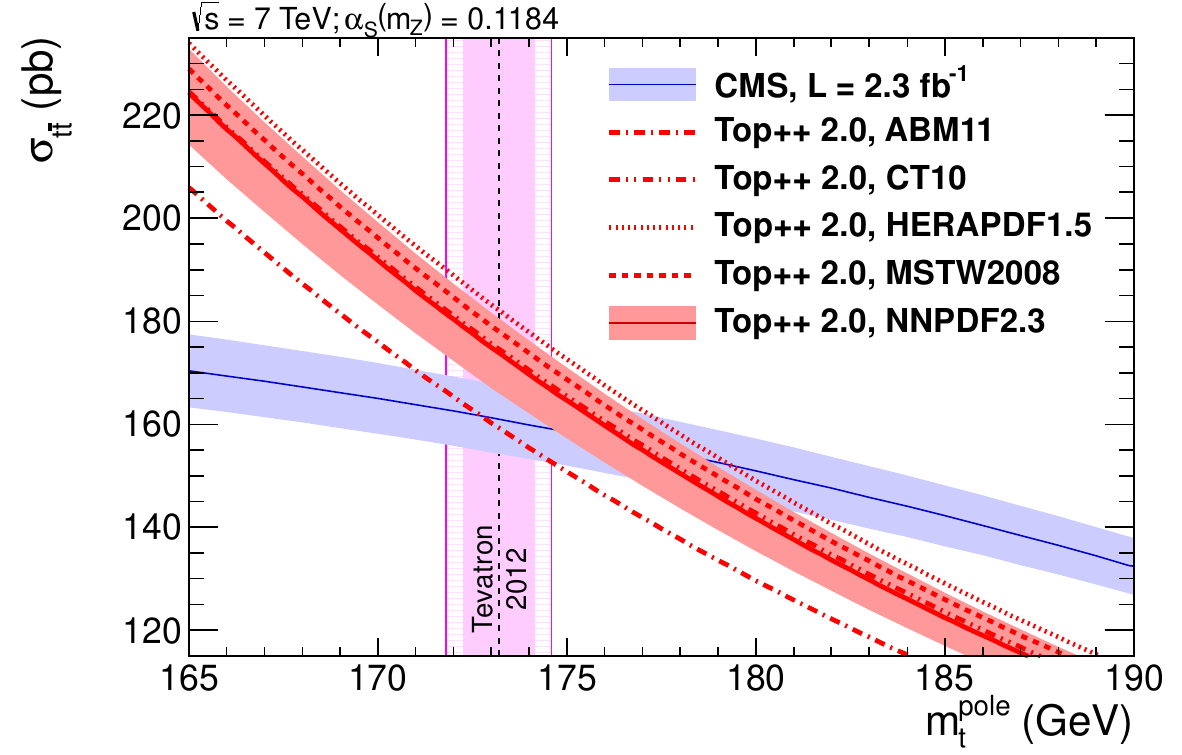}
\caption{%
    Predicted \stt as a function of the top quark pole mass, using different PDF sets (red shaded band and red lines of different styles), compared to the cross section measured by CMS assuming $\mtmc=\mtp$ (blue shaded band). The uncertainties in the
    measured \stt as well as the scale and PDF uncertainties in the prediction with \NNPDF{2.3}~\cite{Ball:2012cx} are illustrated by the filled band. The \mtmc result obtained in direct measurements to that date is shown as hatched area. The inner (solid) area of the vertical band corresponds to the quoted experimental uncertainty in \mtmc, while the outer (hatched) area additionally accounts for a possible difference between this value and \mtp.
    Figure taken from Ref.~\cite{CMS:2014rml}.
}
\label{indirect:fig1}
\end{figure}

In the CMS work~\cite{CMS:2014rml}, the predicted inclusive \stt at NNLO+NNLL~\cite{Czakon:2011xx} was compared to the most precise single measurement at \sqrtseq{7} at CMS to that date~\cite{CMS:2012exf}, using an integrated luminosity of 2.3\fbinv of the data in the dilepton decay channel. The values of \mtp and, alternatively, of \asmz were obtained. In Fig.~\ref{indirect:fig1}, the dependence of the predicted \stt cross section on the value of \mtp is shown.

Besides the value of \mtp, the predicted cross section depends on the value of \alpS. A simultaneous extraction of \mt and \asmz from the inclusive \stt alone is not possible since both parameters alter the predicted \stt in such a way that any variation of one parameter can be compensated by a variation of the other. In cross section calculations, \asmz appears not only in the expression for the parton-parton interaction but also in the QCD evolution of the PDFs. Varying the value of \asmz in the \stt calculation therefore requires a consistent modification of the PDFs. Consequently, to extract the value of \mtp, a choice of the PDFs and of \asmz has to be made. The interplay of \mt, \asmz, and the proton PDFs in the predicted \stt was studied for the first time by using 5 different PDF sets available to that date at NNLO, and for each set a series of different choices of \asmz was considered.

The cross section was measured to be $\stt=161.9\pm2.5\stat\,^{+5.1}_{-5.0}\syst\pm3.6\lum\unit{pb}$~\cite{CMS:2012exf} using the profile likelihood ratio method, where the minimum value of a function $-2\ln[R(\stt)]$ is determined. The ratio $R$ is composed of the likelihood functions depending on \stt and the maximum likelihood estimates of \stt, as well as the sets of nuisance parameters describing the systematic uncertainties in the measurement. The likelihoods are defined by a probability density function binned in a 2-dimensional space of jet multiplicity and the multiplicity of \PQb-tagged jets~\cite{CMS:2012exf}.  The acceptance for \ttbar and, in turn, the measured \stt depend on the value of \mtmc that is used to simulate \ttbar events. The central value of \stt is obtained by assuming $\mtmc=172.5\GeV$, while the dependence of \stt on \mtmc is studied by varying \mtmc in the MC simulation in the range 160--185\GeV and parameterised, as shown in Fig.~\ref{indirect:fig1} by a blue shaded band.

\begin{figure}[!ht]
\centering
\includegraphics[width=0.6\textwidth]{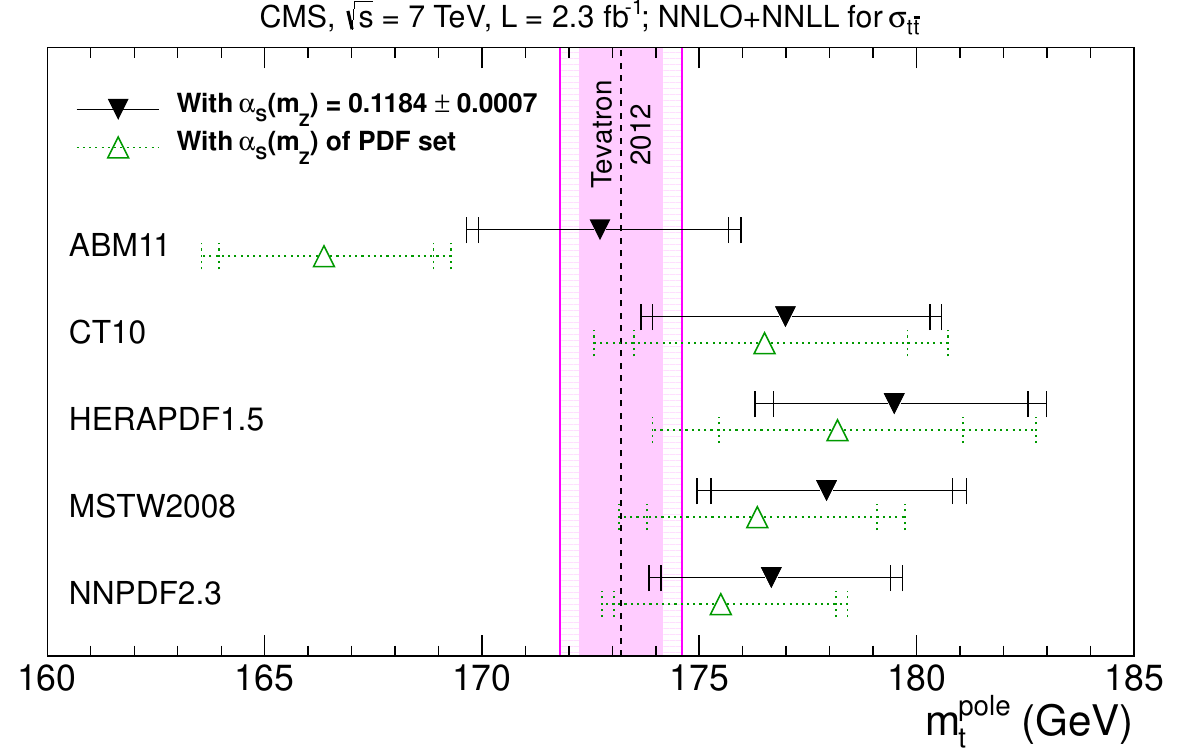}
\caption{%
    Values of \mtp obtained by using measured \stt together with the prediction at NNLO+NNLL using different NNLO PDF sets. The filled symbols represent the results
    obtained when using the world average of \asmz, while the open symbols indicate the results
    obtained with the default \asmz value of the respective PDF set. The inner error bars include the uncertainties in the measured cross section and in the LHC beam energy, as well as the PDF and scale uncertainties in the predicted cross section. The outer error bars additionally account for the uncertainty in the \asmz value used for a specific prediction. For comparison, the most precise \mtmc to that date is shown as vertical band, where the inner (solid)
    area corresponds to the original uncertainty of the direct \mt average, while the outer (hatched) area additionally accounts for the possible difference between \mtmc and \mtp.
    Figure taken from Ref.~\cite{CMS:2014rml}.
}
\label{mtp7tev}
\end{figure}

{\tolerance=1200
The extraction of \mtp was performed through the so-called probabilistic approach by maximising the marginalised posterior
\begin{equation}
    P\big(\mtp\big)=\int_{\mtp} f_{\text{exp}}\Big(\stt\big(\mtp\big)\Big) f_{\text{th}}\Big(\stt\big(\mtp\big)\Big)\,\rd{\stt}.
\end{equation}
The measured cross section and its uncertainty are represented by a Gaussian probability $f_{\text{exp}}(\stt)$.
The probability function for the predicted cross section, $f_{\text{th}}(\stt)$, was obtained through an analytic
convolution of two probability distributions, one accounting for the PDF uncertainty and the
other for scale uncertainties. A Gaussian distribution is used to describe the PDF
uncertainty. Given that no particular probability distribution is known to be adequate
for the confidence interval obtained from the variation of the factorisation, \muf, and renormalisation, \mur, scales, the corresponding uncertainty in the \stt prediction is approximated using a flat prior. The posterior $P\big(\mtp\big)$ is marginalised by integration over \stt and a Bayesian credible interval for \mtp is computed, based on the external constraint for \asmz. The results using different sets of PDF are presented in Fig.~\ref{mtp7tev}. The top quark pole mass is determined to be $\mtp=176.7\,^{+3.0}_{-2.8}\GeV$ using the theoretical prediction based on the \NNPDF{2.3} PDF~\cite{CMS:2014rml}.
The experimental and theoretical uncertainties equally contribute to the final precision of 1.7\%. The theoretical precision is limited by the PDF uncertainties (0.8\%) and the variation of the QCD scales in the theoretical prediction at NNLO+NLL (0.5\%), followed by the uncertainty in the assumption $\mtp=\mtmc$, for which 1\GeV was assumed.
This first LHC measurement of \mtp, although inferior in precision compared to the direct measurements, has set an important milestone in the extraction of the Lagrangian mass of the top quark. The correlations between \mtp, \asmz, and PDFs were for the first time quantified and the remaining dependence of \stt on \mtmc was pointed out.
\par}

In a later work~\cite{CMS:2016yys}, the analysis strategy to measure the \stt was significantly improved. The cross sections were measured through a template fit of the signal and background contributions to multi-differential distributions, binned in the multiplicity of \PQb quark jets and the multiplicity of the other jets in the event. First, the cross section in a fiducial region, \sttvis, was determined, defined by the requirements on the transverse momenta and pseudorapidities of the final-state leptons.
The expected signal and background distributions were modelled in the fit by template histograms, constructed from the simulated samples. The free parameters in the fit were \sttvis, the normalisation for different background contributions, and the nuisance parameters representing other sources of systematic uncertainties, such as the JES and the trigger efficiency. All systematic uncertainties were implemented in the likelihood as nuisance parameters with Gaussian constraints. Each systematic uncertainty was assessed individually by relevant variations in MC simulations or by varying parameter values within their estimated uncertainties in the analysis.
Each source was represented by a nuisance parameter, which was fitted together with \sttvis. The impact of theoretical assumptions in the modelling was determined by repeating the analysis and replacing the signal \ttbar simulation by dedicated simulation samples with varied parameters affecting, \eg the scales for the hard process and for matching to the parton shower, the hadronisation, the colour-reconnection, the underlying event, and PDFs.

The fiducial results were then extrapolated to obtain the value of \stt in the full phase space, by dividing \sttvis by the acceptance, determined from the \ttbar signal MC simulation. Since the acceptance depends on the theoretical model used in the MC event generator, it was parameterised as a function of the same nuisance parameters that were used for the modelling uncertainties in the binned likelihood fit of \sttvis. For the extrapolation of the fitted \sttvis to the full phase space, the full unconstrained variations of the relevant modelling uncertainties were applied.

The \stt measurements at 7 and 8\TeV centre-of-mass energies were simultaneously used to extract \mtp while the correlation between the two measurements for the systematic uncertainties was taken into account.
The cross section fit and the extrapolation to the full phase space were repeated for $\mtmc=169.5$, 172.5, and 175.5\GeV. For each case, a sample of simulated \ttbar events, generated with the corresponding \mtmc value, was used in the fit as a signal model. The dependence of the distributions used in the fit on detector effects
and model variations was evaluated individually and the parameterisation of \stt dependence on \mtmc was obtained. To express the measured dependence as a function of \mtp instead of \mtmc, an additional uncertainty in the measured cross section, \delmtpm, was evaluated by varying \mtmc by $\pm1\GeV$ and reevaluating \stt. The dependence of the \stt measurements on \mtp was modelled by Gaussian likelihoods as
\begin{equation}
    \likelihoodexp\big(\mtp,\stt\big)=\exp\Bigg[\frac{\big(\stt(\mt)-\stt\big)^2}{-2\big(\Delta^2+\delmtpm^2\big)}\Bigg],
\end{equation}
where $\Delta$ is the total uncertainty in each of the \stt measurements, considering
the measured dependence of $\stt\big(\mtp\big)$.

\begin{figure}[!ht]
\centering
\includegraphics[width=0.6\textwidth]{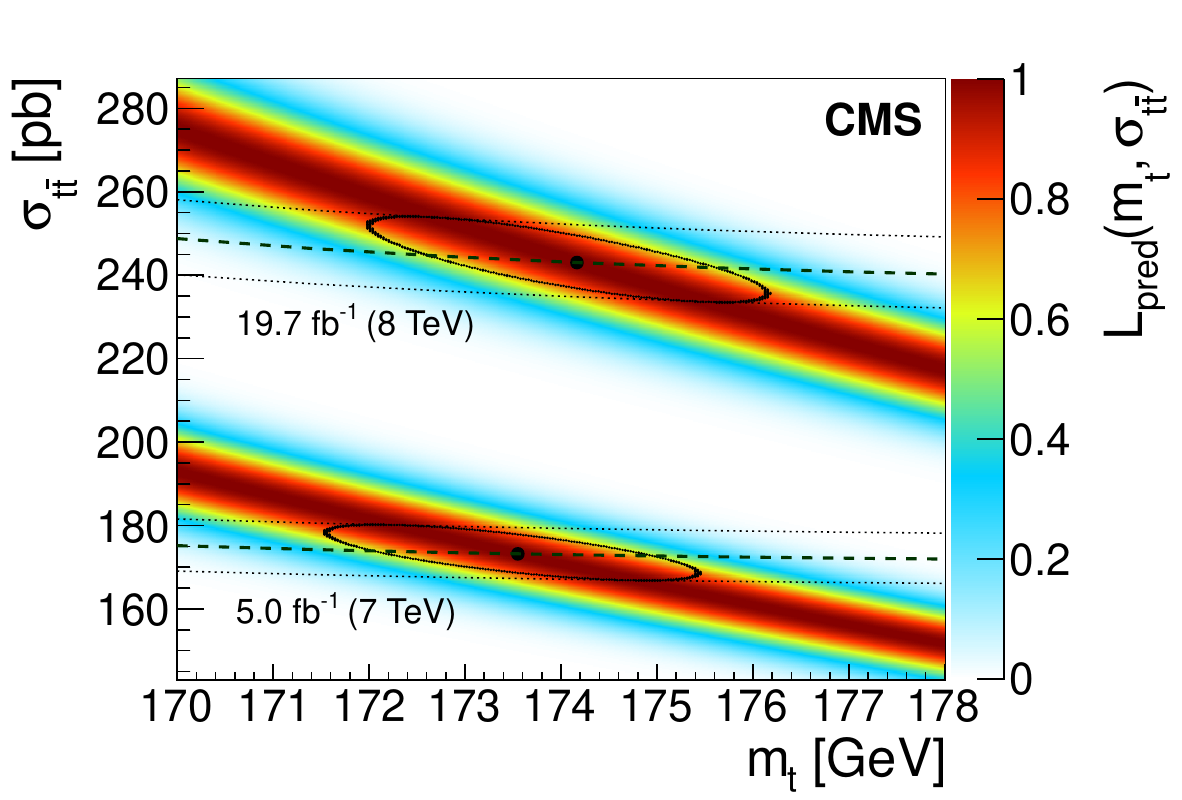}
\caption{%
    Likelihood for the predicted dependence of \stt on \mtp for 7 and 8\TeV determined with \TOPpp, using the \NNPDF{3.0} PDF set. The measured dependencies on the mass are given by the dashed lines, their $1\sigma$ uncertainties are represented by the dotted lines. The extracted mass at each value of \sqrts is indicated by a black point, with its $\pm1$  standard deviation uncertainty constructed from the continuous contour, corresponding to $-2\Delta\log(\likelihoodpred\likelihoodexp)=1$.
    Figure taken from Ref.~\cite{CMS:2016yys}.
}
\label{mtp_78tev}
\end{figure}

The predicted dependence of \stt on \mtp at NNLO+NNLL was determined with  \TOPpp~\cite{Czakon:2011xx}, employing 3 different PDF sets and $\asmz=0.118\pm0.001$. The predicted \stt was represented by an asymmetric Gaussian function with width $\Delta_{p,\pm}$, comprising uncertainties in PDF, \asmz, and the uncertainty in the LHC beam energy, summed in quadrature. This function is convolved with a box function to account for the uncertainty arising from variations of \mur and \muf in the theoretical prediction,
\begin{equation}
    \likelihoodpred\big(\mtp,\stt\big)=\frac{1}{C\big(\mtp\big)}\Bigg(\erf\Bigg[\frac{\stth\big(\mtp\big)-\stt}{\sqrt{2}\Delta_{p,+}}\Bigg]-\erf\Bigg[\frac{\sttl\big(\mtp\big)-\stt}{\sqrt{2}\Delta_{p,-}}\Bigg]\Bigg).
\end{equation}
Here, \stth and \sttl denote the upper and lower predicted cross section values, respectively,
from variations of \mur and \muf. The normalisation factor $C\big(\mtp\big)$
assures that $\max(\likelihoodpred)=1$ for any fixed \mtp.
The value of \mtp is extracted by using the product of the two likelihoods, \likelihoodexp and \likelihoodpred, maximised simultaneously with respect to \mtp and \stt.
The likelihoods for the predicted \stt obtained using the \NNPDF{3.0} PDF set, and the
measurement of \stt at $\sqrts=7$ and 8\TeV as a function of \mtp are shown in Fig.~\ref{mtp_78tev}. As a result, the value of $\mtp=173.8\,^{+1.7}_{-1.8}\GeV$ was obtained~\cite{CMS:2016yys}, with the uncertainty of 1\%.

From the experimental perspective, the remaining dependence of \stt on the assumed \mtmc and the related additional uncertainty in \mtp seemed yet unsatisfactory.
This issue was addressed in later analyses by introducing novel observables in \ttbar production, sensitive to \mtmc, into the template fit in the \stt measurement.

\subsection{Mitigating the dependence of the measured cross section on \texorpdfstring{\mtmc}{mtMC}}
\label{sec:mlb_mtmt}

Beyond the inclusive cross section, the top quark mass can be extracted from \mt-sensitive kinematic distributions. However, the reliability of the precision of the respective results obtained using parton-shower event generators suffers from the aforementioned \mtmc interpretation. Alternative ways to estimate theoretical uncertainties in the description of relevant kinematic distributions and specific observables were investigated. Several kinematic distributions, typically involving top quark decay products were suggested, \eg in  Ref.~\cite{Biswas:2010sa}. The NLO QCD corrections to \ttbar production and decay considering the spin correlations became available at the same time, \eg Refs.~\cite {Melnikov:2009dn, Bernreuther:2010ny}. In particular, the higher order corrections  were important since those allow the distinction between the mass parameters defined in different renormalisation schemes. In Ref.~\cite{Biswas:2010sa}, several observables relevant for the \mt extraction at LO and NLO QCD were studied, and their sensitivity to input parameters was investigated. One of the most promising observables was found to be the invariant mass of the lepton and the \PQb jet, \mlb, in dilepton \ttbar events. Considering the top quark decay \ttobW, $\PW\to\Pell\PGn$ at LO and neglecting the masses of leptons and \PQb quark,
\begin{equation}
    \mlb^2=\frac{\mt^2-\mW^2}{2}(1-\cos\theta_{\Pell\PQb}),
\end{equation}
so the dependence of \mlb on \mt is precisely known, given a value of the \PW boson
mass \mW. Here, $\theta_{\Pell\PQb}$ is the angle between the lepton and the \PQb quark in the  \PW boson rest frame. At maximum, the value of \mlb approaches $\sqrt{\smash[b]{\mt^2-\mW^2}}$.
Experimentally, there is an ambiguity in which of the two \PQb jets should be combined with the chosen lepton of a certain charge. Therefore, the lepton is associated with the \PQb jet resulting in the smallest value of \mlb, \mlbmin. The \mlbmin distribution was shown to be under good theoretical control, but the way higher-order effects are considered appeared important~\cite{Moch:2014tta, Heinrich:2013qaa}. For the experimental extraction of \mt using \mlbmin, however, the respective NLO calculation would need to be implemented in the MC simulations used in the measurement of \stt. In the absence of those, \mlbmin appeared to be a promising observable in the determination of \mtmc and in the mitigation of the \mtmc-dependence of the \stt measurement.

\begin{figure}[!ht]
\centering
\includegraphics[width=0.48\textwidth]{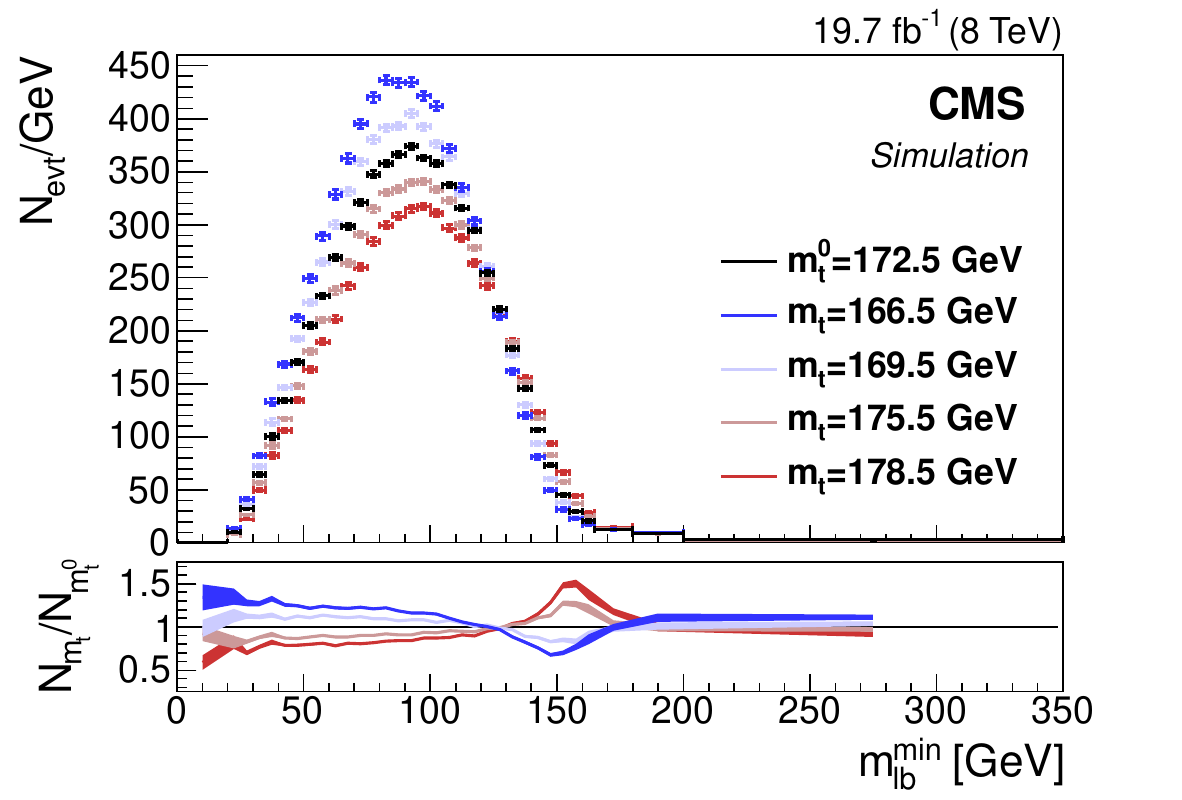}%
\hfill%
\includegraphics[width=0.48\textwidth]{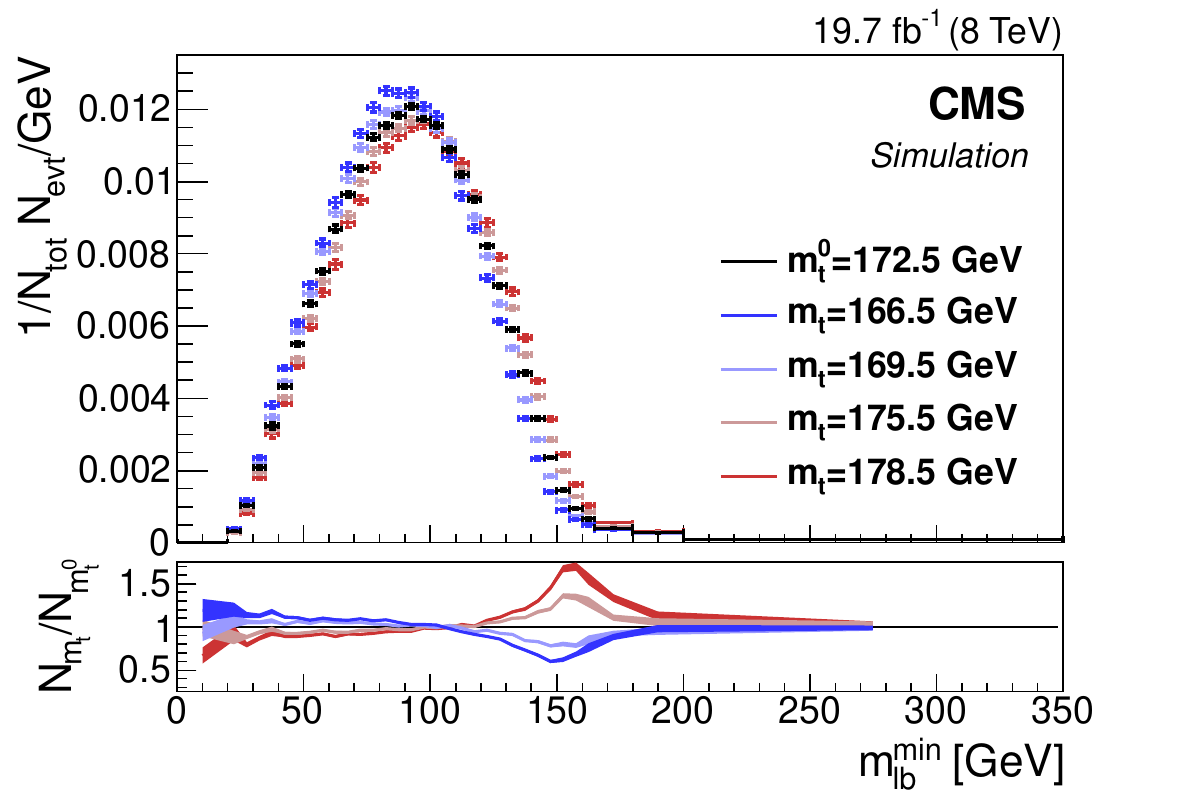}
\caption{%
    Absolute (left) and shape (right) distributions of \mlbmin for \ttbar production at the LHC at \sqrtseq{8} after detector simulation and event selection in the $\Pe\PGm$ channel. The central prediction (black symbols) is obtained at the value of \mtmc of 172.5\GeV, denoted as $\mt^0$. Predictions assuming different \mtmc values are shown by different colours.
}
\label{mlb_mt_dependence}
\end{figure}

The \mlbmin distribution provides strong sensitivity to the choice of \mtmc at values of \mlbmin close to the top quark mass, as demonstrated in Fig.~\ref{mlb_mt_dependence}.

A generic approach to measure any observed distribution $\xi$ sensitive to \mt in a particular renormalisation scheme without any prior assumptions on \mtmc, or its relation to \mt, was suggested in Ref.~\cite{Kieseler:2015jzh}. The method employs a simultaneous likelihood fit of \mtmc and $\xi$, comparing an observed distribution in data to its MC prediction. In later CMS analyses, \mlbmin is chosen as such an observable.

In the view of precision measurements of \mt, the fundamental issue of \mtp is the infrared-sensitivity, also known as the renormalon problem, which leads to poor perturbative behaviour. Alternative renormalisation schemes~\cite{Dowling:2013baa, Hoang:2017suc} were explored in the context of \mt measurements at the LHC, and better perturbative convergence by using the \msbar scheme was demonstrated~\cite{Dowling:2013baa}. Using the higher-order calculations for inclusive and differential \stt in the \msbar scheme, extraction of the running mass of the top quark, \mtmt, and of its scale-dependence becomes possible.

In the CMS analysis~\cite{CMS:2018fks} based on the LHC data collected at a centre-of-mass energy of 13\TeV, the top quark mass is extracted in both the on-shell and the \msbar mass schemes.
In Ref.~\cite{CMS:2018fks}, the \stt measurement was performed using a template fit to multidifferential distributions, similar to the measurement~\cite{CMS:2016yys} at $\sqrts=7$ and 8\TeV.  First, a visible \ttbar cross section \sttvis in the experimentally accessible fiducial volume is determined, using the fit to constrain the systematic uncertainties from the data.
The measured \sttvis is then extrapolated to the full phase space to obtain \stt, which introduces a residual dependence of \stt on \mtmc, due to the impact of \mtmc on the simulated detector acceptance. In contrast to previous measurements, where this dependence was determined by repeating the analysis with varied \mtmc, the approach of Ref.~\cite{Kieseler:2015jzh} is followed and \mtmc is introduced in the fit as an additional free parameter. The sensitivity to \mtmc is enhanced by introducing the \mlbmin distribution in the fit. In the simultaneous fit, \stt and \mtmc are directly constrained
from the data. The resulting \stt and its uncertainty therefore account for the dependence on \mtmc, irrespective of its physics interpretation, and are used for the extraction of \mtp and \mtmt, or alternatively, of \asmz.

While Ref.~\cite{CMS:2018fks} contains \stt measurements obtained in the \EE, \MM, and \EM channels, to minimise the impact from background, only the \EM channel was used for the simultaneous \stt and \mtmc measurement.
The templates describing the distributions for the signal and background events were taken from the simulation and their statistical uncertainty was accounted for by using pseudo-experiments. To construct the templates describing the dependence of the final-state distributions on \mtmc, separate MC simulation samples of \ttbar and \tW production were used, in which \mtmc is varied in the range 169.5--175.5\GeV.

\begin{figure}[!ht]
\centering
\includegraphics[width=0.48\textwidth]{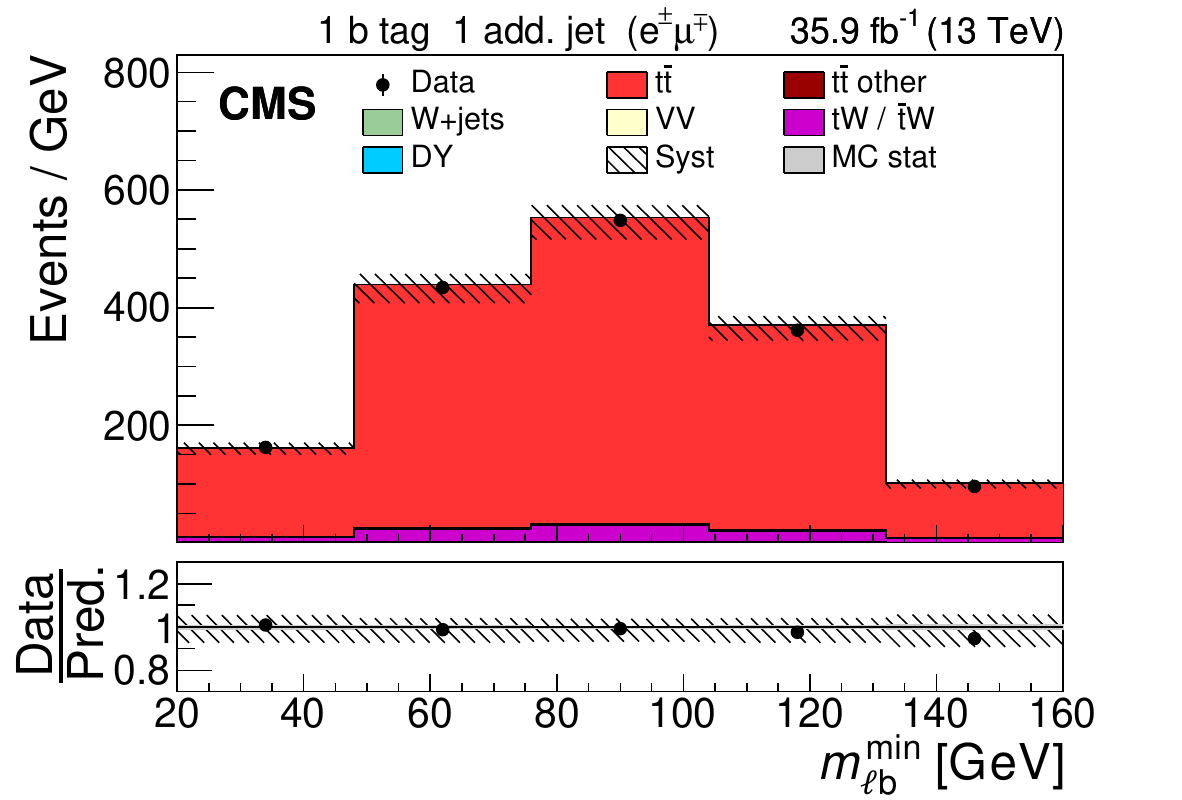}%
\hfill%
\includegraphics[width=0.48\textwidth]{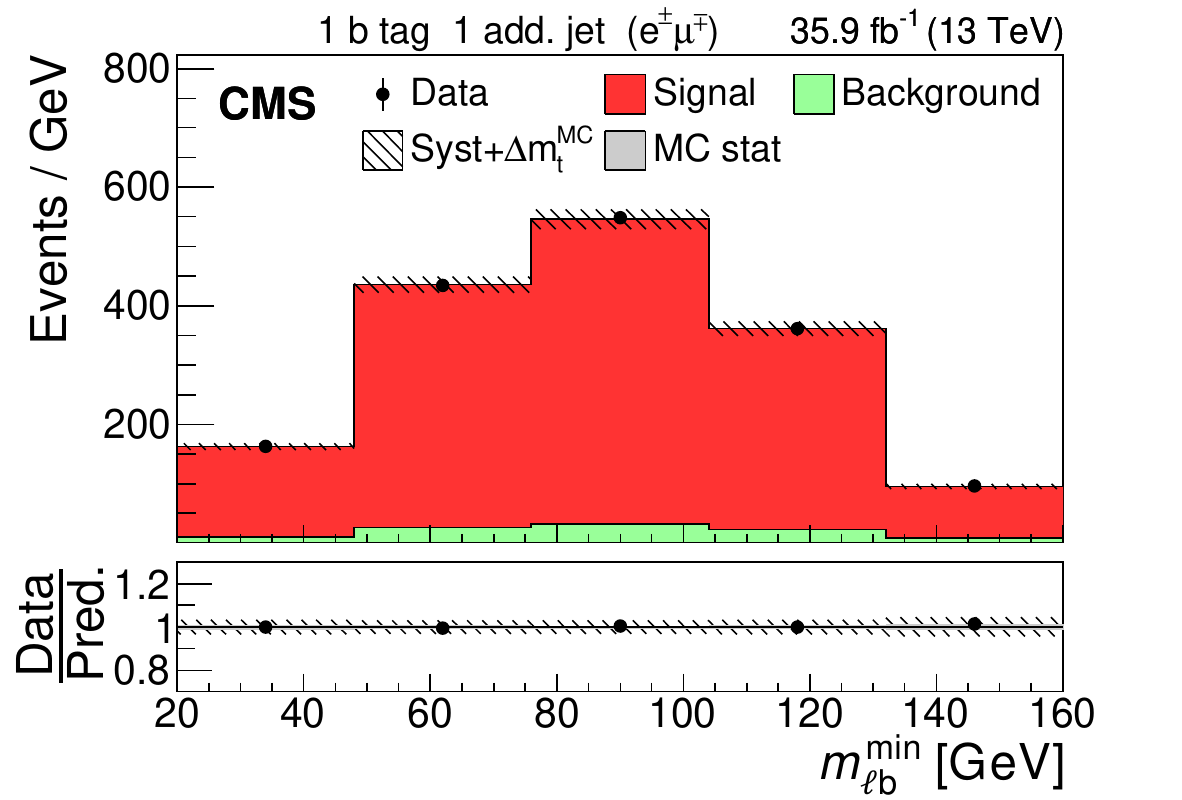}
\caption{%
    Data (points) compared to pre-fit (left) and post-fit (right) \mlbmin distributions of the expected signal and backgrounds from simulation (shaded histograms) used in the simultaneous fit of \stt and \mtmc. Events with exactly one \PQb-tagged jets are shown. The hatched bands correspond to the total uncertainty in the sum of the predicted yields. The ratios of data to the sum of the predicted yields are shown
    in the lower panel. Here, the solid grey band represents the contribution of the statistical uncertainty.
    Figures taken from Ref.~\cite{CMS:2018fks}.
}
\label{top-17-001-mlb}
\end{figure}

The fit was performed in twelve mutually exclusive categories, according to the number of \PQb-tagged jets and of additional non-\PQb-tagged jets in the event. Categorising the events by their \PQb-tagged jet multiplicity allows to constrain the efficiency to select and identify a \PQb jet. Besides \sttvis, the free parameters of the fit are the nuisance parameters \nuisancevec corresponding to the various sources of systematic uncertainty. The function $-2\ln(L)$ was minimised, with likelihood $L$ based on Poisson statistics:
\begin{equation}
    L=\prod_i \frac{\mathrm{e}^{-v_i}v_i^{n_i}}{n_i!}\prod_j\pi(\lambda_j).
\end{equation}
Here, $i$ denotes the bin of the respective final-state distribution, while $v_i$ and $n_i$ are the expected and observed number of events in bin $i$, respectively. The terms $\pi(\lambda_j)$ account for deviations of the nuisance parameters $\lambda_j$ from their nominal values according to their prior density distributions, which are assumed to be Gaussian. In the fit, the expected number of events in each bin $i$, $v_i$, is parameterised as
\begin{equation}
    v_i=s_i\big(\sttvis,\nuisancevec\big)+\sum_k b^{\text{MC}}_{k,i}\big(\nuisancevec\big),
\end{equation}
where $s_i$ is the expected number of \ttbar signal events in bin $i$, and $b_{k,i}^{\text{MC}}$ represents
the predicted number of background events in bin $i$ from a source $k$. Comparisons of the data and the prediction from the MC simulation before and after the fit are presented in Fig.~\ref{top-17-001-mlb} for the \mlbmin distribution.

\begin{figure}[!ht]
\centering
\includegraphics[width=0.8\textwidth]{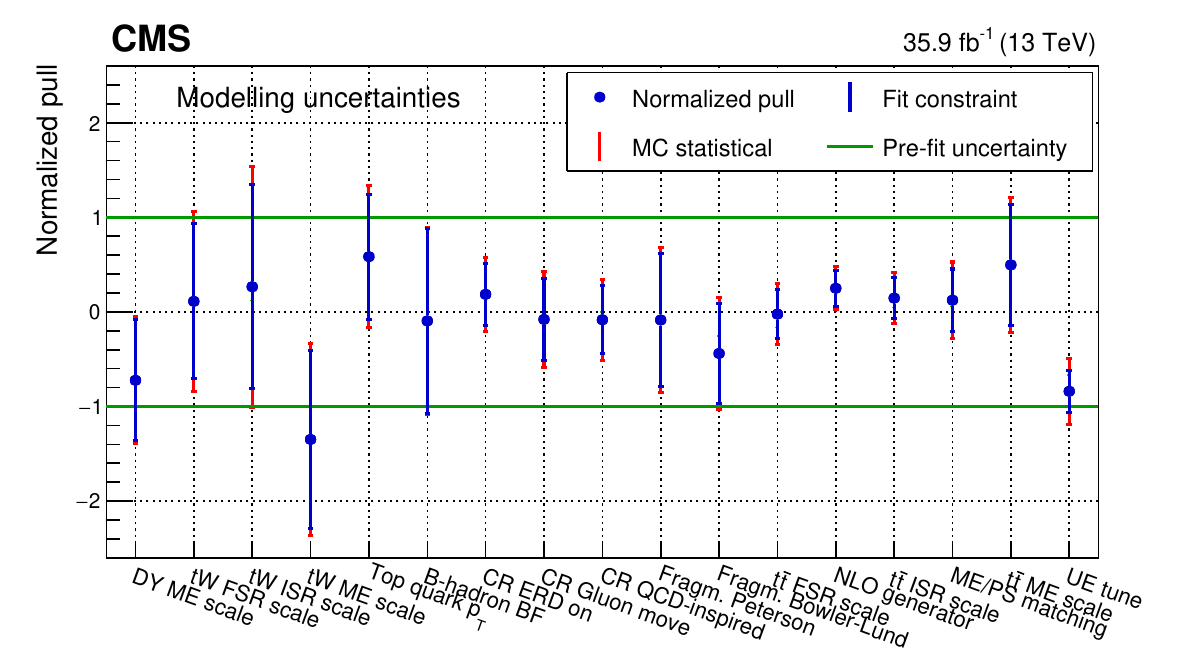}
\caption{%
    Normalised pulls and constraints of the nuisance parameters related to the modelling uncertainties for the simultaneous fit of \stt and \mtmc. The markers denote the fitted value, while the inner vertical bars represent the constraint and the outer vertical bars denote the additional uncertainty as determined from pseudo-experiments. The constraint is defined as the ratio of the post-fit uncertainty to the pre-fit uncertainty of a given nuisance parameter, while the normalised pull is the difference between the post-fit and the pre-fit values of the nuisance parameter normalised to its pre-fit uncertainty. The horizontal lines at $\pm1$ represent the pre-fit uncertainty.
    Figure taken from Ref.~\cite{CMS:2018fks}.
}
\label{top-17-001-constraints}
\end{figure}

The fit impact on the uncertainties can be quantified by the pulls and constraints of the corresponding nuisance parameters. The constraint is defined as the ratio of the post-fit uncertainty to the pre-fit uncertainty of a given nuisance parameter, while the normalised pull is the difference between the post-fit and the pre-fit values of the nuisance parameter normalised to its pre-fit uncertainty.
The normalised pulls and constraints of the nuisance parameters related to the modelling uncertainties for the simultaneous fit of \stt and \mtmc in the CMS analysis~\cite{CMS:2018fks} are shown in Fig.~\ref{top-17-001-constraints}.

As a result of the simultaneous fit, the values of $\stt=815\pm2\stat\pm29\syst\pm20\lum\unit{pb}$, and $\mtmc=172.33\pm0.14\stat\,^{+0.66}_{-0.72}\syst\GeV$ are obtained~\cite{CMS:2018fks}, with 12\% correlation between the two.

The result on \stt is used together with the QCD prediction~\cite{Aliev:2010zk} at NNLO in the \msbar scheme to extract the value of \mtmt. For this purpose, the measured and the predicted cross sections are compared via a \chisq minimisation, using the open-source QCD analysis framework  \xfitter~\cite{Alekhin:2014irh}. For a measurement $\mu$, a corresponding theoretical prediction $m$, and the set of systematic nuisance parameters $\vec{b}$, the following \chisq definition is used:
\begin{equation}
    \chisq(m,\vec{b})=\frac{\big[\mu-m\big(1-\sum_j\gamma_j b_j\big)\big]^2}{\delta^2_{\text{unc}}m^2+\delta^2_{\text{stat}}\,\mu\,m\big(1-\sum_j\gamma_j b_j\big)}+\sum_j b^2_j.
\end{equation}
Here, $\delta_{\text{stat}}$ and $\delta_{\text{unc}}$ are relative statistical and uncorrelated systematic uncertainties of the measurement, $\gamma_j$ quantifies the sensitivity of the measurement to the correlated systematic source $j$. This definition of the \chisq function assumes that
systematic uncertainties are proportional to the values of the central prediction (multiplicative uncertainties, $m_i(1-\sum_j\gamma_j b_j)$), whereas the statistical uncertainties scale with the square root of the expected number of events.

{\tolerance=1200
The four most recent PDF sets available at NNLO to that date were used: ABMP16nnlo, CT14nnlo, MMHT14nnlo, and NNPDF3.1nnlo. Unlike other PDF sets, the ABMP16nnlo employs the \msbar scheme for the heavy quarks in the theoretical predictions used in the PDF determination. For the other PDFs, values of \mtp are assumed and are converted to \mtmt using the number of \alpS loops according to the individual prescription by the corresponding PDF group (as shown in Table~4 of Ref.~\cite{CMS:2018fks}). Because of the strong correlation between \asmz and \mtmt in the prediction of \stt, for the \mtt extraction, the value of \asmz in the theoretical prediction is set to that of the particular PDF set.
\par}

{\tolerance=800
The fit is performed by varying \mtmt in the theoretical prediction in the range $158<\mtmt<163\GeV$ for ABMP16nnlo PDF and in the range $162<\mtmt<167\GeV$ for the other PDFs. The uncertainties related to the variation of \asmz in the PDFs are estimated by repeating the fit using the PDF eigenvectors with \asmz varied within its uncertainty as provided by each PDF, except for ABMP16nnlo, where the value of \asmz is a free parameter in the PDF fit and its uncertainty is included in the eigenvectors.
\par}

\begin{figure}[!t]
\centering
\includegraphics[width=0.48\textwidth]{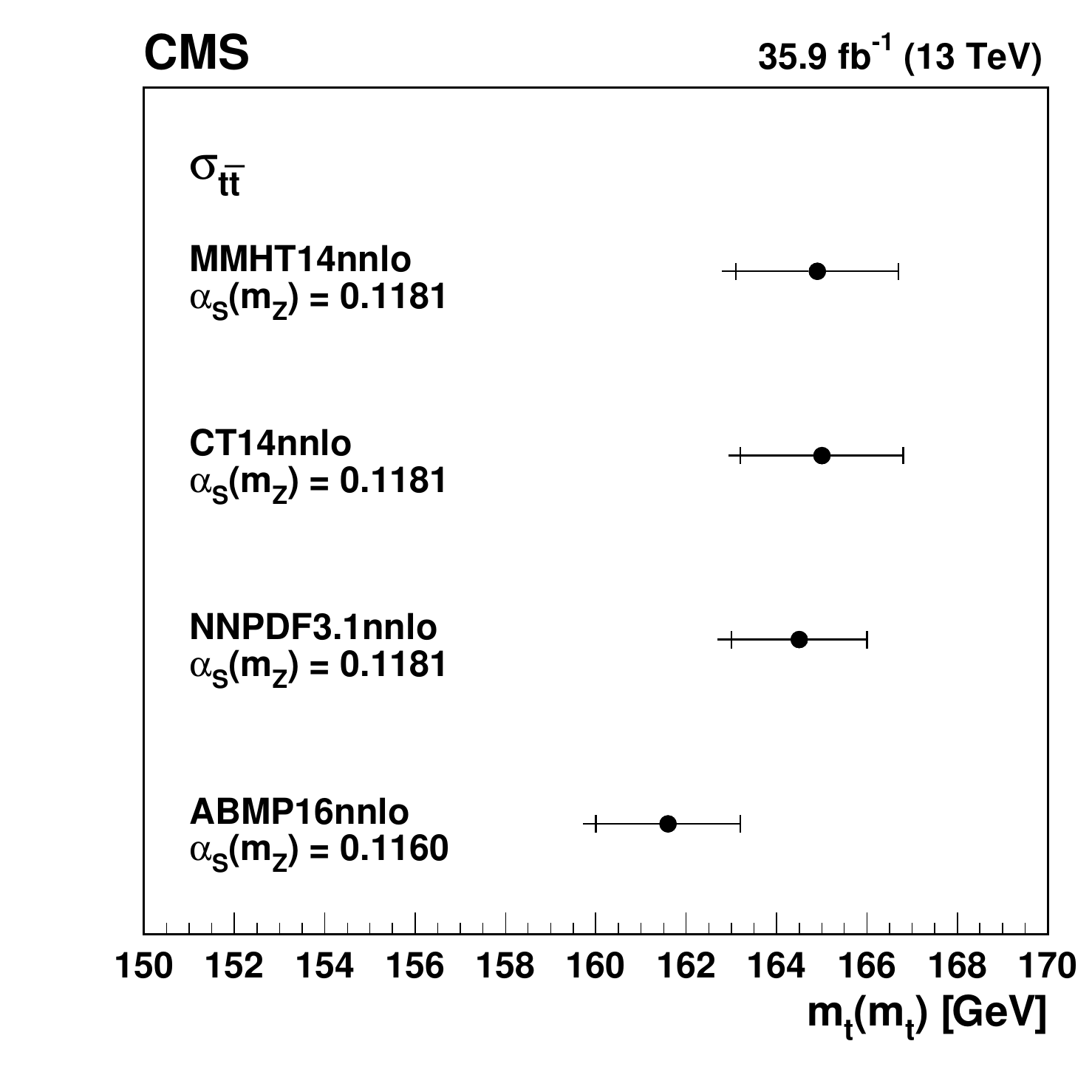}
\caption{%
    Values of \mtmt obtained from comparing the \stt measurement to the theoretical NNLO predictions using different PDF sets. The inner horizontal bars on the points represent the quadratic sum of the experimental, PDF, and \asmz uncertainties, while the outer horizontal bars give the total uncertainties.
    Figure taken from Ref.~\cite{CMS:2018fks}.
}
\label{top-17-001_mtmt}
\end{figure}

Instead of assuming a prior for the scale variation uncertainty, in the analysis~\cite{CMS:2018fks}, the variation of \mur and \muf was externalised, by repeating the \chisq fit independently for different choices of the \mur and \muf in the predicted \stt. The nominal values of these scales were set to \mtmt and varied by a factor of two up and down, independently. The largest differences of the results to the nominal one was considered as scale uncertainty. The results on \mtmt are illustrated in Fig.~\ref{top-17-001_mtmt}.

The results obtained with different PDF sets are in agreement, although the ABMP16nnlo PDF set yields a systematically lower value. This difference is expected and has its origin in a larger value of $\asmz=0.118$ assumed in the \NNPDF{3.1}, MMHT2014, and CT14 PDFs. The result obtained by using ABMP16 PDF,
$\mtmt=161.6\pm1.6\fitPDFas\,^{+ 0.1}_{- 1.0}\scale\GeV$~\cite{CMS:2018fks}, with its total uncertainty of about 1.2\%,
should be considered as the most theoretically consistent, since only ABMP16 PDF implies a heavy quark treatment in the \msbar scheme and considers the correlation between the \asmz and PDF. Using the same theoretical prediction consistently in the pole mass scheme, results in $\mtp=169.1\pm1.8\fitPDFas\,^{+1.3}_{-1.9}\scale\GeV$~\cite{CMS:2018fks} using the ABMP16 PDF.
The shift between the pole and the running mass values is expected, but the significantly smaller scale uncertainties in the case of the \msbar scheme arises from significantly better perturbative convergence in this scheme.

\begin{figure}[!t]
\centering
\includegraphics[width=0.48\textwidth]{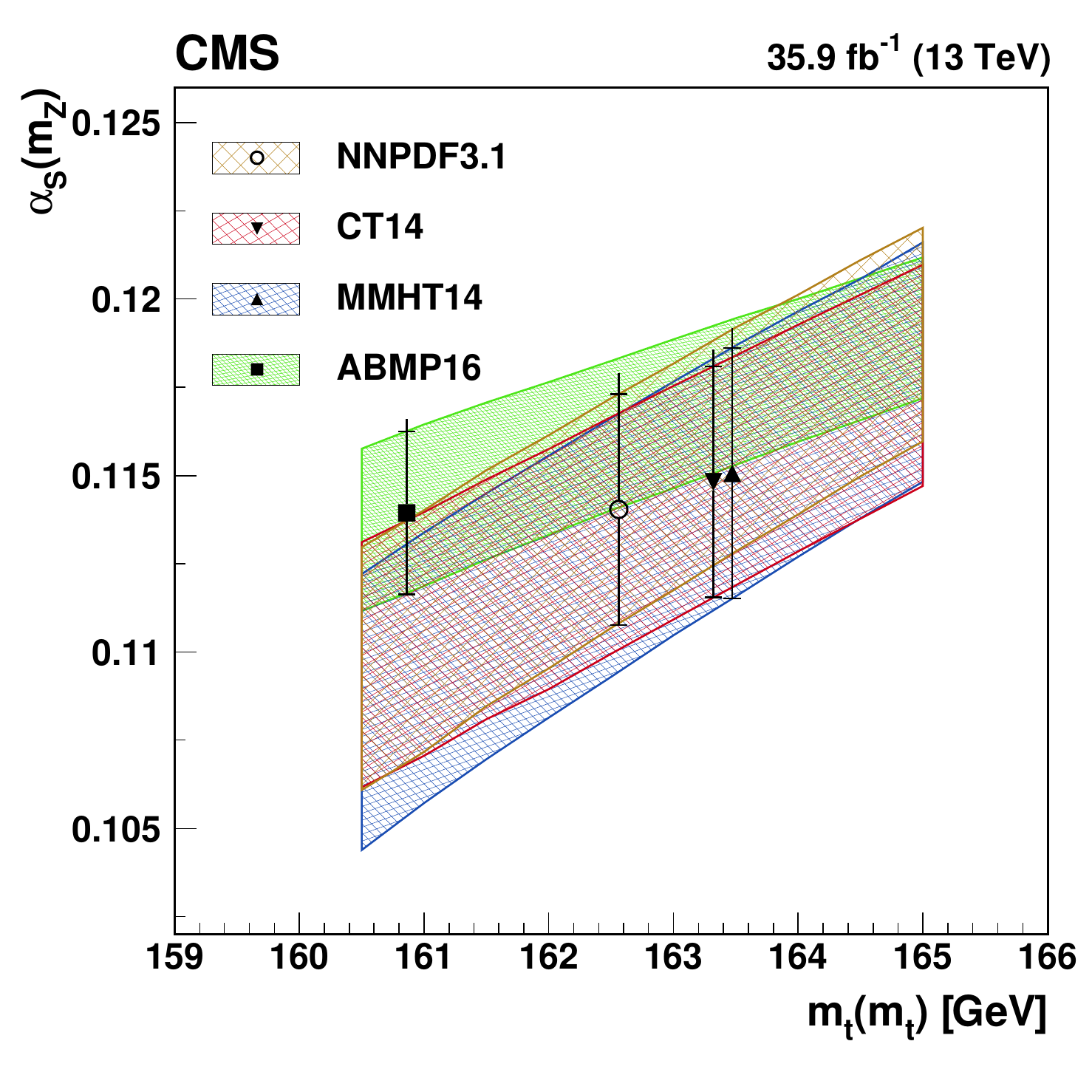}
\caption{%
    Values of \asmz obtained in the comparison of the \stt measurement to the NNLO prediction using different PDFs, as functions of the \mtmt value used in the theoretical calculation. The results from using the different PDFs are shown by the bands with different shadings, with the band width corresponding to the quadratic sum of the experimental and PDF uncertainties in \asmz. The resulting measured values of \asmz are shown by the different style points at the \mtmt values used for each PDF. The inner vertical bars on the points represent the quadratic sum of the experimental and PDF uncertainties in \asmz, while the outer vertical bars show the total uncertainties.
    Figure taken from Ref.~\cite{CMS:2018fks}.
}
\label{top-17-001_mt_as}
\end{figure}

While higher experimental precision is achieved in the 13\TeV analysis as compared to 7 and 8\TeV measurements, the full consideration of the PDF eigenvectors in \stt calculation and, in turn, in the \chisq minimisation procedure, and externalising the scale variations leads to an increased uncertainty with respect to the combined 7 and 8\TeV result. Therefore, the extraction of \mt through comparison of measured and predicted \stt has the limitation by the PDF uncertainty, and aforementioned correlation of PDF, \alpS, and \mt in the prediction of \stt. The correlations between the \mtmt with the assumption on \asmz was investigated in detail for each PDF
by performing a \chisq scan in \asmz for ten different assumptions of \mtmt, varied from 160.5 to 165.0\GeV. A
linear dependence is observed, as shown in Fig.~\ref{top-17-001_mt_as}, illustrating the strong correlation of the PDF, \asmz and \mtmt in the \stt prediction and the related ambiguity in the extraction of one parameter by fixing the others.

\subsection{The first illustration of the running of the top quark mass}
\label{sec:running}

In Section~\ref{sec:mlb_mtmt}, the inclusive measurement of the \ttbar production cross section is used to extract the value of the top quark mass in the \msbar scheme at the top quark mass scale, \mtmt. In the \msbar scheme, which is the standard scheme used to renormalise \alpS, the top quark mass depends on an additional scale \mumass. As already mentioned in Section~\ref{sec:introschemes}, the scale \mumass sets the lower bound of the self-energy contributions absorbed in the \msbar mass and should be chosen close to the dynamical scale governing the \mt sensitivity of the cross section. This scale setting ensures the absence of large logarithmic corrections as far as the mass dependence of the theoretical prediction is concerned and thus ensures an adequate treatment of quantum corrections related to the mass sensitivity. The \msbar mass is adequate for cross sections where this dynamical scale is close to or larger than the top quark mass, \ie $\mumass\gtrsim\mt$. For the inclusive cross section measurement described in Section~\ref{sec:mlb_mtmt} this dynamical scale is set by typical transverse momentum of the produced top quarks which is around the top quark mass, justifying the use of \mtmt.

As in the case of \alpS, the scale evolution (often referred to as ``running'') of \mtmum is described by the renormalisation group equation (RGE):
\begin{equation}
    \mumass^2\frac{\rd\mtmum}{\rd\mumass^2}=-\gamma\big(\alpS(\mumass)\big)\mtmum,
\label{eq:RGE}
\end{equation}
where $\gamma\big(\alpS(\mumass)\big)$ is known as the mass anomalous dimension. This quantity can be calculated in perturbation theory, and the coefficients are currently known up to order $\alpS^5$~\cite{Baikov:2014qja,Luthe:2016xec}. Measuring the running of \mtmu is not only a fundamental test of the validity of perturbative QCD, but also an indirect probe of BSM physics scenarios that can modify the RGE running, \eg supersymmetric theories~\cite{Mihaila:2013wma} or models based on the dynamic mass generation of fermions~\cite{Christensen:2005hm}.

Measuring cross sections where the top quark mass sensitivity is governed at widely different energy scales $Q$ allows the running of the \msbar top quark mass to be measured by extracting the value of $\mt(\mumass=Q)$. This is in close analogy to measurements of the running strong coupling \alpS. In Ref.~\cite{CMS:2019jul}, where the first measurement of the running of the top quark mass is presented, this is achieved by comparing a measurement of the \ttbar production cross section as a function of \mtt to the QCD predictions at NLO. The analysis of Ref.~\cite{CMS:2019jul} makes use of the same data as in Ref.~\cite{CMS:2018fks}, addressing the \ttbar production with the \EM final state. The differential cross section, \dstt, is measured by means of a profile likelihood unfolding of multi-differential distributions, extending the method of Ref.~\cite{CMS:2018fks} presented in Section~\ref{sec:inclusive}.
The investigation of the \msbar mass running adopts $\mtt/2$ as the scale \mumass, which quantifies the energy scale of the hard \ttbar production process.

In order to measure the \ttbar cross section differentially, the \ttbar simulation is split into bins of \mtt at the generator level, and each sub-sample is treated as an independent signal process in the likelihood fit, while preserving the correlation between the systematic uncertainties. This procedure is commonly known as maximum likelihood unfolding.
The expected number of events in each bin is parameterised as:
\begin{equation}
    \nu_i=\sum_{k=1}^{4}s_i^k(\sttk,\mtmc,\nuisancevec) + \sum_j b^j_i(\mtmc,\nuisancevec),
\label{eq:expectev}
\end{equation}
where \sttk is the total cross \ttbar cross section in bin $k$ of \mtt, $s_i^k$ represents the contribution of bin $k$ in \mtt to bin $i$, $b^j_i$ is the contamination from background $j$ in that bin, and \nuisancevec are the nuisance parameters that parameterise the effects of the systematic uncertainties. As in the analysis of Ref.~\cite{CMS:2018fks}, the effect of \mtmc is profiled in the likelihood. This expression incorporates the effect of the detector response and of the signal acceptance, and directly connects parton-level quantities to measurable detector-level distributions. Therefore, the likelihood fit
provides directly the unfolded results at the parton level. In order to allow for a comparison to fixed-order theoretical predictions, in this analysis the parton level is defined as the matrix-element level, \ie before parton showering, assuming stable top quarks. Details on the MC simulation are given in Section~\ref{sec:mcsetup}.

In order to enhance the sensitivity to each individual bin of \mtt, the invariant mass of the \ttbar system is reconstructed at the detector level (\mttreco) using the full kinematic reconstruction described in Section~\ref{sec:kinreco}. The additional dependence on the value of \mt assumed in the kinematic reconstruction is fully parameterised in the likelihood via the parameter \mtmc. As in Ref.~\cite{CMS:2018fks}, this parameter is treated as freely floating in the fit, and is constrained via the \mlbmin distribution.

\begin{figure}[!ht]
\centering
\includegraphics[width=0.48\textwidth]{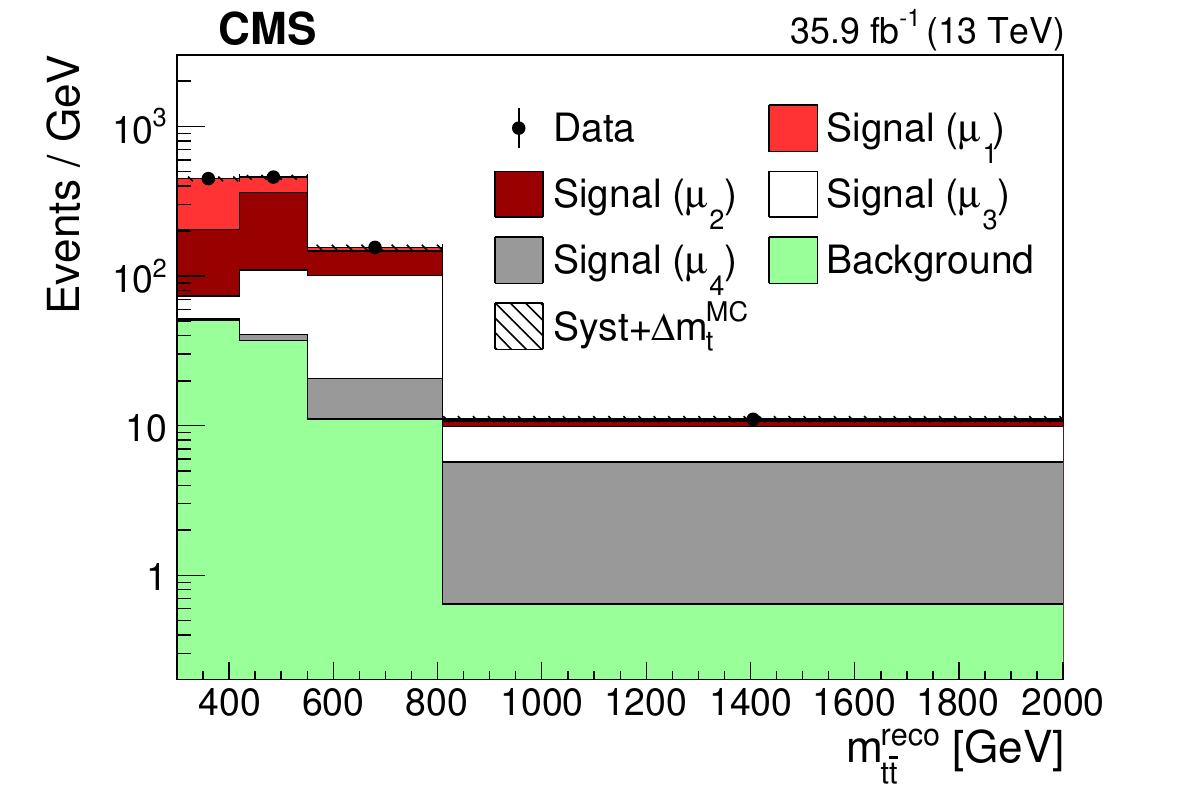}%
\hfill%
\includegraphics[width=0.48\textwidth]{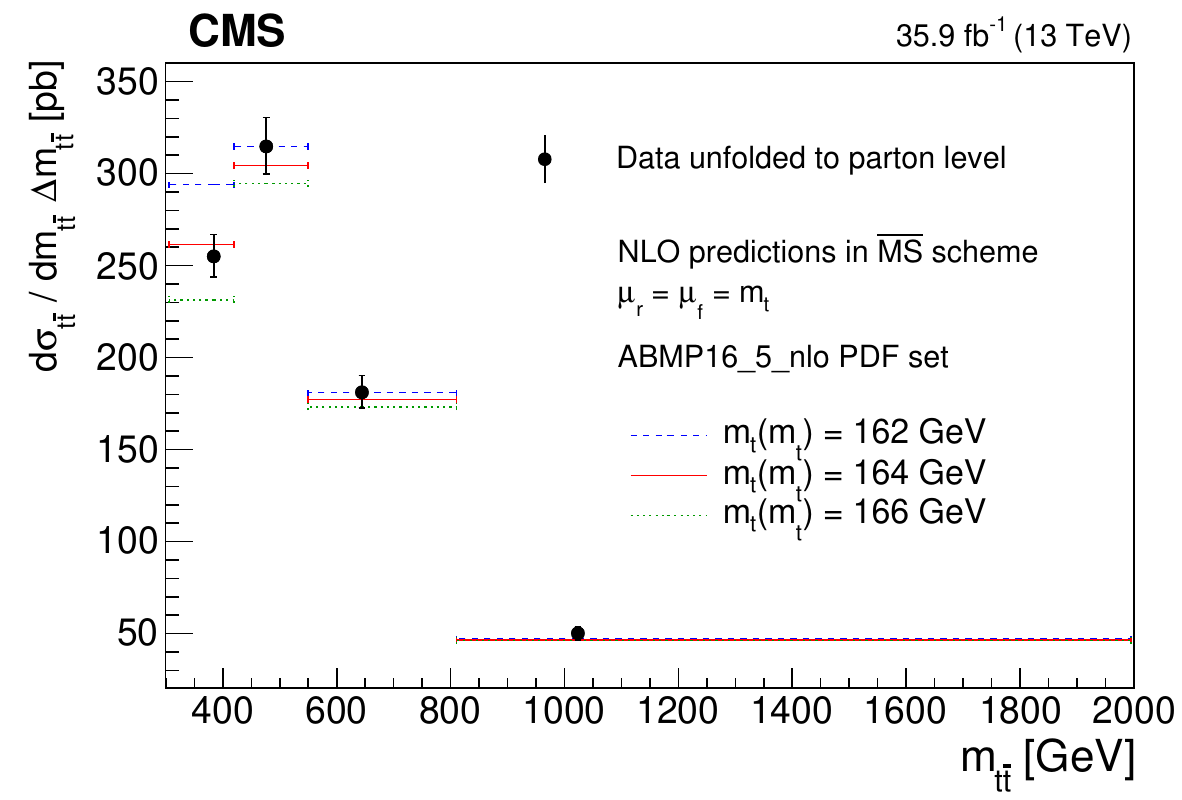}
\caption{%
    Left: profile likelihood unfolding of the \mtt distribution. The signal sample is split into subprocesses in bins of parton-level \mtt, and the signal corresponding to bin $k$ in \mtt is denoted with ``Signal (\muk)''. The vertical bars represent the statistical uncertainty in the data, while the hashed band is the total uncertainty in the MC simulation. Right: unfolded \ttbar cross section as a function of \mtt, compared to theoretical predictions in the \msbar scheme for different values of \mtmt. The vertical bars correspond to the total uncertainty in the unfolded cross section. Here, the bin centres for the unfolded cross section are defined as the average \mtt in the \POWHEGPYTHIAEight simulation.
    Figures taken from Ref.~\cite{CMS:2019jul}.
}
\label{fig:mt_running_unfold}
\end{figure}

The fit is performed in categories of \PQb-tagged jet multiplicity and in bins of \mttreco, while all events with less than two jets in the final state, for which no kinematic reconstruction is possible, are assigned to separate categories. The \mttreco distribution after the fit to the data, which illustrates the likelihood unfolding procedure, is shown in Fig.~\ref{fig:mt_running_unfold} (left). In Fig.~\ref{fig:mt_running_unfold} (right), instead, the unfolded \dstt is compared to the NLO theoretical predictions used in Ref.~\cite{CMS:2019jul} to extract the running of \mt. The bin centres are chosen as the average value of \mtt in each bin according to the \POWHEGPYTHIAEight simulation, and are considered as the representative energy scale of each \mtt bin. As illustrated in Fig.~\ref{fig:mt_running_unfold} (right), the dependence of the \ttbar production cross section on the value of \mt decreases rapidly with increasing \mtt.

An updated extraction of the running of \mt is obtained in the scope of this article, with a similar theoretical setup as the one suggested in Ref.~\cite{Catani:2020tko}, where differential calculations in the \msbar scheme are obtained at NNLO and compared to the results of Ref.~\cite{CMS:2019jul}. Here, unlike in the original result of Ref.~\cite{CMS:2019jul}, a bin-by-bin dynamic scale is implemented in the NLO calculation, which allows the direct extraction of the value of \mtmum. A dynamic scale choice is also favoured from the theoretical point of view, as it accounts for the summation of higher-order QCD corrections. This approach has also been used in the improved analysis of Ref.~\cite{Defranchis:2022nqb}, where the running of \mt is extracted at NNLO in QCD.

The measured cross section of Ref.~\cite{CMS:2019jul} is also updated according to the new luminosity measurement of the 2016 data set~\cite{CMS:2021xjt}, which leads to a significant improvement in the uncertainty in the measured cross section. Following the approach of Ref.~\cite{CMS:2019jul}, the value of \mtmum is extracted in each bin of \mtt separately. Here, \mumass is chosen to be $\muk/2$, where \muk is the representative scale of bin $k$ in \mtt, corresponding to the bin centre in Fig.~\ref{fig:mt_running_unfold} (right). The measured values of \mtmum are normalised to the value of \mtmuref, where \muref is arbitrarily chosen as the scale of the second bin in \mtt, in order to profit from the cancellation of correlated systematic uncertainties.

\begin{figure}[!ht]
\centering
\includegraphics[width=0.65\textwidth]{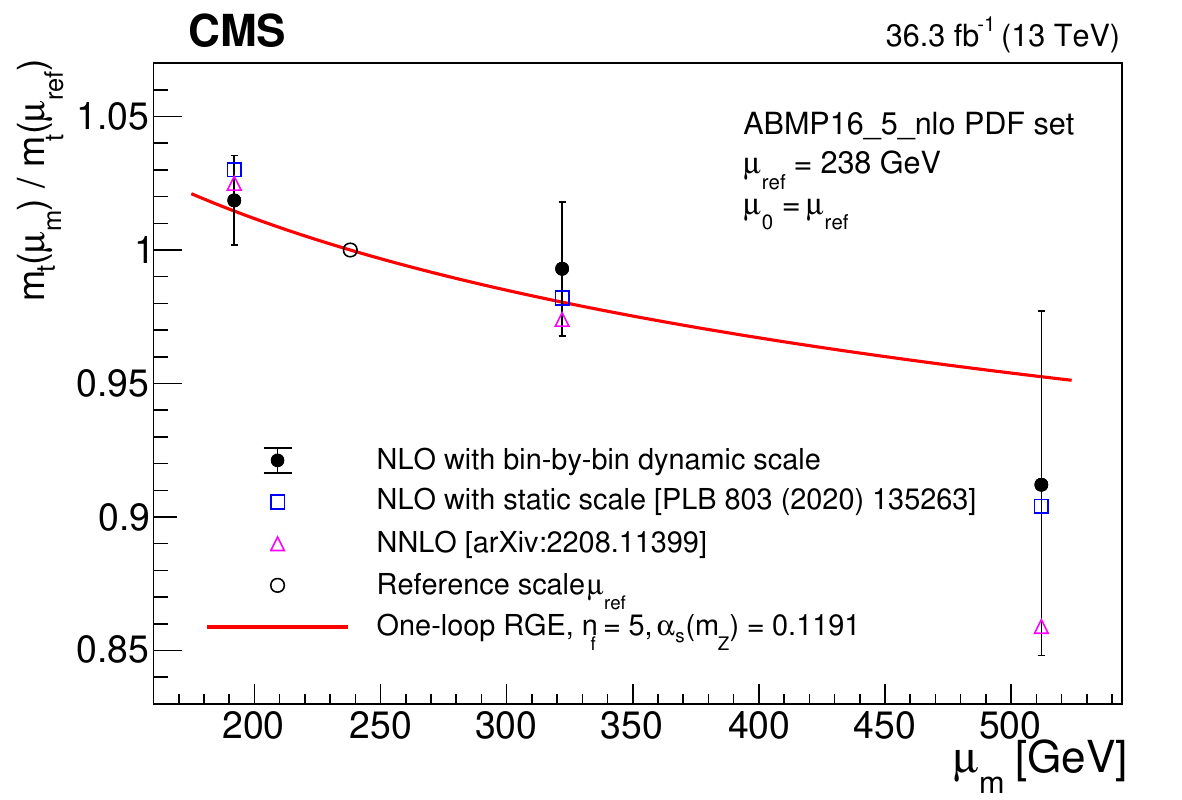}
\caption{%
    Running of the top quark mass as a function of $\mumass=\mtt/2$ obtained with a bin-by-bin dynamic scale $\muk/2$ (full circles), compared to the central values of the results of Ref.~\cite{CMS:2019jul} obtained with a constant scale $\mumass=\muk$ (hollow squares) and to those of the NNLO results of Ref.~\cite{Defranchis:2022nqb} (hollow triangles). As in Ref.~\cite{CMS:2019jul}, the error bars indicate the combination of experimental, extrapolation, and PDF uncertainties in the NLO extraction with bin-by-bin dynamic scale. The full treatment of the QCD scale variations can be found in Ref.~\cite{Defranchis:2022nqb}. The assumptions on the renormalisation and factorisation scales adopted in the different interpretations are summarised in Table~\ref{tab:running}. The uncertainties in the three results, which are mostly correlated, are given in the respective references and are of comparable size.
}
\label{fig:mt_running_result}
\end{figure}

The result is shown in Fig.~\ref{fig:mt_running_result}, where it is compared to the one-loop solution of the QCD RGE, to the original result of Ref.~\cite{CMS:2019jul}, and to the more recent re-interpretation at NNLO in QCD described in Ref.~\cite{Defranchis:2022nqb}. However, it has to be noted that the results are not directly comparable to each other, as they differ not only for the perturbative order in QCD, but also for the choice of the renormalisation and factorisation scales in the fixed-order calculations, as summarised in Table~\ref{tab:running}. Nonetheless, in all cases the RGE running scenario is favoured by the data compared to a hypothetical no-running scenario in which $\rd\mtmum/\rd\mumass=0$.

\begin{table}[!ht]
\centering
\topcaption{%
    Summary of scale choices for \mur, \muf, and \mumass for the three different extractions of the running of the top quark mass. The NLO fixed scale corresponds to the result of Ref.~\cite{CMS:2019jul}, while the NNLO result is described in Ref.~\cite{Defranchis:2022nqb}. The NLO bin-by-bin dynamic result, instead, is obtained in the scope of this review work.
}
\renewcommand{\arraystretch}{1.1}
\begin{tabular}{ccc}
    Fixed-order theory model & \mumass [{\GeVns}]  & \mur, \muf [{\GeVns}]  \\ \hline
    NLO fixed scales  & \mt   & \mtmt \\
    NLO bin-by-bin dynamic scale & $\mtt/2$ & \mtmum \\
    NNLO bin-by-bin dynamic scale & $\mtt/2$ & \mumass \\
\end{tabular}
\label{tab:running}
\end{table}

\subsection{Resolving correlations of \texorpdfstring{\mt}{mt}, \texorpdfstring{\asmz}{as(mZ)}, and PDFs}
\label{sec:pdf_as_mt}

The correlation among PDFs, \asmz, and \mt in the QCD prediction of \stt was already mentioned in the context of the extraction of \mt using the inclusive \stt.
The origin of this correlation is the fact that \ttbar production in \pp collisions is dominated by the gluon-gluon fusion process (to about 90\%), so that the gluon PDF, \asmz, and \mt alter the normalisation and shape of the \stt prediction. At the same time, it means that any of these parameters can be extracted individually, by using the \ttbar cross sections, only once the other two are fixed.
Therefore, besides extraction of \mt or \asmz by using the measurements of inclusive cross section of \ttbar production, the same measurements can be used to constrain the proton PDFs, by fixing \mt and \asmz.  Due to the large scale, provided by the top quark mass, the \ttbar production is sensitive to the gluon distribution $g(x)$ at large fractions $x$ of the proton momentum, carried by the gluon. Due to lack of other experimental data constraining the gluon distribution at high $x$, $g(x)$ has large uncertainties in this region.

\begin{figure}[!ht]
\centering
\includegraphics[width=0.6\textwidth]{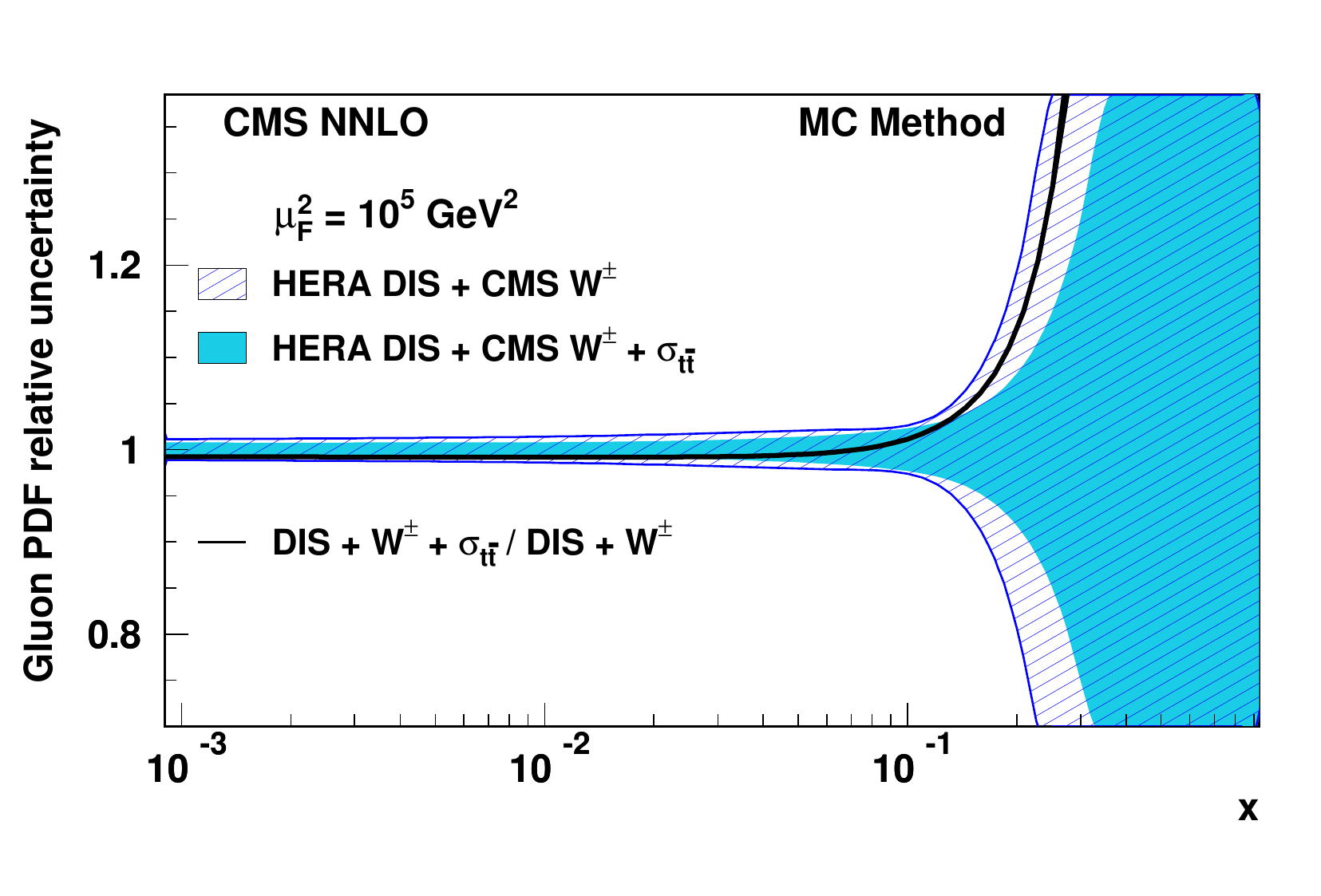}
\caption{%
    The fractional uncertainties in the gluon distribution function of the proton as a function of $x$ at factorisation scale $\muf^2=10^5\GeV^2$ from a QCD analysis using the DIS and CMS muon charge asymmetry measurements (hatched area), and also including the CMS \stt results at \sqrtseq{5.02} (solid area). The relative uncertainties are found after the two gluon distributions have been normalised to unity. The solid line shows the ratio of the gluon distribution function found from the fit with the CMS \stt measurements included to that found without.
    Figure taken from Ref.~\cite{CMS:2017zpm}.
}
\label{stt_5tev_gluon}
\end{figure}

An illustrative example of PDF constraints using the inclusive \stt is the result of the CMS analysis~\cite{CMS:2017zpm}. In this work, the \stt measurement at \sqrtseq{5.02} based on the integrated luminosity of 24.4\pbinv was included in a PDF fit at NNLO together with the cross sections of \ep deep inelastic scattering (DIS) at HERA~\cite{H1:2015ubc}, and the CMS muon charge asymmetry measurements in \PW boson production~\cite{CMS:2016qqr}.
In the fit, performed by using the open-source QCD analysis platform \xfitter~\cite{Alekhin:2014irh}, the values of $\asmz=0.118$ and $\mtp=172.5\GeV$ are assumed. Already by including a single measurement of \stt at 5.02\TeV, the reduction of the uncertainty in $g(x)$ is observed, as shown in Fig.~\ref{stt_5tev_gluon}.

While the PDF constraints by using inclusive \stt are achieved only through the global normalisation, differential cross sections provide further information about the PDFs, \alpS, and \mt. This was investigated in Ref.~\cite{Guzzi:2014wia}, where the differential cross sections were suggested to be used in a QCD analyses to extract PDFs, \asmz, and \mt.
In particular, the invariant mass \mtt and rapidity \ytt of the \ttbar pair are directly related to $x$ as $x=(\mtt/\sqrts)\exp[\pm y(\ttbar)]$ at LO QCD. In the CMS work~\cite{CMS:2017iqf}, measurements of double-differential \ttbar cross sections as functions of \mtt and \ytt were demonstrated to be most sensitive to $g(x)$, providing more significant constraints than inclusive or single-differential cross sections.

By using multi-differential \ttbar cross sections, it is possible to obtain a good overall constraint on the PDFs, \asmz, and \mt, simultaneously, since the \mtt distribution is driven by the value of \mt. To better access the \ttbar threshold in the final states with two leptons, the LKR algorithm, discussed in Section~\ref{sec:kinreco}, was developed, probing \mtt in a less biased way compared to FKR. However, the limited resolution in \mtt mentioned in Section~\ref{sec:kinrecodillep}, prevents splitting the \mtt distribution in bins narrower than 100--150\GeV, in particular close to the threshold. Further, production of \ttbar associated with jets brings in additional sensitivity to \asmz at the scale of \mt, and enhances sensitivity to \mt, since the gluon radiation depends on \mt through threshold and cone effects~\cite{Alioli:2013mxa}.

First simultaneous determination of the PDFs, \asmz, and \mtp by using multi-differential \ttbar cross sections were carried out by CMS in Ref.~\cite{CMS:2019esx}. In particular, double-differential \ttbar cross sections as functions of \mtt and \ytt were measured in different categories with respect to the number of associated additional particle-level jets in the event, \nj, using two ($\nj=0$ and $\nj\geq1$) and three ($\nj=0$, $\nj=1$, and $\nj\geq2$) bins of \nj. These cross sections are denoted as \njmttytttwo and \njmttyttthree, respectively. To correct for the detector resolution and inefficiency, a regularised unfolding was performed simultaneously in bins of the observables in which \stt were measured. To compare the measured cross sections for \ttbar production with additional jets to NLO QCD predictions, the measured cross sections were further corrected from particle to parton level for MPI, hadronisation, and top quark decay effects, by using the MC simulation. The measured triple-differential cross sections are compared to calculations of the order in \alpS required for NLO accuracy: the inclusive \ttbar production at \orderas3~\cite{Mangano:1991jk}; \ttbar production with one jet at \orderas4~\cite{Dittmaier:2007wz}; and \ttbar production with two additional jets at \orderas5~\cite{Bevilacqua:2010ve,Bevilacqua:2011aa}. In particular, the cross sections for inclusive \ttbar production are calculated from the sum of the measured \stt in the $\nj=0$ and $\nj\geq1$ bins. Thus, the cross sections obtained for inclusive \ttbar and \ttbar+1 jet production are compared to the NLO \orderas3 and NLO \orderas4 calculations, respectively. Similarly, cross sections for inclusive \ttbar, $\ttbar+1$, and $\ttbar+2$ jets production are obtained using the \njmttyttthree measurement and compared to the NLO \orderas3, NLO \orderas4, and NLO \orderas5 calculations, respectively.

\begin{figure}[!htp]
\centering
\includegraphics[width=\textwidth]{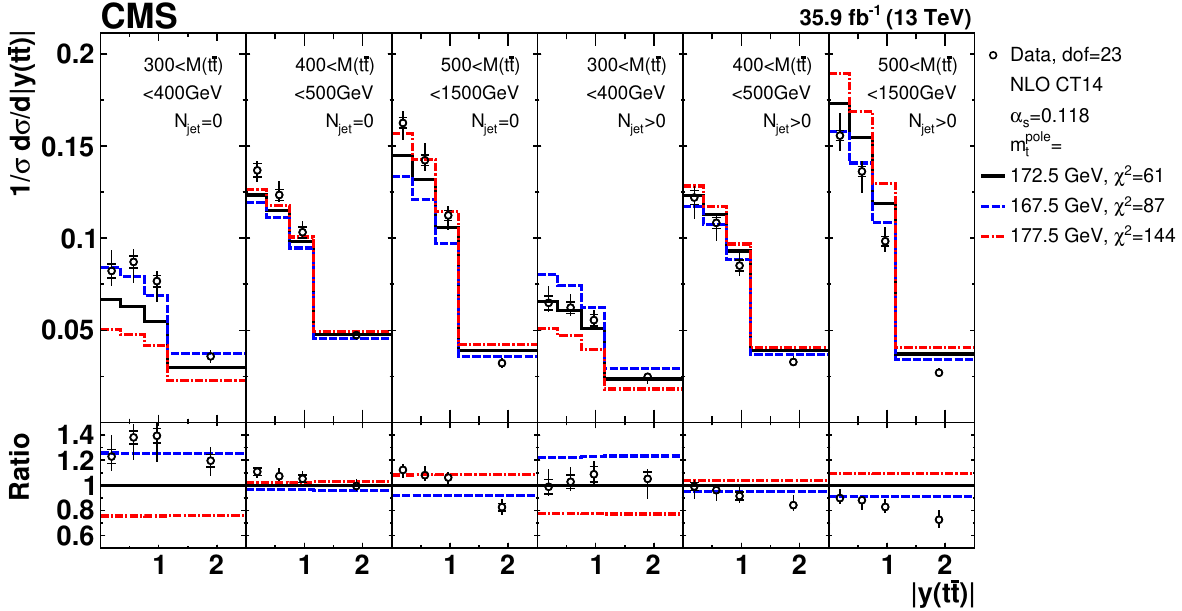}
\caption{%
    Comparison of the measured \njmttytttwo cross sections to NLO predictions obtained using different \mtp values. For each theoretical prediction, values of \chisq and dof for the comparison to the data are reported.
    Figure taken from Ref.~\cite{CMS:2019esx}.
}
\label{fig:xsec-nlo-nj2mttytt-mt}
\end{figure}

\begin{figure}[!htp]
\centering
\includegraphics[width=\textwidth]{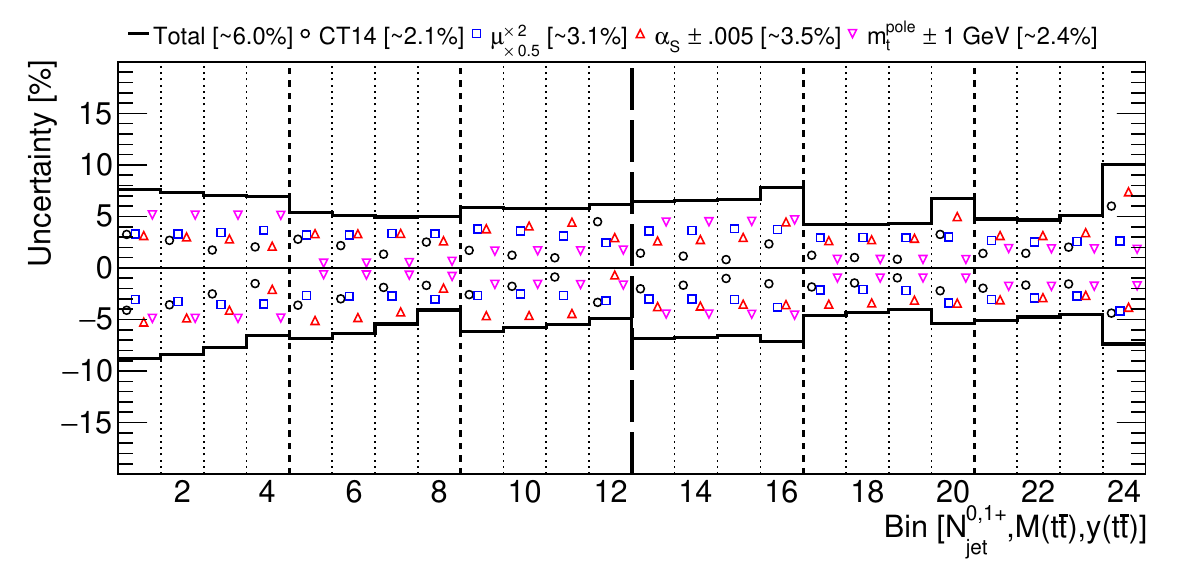}
\caption{%
    The theoretical uncertainties for \njmttytttwo cross sections, arising from the scale, PDF, \asmz, and \mt variations,
    as well as the total theoretical uncertainties obtained from variations in \mur and \muf, with their bin-averaged values shown in brackets. The bins are the same as in Fig.~\ref{fig:xsec-nlo-nj2mttytt-mt}.
    Figure taken from Ref.~\cite{CMS:2019esx}.
}
\label{fig:thunc-nj23mttytt}
\end{figure}

Using the normalised cross sections results in the partial cancellation of experimental and theoretical uncertainties. To demonstrate the sensitivity to \mtp, in Fig.~\ref{fig:xsec-nlo-nj2mttytt-mt}, the data are compared to the predictions obtained with different values of \mtp. The largest sensitivity to \mtp is observed at lower \mtt (indicated as $M_{\ttbar}$ in Fig.~\ref{fig:xsec-nlo-nj2mttytt-mt} and Fig.~\ref{fig:thunc-nj23mttytt}), closest to the \ttbar production threshold, while the sensitivity at higher \mtt occurs mainly because of the cross section normalisation.
To further demonstrate the sensitivity of the theoretical predictions for the measured \njmttytttwo cross sections to different input parameters, in Fig.~\ref{fig:thunc-nj23mttytt}, the contributions arising from the PDF, \asmz ($\pm0.005$), and \mtp ($\pm1\GeV$) uncertainties are shown separately. The total theoretical uncertainties are obtained by adding the uncertainties originating from PDF, \asmz, \mtp, and variations of \mur and \muf, in quadrature.

The normalised triple-differential \njmttytttwo cross sections are used together with the combined HERA DIS data~\cite{H1:2015ubc} in a QCD analysis, where PDF, \asmz, and \mtp are extracted at NLO, using the \xfitter program~\cite{Alekhin:2014irh}. The resulting NLO values of \asmz and \mtp are obtained~\cite{CMS:2019esx} as follows:
\begin{align}
    \asmz&=0.1135\pm0.0016\fit\,^{+0.0002}_{-0.0004}\model\,^{+0.0008}_{-0.0001}\param\,^{+0.0011}_{-0.0005}\scale \nonumber\\
    &= 0.1135\,^{+0.0021}_{-0.0017}, \label{eq:as-nom}\\
    \mtp&=170.5\pm0.7\fit\pm0.1\model\,^{+0.0}_{-0.1}\param\pm0.3\scale\GeV \nonumber\\
    &=170.5\pm0.8\GeV.
\end{align}
Here `fit', `model', and `param' denote the fit, model, and parameterisation uncertainties. The fit uncertainties were obtained using the criterion of $\Delta\chisq=1$. The model uncertainties arise from the variations of assumptions on theoretical inputs, such as masses of \PQc and \PQb quarks or the value of the starting evolution scale. The parameterisation uncertainties originate from the variations of the functional form for the PDFs at the starting scale. In addition, `scale' denotes the uncertainties arising from the scale variations in \stt predictions, which are estimated by repeating the fit using predictions where the values of \mur and \muf are varied by a factor of 2, independently up and down, and taking the differences with respect to the nominal result.

\begin{figure}[!ht]
\centering
\includegraphics[width=0.7\textwidth]{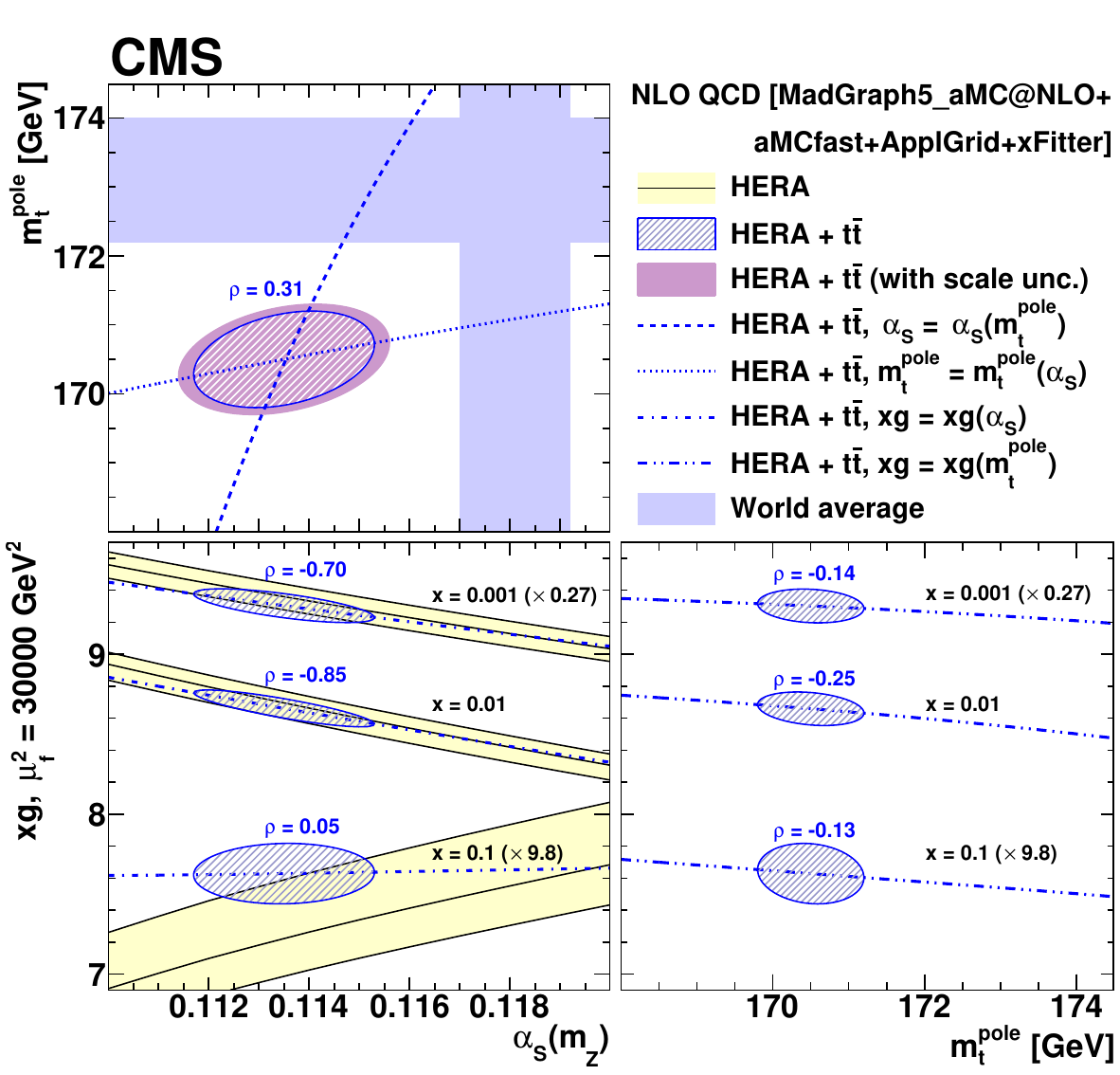}
\caption{%
    The extracted values and their correlations for \alpS and \mtp (upper left), \alpS and gluon PDF (lower left), and \mtp and gluon PDF (lower, right). The gluon PDF is shown at the scale $\muf^2=30\,000\GeV^2$ for several values of $x$.
    For the extracted values of \alpS and \mtp, the additional uncertainties arising from the dependence on the scale are shown.
    The correlation coefficients $\rho$ as defined in Ref.~\cite{CMS:2019esx} are displayed. Furthermore, values of \alpS (\mtp, gluon PDF) extracted using fixed values of $\mtp(\alpS)$ are displayed as dashed, dotted, or dash-dotted lines. The world average values $\asmz=0.1181\pm0.0011$ and $\mtp=173.1\pm0.9\GeV$ from Ref.~\cite{ParticleDataGroup:2018ovx} are shown for reference.
    Figure taken from Ref.~\cite{CMS:2019esx}.
}
\label{fig:allcor}
\end{figure}

In Fig.~\ref{fig:allcor} the extracted \asmz, \mtp, and gluon PDF at the scale $\muf^2=30\,000\GeV^2$ for several values of $x$ are shown, together with their correlations. When using only DIS data, the largest correlation to \asmz is observed in the gluon PDF. Once included in the fit, measurement of the \ttbar production resolves this correlation in the relevant kinematic range, because of its sensitivity to both $g(x)$ and \asmz. In addition, the multi-differential \njmttytttwo cross sections provide constraints on \mtp. As a result, the correlations between $g(x)$, \asmz, and \mtp are significantly reduced in the kinematic range of \ttbar production. This way, the simultaneous QCD analysis of PDFs, \asmz, and \mtp has highest potential to extract \mtp with best precision through mitigating uncertainties in \asmz and $g(x)$. However, an additional theoretical uncertainty in the extracted \mtp value is expected, due to the gluon resummation corrections, and in particular the Coulomb gluon exchange contributions arising from to the toponium quasi bound state dynamics in the small-\mtt region~\cite{Kiyo:2008bv,Piclum:2018ndt}. These corrections are not yet implemented in a form suitable for the \stt analysis in \pp collisions, as discussed in Section~\ref{sec:indirect_outlook}. It was estimated in Ref.~\cite{CMS:2019esx} that this could result in an  uncertainty of $+1\GeV$ in \mtp, in addition to the one quoted in Eq.~\eqref{eq:as-nom}.
Note that the uncertainty in \mt due to the missing Coulomb quasi bound state effects would be considerably smaller, once instead of the pole mass scheme, a renormalisation scheme is chosen, where these Coulomb corrections can be partially absorbed into \mt itself. As shown  in Ref.~\cite{Makela:2023xnt}, this can be achieved by using the MSR mass $\mtmsr(R)$ for a scale $R\approx80\GeV$.

While the resulting values of \mtp and \asmz in Ref.~\cite{CMS:2019esx} are very precise, the central value of \asmz is small in comparison to other extractions at NLO, and to the world average result. In the CMS work~\cite{CMS:2021yzl},
the normalised triple-differential \ttbar cross sections of Ref.~\cite{CMS:2019esx} and further data sets used therein, were included in the QCD fit together with
the double-differential cross section of inclusive jet production at \sqrts of 13\TeV. With increased sensitivity to $g(x)$ and the value of \asmz, provided by the jet production measurements, the simultaneous extraction of PDFs, \asmz, and \mtp could be further refined. The value $\asmz=0.1188\pm0.0031$ is obtained at NLO~\cite{CMS:2021yzl}, in good agreement with the world average, and the value of $\mtp=170.4\pm0.7\GeV$ is obtained with improved precision.

\subsection{Top quark pole mass extracted from \texorpdfstring{\ttbarjet}{tt+jet} events}
\label{sec:ttjetanalysis}

Alternatively to the \mt extraction using inclusive \ttbar production, a novel observable was suggested in Ref.~\cite{Alioli:2013mxa} to extract \mt using events where the \ttbar pair is produced in association with at least one energetic jet (\ttbarjet). Here, the dependence of the gluon radiation on \mt through threshold and cone effects is explored. The observable of interest $\rho$ is defined\footnote{Should not to be confused with correlation coefficients of Ref.~\cite{CMS:2019esx}.} as
\begin{equation}
    \rho=\frac{340\GeV}{\mttbarjet},
\end{equation}
where \mttbarjet is the invariant mass of the \ttbarjet system using the leading additional jet.
By using the \ttbarjet normalised differential cross section as a function of $\rho$, \mt can be extracted. The result of the measurement is independent of the choice of the scaling constant in the numerator, which is introduced to define $\rho$ dimensionless, and is on the order of two times \mt.

\begin{figure}[!ht]
\centering
\includegraphics[width=0.57\textwidth]{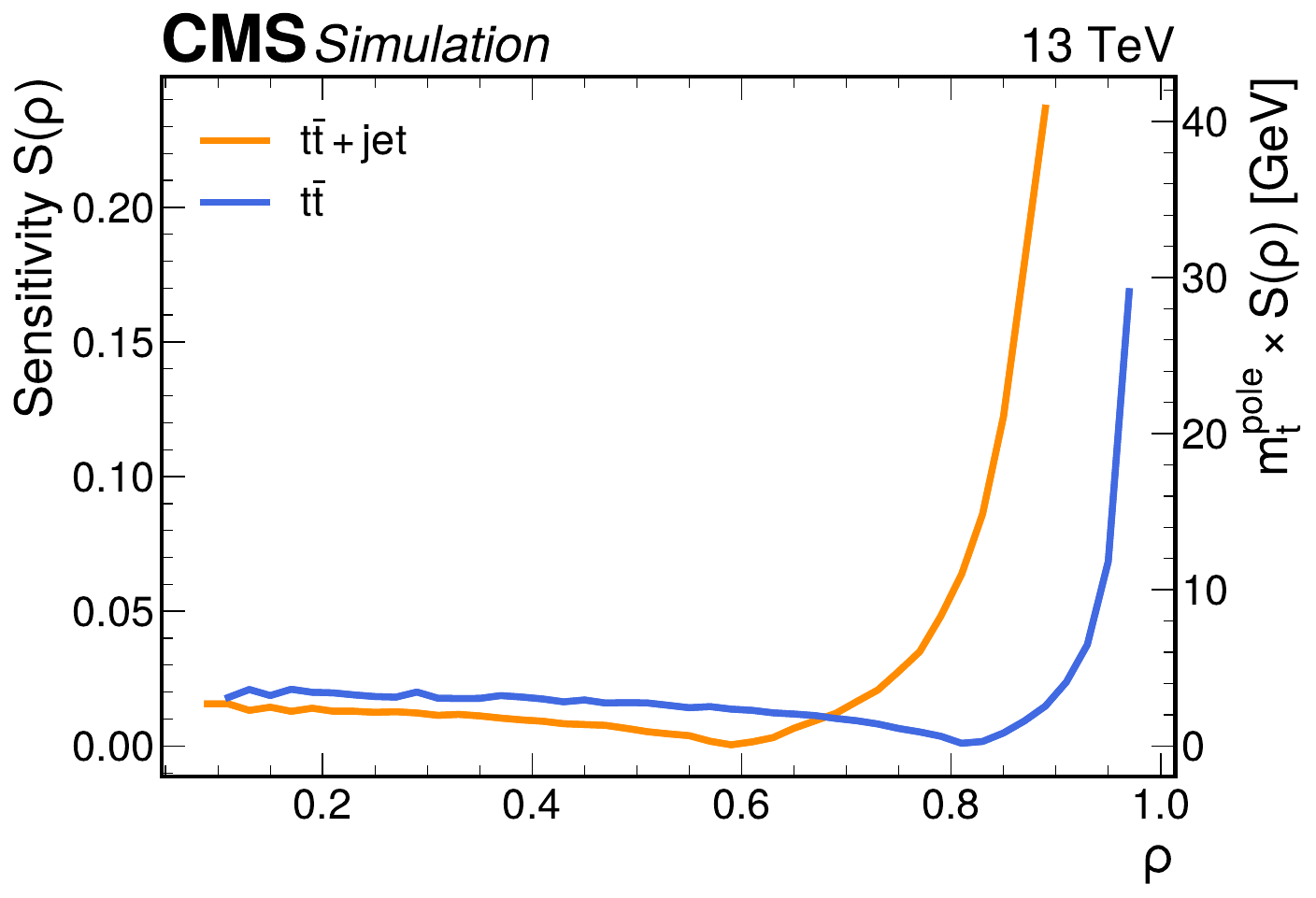}%
\hfill%
\includegraphics[width=0.4\textwidth]{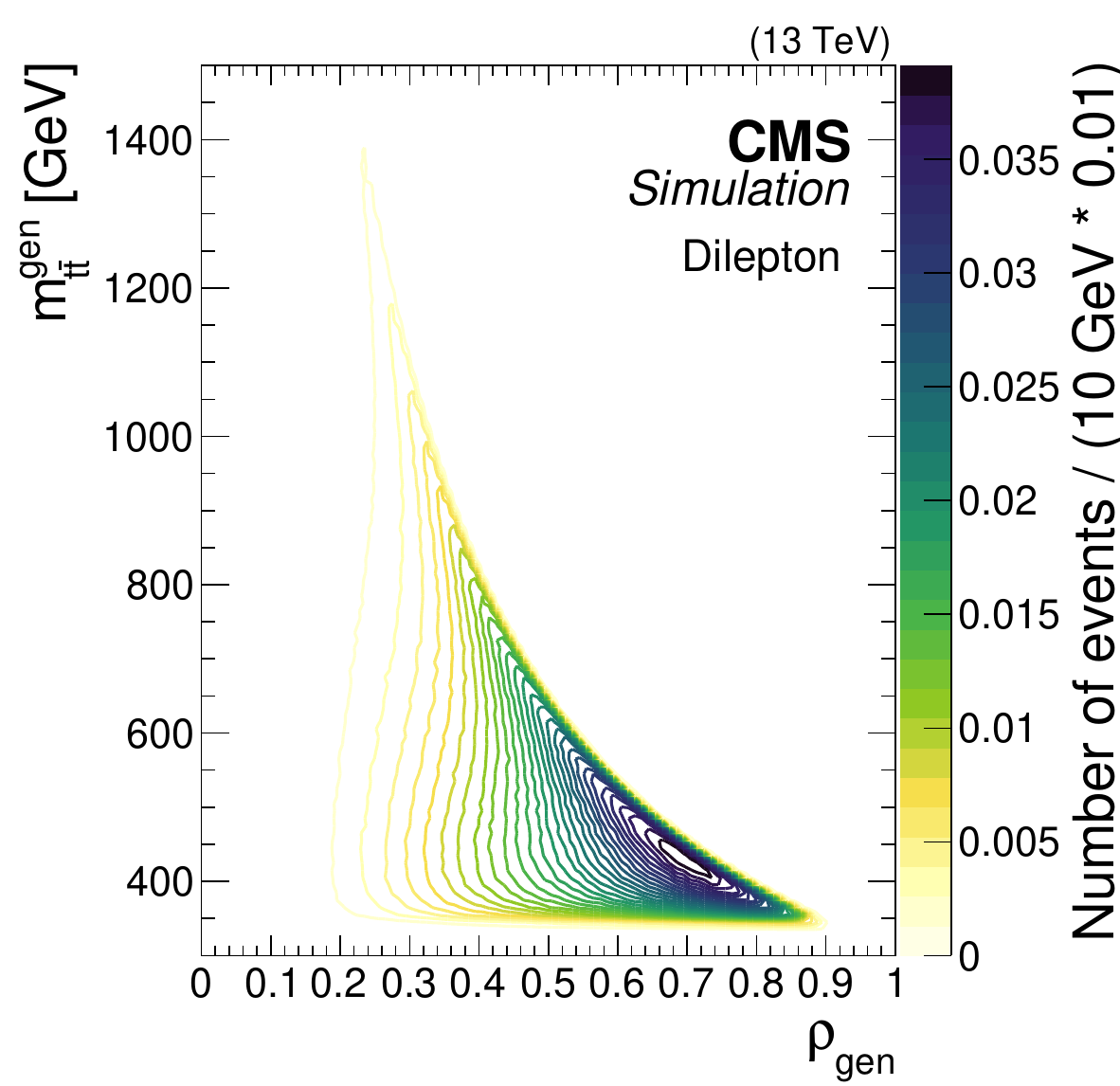}
\caption{%
    Left: Sensitivity \sensitivity to the value of \mtp for \ttbar (blue) and \ttbarjet production (orange).
    Figure taken from Ref.~\cite{CMS:2022emx}.
    Right: The distribution of \mtt at the parton level as given by the \POWHEGPYTHIAEight\ \ttbar simulation as a function of $\rho$ at parton level, obtained in Ref.~\cite{CMS:2022emx}.
}
\label{fig:ttj_rho_vs_mtt}
\end{figure}

A high sensitivity to \mt is expected close to the production threshold, for $\rho>0.65$, while for high \mttbarjet, \eg $\rho<0.55$, this sensitivity is small. The sensitivity \sensitivity is defined as~\cite{Alioli:2013mxa}
\begin{equation}
	\sensitivity(\rho)=\sum_{\delmtp\,=\,\pm3\GeV}
    \frac{\rhodiffxsec(\rho,\mtp) - \rhodiffxsec(\rho,\mtp+\delmtp)}{2\abs{\delmtp}\rhodiffxsec(\rho,\mtp)},
\end{equation}
where \rhodiffxsec is the normalised differential cross section of \ttbarjet production as a function of $\rho$ and \delmtp the variation of \mtp.
The value of \sensitivity quantifies how the differential cross section changes, as a result of the variation in \mtp and is studied in Ref.~\cite{Alioli:2013mxa} by using the \POWHEG generator.
In Fig.~\ref{fig:ttj_rho_vs_mtt} (left), the \mt sensitivities are compared for \ttbarjet and inclusive \ttbar production. For the latter, in the definition of $\rho$, the invariant mass of \ttbarjet is replaced by the invariant mass of the \ttbar pair, \mtt. For both processes, the sensitivity is largest close to the threshold of the \ttbar production, however in the case of \ttbarjet this sensitivity is significantly increased due to the presence of additional gluon radiation. The infrared safety is assured through the requirement for the additional jet in \ttbarjet to have a transverse momentum of at least 30\GeV. As compared to the \ttbar production, the kinematic range accessed by \ttbarjet is shifted further away from the threshold region, where the highest sensitivity to \mt is expected, as shown in Fig.~\ref{fig:ttj_rho_vs_mtt} (right). On the other hand, the reliable theoretical prediction in this region would require resummation of threshold effects and soft-gluon emission, not yet fully available for \ttbar production in \pp collisions.

The first extraction of \mtp using \ttbarjet events in CMS~\cite{CMS:2022emx} was performed at \sqrtseq{13}, using
\pp collision data collected by the CMS experiment in 2016 and corresponding to an integrated luminosity of 36.3\fbinv. Dilepton decays of
\ttbar are used, and a novel method of kinematic reconstruction, based on a NN regression, developed for the purpose of this measurement, is applied, as discussed in details in Section~\ref{sec:kinreco}.
By using a maximum likelihood fit to the final-state distributions of \ttbar and \ttbarjet events, the differential cross section of \ttbarjet production as a function of $\rho$ is measured. The method of Refs.~\cite{CMS:2019jul,CMS:2018fks}, as described above, is extended in order to constrain systematic uncertainties in the visible phase space together with the differential cross section. To mitigate the correlation between the extracted cross section and \mtmc, the latter is treated as an additional free parameter in the fit, by considering the \mlbmin distribution.

The cross section is measured at the parton level, as defined in Section~\ref{sec:particlePartonLevel}. Additional jets are reconstructed using the anti-\kt algorithm with a distance parameter of 0.4, and jets originating from the top quark decay products are removed. At least one such additional jet at the parton level with $\pt>30\GeV$ and $\abseta<2.4$ is required.
This definition allows for the direct comparison of the measurement to the fixed-order theoretical predictions.
The measurement~\cite{CMS:2022emx} is performed in four bins of \rhotrue and \rhoreco: 0--0.3, 0.3--0.45, 0.45--0.7, and 0.7--1.0. Eleven exclusive event categories are introduced, based on the number of \PQb-tagged jets ($\nbj=1$, $\nbj\geq2$), jets ($\nj=1$, $\nj=2$, $\nj\geq3$), and the four bins in \rhoreco, as listed in Table~\ref{tab:categories}.
In the \rhoreco categories, a discriminating variable (\dnnresponse) originating from a NN-based multiclassifier is fitted to maximise the signal sensitivity. The classifier aims to separate events originating from the \ttbarjet, \ttbarnojet, and \Zjets processes, and \dnnresponse is defined such to optimise the \ttbarjet over \ttbarnojet separation. The systematic uncertainties related to the calibration of the JES are constrained by fitting jet \pt distributions.

\begin{table}[!ht]
\centering
\topcaption{%
    A list of the event categories and distributions used in the maximum likelihood fit.
}
\label{tab:categories}
\renewcommand{\arraystretch}{1.4}
\begin{tabular}{ccccc{c}@{\hspace*{5pt}}cc}
    & \multicolumn{4}{c}{Reconstructed $\rho$} && \multicolumn{2}{c}{No reconstructed $\rho$}                           \\
    \cline{2-5}\cline{7-8}
    & \multicolumn{4}{c}{$\nj\geq3$}  && \multirow{2}{*}{$\nj=1$} & \multirow{2}{*}{$\nj=2$} \\
    & $\rho<0.3$    & $0.3<\rho<0.45$    & $0.45<\rho<0.7$  & $\rho>0.7$ &&                           &                         \\\hline
    $\nbj=1$ & \dnnresponse     & \dnnresponse     & \dnnresponse   & \dnnresponse && $\pt^{\text{leading jet}}$                       & $\pt^{\text{subleading jet}}$                    \\
    $\nbj\geq2$ & \dnnresponse     & \dnnresponse     & \dnnresponse  &  \dnnresponse && \NA                         & \mlbmin
\end{tabular}
\end{table}

\begin{figure}[!t]
\centering
\includegraphics[width=0.48\textwidth]{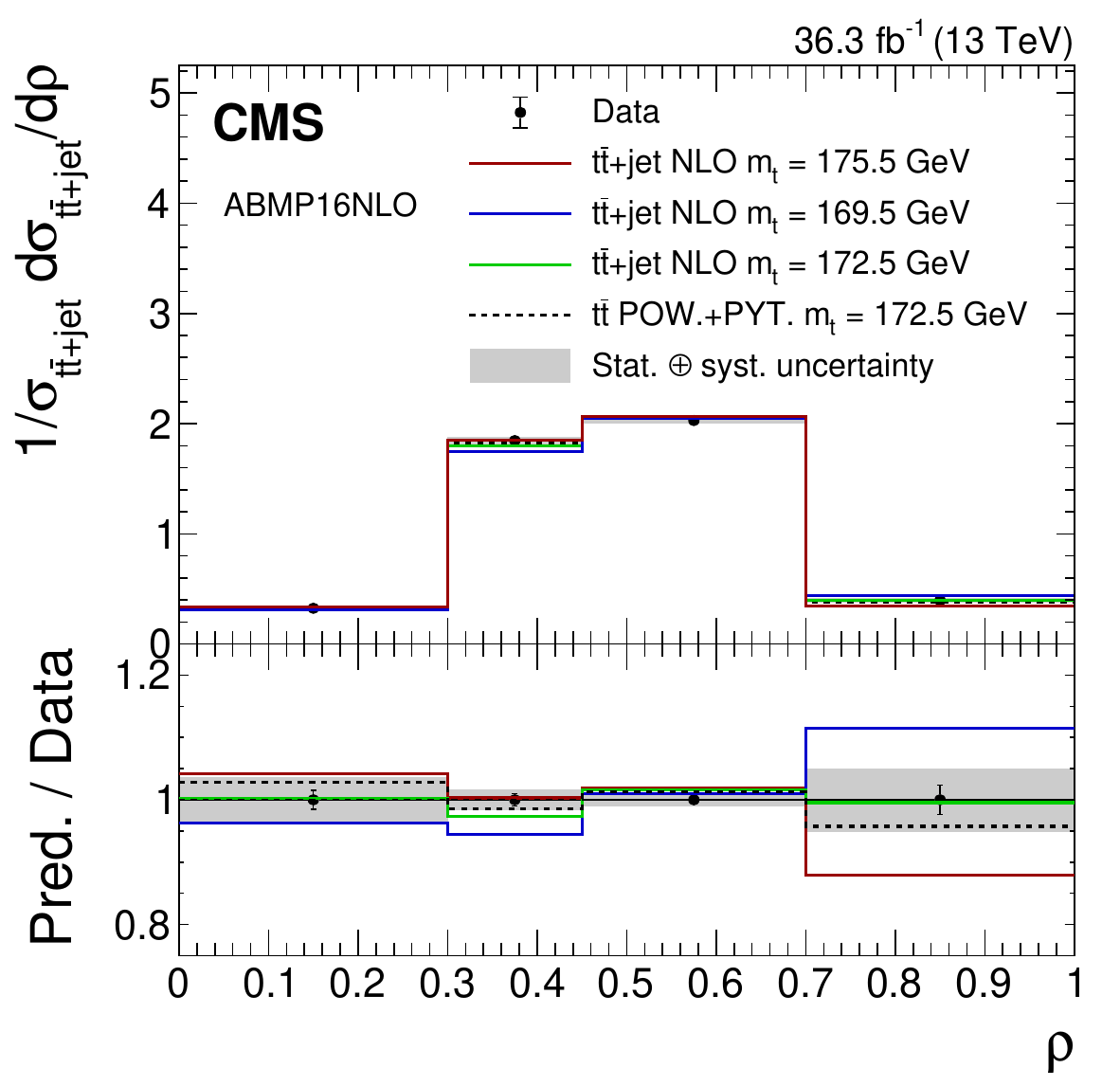}%
\hfill%
\includegraphics[width=0.48\textwidth]{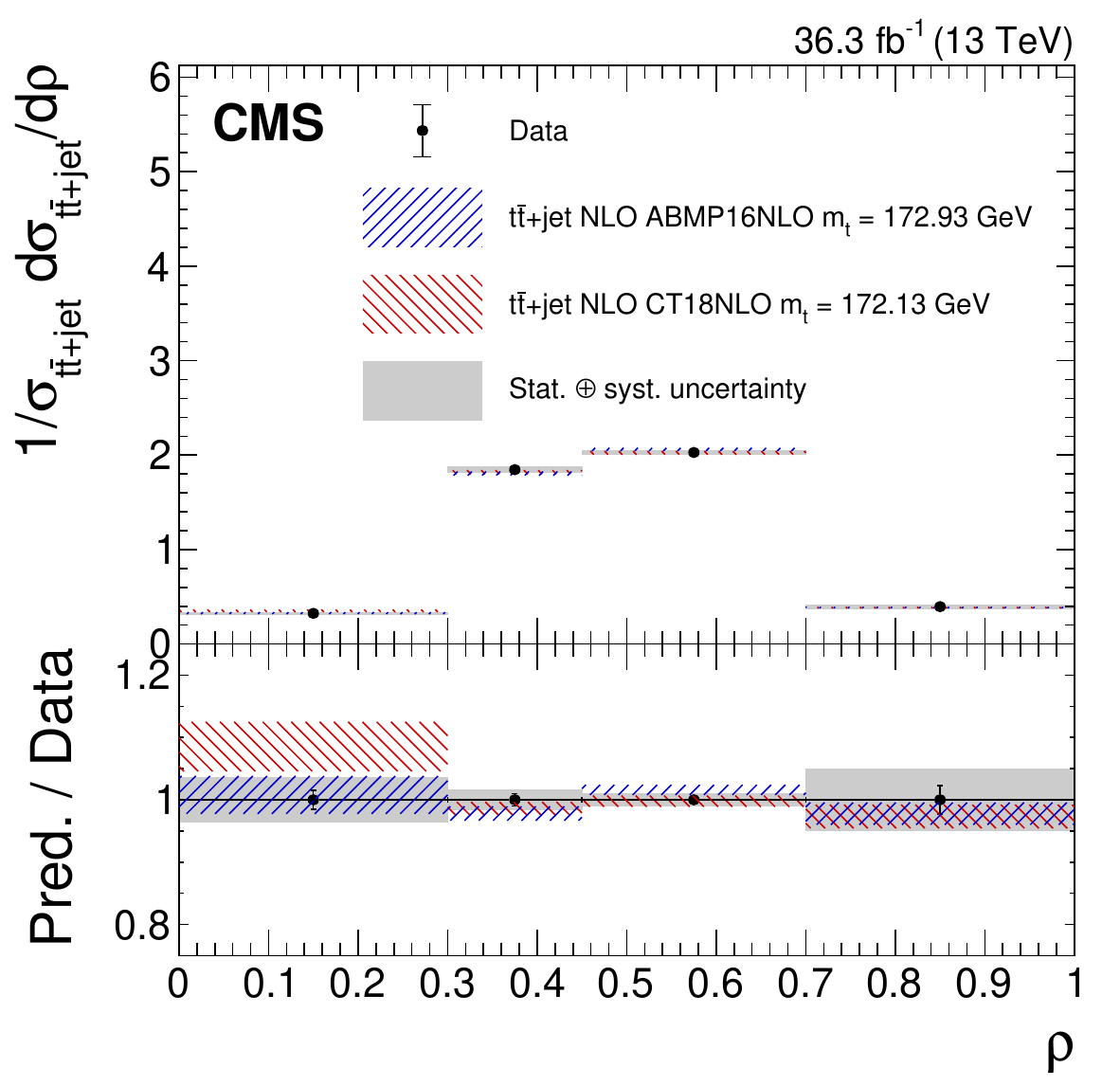}
\caption{%
    The measured normalised \ttbarjet differential cross section (closed symbols) as a function of $\rho$.
    The vertical error bars (shaded areas) show the statistical (statistical plus systematic) uncertainty. The data are
    compared to theoretical predictions and the \POWHEGPYTHIAEight simulation, either using alternative values of \mt (left panel), shown by the solid lines, or two alternative PDF sets (right), shown by the hatched areas.  In the lower panels, the ratio of the predictions to the measurement is shown.
    Figures taken from Ref.~\cite{CMS:2022emx}.
}
\label{fig:ttj_diffXsec}
\end{figure}

The resulting \ttbarjet cross section is shown in Fig.~\ref{fig:ttj_diffXsec}. It is compared to fixed-order theoretical calculations obtained using the \ttbarjet process implemented in \POWHEG-\textsc{box}~\cite{bib:ttjPowheg} at NLO, with the ABMP16NLO~\cite{bib:ABMP16} PDF set, and assuming \mtp values of 169.5, 172.5, and 175.5\GeV. Alternatively, the CT18NLO PDF set~\cite{bib:CT18} is considered. The NLO calculation benefits from the implementation of a dynamical scale, as discussed in Ref.~\cite{bib:ttjPheno}, which depends on the scalar sum of the top quark and antiquark transverse masses and the \pt of the additional jet.

{\tolerance=3100
The value for \mtp is extracted using a \chisq fit of the theoretical predictions to the measured normalised \ttbarjet cross section, taking into account its full covariance obtained from the likelihood fit. The PDF uncertainties are evaluated in each bin and included in the total covariance matrix.
For CT18NLO, the uncertainties evaluated at 90\% confidence level (\CL), are symmetrised and rescaled to the 68\% \CL to be consistent with the precision of the ABMP16NLO PDF. To estimate the scale variation uncertainty, the fit is repeated for each choice of \mur and \muf and the maximum difference in the results to the nominal one was considered as the total uncertainty.
Using the ABMP16NLO PDF set, the resulting \mtp value is obtained as
\begin{equation}
    \mtp=172.93\pm1.26\fit\,^{+0.51}_{-0.43}\scale\GeV.
\end{equation}
Using the CT18NLO PDF set instead, this results in
\begin{equation}
    \mtp=172.13\pm1.34\fit\,^{+0.50}_{-0.40}\scale\GeV.
\end{equation}
The total uncertainty in \mtp corresponds to 1.37 (1.44)\GeV for the ABMP16NLO (CT18NLO) PDF set. The comparison of the predictions using the best fit top quark mass value to the unfolded data is shown in the right panel of Fig.~\ref{fig:ttj_diffXsec}.
The impact of the individual PDF uncertainties is estimated to be 0.35 (0.27)\GeV for the CT18NLO (ABMP16NLO) PDF set by excluding the effect of the PDF uncertainties in a \chisq fit and replacing the central values of the measured cross section with the ones obtained from the theoretical prediction.
\par}

\subsection{Problems and prospects for indirect top quark mass extraction}
\label{sec:indirect_outlook}

The described indirect methods to extract \mt from \pp collision data using \ttbar and \ttbarjet production result in an uncertainty of about 1\GeV.

The experimental uncertainties in \mt, obtained indirectly by using inclusive \stt are limited by the uncertainty associated with the integrated luminosity, which itself is a subject of careful refinements and improvements~\cite{Babaev:2023fim}. The main limitation of such measurements, however, arises from the correlations of PDFs, \asmz, and \mt in the theoretical predictions for \stt and resulting theoretical uncertainty.

Therefore, the most precise \mt results are obtained in analyses, where together with \mt, the PDFs and \alpS are extracted, based on normalised multi-differential \stt measurements, so that the respective correlations are mitigated. To ensure minimal uncertainty in the theoretical prediction, calculations at NNLO or higher order are of an advantage. The presence of a reconstructed jet in the final state makes the computation of NNLO QCD correction more involved so that in the foreseeable future only theoretical predictions at NLO may be available for the \mttbarjet analysis. Therefore, the extraction of \mt by using \ttbar production seems currently more preferable, which makes \mtt and \ytt most promising observables of interest. In the HL-LHC scenario, improvements in experimental precision in the measurement of \mtt or \ytt distributions, and in turn of \mt or \mtmu are expected from better population of the respective spectra~\cite{Azzi:2019yne}.

Further improvements in the precision in \mt would require several important developments in the theoretical predictions that can be used for the experimental analyses: improved description of the threshold of \ttbar production; implementation of scale-dependent and renormalon-free mass schemes with suitable scale choice prescriptions for the different observables; availability of open-source, fast, and numerically precise multi-differential calculations of \ttbar and \ttbarjet production to at least NNLO in QCD with fast-grid interface to PDF convolution; and availability of electroweak corrections to at least NLO with a systematic treatment of finite-width and off-shell effects. In the following, the need for these improvements is discussed in more details.

In \ttbar production, calculated recently at NNLO in QCD~\cite{Kidonakis:2014pja, Kidonakis:2019yji,Czakon:2015owf, Czakon:2017wor, Czakon:2018nun, Catani:2019hip, Czakon:2020qbd}, the strongest sensitivity to \mt arises from the threshold \ttbar region, \ie, where \mtt is in the range from 340 to 360\GeV. However, in this region, the fixed-order perturbative calculations become insufficient and the theoretical uncertainty cannot be estimated reliably through the common normalisation scale variations. Here, nonrelativistic quasi-bound state QCD corrections become important since the produced top quarks attain small nonrelativistic velocities in the \ttbar centre-of-mass frame, and the dynamics of the \ttbar system is governed by \mt, relative momentum, and kinetic energy of the top quark. Appearance of ratios involving the masses, momenta, and kinetic energy of the top quark makes the standard fixed-order expansion in powers of \alpS unreliable and, in contrast to the simpler situation at \EE linear colliders~\cite{Hoang:2000yr}, colour singlet as well as colour octet \ttbar states need to be described systematically.
The most pronounced quasi-bound state effects arise from the Coulomb corrections due
to the exchange of gluons between the produced \PQt and \PAQt. There are a number of predictions available for the Coulomb corrections~\cite{Hagiwara:2008df,Kiyo:2008bv,Ju:2020otc}, suitable for the threshold region and provided in the pole mass scheme. It was shown in the NLO analysis of Ref.~\cite{Makela:2023xnt} that the fixed-order corrections in the threshold region are significantly smaller if the MSR mass at an intermediate scale $R\approx80\GeV$ is employed, since this choice partially sums bound state binding energy effects that lower the threshold value of \mtt.
However, none of the current theoretical predictions provides an adequate description of the entire lowest \mtt interval between 300\GeV and the quasi-bound state region, where the imaginary energy and the optical theorem approach to account for the top quark width~\cite{Hoang:2000yr} used in Refs.~\cite{Hagiwara:2008df,Kiyo:2008bv,Ju:2020otc} is not adequate and yields an unreliable description of the \ttbar production rate (as shown in Ref.~\cite{Ju:2020otc}). Here, a matching to nonresonant production of the top quark related final states as well as a careful account for definition of the reconstructed experimental final state needs to be implemented. Furthermore, a systematic treatment of the intermediate region for \mtt above 360\GeV has to be devised, where the nonrelativistic and relativistic calculations are matched, such that the reliable uncertainty estimates in this region are possible. It should also be mentioned that the foundation of the particle to parton unfolding procedure to determine the momenta of the top quarks and antiquarks in the on-shell approximation that is used in the theoretical differential \ttbar cross section predictions deserves some scrutiny from the theoretical perspective because it is based entirely on the particle picture of the top quark implemented in the simulations.

An important further desired theoretical improvement concerns the implementation of top quark mass renormalisation schemes for the differential cross section, most notably the \msbar mass \mtmum (suitable for scales above \mt) or the MSR mass $\mtmsr(R)$ (suitable for scales below \mt) with adaptable choice of the mass renormalisation scales \mumass and $R$, to allow for flexible dynamical scale settings. This also avoids the impact of the pole mass renormalon problem, which will become increasingly relevant for improving precision.
Currently, no open-source code for calculation of differential cross sections at NNLO using an arbitrary short-distance mass scheme is yet available. Further, to perform a full QCD analysis with simultaneous extraction of \mtmt, \asmz, and PDFs, the interpolation of fast-grid techniques (\eg \textsc{fastNLO}~\cite{Britzger:2012bs}, \textsc{APPLgrid}~\cite{Carli:2010rw} or  \textsc{APPLfast}~\cite{Britzger:2022lbf}) to such a theoretical calculation would be necessary. It should also be mentioned that eventually electroweak corrections should be provided in the cross section predictions used for the experimental analyses. This also entails the treatment of off-shell and nonresonant effects and the dependence on the definition of the electroweak vacuum expectation value~\cite{Kataev:2022dua,Dittmaier:2022maf} that affects the relation of the pole or the MSR mass, both of which can be defined in theories where all massive boson effects are integrated out, with the \msbar mass and the top quark Yukawa coupling relevant for applications at the electroweak scale and above. Furthermore, the availability of off-shell theoretical calculations, implying only top quark decay products in the final state, would imply changes in the experimental analysis strategy, since no unfolding to the parton level would be required.

\section{Measurements in the Lorentz-boosted regime}
\label{sec:boosted}

Measurements of the jet mass in decays of Lorentz-boosted top quarks provide an alternative approach to \mt measurements in a phase space region where the top quarks are produced at very high \pt, dominated by different systematic uncertainties than direct \mt measurements and indirect extractions of the top quark mass.
As is the case for the direct measurements presented in Sec.~\ref{sec:direct}, the measurements in the boosted top quark regime aim to extract \mt from the invariant mass of the top quark decay products. Ultimately, the goal is to use theoretical calculations of the top jet mass at particle level, contained in a single jet, to extract the top quark mass from the measured, unfolded, jet mass distribution. However, currently such calculations do not yet exist in the same phase space that is probed experimentally. The measurements presented in this section therefore rely on MC simulation to extract the \mtmc parameter in a manner analogous to the direct measurements at low top quark boosts, where the top quark decays can be resolved in separate jets using traditional jet clustering techniques.
As the reconstruction techniques in boosted topologies are affected by systematic uncertainties in different ways, this constitutes a valuable alternative measurement from an experimental point of view.

From the theoretical perspective, the boosted topology where the top quark and antiquark decay products are well separated offers the possibility of analytic and resummed particle-level theory predictions that may eventually lead to alternative measurements of \mt in a well-defined renormalisation scheme.
In this regime, Coulomb effects modifying predictions in the \ttbar threshold region, important for the indirect top quark mass extraction, are irrelevant.
The sensitivity to the top quark mass predominantly comes from the inclusive kinematic properties of the jet initiated by a boosted top quark and its decay products, and subtle effects from the modelling of the inclusive and differential \ttbar production cross sections have a negligible impact.

Although top quarks are dominantly produced at lower \pt, top quarks with large \pt are still abundantly produced at the LHC.
Their decay products receive large Lorentz boosts and are thus strongly collimated, such that the fully hadronic decay \ttobqq can be reconstructed with a single large-$R$ jet, where $R$ is the jet distance parameter and usually lies in the range 0.8--1.2.
The distribution in the invariant mass (\mjet) of these jets features a distinct peak, the position of which is closely related to the value of \mt.
The \mjet measurement is robust against typical uncertainties affecting \ttbar production close to the threshold, such as uncertainties in the proton PDFs, resummation effects, and Coulomb corrections.
In addition to having complementary uncertainties, this measurement is based on high-energy events that have a negligible impact on direct measurements, and thus constitutes an additional independent method, which can readily be combined with other measurements of \mt.

An analysis of the measured distribution of \mjet allows for a precise determination of \mt,
which can be \mtmc in a generator-based analysis or the top quark mass in a well-defined renormalisation scheme in an analysis based on analytic theory calculations.
The jet mass distribution of boosted top quarks has good prospects for systematic analytical
first-principle QCD predictions at the particle level.
The boosted topology allows the application of factorisation and effective theory methods
for hadron-level descriptions that do not rely on multipurpose MC event generators.
Theoretical studies in this direction are based on the strong collimation of the top quark decay products,
such that all relevant QCD radiation can be classified into factorisable soft, collinear,
or collinear-soft radiation (in the directions of the top quark and antiquark) where also jet grooming
techniques can be accounted for~\cite{Fleming:2007qr,Fleming:2007xt,Hoang:2017kmk,Hoang:2019ceu}.
As for observables related to global event shapes used in the conceptual studies of Refs.~\cite{Butenschoen:2016lpz,Hoang:2018zrp}, these analytic computations allow for a consistent
implementation of the top quark mass in well-defined renormalisation schemes.
Unfortunately, because of very limited statistical precision, the phase space with jet $\pt>750\GeV$, for which the theoretical results~\cite{Hoang:2017kmk,Hoang:2019ceu} are currently available, is not experimentally accessible with the LHC \Run2 data.
Still, we perform the extraction of \mtmc based on the predicted \mjet distributions
from simulations by MC event generators in analogy to the direct measurements.
This measurement of \mtmc is, however, quite uncorrelated from direct measurements
and demonstrates the principle capability and precision of this method.
For the time being, this approach also provides an important consistency check of the
direct measurements within the MC simulation framework.
Once the theoretical calculations and experimental measurements are carried out
in a comparable kinematic phase space, the measurement of \mjet may turn into
a precision measurement of a top quark mass in a well-defined mass scheme,
which does not rely on the picture of a top quark particle with a Breit--Wigner distributed mass.

\subsection{Overview of existing jet mass measurements}

All the jet mass measurements by CMS have been performed in the lepton+jets channel of \ttbar production,
where the semi-leptonic top quark decay $\ttobW\to\PQb\Pell\PGnell$ is used to identify \ttbar events, and the measurement is performed on the fully hadronic decay $\ttobW\to\PQb\qqpr$.
The single lepton in this decay mode of the \ttbar system allows the selection of a pure sample with a small background contribution, and is required to be an electron or muon carrying a minimum \pt of approximately 50\GeV.
We require each event to have exactly two large-$R$ jets with high \pt, aiming at reconstructing the hadronic top quark decay \ttobqq in one jet, and the \PQb jet of the leptonic top quark decay in a separate jet with large angular separation.
The jet containing the hadronic top quark decay is identified by the larger distance to the single lepton and is required to have $\pt>400\GeV$.
In addition, \mjet has to exceed the invariant mass of the system composed of the second jet and the single lepton.
The latter criterion should always hold true if all products of the hadronic decay are within the selected jet, since the neutrino from the leptonic decay is not reconstructed.

The CMS Collaboration has carried out three measurements of the jet mass in decays of boosted top quarks. The first measurement has been
performed using 8\TeV data corresponding to an integrated luminosity of 19.7\fbinv~\cite{CMS:2017pcy}.
This measurement has large statistical and modelling uncertainties, with a total uncertainty in the extracted value of \mt of 9\GeV. Nevertheless, it was the first measurement of this kind and showed the possibility of a determination of \mt from the jet mass.
The first \mjet measurement at \sqrtseq{13} used data corresponding to an integrated luminosity of 35.9\fbinv~\cite{CMS:2019fak}.
The increase in centre-of-mass energy, together with the larger data set, resulted in an increase in the number
of selected events by more than a factor of ten with respect to the 8\TeV measurement.
The use of a novel jet reconstruction resulted in a decreased width of the \mjet distribution at the particle level and better experimental resolution in \mjet, which subsequently improved the sensitivity to \mt.
Furthermore, the optimised jet clustering led to a significant reduction in the experimental and modelling uncertainties, resulting in a total uncertainty of 2.5\GeV in \mt.
The most recent measurement used the \Run2 data set corresponding to an integrated luminosity of 138\fbinv~\cite{CMS:2022kqg}. For this measurement, CMS has developed a new method for calibrating the jet mass,
and an auxiliary measurement of the jet substructure of large-$R$ jets has resulted in a smaller
uncertainty from the modelling of final state radiation. These improvements, together with the larger
data set, result in an uncertainty of 0.84\GeV in \mt.

\subsection{The jet mass}

The jet mass is defined as the invariant mass of the sum of all jet constituent four-momenta,
\begin{equation}
    \mjet^2=\bigg(\sum_i^N p_i\bigg)^2,
\end{equation}
where $p_i$ is the four-momentum of constituent $i$ from $N$ jet constituents.
In gluon and light-quark jets, the jet mass is dominantly generated by a series of collinear $1\to2$ splittings.
The invariant mass of two massless particles $i$ and $j$ can be approximated by
$m^2\approx p_{\mathrm{T},i}\,p_{\mathrm{T},j}\,\DRij^2$~\cite{Salam:2010nqg}
and depends on the \pt of both particles and their angular separation \DRij.
This causes \pt-dependent Sudakov peaks~\cite{Dasgupta:2013ihk} in the \mjet distribution
in light-quark and gluon jets.
In the case of on-shell decays of top quarks, the dominant part of the jet mass is generated by
the resonance decay, with corrections from additional radiation.
In order to have a reliable correlation between the peak in the \mjet distribution and the value
of \mt, the precise knowledge of which constituents produced in the event
are included in the calculation of \mjet is mandatory.
Ideally, within the picture of an on-shell decay of a top quark,
all particles from the top quark decay would be included in the large-$R$ jet.
This would only be possible if the size of the jet cone is equal to or larger than the largest angular
distance between the decay products of the top quark, which depends on the top quark \pt.
In the following discussion and in the evaluation of suitable jet algorithms, we use the picture of an on-shell top quark particle decaying via
\ttobqq, as it is implemented in event generators simulating \ttbar production,
where we use the generator information of the three decay quarks at the parton level before PS.
Even though this simplified picture is used to find an optimal jet reconstruction algorithm,
the analysis does not rely on this simplified picture, since the jet mass is defined by the jet constituents
at the particle level as discussed below.
After the unfolding to the particle level,
the data include effects not accounted for in event generators,
such as gluons that provide colour neutralisation and off-shell contributions beyond the Breit--Wigner mass distribution.
For the \mtmc measurement it is implicitly assumed that these effects are small.

\begin{figure}[!ht]
\centering
\includegraphics[width=0.6\textwidth]{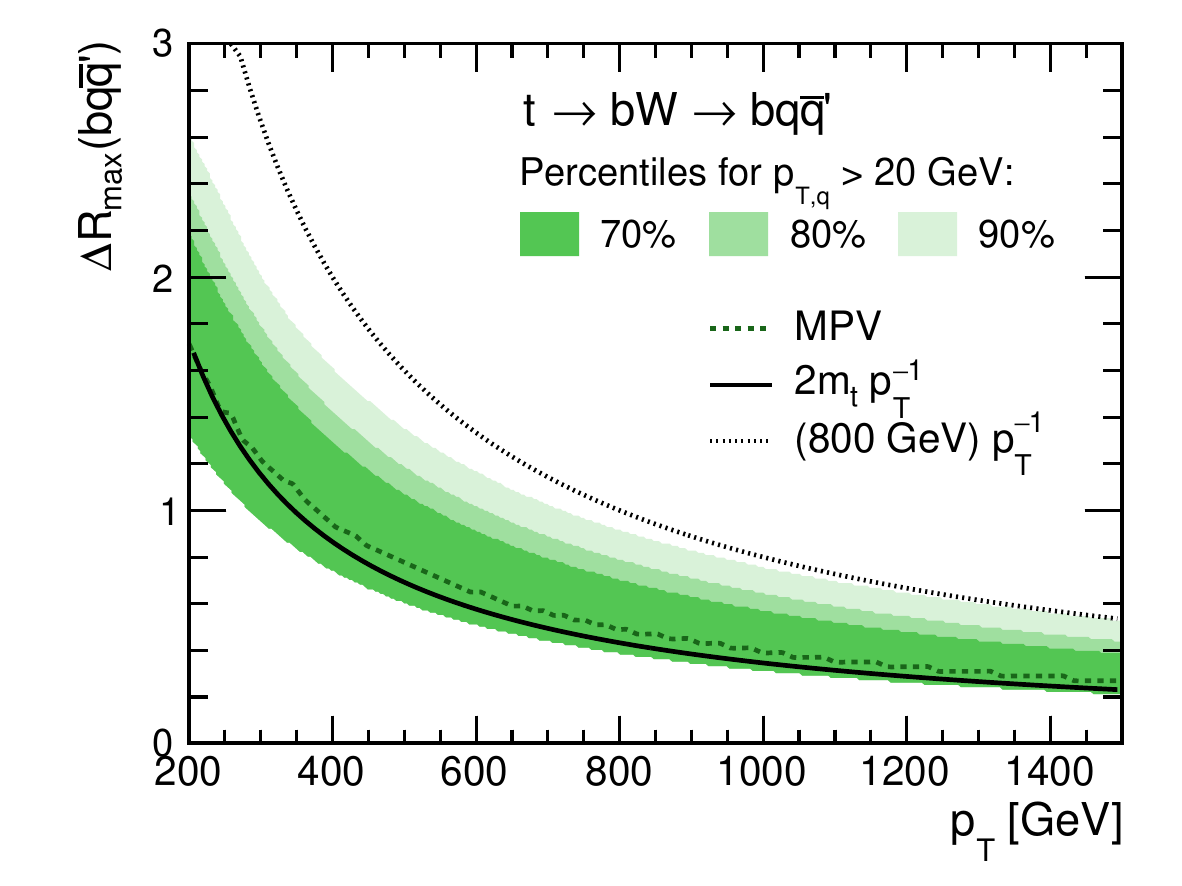}
\caption{%
    Percentiles of maximum angular distance between the top quark decay partons as a function of the top
    quark \pt obtained from \ttbar simulation. The filled bands indicate the areas that are
    populated by 70, 80, and 90\% of all simulated \ttbar events,
    where the decay partons have at least $\pt>20\GeV$. The most probable value (MPV) is
    shown as a dashed line, and two functional forms are shown that approximate the \pt-dependence
    of $\DR_{\text{max}}$.
    Figure taken from Ref.~\cite{Kogler:2021kkw}.
}
\label{boosted:kinematics}
\end{figure}

Figure~\ref{boosted:kinematics} shows the most probable region of maximum distance of the three partons from the decay
\ttobqq, as a function of the top quark \pt.
At \pt larger than 800\GeV, a distance parameter of $R=0.8$ is sufficient to fully reconstruct
the decay products of the top quark in about 80\% of the time.
In order to obtain a similar coverage at lower \pt, the value of $R$ has to be increased
proportionally to approximately $1/\pt$.

The jet mass is affected by additional effects, some of which are not correlated to the top quark decay.
At the particle level, the jet mass receives contributions from ISR, the underlying event, and multi-particle interactions.
Since these processes are not correlated with the production and decay of the top quark,
their effect is independent of the top quark kinematics and scales with $\pt R^4$ because it depends quadratically on the active area of the jet.
The linear dependence in \pt stems from the fact that these contributions
increase the jet \pt, but the leading effect comes from the size of the jet distance parameter.
Since including more particles can only increase the jet mass, the peak position
in the \mjet distribution is shifted towards higher values, and a tail is introduced at large $\mjet\gg\mt$.
The leading power corrections to the jet mass from hadronisation scale as $\pt R$,
and are more than a factor of ten smaller than the effects from the underlying event.
At the detector level, contributions from pileup have a similar effect as the underlying event,
but the effect is larger because of the high energy density of pileup at high instantaneous luminosities.
In the data analysis, several corrections are applied to remove the effects of pileup,
enabled by the possibility to distinguish pileup particles from the hard scattering
and by subtracting on average the pileup contributions from jets, such
that the measured distribution in \mjet at the particle level is free of pileup effects.

The correlation of \mjet to the mass of the particle initiating the jet makes \mjet an important
observable for jet tagging algorithms, where jet substructure information
is used for large-$R$ jet identification~\cite{Larkoski:2017jix, Asquith:2018igt, Kogler:2021kkw}.
In order to increase the tagging performance, grooming or trimming algorithms
are used to remove wide-angle and soft radiation from the jet before calculating \mjet.
Depending on the strength of the grooming algorithm, this largely removes the \pt-dependent
Sudakov peaks in light-quark and gluon jets and leads to a steeply falling \mjet spectrum
with a peak at very small values~\cite{CMS:2018vzn}.
In top quark decays, grooming removes additional particles in the jet from ISR, the underlying event and pileup,
and subsequently improves the jet mass resolution at the detector level and reduces the width of the lineshape
of the \mjet distribution at the particle level, and thus increases the sensitivity to \mt.
For top quark tagging this is an essential tool to increase the separating power
of \mjet in the categorisation into jets initiated by top quarks or light quarks and gluons.
In measurements of \mjet, grooming not only enhances the sensitivity to \mt,
but also removes a large fraction of the nonperturbative effects, particularly arising from ISR and underlying event.
We note that there is no algorithm that removes all nonperturbative effects,
such that these still have to be accounted for in the description of \mjet.

\subsubsection{Theoretical considerations}

An important motivation for these measurements~\cite{CMS:2017pcy} is to confront the measured jet mass distribution directly with theoretical predictions.
The large angular separation between the decay products of the top quark and antiquarks at high
top quark boosts allows for the derivation of factorisation formulae for differential cross sections,
where the scales of the hard interaction, collinear and soft radiation within the jets,
and nonperturbative effects can be separated.
Previous calculations for \EE collisions~\cite{Fleming:2007qr, Fleming:2007xt}, based on soft-collinear effective theory (SCET)~\cite{Bauer:2000ew, Bauer:2000yr, Bauer:2001ct, Bauer:2001yt, Bauer:2002nz} and boosted heavy-quark effective theory~\cite{Fleming:2007qr,Fleming:2007xt}, have been extended to \pp
collisions with the help of light soft-drop grooming~\cite{Hoang:2017kmk,Hoang:2019ceu} to reduce the impact of ISR and the underlying event. Light soft-drop grooming is a less restrictive version of the soft-drop grooming algorithm~\cite{Larkoski:2014wba, Dasgupta:2013ihk} so that the top quark decay products are not affected.
The presented calculation considers top quark jets with $\pt>750\GeV$, where
soft-drop grooming enables the factorisation between the top quark and antiquark,
by removing soft-wide angle radiation, such that the analysis can be carried out in the
lepton+jets channel.
The groomed jet mass is measured on the fully hadronic decay leg of the \ttbar decay,
which has a large angular separation from the semi-leptonic top quark decay,
thanks to the large Lorentz boost.
Light soft-drop grooming, with the soft-drop parameters $z_{\text{cut}}=0.01$ and $\beta=2$~\cite{Hoang:2017kmk},
removes significant nonperturbative contamination from the top quark jet while
retaining collinear radiation associated with the top quark decay products
within the cone defined by the hard jets from the top quark decay.
This allows for a treatment of the top quark and antiquark as individual radiators
and a clear interpretation in terms of a short-distance mass scheme since all radiation that is soft in the top quark (or antiquark) rest frame (called ultracollinear in the laboratory frame) remains ungroomed and is treated inclusively.
A stronger soft-drop grooming, for example with $z_{\text{cut}}=0.1$ and $\beta=0$ as used in many CMS analyses,
would result in a breakdown of the validity of the factorisation formulae since
parts of the ultracollinear radiation would be restricted.
The calculation predicts the jet mass distribution in the MSR and the pole mass schemes, such that
it can be used to determine the MSR mass from a corresponding measurement.
Since nonperturbative effects are not fully removed by the light soft-drop grooming, a
free parameter is introduced in the particle level factorisation formulae to
account for the shift of the \mjet distribution because of the underlying event.
This parameter needs to be obtained from data
and shows a correlation with the value of the top quark mass, which can impact the accuracy of the
\mt determination if not accounted for.
While the requirement of top quark $\pt>750\GeV$ is not yet experimentally accessible
with the present 13\TeV data set because of the small \ttbar production cross section at high \pt,
this measurement will become feasible at the HL-LHC. We also note that the effects from multi-particle interactions and the underlying event are still significant despite grooming,
such that a first-principle description of these effects would be desirable.
The existing calculations provide a tool for the calibration of the top quark mass parameter
in the event generator used for the simulation of \ttbar production, such that a numerical relation between \mtmc and the MSR (or the pole) mass can be determined~\cite{ATLAS:2021urs}.
This is in close analogy to the \mtmc calibration framework proposed in Refs.~\cite{Butenschoen:2016lpz,Dehnadi:2023msm} based on global event shapes in \EE collisions.
Calculations for moderate top quark \pt starting at 400\GeV will need considerable theoretical work,
because the three decay quarks cannot be considered as a single radiator anymore, but a factorisation
theorem needs to be developed taking into account the dynamics of three separate colour-charged radiators.

Finally we note that the \mjet distribution in boosted top quark decays shares many physical aspects with the \EE shape observables~\cite{Hoang:2018zrp}---such as the 2-jettiness---for which some concrete insights concerning the interpretation of the MC top quark mass parameter \mtmc exist. Similar insights do not yet exist for observables close to the ones used for the direct measurements.

\subsubsection{Experimental methods}

The most important experimental elements of this measurement are well reconstructed and calibrated large-$R$ jets.
Jets are clustered from the list of PF candidates as described in Section~\ref{sec:reconstruction}.
In addition to the commonly used anti-\kt jets, large-$R$ jets are clustered for measurements of boosted heavy objects.

In the presented \mjet measurements, all ingredients to jet clustering play a crucial role
since the width of the peak in the \mjet distribution, possible shifts from pileup and the
underlying event, and the jet mass scale (JMS) and resolution directly translate to the sensitivity to \mt.
All three existing measurements of \mjet~\cite{CMS:2017pcy, CMS:2019fak, CMS:2022kqg} make use of jets clustered from a list of PF particles.
The 8\TeV measurement~\cite{CMS:2017pcy} did not use any pileup mitigation technique, while
the measurements at 13\TeV~\cite{CMS:2019fak, CMS:2022kqg} use the CHS algorithm.
A specialised two-step jet clustering was introduced with the first measurement at 13\TeV~\cite{CMS:2019fak}, using the XCone algorithm~\cite{Stewart:2015waa}.
The clustering procedure acts as a grooming algorithm on the large-$R$ jets.
It improves both the peak width and the jet mass resolution by factors of two compared
to the initial measurement at 8\TeV~\cite{CMS:2017pcy} and reduces the
shift of the peak due to additional particles from pileup and the underlying event.
In the future, the measurement of \mjet will also profit from studies in the context of
jet substructure tagging, where PUPPI and soft-drop grooming have been calibrated with sufficient precision.

Another crucial aspect of the \mjet measurement regards an optimal selection of the jet including the hadronic top quark decay.
High-energy ISR and FSR cannot only affect the \mjet distribution
of the top quark jet, but can also lead to the selection of a wrong jet that reconstructs radiation
uncorrelated with the top quark decay.
This leads to enhanced tails to both sides of the \mjet peak and degrades the sensitivity to \mt by
shifting the peak position.
Thus, the jet definition and the selection of the jet that fully contains the \ttobqq decay has to be carefully optimised in order to reduce the
influence of radiation not connected with the top quark decay.

\subsection{Optimising the jet definition for jet mass measurements}

Measurements of the jet mass aim to reconstruct all particles associated with the top quark decay in a single large-$R$ jet.
In \pp collisions at the LHC, additional particles arise from various sources such as pileup, underlying event, and final-state radiation.
Since all these effects can change the jet mass and might even affect the identification of the jet that contains the hadronic top quark decay, a suitable jet algorithm is crucial for measurements of \mjet.
In commonly used jet clustering algorithms the distance parameter $R$ controls the largest distance at which particles are combined to form a jet.
The Lorentz boost that subsequently defines the opening angle of the decay in the lab frame depends on the top quark \pt.
Thus, an optimal value of $R$ has to be chosen such that the jet cone is large enough for a given top quark momentum in order to catch all products of the hadronic top quark decay.
On the other hand, effects from pileup and the underlying event are enhanced with a larger jet size, such that a compromise needs to be made for $R$ sufficiently large, but just large enough.

{\tolerance=800
In the measurement using the LHC 8\TeV data~\cite{CMS:2017pcy}, Cambridge--Aachen (CA)~\cite{Dokshitzer:1997in,Wobisch:1998wt} jets with $R=1.2$ were chosen.
At 8\TeV, this decision was driven by the available size of the selected data set.
A smaller value of $R$ would have improved the experimental resolution but also leads to a larger fraction of top quark decays that are not fully reconstructed within the jet or the need to require a minimum jet \pt larger than 400\GeV.
While the former would have decreased the sensitivity to the top quark mass, the latter would have drastically reduced the already limited statistical precision of the measurement because of the
steeply falling top quark \pt spectrum.
No grooming was applied in this measurement and although the statistical uncertainty dominates the extraction of \mt, the effects of additional particles from the underlying event and pileup are visible in a \pt-dependent shift of the peak in the \mjet distribution.
\par}

\begin{figure}[!t]
\centering
\includegraphics[width=0.48\textwidth]{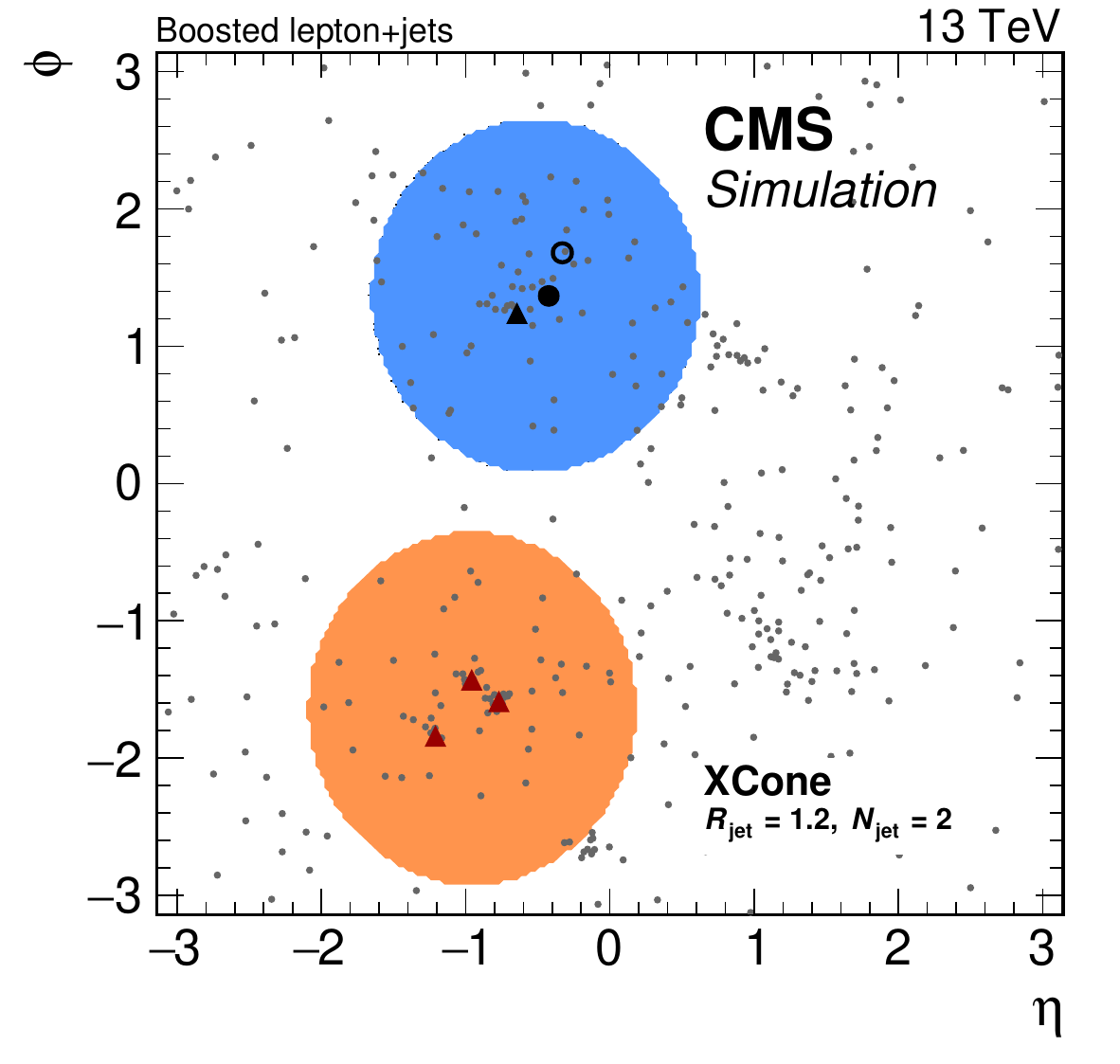}%
\hfill%
\includegraphics[width=0.48\textwidth]{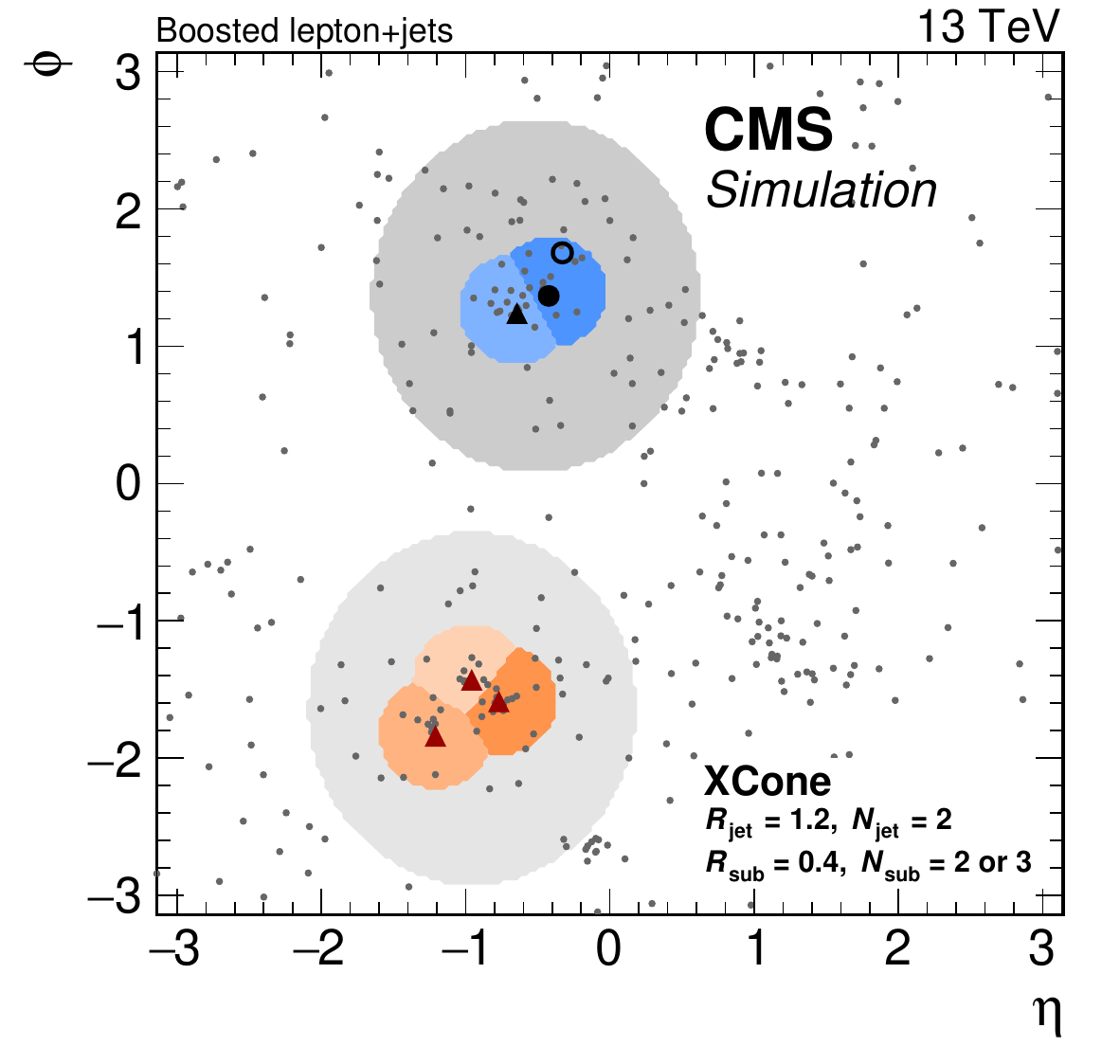}
\caption{%
    Display of a simulated \ttbar event. Each point marks the position of a particle at the particle level in the \etaphi plane. Decay products of the top quarks are highlighted with triangles or larger circles. The red triangles mark the three quarks from the hadronic decay; the black triangle, black circle, and open circle correspond to the \PQb quark, charged lepton, and neutrino from the leptonic top quark decay, respectively. The jet areas are shown as coloured shapes. The left panel shows the first clustering step with $N=2$ and $R=1.2$, while the right panel shows the subjet clustering.
}
\label{boosted:xcone}
\end{figure}

For the first \mjet measurement with 13\TeV data~\cite{CMS:2019fak}, the jet reconstruction was changed from CA jets to a two-step clustering~\cite{Thaler:2015xaa} using XCone~\cite{Stewart:2015waa}.
First, XCone is run with $R=1.2$ and $N=2$ using all CHS PF candidates as input particles.
As an exclusive jet algorithm, XCone returns exactly two large-$R$ jets, where the jet axes are found by minimising the $N$-jettiness~\cite{Stewart:2010tn}.
This setup is optimised to include all partons from the two top quark decays in a phase space where the jet \pt is larger than 400\GeV.
Subsequently, XCone is run again separately for the constituents of each large-$R$ jet, now with $R=0.4$ and $N=3$, which aims at reconstructing the three-prong top quark decay.
All particles that are not part of one of the three subjets are removed from the jet.
In this way, the two-step procedure acts as a grooming algorithm and the effects of additional and soft radiation are mitigated.
A display of the clustering procedure in a simulated \ttbar event is shown in Fig.~\ref{boosted:xcone}.
In this example, the first clustering step reconstructs both top quarks.
In the next step, soft and wide angle radiation is removed by reconstructing three subjets. Ideally, the subjets match the three-prong structure of the hadronic top quark decay.
On the leptonic side, we aim at a two-prong decay and run XCone with $N=2$, since the lepton is part of the clustering and the neutrino cannot be detected. However, the measurement is performed using the hadronic jet only and it was verified that the details of the clustering procedure of the leptonic side do not change the measurement.
In Fig.~\ref{boosted:xcone} another feature of the XCone algorithm becomes visible.
The XCone subjets can be arbitrarily close and form a straight border separating the jets.
In contrast, the anti-\kt algorithm commonly used in other analyses would result in an approximately circular high-energy jet at the centre of the overlap of two jets and lower-energy jets clustering the remnants around the jet in the centre.
This feature of the XCone algorithm allows a reconstruction of the three-prong structure of the top quark decay despite an angular overlap of size $R=0.4$ of the subjets at large Lorentz boosts.
A distinct advantage of this approach is that the two-prong \PW boson decay can be identified and reconstructed from two XCone subjets, which is subsequently used in the calibration of the JMS.

\begin{figure}[!t]
\centering
\includegraphics[width=0.48\textwidth]{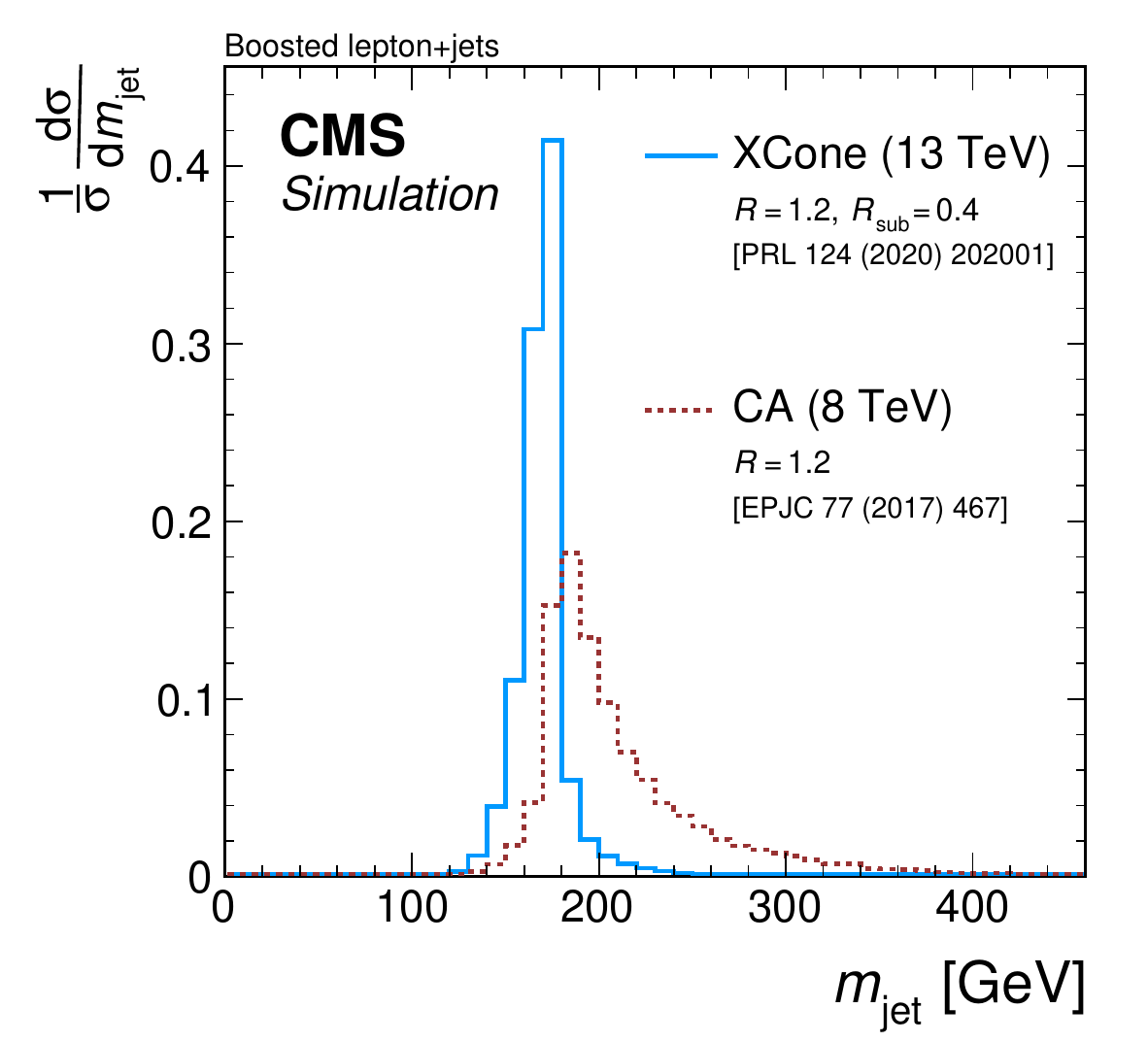}
\caption{%
    Normalised jet mass distribution at the particle level for the two-step XCone clustering (blue solid) used in Ref.~\cite{CMS:2019fak,CMS:2022kqg} and CA jets with $R=1.2$ (red dotted) used in Ref.~\cite{CMS:2017pcy}. Only events where all top quark decay products are within $\DR=0.4$ to any XCone subjet or within $\DR=1.2$ to the CA jet are shown.
}
\label{boosted:particle_level}
\end{figure}

A comparison of this approach to the CA jets used for the 8\TeV measurement
is shown in Fig.~\ref{boosted:particle_level}, displaying the normalised \mjet distribution
for the fraction of ``matched'' events.
The width of the distribution around the peak in \mjet reduces by a factor of two with the
two-step clustering, and the shift of the peak position towards larger values is  strongly reduced.
While the performance is comparable to jets with $R=0.8$, the first step in the XCone clustering with $R=1.2$ maintains high reconstruction
efficiencies also for jets close to the selection threshold of 400\GeV and improves the statistical precision in the measurement.
In this way, the two-step clustering allows a smoother transition between moderately and highly boosted top quark jets.

\subsection{Reconstruction effects in the jet mass}

The event selection at the detector level is very similar to the particle level phase space detailed above in order to minimise migrations in the detector response matrix used in the unfolding, such that the respective corrections are small.
The data are selected with a single-lepton trigger, which usually provides high efficiency in the selection of high-energy \ttbar events in the lepton+jets channel.
Moreover, a few well known and understood selection criteria, such as \PQb jet tagging, a customised lepton isolation, and a cut on \ptmiss, are used in order to reduce backgrounds and select a pure \ttbar sample.

Pileup effects play a role at the detector level, but are absent at the particle level.
Together with detector resolution effects, this leads to a finite jet mass resolution that
highly depends on the jet reconstruction.
Here we define the resolution as the width of the distribution in $(\mjetrec-\mjetgen)/\mjetgen$, where \mjetrec and \mjetgen are the jet mass at the detector and particle levels, respectively.
The specialised XCone reconstruction, because of its grooming, results in a resolution of 7--8\%.
This translates to an improvement by a factor of 2 compared to 14\%, obtained for nongroomed CA jets.
Furthermore, we only observe a very small dependence on the number of reconstructed primary vertices, which indicates a significant reduction of pileup effects.

At detector level, the calibration of physics objects is a crucial aspect of the measurement.
The connected uncertainties are grouped into experimental uncertainties and are dominated by uncertainties in the jet calibration.
Variations in the JES shift the peak in the \mjet distribution
and thus lead to large uncertainties in the extraction of \mt.
At 8\TeV, the statistical uncertainty was very large, such that a reduction of the JES uncertainty would not have improved the measurement precision.
For the first measurement at 13\TeV~\cite{CMS:2019fak}, a dedicated calibration for XCone subjets was derived to correct for differences in the reconstruction compared to anti-\kt jets with $R=0.4$, which are used
to derive JES corrections.
The improvements introduced with the first measurement at 13\TeV, most importantly the two-step jet clustering with XCone which results in an improved line shape of the \mjet distribution, improved jet mass resolution, and pileup stability, and the large gain in statistical precision, resulted
in the JES uncertainty becoming the dominant experimental uncertainty.
Therefore, in the measurement with the full \Run2 data set~\cite{CMS:2022kqg}, a dedicated calibration of the JMS was introduced.
The centrally provided JES corrections are derived by calibrating the jet with \pt- and $\eta$-dependent correction factors that scale the full jet four-momentum.
However, the jet mass is not necessarily affected in the same way as the jet three-momentum,
calling for a technique to calibrate the JMS.
The method developed for this measurement uses the distribution in the reconstructed \PW boson mass
for the JMS calibration, similar to JEC constraints from \mW in the direct measurements discussed in Sec.~\ref{sec:direct}.
The \PW boson decay is reconstructed by selecting the two XCone subjets that are not
associated with the \PQb quark from the top quark decay,
which is identified by using the \PQb tagging score.
The JMS response is parameterised as a function of two parameters, which affect the JES and XCone
corrections. These parameters are obtained from a fit to data in the reconstructed \mW distributions.
The jet four-momentum is then constructed such that the JES only changes the jet three-momentum, while the JMS acts on \mjet.
Since the \PW boson decay results in a sample of light-flavour jets,
there is an additional uncertainty connected to the jet response to heavy-flavour jets,
estimated from a comparison of \PYTHIA and \HERWIG.
The dedicated JMS calibration reduces the effect of the uncertainty in the JES from $\delmt=1.47\GeV$ in the \mt extraction to $\delmt=0.37\oplus0.26\oplus0.07\GeV=0.46\GeV$, where the uncertainty is split into the contributions from the JMS, JMS flavour, and JES, respectively.

\subsection{Uncertainties from the modelling of the jet mass}

Modelling uncertainties arise from potential differences of the data compared to the simulation used to construct the response matrix in the unfolding.
These differences can introduce a model dependence in the unfolding and subsequently lead to a bias in the unfolded distribution.
Thus, all theoretical uncertainties enter this measurement twice: as biases in the unfolding and through the prediction of the \mjet distribution when extracting the top quark mass.
The modelling uncertainties are estimated by varying the simulation within theoretical uncertainties, unfolding the detector level distribution of the varied simulation and comparing the unfolded result to the true particle-level distribution. Any difference points to a potential bias due to the modelling and is accounted for as a model uncertainty.
A full list of modelling uncertainties that are considered in top quark mass measurements in CMS can be found in Section~\ref{sec:mcsetup}.

By focusing on the jet mass in hadronic decays of boosted top quarks rather than on the
reconstructed top quark mass in resolved decays or on \ttbar production rates,
many uncertainties relevant for the latter are small in jet mass measurements.
This includes uncertainties in the factorisation and renormalisation scales,
choice of PDFs, and \PQb fragmentation model.
The uncertainty in the colour reconnection model is estimated as non-negligible in the latest measurements at 13\TeV, but includes a significant statistical uncertainty due to the limited statistical precision in the simulated samples that are used for these variations.
In addition, our studies show that uncertainties in the underlying event tune are small in \mjet measurements
when using the XCone jet clustering. This can be understood by the jet grooming properties of the two-step XCone
clustering, which removes additional particles in the large-$R$ jet that are not connected to the top quark decay.

However, uncertainties in the parton shower model are very relevant for the measurement of \mjet.
Since the precision in the \mt extraction at 8\TeV was limited by the statistical uncertainty, a simple comparison of the \mjet distribution between simulated \ttbar samples using \POWHEGPYTHIA and \POWHEGHERWIG was used as an estimate of the uncertainty in the parton shower and hadronisation modelling.
With increasing precision in the first measurement at 13\TeV, the parton shower uncertainty was studied in more detail by evaluating variations of single model parameters that vary ISR, FSR, and the parameter \hdamp, that steers the matching between matrix element and parton shower.
The uncertainties in the scale choice of FSR modelling turned out to be the dominant modelling uncertainty in the 13\TeV measurement using data collected in 2016~\cite{CMS:2019fak}.
Already then it was assumed that the variations by a factor of 2 in the FSR energy scale in the CUETP8M2T4~\cite{CMS:2019csb} tune was overestimating this uncertainty.
With the switch to the CP5~\cite{CMS:2019csb} tune for the simulated samples for the data-taking periods of 2017 and 2018, this uncertainty is already much reduced, which is directly visible in the decreasing theoretical uncertainties of the latest \Run2 measurement~\cite{CMS:2022kqg} compared to the measurement with 2016 data~\cite{CMS:2019fak}, where the FSR uncertainty is the dominant source.
In addition, the latest \mjet measurement makes use of jet substructure observables in order to constrain the FSR modelling uncertainty.
The $N$-subjettiness ratio $\tau_{32}=\tau_3/\tau_2$~\cite{Thaler:2010tr,Thaler:2011gf} is sensitive to the amount of additional radiation that affects the three-prong top quark decay and is thus used to tune the FSR modelling in \ttbar simulation and consequently reduce the corresponding uncertainties.

With the FSR uncertainty being under control, the uncertainty in the choice of the value of \mtmc is the dominant modelling uncertainty.
This uncertainty reproduces a possible bias when unfolding a distribution that corresponds to a different value of \mt compared to the one used in the simulation that populates the response matrix.
In order to estimate this effect, we unfold the \mjet distribution of alternative simulated samples with different values of \mtmc with the nominal response matrix and compare the result to the \mjet particle-level distribution of the alternative samples.
Unfortunately, the available simulated samples with different values of \mtmc are very limited in statistical precision, especially at high top quark energies. Thus, a substantial fraction of this estimated uncertainty is caused by statistical effects.

\subsection{Aspects in the unfolding of the data}

The data are unfolded using regularised unfolding as implemented in the \textsc{TUnfold} software package~\cite{Schmitt:2012kp}.
We unfold the data to the particle level, which differs from the procedure in indirect top quark mass extractions, where one unfolds to the level of stable on-shell top quarks.
The response matrix, which contains the information about the transition from the particle to the detector levels, is filled using simulated \ttbar events, where each event contributes with the value of \mjet at the particle level and the \mjet at the detector level.
Although the response matrix is created from a \ttbar sample that simulates on-shell top quarks that further decay, the unfolding procedure in this measurement does not rely on a definition of an on-shell top quark, since all information is extracted from jets at the particle and detector level.

Another key feature of the unfolding setup in the jet mass measurement is the inclusion of events into and out of the measured phase space by adding multiple sideband regions to the response matrix.
Furthermore, the response matrix is built differentially in jet mass and jet \pt.
The high granularity is crucial in order to make the unfolding more independent from the model chosen in the simulation and subsequently reduce modelling uncertainties.
Thus, the increase in the number of selected events by collecting more data and the growth of the \ttbar production cross section---especially at high top quark energies---with the LHC upgrade from $\sqrts=8$ to 13\TeV did not only increase the statistical precision but also allowed the response matrix to be more granular and reduced modelling uncertainties.
The smaller jet mass resolution in the two-step XCone jet clustering enables smaller bin sizes at the particle level that help the unfolding to disentangle modelling differences and increases sensitivity to the later extracted top quark mass.
Furthermore, the binning is set up such that the purity and stability---defined as the fraction of events that are reconstructed in the same bin as they are generated and the fraction of events that are generated in the same bin as they are reconstructed---surpass 40\% over the full range of the particle-level phase space.
We also split the \mjet bins in the peak region in the unfolding in order to increase the sensitivity to model differences and retain the statistical precision by recombining them after the procedure.
With the currently available data set after \Run2, this results in a response matrix consisting of 200 bins at the detector level and 72 bins at the particle level.

\subsection{Top quark mass from jet mass}

\begin{figure}[!tp]
\centering
\includegraphics[width=0.48\textwidth]{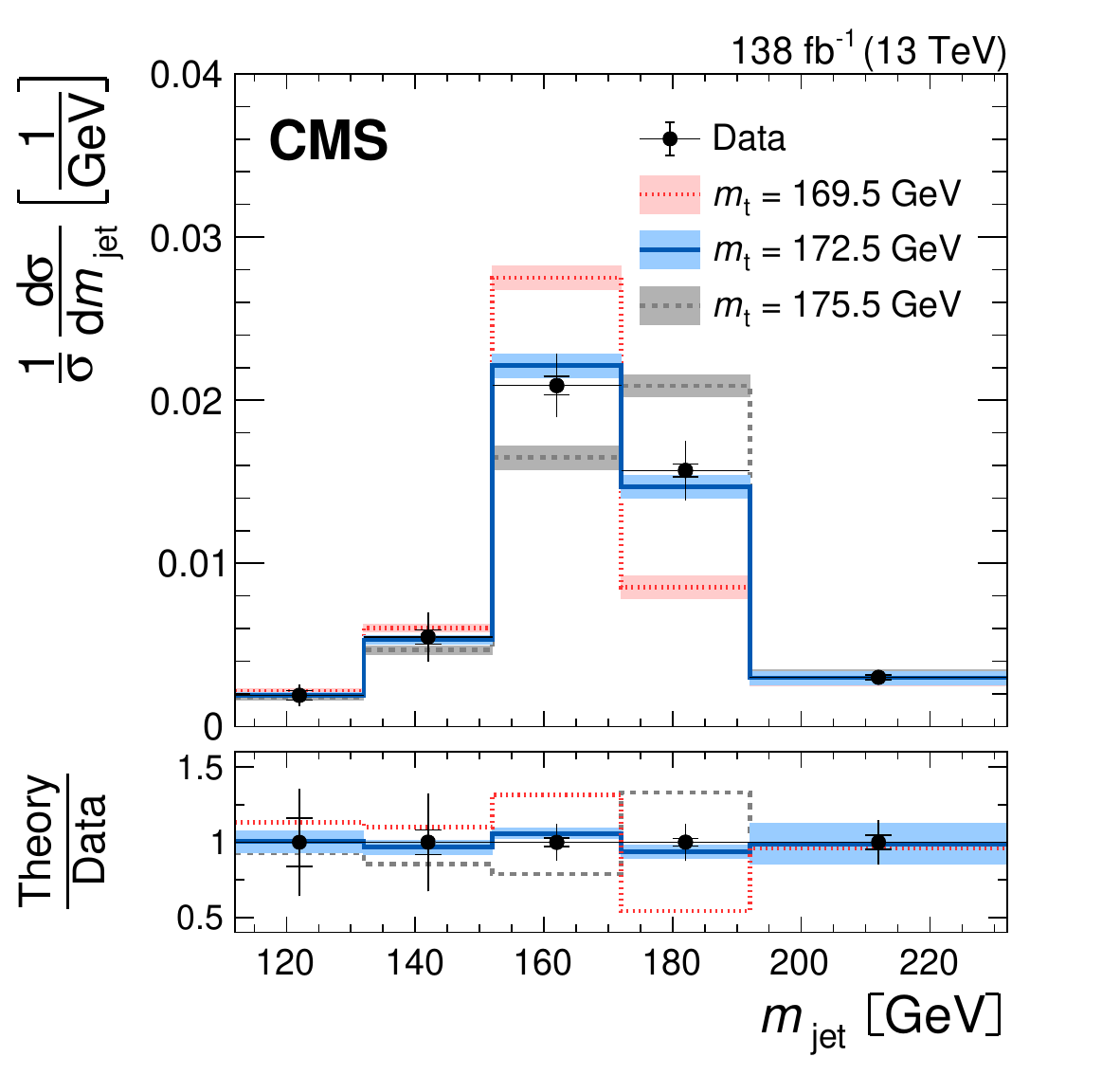}
\caption{%
    Normalised differential \ttbar production cross section as a function of \mjet. Data (markers) are compared to predictions for different \mt obtained from simulation (lines). The bars on the markers display the statistical (inner bars) and total (outer bars) uncertainties. The theoretical uncertainty is shown as coloured area.
    Figure taken from Ref.~\cite{CMS:2022kqg}.
}
\label{boosted:xs_norm}
\end{figure}

The top quark mass has been extracted from the normalised differential \ttbar cross section as a function of \mjet in order to be insensitive to normalisation effects.
Figure~\ref{boosted:xs_norm} shows the normalised measurement with the full \Run2 data set~\cite{CMS:2022kqg}.
So far, no analytical calculations are available for the selected phase space, thus we have extracted \mt using the \POWHEGPYTHIA simulation (detailed in Section~\ref{sec:mcsetup}), resulting in a value of $\mtmc=173.06\pm0.84\GeV$, which is compatible with direct measurements at moderate top quark energies.

\begin{figure}[!tp]
\centering
\includegraphics[width=\textwidth]{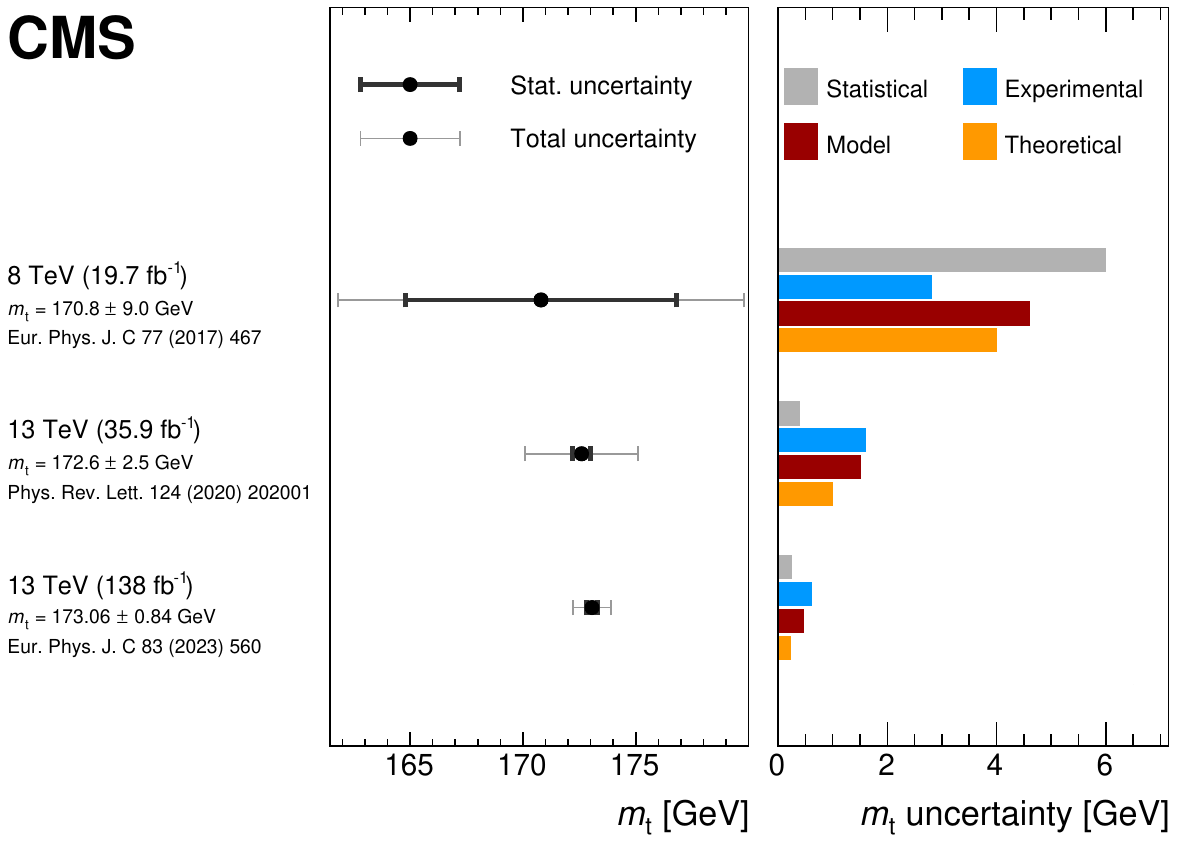}
\caption{%
    Summary of the \mt extraction in \mjet measurements. The left panel shows the extracted value of \mt (marker) with statistical (thick bars) and total (thin bars) uncertainties. The right panel displays a breakdown of contributing uncertainty groups and their impact on the uncertainty in the \mt extraction. The figure is compiled from Refs.~\cite{CMS:2017pcy,CMS:2019fak,CMS:2022kqg}.
}
\label{boosted:sumary}
\end{figure}

The resulting values and uncertainties in the extraction of \mt in the three \mjet measurements~\cite{CMS:2017pcy,CMS:2019fak,CMS:2022kqg} are summarised in Fig.~\ref{boosted:sumary}. The uncertainties are broken down into statistical, experimental, model, and theoretical contributions. The statistical uncertainty accounts for the finite statistical precision in the available data set. Experimental uncertainties arise from the calibration of physics objects. Model uncertainties and theoretical uncertainties both originate from choices of modelling parameters in the simulation. While theoretical uncertainties are taken into account on the particle-level predictions for the \mt hypotheses, model uncertainties arise from the potential bias in the unfolding that can be introduced by differences between data and the \ttbar simulation.

After the first measurement at \sqrtseq{8} with an initial statistical uncertainty of 6\GeV, the extraction of the top quark mass from the jet mass has largely profited from the increased production cross section of boosted top quarks at \sqrtseq{13} and the vast amount of data collected during \Run2.
Already with the data collected during 2016, the statistical uncertainty was no longer dominant.
The sensitivity to \mt was improved by the specialised two-step jet clustering procedure using XCone. The width of the peak in the \mjet distribution and jet mass resolution could both be reduced by a factor of two.
The significantly larger data set allowed the use of a much more granular response matrix that leads to smaller biases in the unfolding and subsequently reduced modelling uncertainties.
Better knowledge of the data also led to improved \ttbar modelling through constraining the variations in the choice of tuning parameters, which reduced the size of modelling variations and theoretical uncertainties.
Furthermore, parton shower uncertainties were no longer estimated by a comparison of \PYTHIA to \HERWIG but by a variation of dedicated parameters, which allows for a more detailed breakdown of systematic sources.
With the full \Run2 data set and dedicated calibrations of the JMS and FSR modelling in \ttbar simulation, the dominant sources of experimental and modelling uncertainties were reduced.
In addition, the newly introduced CP5 tune (see Section~\ref{sec:mcsetup}) featured reduced variations of the value of \alpS that controls the amount of FSR, which directly translates to reduced theoretical uncertainties.
For the increased data set also the number of simulated events was substantially increased.
This led to a decrease of the statistical part in the estimation of modelling and theoretical uncertainties.
Especially the estimation of uncertainties that rely on an additional sample and led to artificially large theoretical uncertainties in the first measurement at 13\TeV are now reduced with the increased statistical precision in the simulation for the full \Run2 data.

\section{Summary and outlook}
\label{sec:conclusions}

To date, the most precise measurements of the mass of the top quark \mt reach a relative precision of approximately 0.2\%. And still, the value of \mt and its uncertainty remain a focal point in particle physics, because of the central role of \mt in the electroweak symmetry breaking and fermion mass generation, and in probing physics beyond the standard model, where it enters as an essential parameter for the theoretical predictions and their quantum corrections. This makes the determination of \mt a compelling topic for both experimental and theory communities.

\subsection{Summary of the top quark mass results}

\begin{figure}[!p]
\centering
\includegraphics[width=0.98\textwidth]{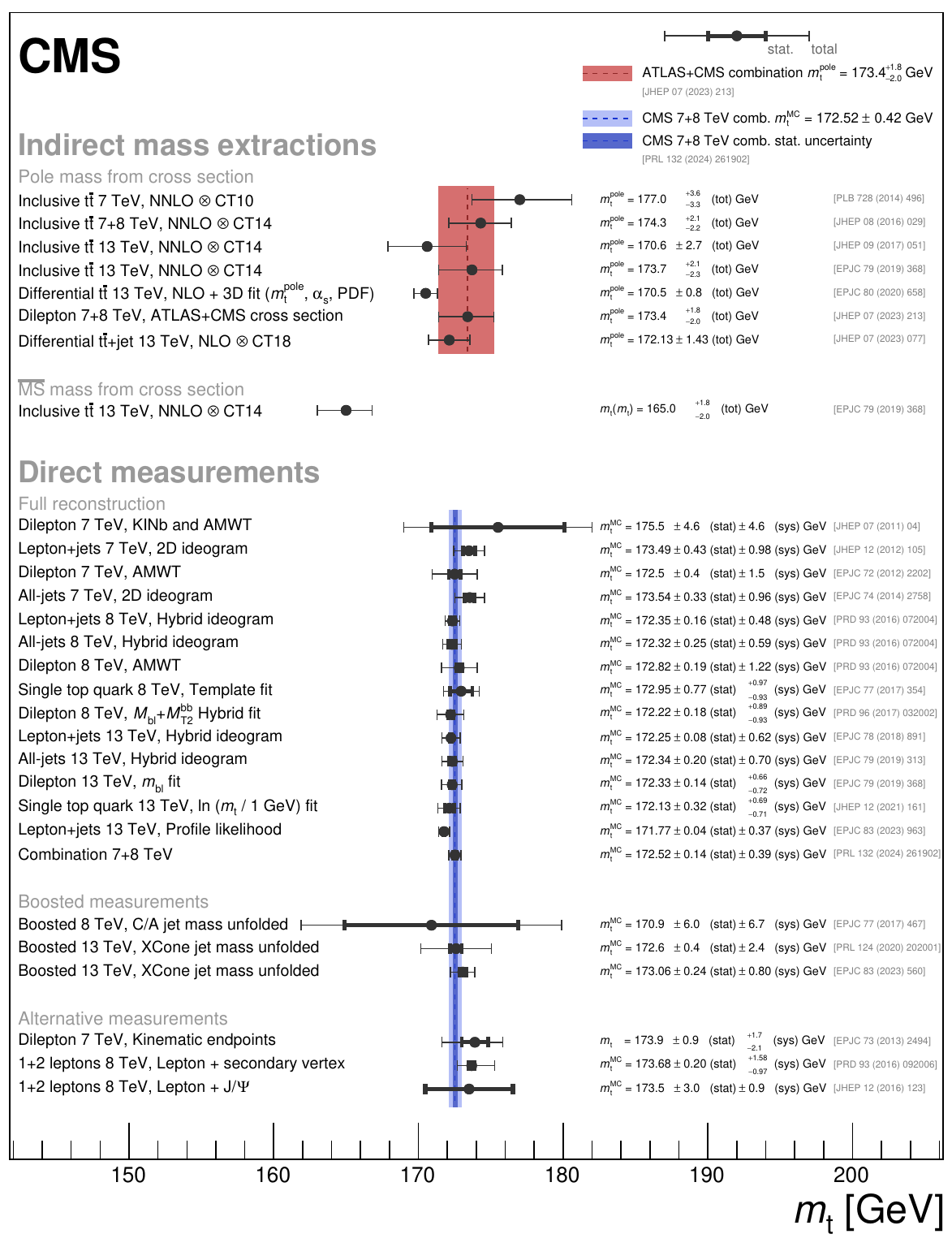}
\caption{%
    Overview of top quark mass measurement results published by the CMS Collaboration. The markers display the respective measured value of \mt with the statistical (inner) and total (outer) uncertainties shown as horizontal error bars. The measurements are categorised into indirect extractions from cross section measurements and direct measurements of \mtmc and are compared to the combined cross section measurement of the ATLAS and CMS Collaborations (red) and a CMS combination of \Run1 results (blue). Similar labelling as in Table~\ref{tab:all_measurements} is used. The figure is compiled from Refs.~\cite{CMS:2011acs,CMS:2012sas,CMS:2012tdr,CMS:2013wbt,CMS:2013lqq,CMS:2014rml,CMS:2015lbj,CMS:2016yys,CMS:2016iru,CMS:2016ixg,CMS:2017xrt,CMS:2017mpr,CMS:2017pcy,CMS:2017znf,CMS:2018quc,CMS:2018tye,CMS:2018fks,CMS:2019esx,CMS:2019fak,CMS:2021jnp,ATLAS:2022aof,CMS:2022emx,CMS:2022kqg,CMS:2023ebf, CMS:2023wnd}.
}
\label{conclusions:MtopResults}
\end{figure}

The CMS Collaboration embarked on an extensive and diverse program of \mt measurements. Some of the most recent results were highlighted in Sections~\ref{sec:direct}, \ref{sec:indirect}, and \ref{sec:boosted}, for direct measurements, indirect \mt extraction in different renormalisation schemes, and analyses in the boosted top quark regime, respectively, together with their historical development. In Fig.~\ref{conclusions:MtopResults}, the summary of \mt results published by the CMS Collaboration to date, also listed in Table~\ref{tab:all_measurements}, is shown.
The measurements are sorted in different groups, according to their traditional classification into the three approaches and \mt definitions used. Also, the results obtained in the alternative measurements mentioned in Section~\ref{sec:intro} are shown.
So far, the results on \mt from indirect extractions have been obtained for the pole mass \mtp and the \msbar mass \mtmt.
Note that the QCD conversion between the pole mass and the \msbar mass schemes yields a value of \mtmt of about 9\GeV lower than corresponding \mtp, as discussed in Section~\ref{sec:introschemes}, which is consistent with the difference found between the \mtp and \mtmt determinations.
Although the results obtained in direct measurements of the top quark mass \mtmc and from indirect extractions of the Lagrangian parameter \mt might be numerically similar, it is important to consider ambiguities in the relation between them, originating from theoretical uncertainties and limitations of the current Monte Carlo (MC) simulations.

The measurements collectively indicate results that are consistent with each other, whether considering top quark pole mass \mtp or direct \mtmc measurements. Nevertheless, it is crucial to acknowledge that while the methods employed may vary, there are notable correlations of systematic and statistical nature among many of the measurements. These correlations arise from common sources of systematic uncertainties and, in certain instances, event overlap. To accurately assess compatibility and to consolidate results into a unified top quark mass extraction, it is essential to consider these correlations. However, this detailed task falls beyond the scope of the present review.

\subsection{Evolution of analysis methods in CMS}

The development of the analysis strategies for the \mt measurements at the LHC in the last decade has resulted in significant advancements in precision.

In the case of direct \mt measurements using \ttbar production, the evolution of the analysis methods has led to a yet unprecedented experimental precision of less than 400\MeV. Direct measurements of \mt using single top quark production allow
the probing of \mt
in a different process and event topology compared to \ttbar events, and thus provide different sensitivity to systematic uncertainties which can be beneficial in mass combinations~\cite{CMS:2023wnd}.
However, any of these direct \mt measurements rely to large extent on MC simulations. This fact complicates the interpretation of the resulting MC parameter, \mtmc, in terms of a Lagrangian \mt defined in a certain renormalisation scheme of Quantum Chromodynamics (QCD). In the face of the high experimental precision, the adequate theoretical interpretation of \mtmc remains an active area of research. In fact, a deeper understanding of both perturbative and nonperturbative effects in MC simulations is required in order to relate the value of \mtmc to that of a Lagrangian mass \mt with reliable uncertainty estimates.

For the indirect \mt extractions, performed by comparing the measured cross sections of top quark-antiquark pair (\ttbar) production or \ttbarjet to theoretical predictions obtained in perturbative QCD, the current uncertainties in \mt are larger by a factor of about two, as compared to direct measurements.
The theoretical uncertainty is dominated by the missing higher-order corrections, estimated by variation of the renormalisation and factorisation scales, and the uncertainties in the strong coupling constant \asmz and parton distribution functions (PDFs).
Experimentally, an unfolding procedure is necessary in order to relate observed detector-level variables with the theoretical calculations involving on-shell top quarks and antiquarks. Analysis strategies for measurements of cross sections of \ttbar and \ttbarjet production,  \stt and \sttjet, have seen improvements both from the experimental and phenomenological side. Template fits to multidifferential distributions considering both signal and background topologies are utilised. Techniques for the reconstruction of \ttbar pairs have also been substantially advanced. Conceptually, using normalised multidifferential cross sections in an analysis, where \asmz, PDFs and \mt can be extracted simultaneously, helps to mitigate their correlation in the theoretical predictions of \stt and leads to reduction of the uncertainties due to missing higher-order corrections.
All these improvements lead to a precision in \mt of about 1\GeV. These results must be further refined by improvements in theoretical calculations, \eg consideration of the Coulomb and off-shell effects.

Boosted topology measurements make use of top quarks that are produced at transverse momenta higher than about 400\GeV, where the decay products can be reconstructed in single jets of large distance parameter $R$, and \mt can be extracted from the mass of the jet \mjet.
This is in contrast to both of the aforementioned approaches, dominated by events where the \ttbar system is produced at transverse momenta of about 100\GeV, and with top quark decay products that are well resolved in the measurement.
Significant progress has been made experimentally in boosted measurements, achieving sub-\GeVns precision in \mt.
This progress involves a dedicated calibration of the jet mass scale and a thorough investigation of the impact of final-state radiation within large-$R$ jets.
Measurements utilising boosted topologies are of particular interest, as the \mjet distribution is calculable within the framework of soft collinear effective theories.
When such theoretical calculations become available, they can be used for Lagrangian \mt measurements, with the unfolded \mjet distribution serving as a means to extract \mt in a well-defined renormalisation scheme.
Such measurements could be compared to those of \mtmc obtained using the same data, offering not only an alternative method for measuring \mt but also an experimental input for the interpretation of \mtmc.
The precision of these measurements is anticipated to improve further with a larger number of \ttbar events at high transverse momenta.

As discussed in Section~\ref{sec:extraction}, studies are in progress to further refine the understanding of the systematic uncertainties related to experimental effects, the modelling of \ttbar events in MC simulation using the latest generators and tunes, and theoretical calculations of differential \ttbar cross sections. Further improvements in precision can therefore be expected from new \mt measurements in the coming years, based on full \Run2 and \Run3 data. Early data from the \Run3 of the LHC has already led to the first inclusive \stt measurement~\cite{CMS:2023qyl}, also shown in Fig.~\ref{fig:tt_xsection_vs_s}.
Moreover, the forthcoming full \Run3 holds the promise of increasing the recorded top quark data set by more than twice its current size.
This increase in the size of the data set, together with improvements in systematic treatment should allow for relevant advances in all the top quark studies.

In the following section, the prospects for the future \mt measurements beyond \Run3 are discussed in the context of the upcoming HL-LHC, which will bring the next big step in integrated luminosity and detector performance improvements.

\subsection{Prospects at the HL-LHC}
\label{sec:ProspectsHiLumi}

The High-Luminosity LHC (HL-LHC) upgrade~\cite{Apollinari:2015bam} has the goal of accumulating data corresponding to an integrated luminosity of up to 3\abinv at a centre-of-mass energy of 14\TeV. The average number of simultaneous \pp collisions bunch crossings is expected to reach nominal values up to 200. To mitigate the effect of this challenging environment, and since some detector components will have suffered from too much radiation damage, several detector components will be replaced, introducing new technology and capability into the CMS detector (Phase-2 upgrade). Among these upgrades, significant improvements are being made in the tracker and muon resolution and coverage~\cite{CMS:2017lum,Hebbeker:2017bix}, dedicated timing detectors~\cite{Butler:2019rpu}, and highly granular endcap calorimeters~\cite{CMS:2017jpq}, as well as improved barrel calorimeters~\cite{CMS:barrelTDR_no_inspire}.

\begin{figure}[!ht]
\centering
\includegraphics[width=0.55\textwidth]{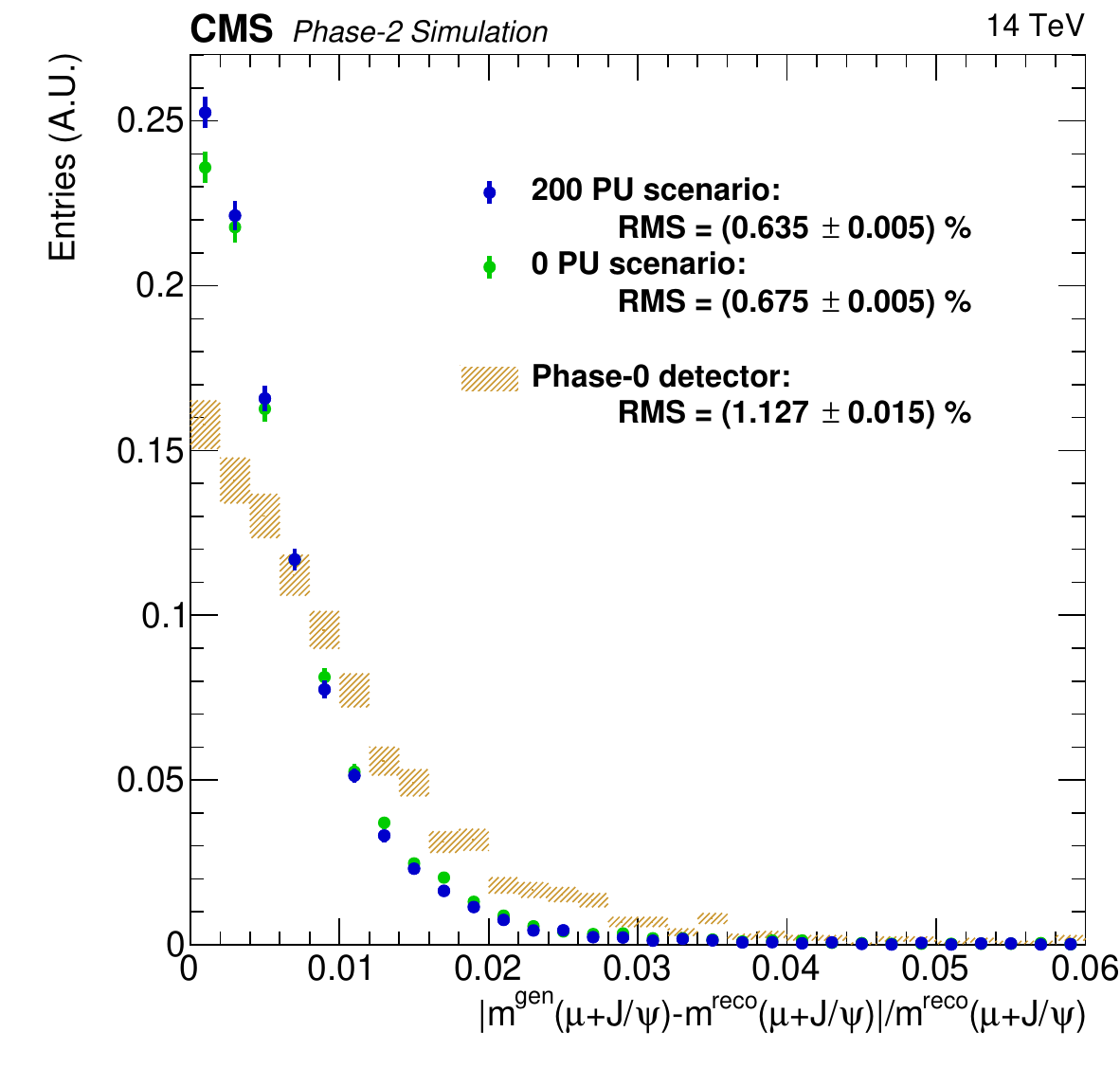}
\caption{%
    The resolution of the $\PGm+\PJGy$ mass for the CMS Phase-2 upgraded detector, for the two PU scenarios, and for the \Run2 (Phase-0) detector.
    Figure taken from Ref.~\cite{Hebbeker:2017bix}.
}
\label{conclusions:JPsi}
\end{figure}

Measurements of \mt will profit twofold from the HL-LHC upgrade. The larger data sample will enable measurements in currently less populated areas of the phase space, and will allow the application of methods exploiting
processes with small branching fractions.
Also, the detector upgrades can lead to more accurate measurements of the physics objects, subsequently providing the basis for higher precision \mt measurements.
An illustrative example is \mt extraction from \PJGy meson decays inside \PQb jets~\cite{CMS:2016ixg} accompanied by a lepton from the \PW boson decay. This measurement is less affected by the jet energy scale uncertainty than classical direct \mt measurements, but suffers from large statistical uncertainties and uncertainties in \PQb quark fragmentation.
The core of this analysis relies on an accurately measurable peak in the $\PJGy\to\PGm\PGm$ invariant mass distribution, and subsequent determination of the $\PGm+\PJGy$ mass.
With the new higher-resolution tracker and with the improvements in the muon system for the HL-LHC, the resolution of this peak will improve by almost a factor of two, as shown in Fig.~\ref{conclusions:JPsi}.

\begin{figure}[!t]
\centering
\includegraphics[width=0.6\textwidth]{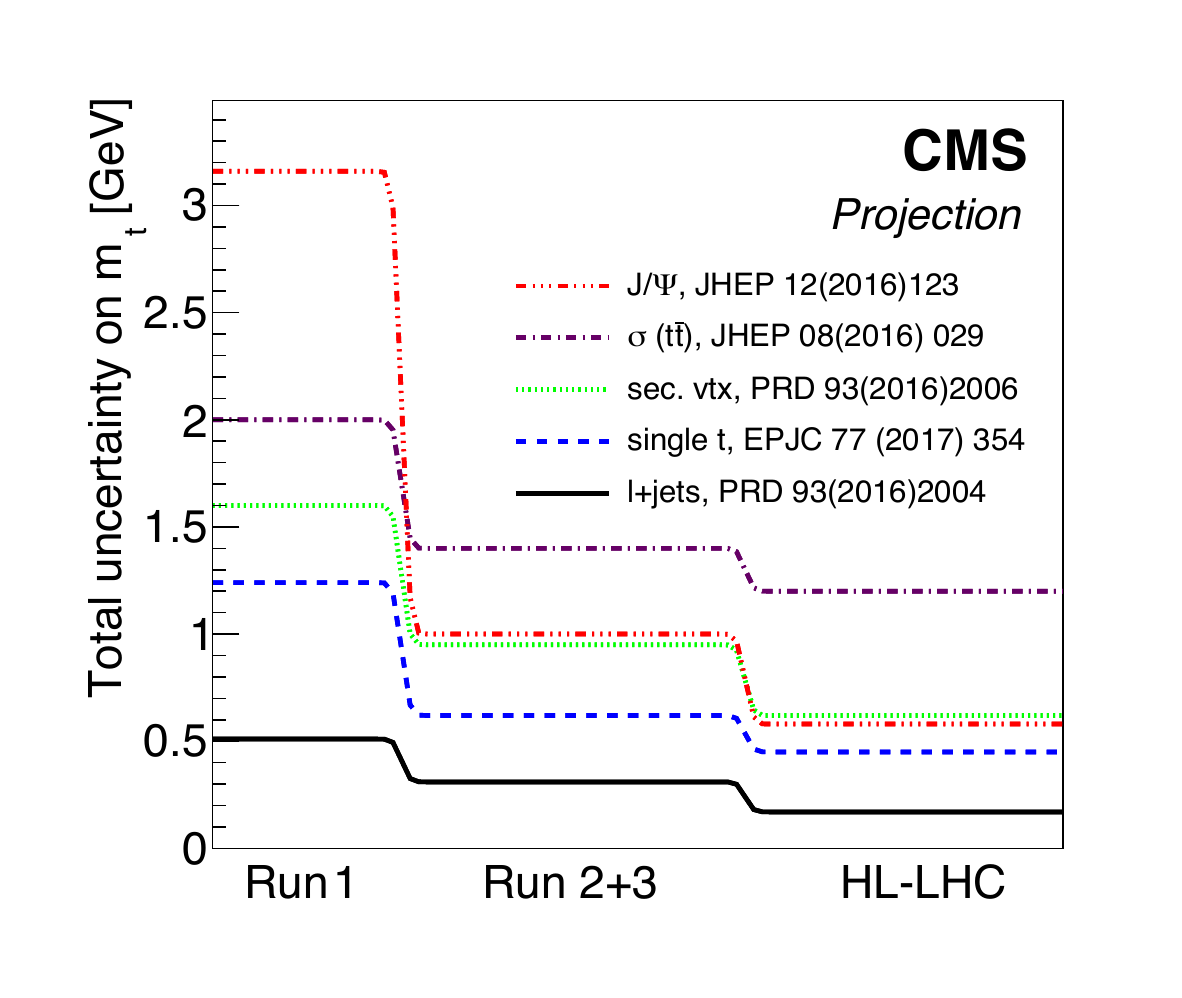}
\caption{%
    Total uncertainty in \mt obtained with a selection of different measurement methods and their projections for expected running conditions in \Run2 + \Run3 and at the HL-LHC. The projections are based on \mt measurements performed during the LHC \Run1, also listed in Table~\ref{tab:all_measurements}: the \PJGy~\cite{CMS:2016ixg}, total \ttbar cross section~\cite{CMS:2016yys} in the dilepton channel, secondary vertex~\cite{CMS:2016iru}, single top quark~\cite{CMS:2017mpr}, and lepton+jets direct~\cite{CMS:2015lbj} measurements. These projections do not fully account for improvements in the performance of the upgraded CMS detector.
    Figure taken from Ref.~\cite{CMS:2017gvo}.
}
\label{conclusions:ftr}
\end{figure}

Most \mt measurements are limited by the systematic uncertainties. Approximate studies to obtain HL-LHC projections for the \mt measurements were performed and are shown in Fig.~\ref{conclusions:ftr}. These do not fully account for improvements in the performance of the upgraded CMS detector. An ultimate relative precision of direct \mt measurement better than 0.1\% is expected. But also other methods profit significantly from the HL-LHC data and will continue to provide complementary information.
To estimate the HL-LHC prospects for these analyses, the systematic uncertainties are assumed to decrease, as expected considering the detector upgrades, developments of the reconstruction algorithms, refinements in the theoretical predictions, and improvements in the modelling from ancillary measurements~\cite{CMS:2017gvo}. In particular, the effect of the increased pileup is expected to be controllable for all objects, given higher detector granularity, timing capabilities of subdetectors, dedicated timing detectors, and exploiting the potential of pileup mitigation algorithms such as PUPPI~\cite{Bertolini:2014bba}. A moderate increase in the production cross section is expected to compensate possible losses in selection and trigger efficiencies. Furthermore, an increase in the acceptance of the upgraded detectors is expected.

Significant reduction of the systematic uncertainties in the signal modelling is expected too. Ancillary studies are being performed for the modelling of colour reconnection and the underlying event tunes, as outlined in previous sections. These are partially limited by statistical effects, and are therefore assumed to improve under HL-LHC conditions. These improvements are expected to reduce the corresponding uncertainties by about a factor of two. Further, the precision of modelling QCD and fragmentation effects is expected to increase, by using new MC generators at next-to-leading (NLO) and next-to-next-to-leading order (NNLO) QCD, improvements in the parton-shower simulation, and a fine-grained tuning of their parameters by exploiting larger data sets.
While the choice of the PDF set and the PDF uncertainties typically only have a small effect in direct \mt measurements, these are of high importance in the indirect extraction of \mt using QCD predictions in well-defined renormalisation schemes. For the HL-LHC projections, the contribution of the PDF uncertainty is usually assumed to be reduced by a factor of two.
The experimental uncertainties, often dominated by the jet energy scale, are also expected to be reduced by approximately a factor of two by the end of the HL-LHC running. However, the relative importance of the individual effects differs between the various \mt measurement methods~\cite{CMS:EDQ_note_no_inspire,CMS:2017gvo}.
The flavour-dependent components of the jet energy scale and the corresponding modelling of the \PQb quark fragmentation and the hadronisation model limit the precision of the direct measurements of \mt in \ttbar production. With dedicated measurements and improvements in the modelling, these contributions are expected to reduce.
The projected uncertainty reduction does not yet account for in-situ constraints for fits to multi-dimensional final-state distributions, introduced in Refs.~\cite{Kieseler:2015jzh,CMS:2018fks,CMS:2019jul} and used successfully for the most precise single measurement to date~\cite{CMS:2023ebf}.

In measurements that exploit the electroweak production modes in single top quark events, the background modelling is among the dominant sources of systematic uncertainties. With increasing centre-of-mass energy, the cross section of the leading contributions from \Wjets production increases more slowly than for top quark production, in particular compared to \Run1. Moreover, due to the large data sample, fine-grained regions can be used to constrain the background processes, which is why finally their contribution to the uncertainty is expected to be reduced by a factor of three with respect to \Run1.

As mentioned earlier, \mt analyses relying on secondary vertices in the \PQb jets or a full reconstruction of particles therein, \eg the \PJGy meson, will profit from the upgraded tracking detector. The dominant systematic uncertainties remain related to the modelling of the \PQb quark hadronisation. These effects are studied through dedicated analyses, and could be constrained in situ, given the improved vertex resolution, leading to the assumption that their impact on the precision of \mt will be reduced significantly.

\begin{figure}[!ht]
\centering
\includegraphics[width=0.48\textwidth]{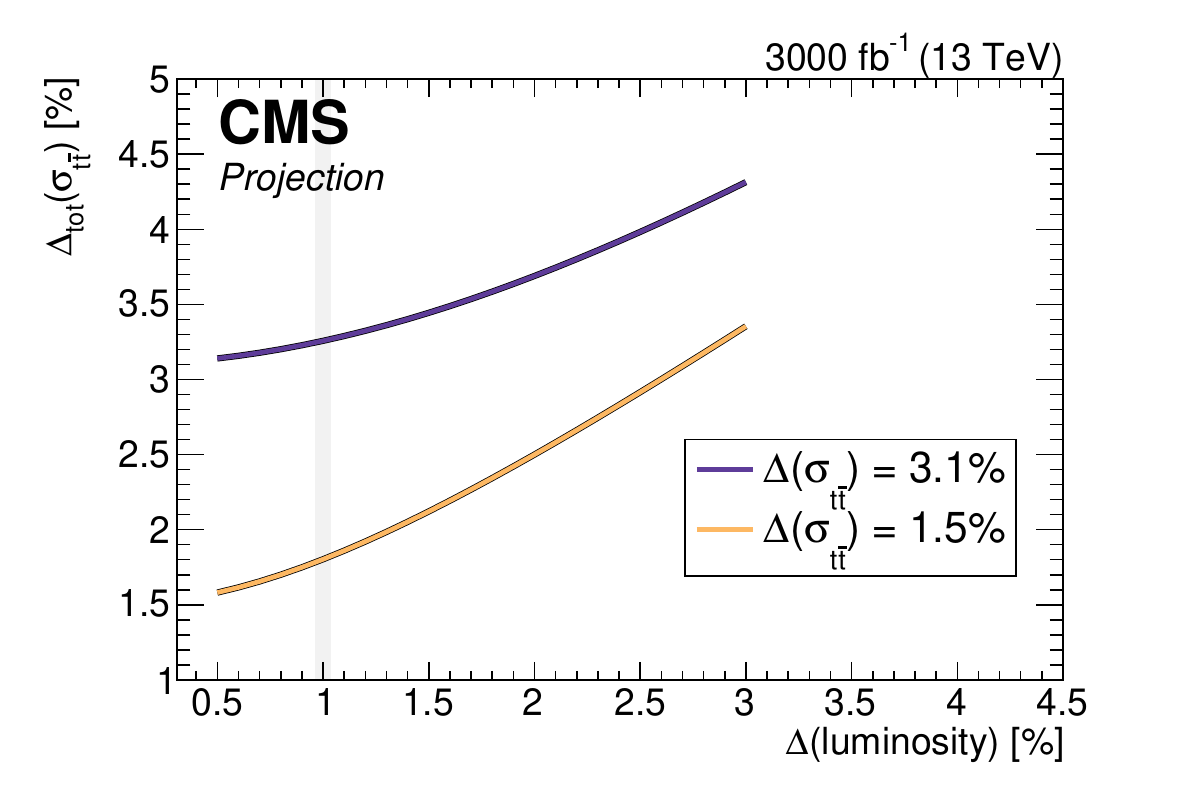}%
\hfill%
\includegraphics[width=0.48\textwidth]{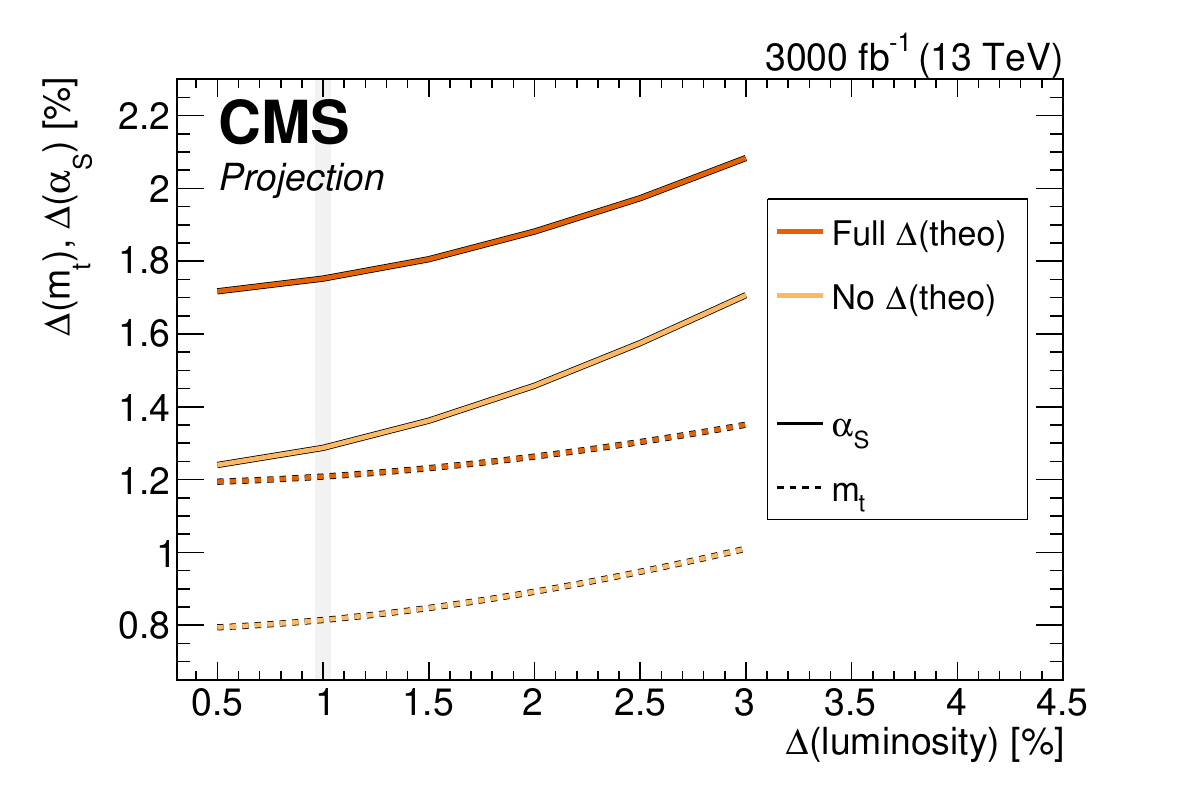}
\caption{%
    Left: The projected total experimental uncertainty in the top quark pair production cross section as a function of the uncertainty in the integrated luminosity, for two experimental scenarios, assuming no reduction of the experimental uncertainties with respect to \Run2 and a reduction of the uncertainties following the recommendations outlined in Ref~\cite{CMS:EDQ_note_no_inspire}.
    Right: The projected relative uncertainties in the extracted values of \mt (dotted lines) and \alpS (solid lines) as a function of the uncertainty in the integrated luminosity, comparing the case of the full uncertainty in the prediction and no uncertainty in the prediction. The results are obtained assuming a reduction of the uncertainties in the measurement to 1.5\%.
    Figure taken from Ref.~\cite{CMS:BRILTDR_no_inspire}.
}
\label{conclusions:lumi}
\end{figure}

The indirect extractions of \mt, \eg from the inclusive \ttbar production cross section, are expected to become more precise. Besides the conceptual issue of correlation of PDF, \asmz and \mt in the \stt prediction, the extraction of \mtp from the inclusive \stt is limited in almost equal parts by uncertainties in the theoretical prediction, currently available up to NNLO in QCD, and the experimental precision of the \stt measurement.
With several improvements in the analysis techniques~\cite{CMS:2016yys,CMS:2018fks}, the experimental precision of the inclusive \stt measurement is already mostly limited by the uncertainty in the integrated luminosity. A projection~\cite{CMS:2018fks} of the \Run2 measurement is shown in Fig.~\ref{conclusions:lumi}. It has been obtained in the context of the CMS beam and radiation monitoring system upgrade studies~\cite{CMS:BRILTDR_no_inspire}. The systematic uncertainties are scaled according to the assumptions outlined above, and the fit to the measured distributions has been repeated. In order to show their impact, the uncertainties in the NNLO prediction are assumed to remain at the current level and compared to a scenario with no uncertainties.
Depending on the scenario, a precision of up to 1.3\GeV in the \mtp can be reached.

This is approximately consistent with the projection from Ref.~\cite{CMS:2017gvo} shown in Fig.~\ref{conclusions:ftr}, where a reduction of the uncertainty in the integrated luminosity down to 0.5\% is expected. Furthermore, a reduction of theoretical uncertainties in \stt is assumed, originating from uncertainties in PDFs, \asmz, and from missing higher-order corrections. With additional measurements, the PDF and \alpS uncertainty are assumed to be reduced by a factor of two by the end of the HL-LHC phase. However, it is uncertain whether QCD predictions beyond NNLO will become available. Therefore, the uncertainties from the scale variations are assumed to be constant.

\begin{figure}[!ht]
\centering
\includegraphics[width=0.7\textwidth]
{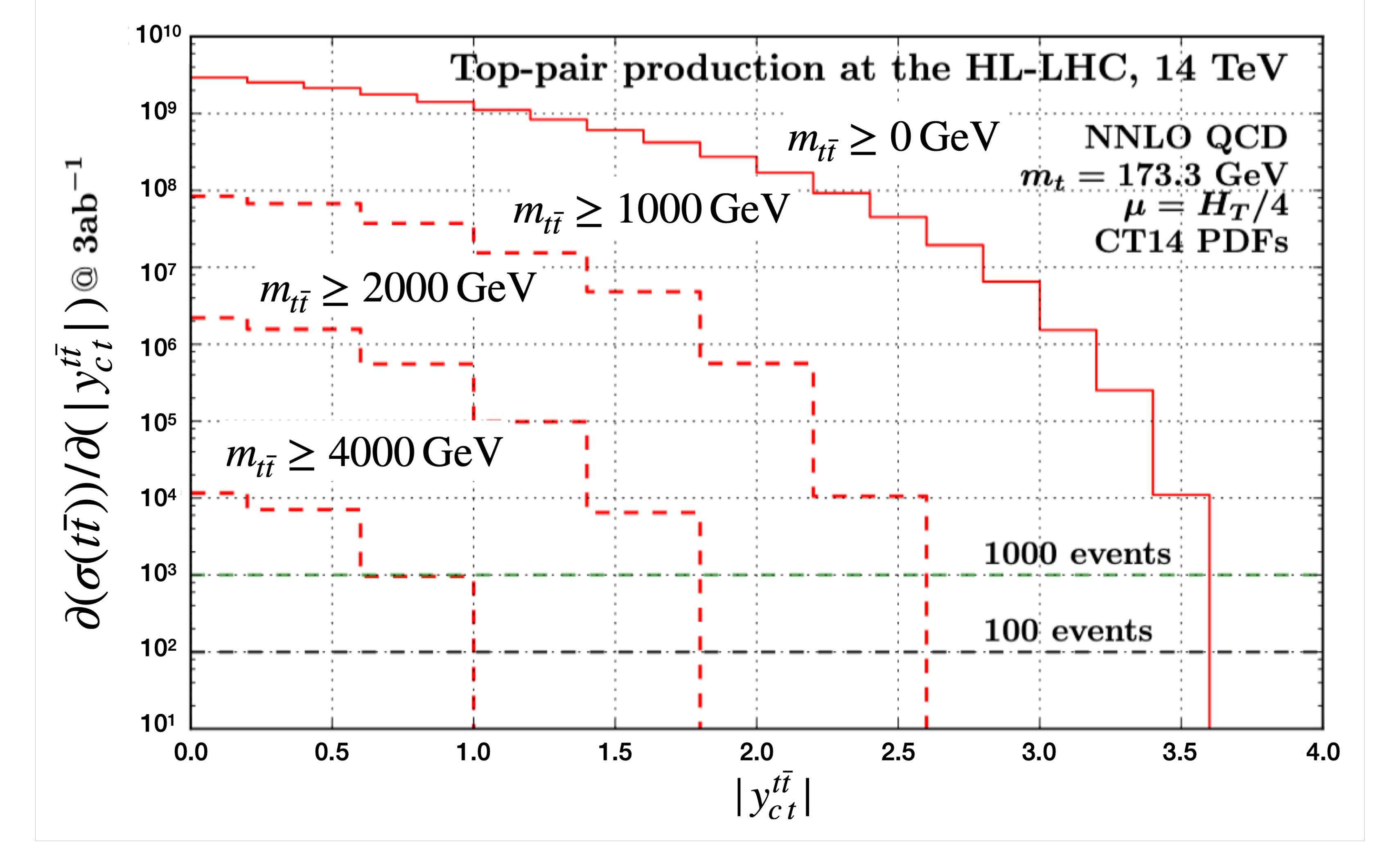}
\caption{%
    Projected cumulative differential \ttbar distributions for HL-LHC scenario as functions of rapidity and invariant mass of the \ttbar pair.
    Figure taken from Ref.~\cite{Dainese:2019rgk}.
}
\label{conclusions:yr_68}
\end{figure}

In the HL-LHC phase, the precision of the differential \ttbar cross section measurements and, in turn, the experimental accuracy of extraction of \mt, \asmz and of PDFs will profit from both the increased amount of data and the extended rapidity reach of the HL-LHC CMS detector. The projection study of Ref.~\cite{Dainese:2019rgk} demonstrated that despite the significantly higher pileup, the performance of the \ttbar reconstruction in the HL-LHC phase is expected to remain similar to the one of analyses based on data taken in 2016. The measurable phase space will increase due to the extended rapidity range, allowing for finer binning of double-differential measurements of \mtt and \ytt in a phase space not accessible in current measurements, as illustrated in Fig.~\ref{conclusions:yr_68}.

\begin{figure}[!ht]
\centering
\includegraphics[width=0.32\textwidth]{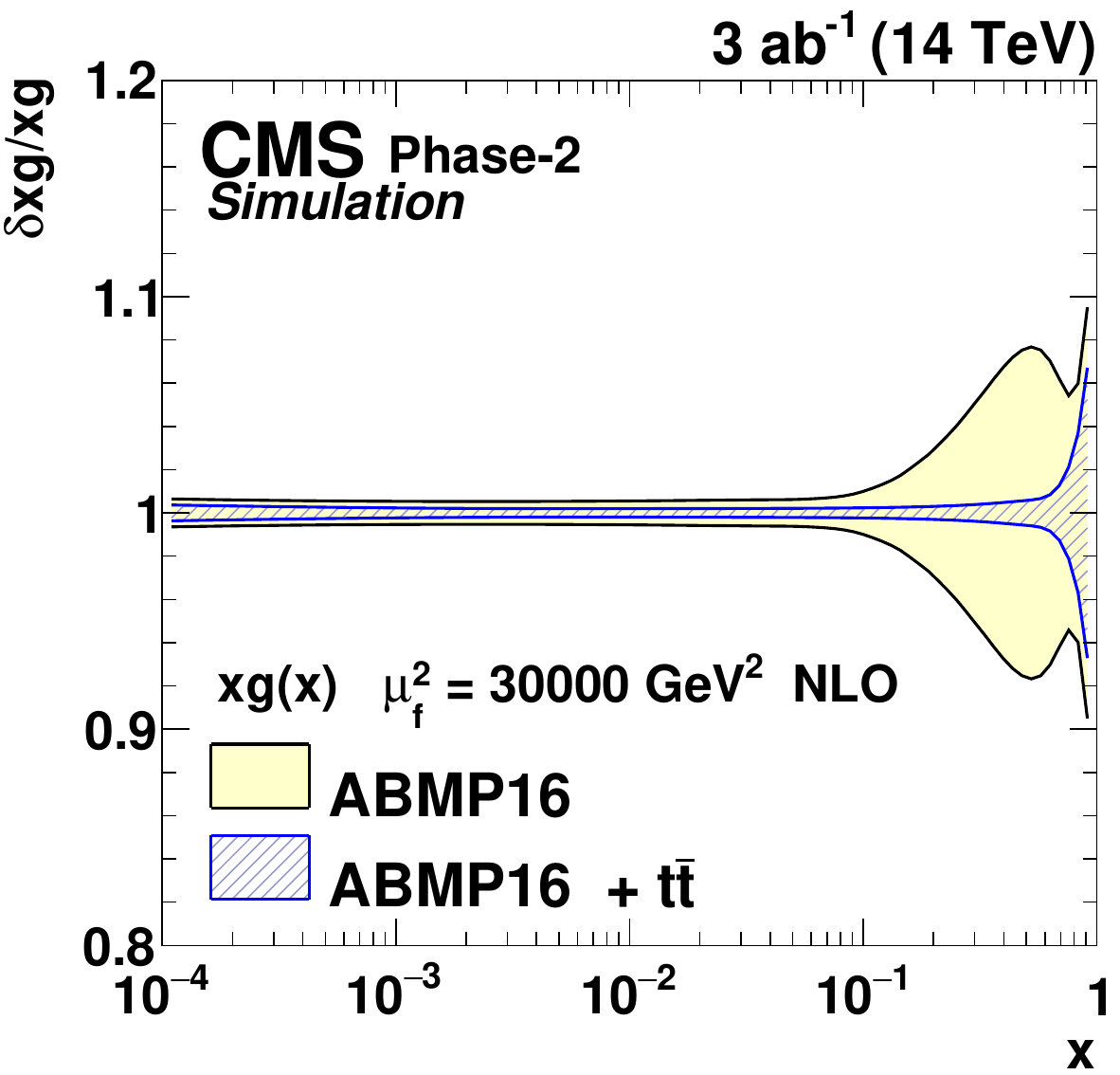}%
\hfill%
\includegraphics[width=0.32\textwidth]{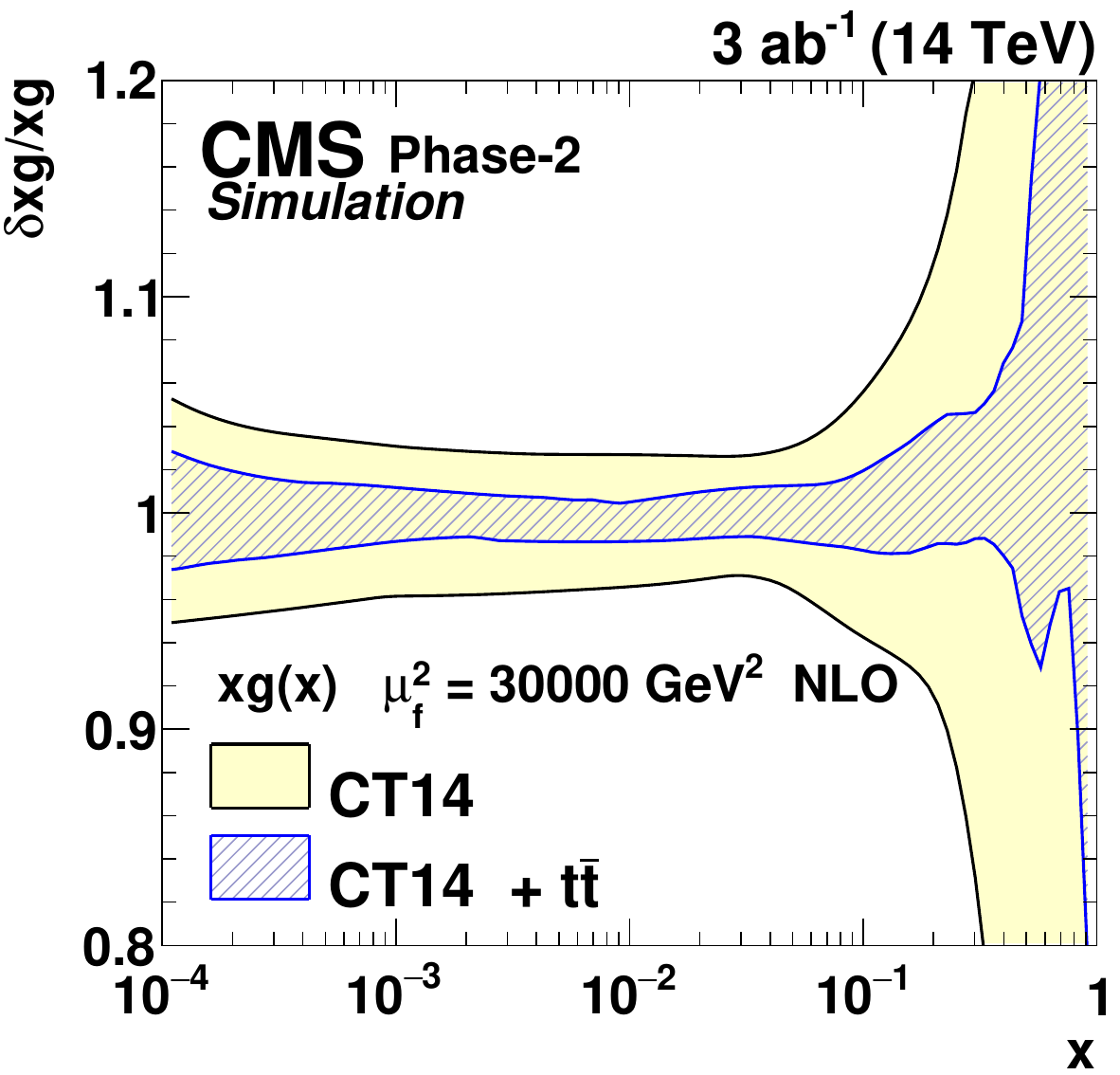}%
\hfill%
\includegraphics[width=0.32\textwidth]{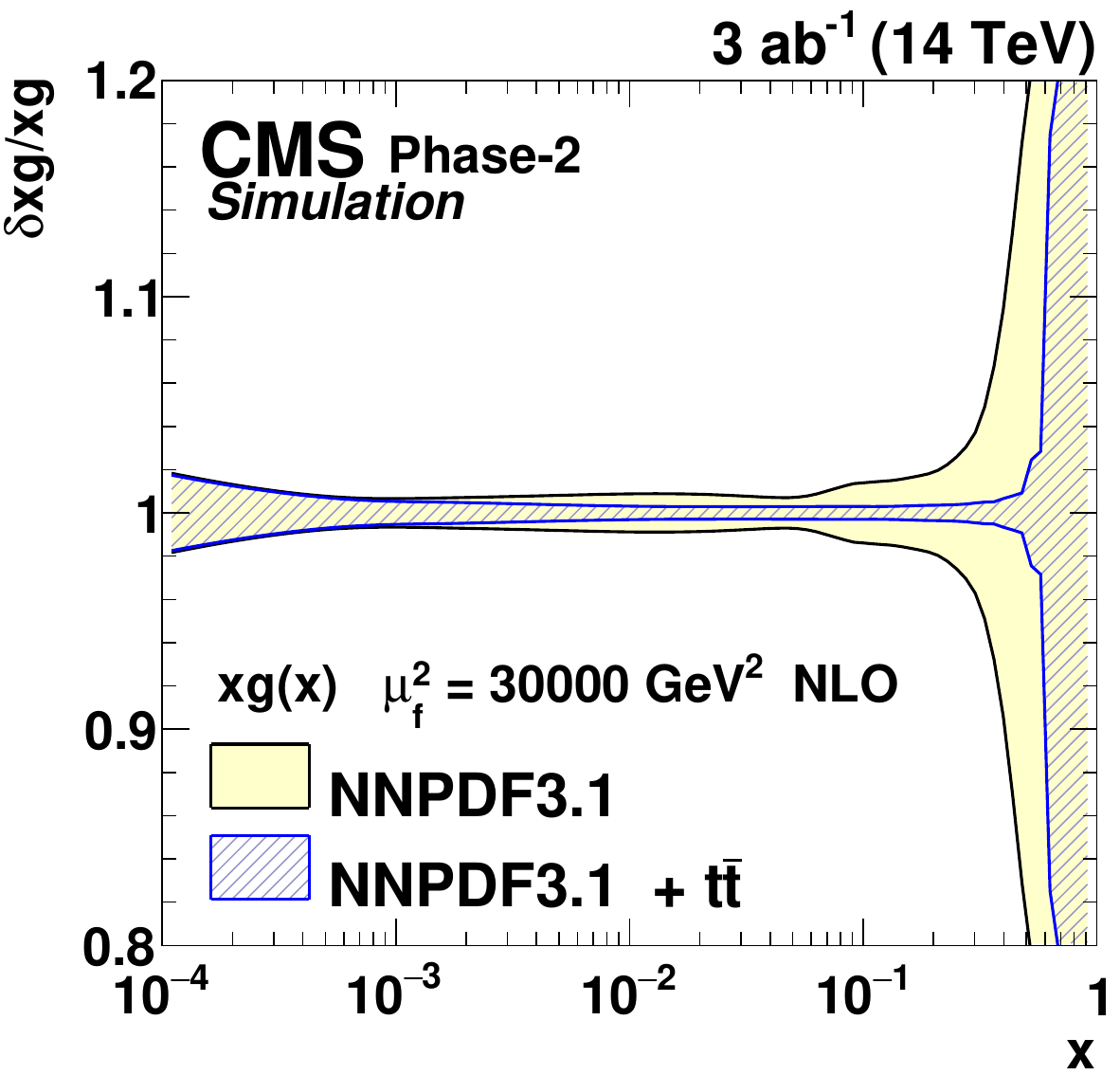}
\caption{%
    The relative gluon PDF uncertainties of the original and profiled ABMP16 (left), CT14 (middle), and \NNPDF{3.1} (right) sets.
    Figure taken from Ref.~\cite{Dainese:2019rgk}.
}
\label{conclusions:yr_73}
\end{figure}

While no projection is available for the precision of \mt when extracted from the differential cross sections, the projected precision in the PDF extraction from \ttbar multi-differential measurements is investigated in Ref.~\cite{Dainese:2019rgk}.
The inclusion of \ttbar cross section measurements is found to significantly improve the precision in PDF extraction.
In particular, the uncertainties in $g(x)$ could be reduced by a factor of 5--10 at high $x$, as illustrated in Fig.~\ref{conclusions:yr_73}, obtained using a profiling technique~\cite{Paukkunen:2014zia}. The latter is based on minimising the \chisq function between the data and theoretical predictions using available PDFs and taking into account both experimental and theoretical uncertainties arising from the PDF variations. As discussed in Section~\ref{sec:pdf_as_mt}, this significant reduction in the $g(x)$ would immediately translate in reduction of related uncertainty in \mt due to large correlations of both in theoretical predictions of \stt.
Beyond these projections, further improvement is expected from higher-order calculations of double-differential distributions and the use of dynamical \mt renormalisation schemes such as the MSR mass scheme~\cite{Makela:2023xnt}, which should be provided with fast interpolation grids in the future. By performing the full QCD analysis of PDFs, \mt and \asmz, the correlation between those is expected to be diminished, so that ultimate precision of \mt unambiguously defined in a particular renormalisation scheme can be achieved.
Furthermore, QCD corrections from resummations beyond the fixed-order approach, off-shell corrections, and flexible implementations of different \mt renormalisation schemes should be accounted for in these analyses once available. These are essential to achieve the ultimate theoretical accuracy.

\begin{figure}[!ht]
\centering
\includegraphics[width=0.48\textwidth]{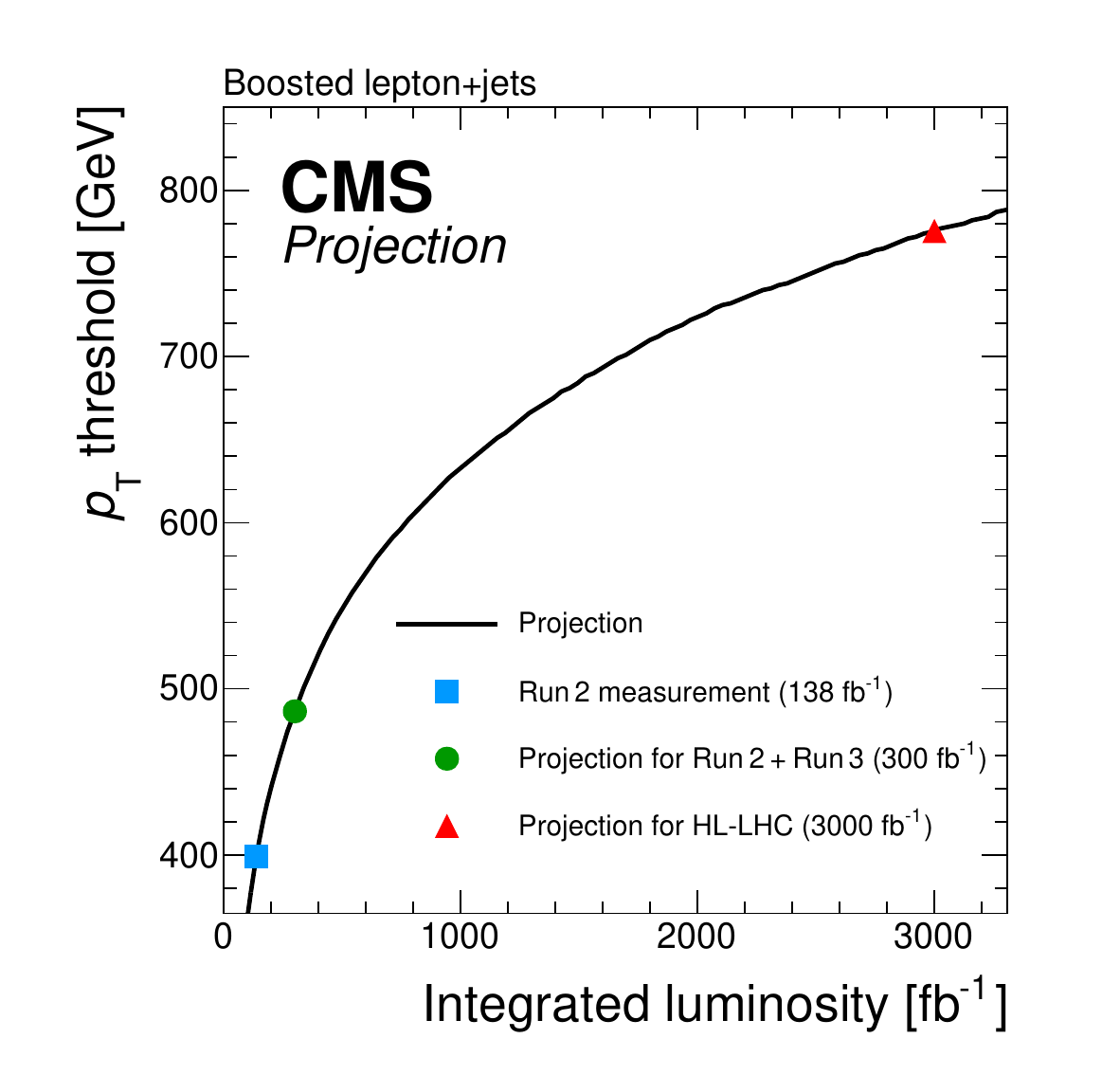}
\caption{%
    Scan of the jet \pt threshold in the measurements of the jet mass against integrated luminosity resulting in the same event yield in data after the full selection as in the most recent measurement~\cite{CMS:2022kqg}. The projection is obtained by scanning the jet \pt spectrum observed in data. The markers correspond to 138\fbinv of LHC \Run2 data used in Ref.~\cite{CMS:2022kqg}, to an estimated data set for the combination of \Run2 and \Run3, and to the HL-LHC scenario. For simplicity a constant centre-of-mass energy of 13\TeV and a similar detector acceptance to \Run2 are assumed in all scenarios.
}
\label{boosted:projection}
\end{figure}

The extraction of \mt from the \mjet distribution in decays of Lorentz-boosted top quarks will also benefit from the increased centre-of-mass energy and the large data set expected after the HL-LHC upgrade.
While the possibility of a precision \mt measurement from high-energy top quarks has been demonstrated with the data collected already today, the full potential of this measurement is not reached yet.
Already for the generator based extraction of \mtmc more data will allow to make the unfolding more granular and even to perform the measurement differentially in jet \pt.
With the CMS \Run2 data set, corresponding to an integrated luminosity of 138\fbinv, about 52\,000 events were selected in the measurement region.
This number is reduced to 21\,500 when requiring jets to have $\pt>500\GeV$ and even drops to below 3000 events for $\pt>750\GeV$, which would coincide with the space for which analytical calculations exist.
Figure~\ref{boosted:projection} shows a study where the possible jet \pt threshold is calculated as a function of integrated luminosity in order to achieve the same statistical precision as in the latest \Run2 measurement~\cite{CMS:2022kqg}.
After the HL-LHC upgrade, a data set corresponding to 3000\fbinv in combination with a slightly increased \ttbar production cross section at higher \sqrts is expected.
Thus, the phase space at very high \pt becomes available experimentally.

In addition, systematic uncertainties can be further reduced. On the experimental side, the calibration of the jet mass scale can be extended to include a measurement of the jet mass resolution in order to constrain this dominant uncertainty and become independent from the \pt driven calibration of the jet energy resolution.
Modelling uncertainties will benefit from a more granular unfolding process. This involves increasing the number of bins in the \mjet and jet \pt measurements, as well as incorporating additional observables. These steps will help to separate the model dependencies more effectively. This is particularly relevant for reducing uncertainties related to the choice of \mtmc in simulations. By adopting a more detailed approach, we can better distinguish between the correlations of jet \pt and \mjet, thus reducing this uncertainty.
With more data available, one cannot only increase the jet \pt threshold to a higher value but also perform the \mjet measurement differentially in jet \pt.
This could be used to dampen any \pt-dependent effects in the \mjet distribution and further increase the sensitivity to \mt.
Furthermore, a precise test of \pt independence of the measured \mtmc would provide an important consistency check of the generator-based measurement.

Already now, the measurement of \mjet provides a precise determination of \mtmc at energy scales not probed before.
However, the full potential of these measurements can only be reached once the definitions in calculations and the experimental analysis are brought into concordance, requiring developments from both sides.
At this point, these will become a powerful tool not only for precisely measuring \mt in a well-defined theoretical scheme but also for resolving the ambiguities in relation to \mtmc.

\subsection{Conclusions}
\label{sec:Conclusions}

Measurements of the top quark mass have been an essential part of the CMS research programme since the first data were recorded in 2010, with more than 20 journal publications that reveal different aspects related to this fundamental parameter of the standard model.
A growing understanding of theoretical and experimental issues on the way towards increasing precision in \mt, demanded by matching the accuracy of other electroweak parameters, were followed by steady improvements in analysis techniques. Different complementary methods have been used for measurements of \mt, affected by different sources of theoretical and experimental systematic uncertainties. An impressive sub-GeV precision has been achieved, despite the challenging environment of high-energy \pp collisions at the LHC, where events are affected by QCD and electroweak radiation, the underlying event and an unprecedented level of pileup interactions.

This success, and a clear perspective of experimental improvements envisaged for the HL-LHC, give confidence in reaching the ultimate precision in \mt achievable at a hadron collider in the next decade. This experimental goal requires that the necessary theoretical developments will take place, including advancements in the description of the top quark beyond the picture of a free particle, matching higher-order calculations to resummations and hadronisation models, and calculating corrections at the threshold of \ttbar production. The precise determination of \mt is an ongoing endeavour that fosters a close collaboration of the experimental and theoretical communities, with bright prospects in the coming years.

\begin{acknowledgments}
\hyphenation{Bundes-ministerium Forschungs-gemeinschaft Forschungs-zentren Rachada-pisek} We congratulate our colleagues in the CERN accelerator departments for the excellent performance of the LHC and thank the technical and administrative staffs at CERN and at other CMS institutes for their contributions to the success of the CMS effort. In addition, we gratefully acknowledge the computing centres and personnel of the Worldwide LHC Computing Grid and other centres for delivering so effectively the computing infrastructure essential to our analyses. Finally, we acknowledge the enduring support for the construction and operation of the LHC, the CMS detector, and the supporting computing infrastructure provided by the following funding agencies: the Armenian Science Committee, project no. 22rl-037; the Austrian Federal Ministry of Education, Science and Research and the Austrian Science Fund; the Belgian Fonds de la Recherche Scientifique, and Fonds voor Wetenschappelijk Onderzoek; the Brazilian Funding Agencies (CNPq, CAPES, FAPERJ, FAPERGS, and FAPESP); the Bulgarian Ministry of Education and Science, and the Bulgarian National Science Fund; CERN; the Chinese Academy of Sciences, Ministry of Science and Technology, the National Natural Science Foundation of China, and Fundamental Research Funds for the Central Universities; the Ministerio de Ciencia Tecnolog\'ia e Innovaci\'on (MINCIENCIAS), Colombia; the Croatian Ministry of Science, Education and Sport, and the Croatian Science Foundation; the Research and Innovation Foundation, Cyprus; the Secretariat for Higher Education, Science, Technology and Innovation, Ecuador; the Estonian Research Council via PRG780, PRG803, RVTT3 and the Ministry of Education and Research TK202; the Academy of Finland, Finnish Ministry of Education and Culture, and Helsinki Institute of Physics; the Institut National de Physique Nucl\'eaire et de Physique des Particules~/~CNRS, and Commissariat \`a l'\'Energie Atomique et aux \'Energies Alternatives~/~CEA, France; the Shota Rustaveli National Science Foundation, Georgia; the Bundesministerium f\"ur Bildung und Forschung, the Deutsche Forschungsgemeinschaft (DFG), under Germany's Excellence Strategy -- EXC 2121 ``Quantum Universe" -- 390833306, and under project number 400140256 - GRK2497, and Helmholtz-Gemeinschaft Deutscher Forschungszentren, Germany; the General Secretariat for Research and Innovation and the Hellenic Foundation for Research and Innovation (HFRI), Project Number 2288, Greece; the National Research, Development and Innovation Office (NKFIH), Hungary; the Department of Atomic Energy and the Department of Science and Technology, India; the Institute for Studies in Theoretical Physics and Mathematics, Iran; the Science Foundation, Ireland; the Istituto Nazionale di Fisica Nucleare, Italy; the Ministry of Science, ICT and Future Planning, and National Research Foundation (NRF), Republic of Korea; the Ministry of Education and Science of the Republic of Latvia; the Research Council of Lithuania, agreement No.\ VS-19 (LMTLT); the Ministry of Education, and University of Malaya (Malaysia); the Ministry of Science of Montenegro; the Mexican Funding Agencies (BUAP, CINVESTAV, CONACYT, LNS, SEP, and UASLP-FAI); the Ministry of Business, Innovation and Employment, New Zealand; the Pakistan Atomic Energy Commission; the Ministry of Education and Science and the National Science Centre, Poland; the Funda\c{c}\~ao para a Ci\^encia e a Tecnologia, grants CERN/FIS-PAR/0025/2019 and CERN/FIS-INS/0032/2019, Portugal; the Ministry of Education, Science and Technological Development of Serbia; MCIN/AEI/10.13039/501100011033, ERDF ``a way of making Europe", Programa Estatal de Fomento de la Investigaci{\'o}n Cient{\'i}fica y T{\'e}cnica de Excelencia Mar\'{\i}a de Maeztu, grant MDM-2017-0765, projects PID2020-113705RB, PID2020-113304RB, PID2020-116262RB and PID2020-113341RB-I00, and Plan de Ciencia, Tecnolog{\'i}a e Innovaci{\'o}n de Asturias, Spain; the Ministry of Science, Technology and Research, Sri Lanka; the Swiss Funding Agencies (ETH Board, ETH Zurich, PSI, SNF, UniZH, Canton Zurich, and SER); the Ministry of Science and Technology, Taipei; the Ministry of Higher Education, Science, Research and Innovation, and the National Science and Technology Development Agency of Thailand; the Scientific and Technical Research Council of Turkey, and Turkish Energy, Nuclear and Mineral Research Agency; the National Academy of Sciences of Ukraine; the Science and Technology Facilities Council, UK; the US Department of Energy, and the US National Science Foundation.

Individuals have received support from the Marie-Curie programme and the European Research Council and Horizon 2020 Grant, contract Nos.\ 675440, 724704, 752730, 758316, 765710, 824093, 101115353, and COST Action CA16108 (European Union) the Leventis Foundation; the Alfred P.\ Sloan Foundation; the Alexander von Humboldt Foundation; the Belgian Federal Science Policy Office; the Fonds pour la Formation \`a la Recherche dans l'Industrie et dans l'Agriculture (FRIA-Belgium); the Agentschap voor Innovatie door Wetenschap en Technologie (IWT-Belgium); the F.R.S.-FNRS and FWO (Belgium) under the ``Excellence of Science -- EOS" -- be.h project n.\ 30820817; the Beijing Municipal Science \& Technology Commission, No. Z191100007219010; the Ministry of Education, Youth and Sports (MEYS) of the Czech Republic; the Shota Rustaveli National Science Foundation, grant FR-22-985 (Georgia); the Hungarian Academy of Sciences, the New National Excellence Program - \'UNKP, the NKFIH research grants K 124845, K 124850, K 128713, K 128786, K 129058, K 131991, K 133046, K 138136, K 143460, K 143477, 2020-2.2.1-ED-2021-00181, and TKP2021-NKTA-64 (Hungary); the Council of Scientific and Industrial Research, India; ICSC -- National Research Centre for High Performance Computing, Big Data and Quantum Computing, funded by the EU NexGeneration program, Italy; the Latvian Council of Science; the Ministry of Education and Science, project no. 2022/WK/14, and the National Science Center, contracts Opus 2021/41/B/ST2/01369 and 2021/43/B/ST2/01552 (Poland); the Funda\c{c}\~ao para a Ci\^encia e a Tecnologia, grant FCT CEECIND/01334/2018; the National Priorities Research Program by Qatar National Research Fund; the Programa Estatal de Fomento de la Investigaci{\'o}n Cient{\'i}fica y T{\'e}cnica de Excelencia Mar\'{\i}a de Maeztu, grant MDM-2017-0765 and projects PID2020-113705RB, PID2020-113304RB, PID2020-116262RB and PID2020-113341RB-I00, and Programa Severo Ochoa del Principado de Asturias (Spain); the Chulalongkorn Academic into Its 2nd Century Project Advancement Project, and the National Science, Research and Innovation Fund via the Program Management Unit for Human Resources \& Institutional Development, Research and Innovation, grant B37G660013 (Thailand); the Kavli Foundation; the Nvidia Corporation; the SuperMicro Corporation; the Welch Foundation, contract C-1845; and the Weston Havens Foundation (USA).
\end{acknowledgments}

\bibliography{auto_generated}

\providecommand{\href}[2]{#2}\begingroup\raggedright\begin{thebibliography}{100}%
\makeatletter
\providecommand{\hrefCMSnoop }[0]{\@secondoftwo}%
\makeatother
\providecommand{\doi}{\texttt{doi:}\begingroup \urlstyle{tt}\Url}

\bibitem{ParticleDataGroup:2022pth}
\hrefCMSnoop {}{{Particle Data Group}, R.~L. Workman { et~al.}, ``Review of
  particle physics'',} \textit{ Prog. Theor. Exp. Phys.} \textbf{ 2022} (2022)
  083C01,
  \href{http://dx.doi.org/10.1093/ptep/ptac097}{\doi{10.1093/ptep/ptac097}}.

\bibitem{Mahlon:2010gw}
\hrefCMSnoop {}{G.~Mahlon and S.~J. Parke, ``Spin correlation effects in top
  quark pair production at the {LHC}'',} \textit{ Phys. Rev. D} \textbf{ 81}
  (2010) 074024,
  \href{http://dx.doi.org/10.1103/PhysRevD.81.074024}{\doi{10.1103/PhysRevD.81.074024}},
  \href{http://www.arXiv.org/abs/1001.3422}{\texttt{arXiv:1001.3422}}.

\bibitem{Azzi:2019yne}
\hrefCMSnoop {}{P.~Azzi { et~al.}, ``Report from working group 1: Standard
  model physics at the {HL-LHC} and {HE-LHC}'',} CERN Report CERN-LPCC-2018-03,
  2019.
\newblock
  \href{http://dx.doi.org/10.23731/CYRM-2019-007.1}{\doi{10.23731/CYRM-2019-007.1}},
  \href{http://www.arXiv.org/abs/1902.04070}{\texttt{arXiv:1902.04070}}.

\bibitem{Hoang:2020iah}
\hrefCMSnoop {}{A.~H. Hoang, ``What is the top quark mass?'',} \textit{ Ann.
  Rev. Nucl. Part. Sci.} \textbf{ 70} (2020) 225,
  \href{http://dx.doi.org/10.1146/annurev-nucl-101918-023530}{\doi{10.1146/annurev-nucl-101918-023530}},
  \href{http://www.arXiv.org/abs/2004.12915}{\texttt{arXiv:2004.12915}}.

\bibitem{Kobayashi:1973fv}
\hrefCMSnoop {}{M.~Kobayashi and T.~Maskawa, ``${CP}$-violation in the
  renormalizable theory of weak interaction'',} \textit{ Prog. Theor. Phys.}
  \textbf{ 49} (1973) 652,
  \href{http://dx.doi.org/10.1143/PTP.49.652}{\doi{10.1143/PTP.49.652}}.

\bibitem{Ginsparg:1983zc}
\hrefCMSnoop {}{P.~H. Ginsparg, S.~L. Glashow, and M.~B. Wise, ``Top-quark mass
  and bottom-quark decay'',} \textit{ Phys. Rev. Lett.} \textbf{ 50} (1983)
  1415,
  \href{http://dx.doi.org/10.1103/PhysRevLett.50.1415}{\doi{10.1103/PhysRevLett.50.1415}}.
  [Erratum: \DOI{10.1103/PhysRevLett.51.1395.2}].

\bibitem{Buras:1983ap}
\hrefCMSnoop {}{A.~J. Buras, W.~Slominski, and H.~Steger, ``{\PB} meson decay,
  ${CP}$ violation, mixing angles and the top quark mass'',} \textit{ Nucl.
  Phys. B} \textbf{ 238} (1984) 529,
  \href{http://dx.doi.org/10.1016/0550-3213(84)90334-1}{\doi{10.1016/0550-3213(84)90334-1}}.

\bibitem{JADE:1984bmx}
\hrefCMSnoop {}{{JADE} Collaboration, ``A measurement of the electroweak
  induced charge asymmetry in ${\EE\to\bbbar}$'',} \textit{ Phys. Lett. B}
  \textbf{ 146} (1984) 437,
  \href{http://dx.doi.org/10.1016/0370-2693(84)90156-4}{\doi{10.1016/0370-2693(84)90156-4}}.

\bibitem{Glashow:1970gm}
\hrefCMSnoop {}{S.~L. Glashow, J.~Iliopoulos, and L.~Maiani, ``Weak
  interactions with lepton-hadron symmetry'',} \textit{ Phys. Rev. D} \textbf{
  2} (1970) 1285,
  \href{http://dx.doi.org/10.1103/PhysRevD.2.1285}{\doi{10.1103/PhysRevD.2.1285}}.

\bibitem{Adeva:1982bs}
\hrefCMSnoop {}{B.~Adeva { et~al.}, ``Search for top quark and a test of models
  without top quark up to {38.54\GeV} at {PETRA}'',} \textit{ Phys. Rev. Lett.}
  \textbf{ 50} (1983) 799,
  \href{http://dx.doi.org/10.1103/PhysRevLett.50.799}{\doi{10.1103/PhysRevLett.50.799}}.

\bibitem{TOPAZ:1987pqb}
\hrefCMSnoop {}{{TOPAZ} Collaboration, ``Search for top quark in {\EE}
  collisions at $\sqrt{s}={52\GeV}$'',} \textit{ Phys. Rev. Lett.} \textbf{ 60}
  (1988) 97,
  \href{http://dx.doi.org/10.1103/PhysRevLett.60.97}{\doi{10.1103/PhysRevLett.60.97}}.

\bibitem{UA1:1990rck}
\hrefCMSnoop {}{{UA1} Collaboration, ``Search for new heavy quarks in
  proton-antiproton collisions at $\sqrt{s}={0.63\TeV}$'',} \textit{ Z. Phys.
  C} \textbf{ 48} (1990) 1,
  \href{http://dx.doi.org/10.1007/BF01565600}{\doi{10.1007/BF01565600}}.

\bibitem{UA2:1989tae}
\hrefCMSnoop {}{{UA2} Collaboration, ``Search for top quark production at the
  {CERN} ${\PAp\Pp}$ collider'',} \textit{ Z. Phys. C} \textbf{ 46} (1990) 179,
  \href{http://dx.doi.org/10.1007/BF01555995}{\doi{10.1007/BF01555995}}.

\bibitem{ARGUS:1987xtv}
\hrefCMSnoop {}{{ARGUS} Collaboration, ``Observation of {\PBz}-{\PABz}
  mixing'',} \textit{ Phys. Lett. B} \textbf{ 192} (1987) 245,
  \href{http://dx.doi.org/10.1016/0370-2693(87)91177-4}{\doi{10.1016/0370-2693(87)91177-4}}.

\bibitem{CLEO:1989fuk}
\hrefCMSnoop {}{{CLEO} Collaboration, ``${\PBz\PABz}$ mixing at the {\PgUd}'',}
  \textit{ Phys. Rev. Lett.} \textbf{ 62} (1989) 2233,
  \href{http://dx.doi.org/10.1103/PhysRevLett.62.2233}{\doi{10.1103/PhysRevLett.62.2233}}.

\bibitem{Altarelli:1987zf}
\hrefCMSnoop {}{G.~Altarelli and P.~J. Franzini, ``{\PBz}-{\PABz} mixing within
  and beyond the standard model'',} \textit{ Z. Phys. C} \textbf{ 37} (1988)
  271, \href{http://dx.doi.org/10.1007/BF01579913}{\doi{10.1007/BF01579913}}.

\bibitem{ALEPH:1989tsb}
\hrefCMSnoop {}{{ALEPH} Collaboration, ``A search for new quarks and leptons
  from {\PZz} decay'',} \textit{ Phys. Lett. B} \textbf{ 236} (1990) 511,
  \href{http://dx.doi.org/10.1016/0370-2693(90)90392-J}{\doi{10.1016/0370-2693(90)90392-J}}.

\bibitem{OPAL:1989mxj}
\hrefCMSnoop {}{{OPAL} Collaboration, ``A search for the top and {\PQbpr}
  quarks in hadronic {\PZz} decays'',} \textit{ Phys. Lett. B} \textbf{ 236}
  (1990) 364,
  \href{http://dx.doi.org/10.1016/0370-2693(90)90999-M}{\doi{10.1016/0370-2693(90)90999-M}}.

\bibitem{LEP:1991hsu}
\hrefCMSnoop {}{{ALEPH, DELPHI, L3, and OPAL Collaborations}, ``Electroweak
  parameters of the {\PZz} resonance and the standard model'',} \textit{ Phys.
  Lett. B} \textbf{ 276} (1992) 247,
  \href{http://dx.doi.org/10.1016/0370-2693(92)90572-L}{\doi{10.1016/0370-2693(92)90572-L}}.

\bibitem{ALEPH:1995ac}
\href {https://cds.cern.ch/record/293395}{{ALEPH, DELPHI, L3, and OPAL
  Collaborations, and LEP Electroweak Working Group}, ``A combination of
  preliminary {LEP} electroweak measurements and constraints on the standard
  model'',} LEP Note CERN-PPE-95-172, 1995.

\bibitem{Lusignoli:1991bm}
\hrefCMSnoop {}{M.~Lusignoli, L.~Maiani, G.~Martinelli, and L.~Reina, ``Mixing
  and ${CP}$ violation in {\PK}- and {\PB}-mesons: a lattice {QCD} point of
  view'',} \textit{ Nucl. Phys. B} \textbf{ 369} (1992) 139,
  \href{http://dx.doi.org/10.1016/0550-3213(92)90381-K}{\doi{10.1016/0550-3213(92)90381-K}}.

\bibitem{Buras:1993wr}
\hrefCMSnoop {}{A.~J. Buras, ``A 1993 look at the lower bound on the top quark
  mass from ${CP}$ violation'',} \textit{ Phys. Lett. B} \textbf{ 317} (1993)
  449,
  \href{http://dx.doi.org/10.1016/0370-2693(93)91023-G}{\doi{10.1016/0370-2693(93)91023-G}},
  \href{http://www.arXiv.org/abs/hep-ph/9307318}{\texttt{arXiv:hep-ph/9307318}}.

\bibitem{CDF:1995wbb}
\hrefCMSnoop {}{{CDF} Collaboration, ``Observation of top quark production in
  ${\PAp\Pp}$ collisions with the {Collider Detector at Fermilab}'',} \textit{
  Phys. Rev. Lett.} \textbf{ 74} (1995) 2626,
  \href{http://dx.doi.org/10.1103/PhysRevLett.74.2626}{\doi{10.1103/PhysRevLett.74.2626}},
  \href{http://www.arXiv.org/abs/hep-ex/9503002}{\texttt{arXiv:hep-ex/9503002}}.

\bibitem{D0:1995jca}
\hrefCMSnoop {}{{\DZERO} Collaboration, ``Observation of the top quark'',}
  \textit{ Phys. Rev. Lett.} \textbf{ 74} (1995) 2632,
  \href{http://dx.doi.org/10.1103/PhysRevLett.74.2632}{\doi{10.1103/PhysRevLett.74.2632}},
  \href{http://www.arXiv.org/abs/hep-ex/9503003}{\texttt{arXiv:hep-ex/9503003}}.

\bibitem{D0:2009isq}
\hrefCMSnoop {}{{\DZERO} Collaboration, ``Observation of single top-quark
  production'',} \textit{ Phys. Rev. Lett.} \textbf{ 103} (2009) 092001,
  \href{http://dx.doi.org/10.1103/PhysRevLett.103.092001}{\doi{10.1103/PhysRevLett.103.092001}},
  \href{http://www.arXiv.org/abs/0903.0850}{\texttt{arXiv:0903.0850}}.

\bibitem{CDF:2009itk}
\hrefCMSnoop {}{{CDF} Collaboration, ``Observation of electroweak single
  top-quark production'',} \textit{ Phys. Rev. Lett.} \textbf{ 103} (2009)
  092002,
  \href{http://dx.doi.org/10.1103/PhysRevLett.103.092002}{\doi{10.1103/PhysRevLett.103.092002}},
  \href{http://www.arXiv.org/abs/0903.0885}{\texttt{arXiv:0903.0885}}.

\bibitem{CDF:2016vzt}
\hrefCMSnoop {}{{CDF and \DZERO Collaborations}, ``Combination of {CDF} and
  {\DZERO} results on the mass of the top quark using up to 9.7\fbinv at the
  {Tevatron}'',} 2016.
  \href{http://www.arXiv.org/abs/1608.01881}{\texttt{arXiv:1608.01881}}.

\bibitem{Campagnari:1996ai}
\hrefCMSnoop {}{C.~Campagnari and M.~Franklin, ``The discovery of the top
  quark'',} \textit{ Rev. Mod. Phys.} \textbf{ 69} (1997) 137,
  \href{http://dx.doi.org/10.1103/RevModPhys.69.137}{\doi{10.1103/RevModPhys.69.137}},
  \href{http://www.arXiv.org/abs/hep-ex/9608003}{\texttt{arXiv:hep-ex/9608003}}.

\bibitem{Flynn:1989iu}
\hrefCMSnoop {}{J.~M. Flynn and L.~Randall, ``The electromagnetic penguin
  contribution to $\epsilon^\prime/\epsilon$ for large top quark mass'',}
  \textit{ Phys. Lett. B} \textbf{ 224} (1989) 221,
  \href{http://dx.doi.org/10.1016/0370-2693(89)91078-2}{\doi{10.1016/0370-2693(89)91078-2}}.
  [Erratum: \DOI{10.1016/0370-2693(90)91986-L}].

\bibitem{Buchalla:1989we}
\hrefCMSnoop {}{G.~Buchalla, A.~J. Buras, and M.~K. Harlander, ``The anatomy of
  $\epsilon^\prime/\epsilon$ in the standard model'',} \textit{ Nucl. Phys. B}
  \textbf{ 337} (1990) 313,
  \href{http://dx.doi.org/10.1016/0550-3213(90)90275-I}{\doi{10.1016/0550-3213(90)90275-I}}.

\bibitem{Erler:2019hds}
\hrefCMSnoop {}{J.~Erler and M.~Schott, ``Electroweak precision tests of the
  standard model after the discovery of the {Higgs} noson'',} \textit{ Prog.
  Part. Nucl. Phys.} \textbf{ 106} (2019) 68,
  \href{http://dx.doi.org/10.1016/j.ppnp.2019.02.007}{\doi{10.1016/j.ppnp.2019.02.007}},
  \href{http://www.arXiv.org/abs/1902.05142}{\texttt{arXiv:1902.05142}}.

\bibitem{Haller:2018nnx}
J.~Haller\hrefCMSnoop {}{ { et~al.}, ``Update of the global electroweak fit and
  constraints on two-{Higgs}-doublet models'',} \textit{ Eur. Phys. J. C}
  \textbf{ 78} (2018) 675,
  \href{http://dx.doi.org/10.1140/epjc/s10052-018-6131-3}{\doi{10.1140/epjc/s10052-018-6131-3}},
  \href{http://www.arXiv.org/abs/1803.01853}{\texttt{arXiv:1803.01853}}.

\bibitem{ATLAS:2022vkf}
\hrefCMSnoop {}{{ATLAS Collaboration}, ``A detailed map of {Higgs} boson
  interactions by the {ATLAS} experiment ten years after the discovery'',}
  \textit{ Nature} \textbf{ 607} (2022) 52,
  \href{http://dx.doi.org/10.1038/s41586-022-04893-w}{\doi{10.1038/s41586-022-04893-w}},
  \href{http://www.arXiv.org/abs/2207.00092}{\texttt{arXiv:2207.00092}}.
  [Publisher correction: \DOI{10.1038/s41586-022-05581-5}, author correction:
  \DOI{10.1038/s41586-023-06248-5}].

\bibitem{CMS:2022dwd}
\hrefCMSnoop {}{{CMS Collaboration}, ``A portrait of the {Higgs} boson by the
  {CMS} experiment ten years after the discovery'',} \textit{ Nature} \textbf{
  607} (2022) 60,
  \href{http://dx.doi.org/10.1038/s41586-022-04892-x}{\doi{10.1038/s41586-022-04892-x}},
  \href{http://www.arXiv.org/abs/2207.00043}{\texttt{arXiv:2207.00043}}.
  [Author correction: \DOI{10.1038/s41586-023-06164-8}].

\bibitem{CMS:2020djy}
\hrefCMSnoop {}{{CMS Collaboration}, ``Measurement of the top quark {Yukawa}
  coupling from \ttbar kinematic distributions in the dilepton final state in
  proton-proton collisions at $\sqrt{s}={13\TeV}$'',} \textit{ Phys. Rev. D}
  \textbf{ 102} (2020) 092013,
  \href{http://dx.doi.org/10.1103/PhysRevD.102.092013}{\doi{10.1103/PhysRevD.102.092013}},
  \href{http://www.arXiv.org/abs/2009.07123}{\texttt{arXiv:2009.07123}}.

\bibitem{Degrassi:2012ry}
G.~Degrassi\hrefCMSnoop {}{ { et~al.}, ``{Higgs} mass and vacuum stability in
  the standard model at {NNLO}'',} \textit{ JHEP} \textbf{ 08} (2012) 098,
  \href{http://dx.doi.org/10.1007/JHEP08(2012)098}{\doi{10.1007/JHEP08(2012)098}},
  \href{http://www.arXiv.org/abs/1205.6497}{\texttt{arXiv:1205.6497}}.

\bibitem{Alekhin:2012py}
\hrefCMSnoop {}{S.~Alekhin, A.~Djouadi, and S.~Moch, ``The top quark and
  {Higgs} boson masses and the stability of the electroweak vacuum'',} \textit{
  Phys. Lett. B} \textbf{ 716} (2012) 214,
  \href{http://dx.doi.org/10.1016/j.physletb.2012.08.024}{\doi{10.1016/j.physletb.2012.08.024}},
  \href{http://www.arXiv.org/abs/1207.0980}{\texttt{arXiv:1207.0980}}.

\bibitem{Heinemeyer:2013dia}
\hrefCMSnoop {}{S.~Heinemeyer, W.~Hollik, G.~Weiglein, and L.~Zeune,
  ``Implications of {LHC} search results on the {\PW} boson mass prediction in
  the {MSSM}'',} \textit{ JHEP} \textbf{ 12} (2013) 084,
  \href{http://dx.doi.org/10.1007/JHEP12(2013)084}{\doi{10.1007/JHEP12(2013)084}},
  \href{http://www.arXiv.org/abs/1311.1663}{\texttt{arXiv:1311.1663}}.

\bibitem{Bezrukov:2012sa}
\hrefCMSnoop {}{F.~Bezrukov, M.~Y. Kalmykov, B.~A. Kniehl, and M.~Shaposhnikov,
  ``{Higgs} boson mass and new physics'',} \textit{ JHEP} \textbf{ 10} (2012)
  140,
  \href{http://dx.doi.org/10.1007/JHEP10(2012)140}{\doi{10.1007/JHEP10(2012)140}},
  \href{http://www.arXiv.org/abs/1205.2893}{\texttt{arXiv:1205.2893}}.

\bibitem{Dunsky:2020yhv}
\hrefCMSnoop {}{D.~Dunsky, L.~J. Hall, and K.~Harigaya, ``Dark matter
  detection, standard model parameters and intermediate scale supersymmetry'',}
  \textit{ JHEP} \textbf{ 04} (2021) 052,
  \href{http://dx.doi.org/10.1007/JHEP04(2021)052}{\doi{10.1007/JHEP04(2021)052}},
  \href{http://www.arXiv.org/abs/2011.12302}{\texttt{arXiv:2011.12302}}.

\bibitem{Buchmuller:1985jz}
\hrefCMSnoop {}{W.~Buchmuller and D.~Wyler, ``Effective {Lagrangian} analysis
  of new interactions and flavour conservation'',} \textit{ Nucl. Phys. B}
  \textbf{ 268} (1986) 621,
  \href{http://dx.doi.org/10.1016/0550-3213(86)90262-2}{\doi{10.1016/0550-3213(86)90262-2}}.

\bibitem{Giudice:2007fh}
\hrefCMSnoop {}{G.~F. Giudice, C.~Grojean, A.~Pomarol, and R.~Rattazzi, ``The
  strongly-interacting light {Higgs}'',} \textit{ JHEP} \textbf{ 06} (2007)
  045,
  \href{http://dx.doi.org/10.1088/1126-6708/2007/06/045}{\doi{10.1088/1126-6708/2007/06/045}},
  \href{http://www.arXiv.org/abs/hep-ph/0703164}{\texttt{arXiv:hep-ph/0703164}}.

\bibitem{Grzadkowski:2010es}
\hrefCMSnoop {}{B.~Grzadkowski, M.~Iskrzynski, M.~Misiak, and J.~Rosiek,
  ``Dimension-six terms in the standard model {Lagrangian}'',} \textit{ JHEP}
  \textbf{ 10} (2010) 085,
  \href{http://dx.doi.org/10.1007/JHEP10(2010)085}{\doi{10.1007/JHEP10(2010)085}},
  \href{http://www.arXiv.org/abs/1008.4884}{\texttt{arXiv:1008.4884}}.

\bibitem{Gao:2022srd}
J.~Gao\hrefCMSnoop {}{ { et~al.}, ``Simultaneous {CTEQ-TEA} extraction of
  {PDFs} and {SMEFT} parameters from jet and \ttbar data'',} \textit{ JHEP}
  \textbf{ 05} (2023) 003,
  \href{http://dx.doi.org/10.1007/JHEP05(2023)003}{\doi{10.1007/JHEP05(2023)003}},
  \href{http://www.arXiv.org/abs/2211.01094}{\texttt{arXiv:2211.01094}}.

\bibitem{Buras:2012ru}
\hrefCMSnoop {}{A.~J. Buras, J.~Girrbach, D.~Guadagnoli, and G.~Isidori, ``On
  the standard model prediction for
  {$\mathcal{B}(\HepParticle{\PB}{s,d}{}\to\MM)$}'',} \textit{ Eur. Phys. J. C}
  \textbf{ 72} (2012) 2172,
  \href{http://dx.doi.org/10.1140/epjc/s10052-012-2172-1}{\doi{10.1140/epjc/s10052-012-2172-1}},
  \href{http://www.arXiv.org/abs/1208.0934}{\texttt{arXiv:1208.0934}}.

\bibitem{Misiak:2015xwa}
\hrefCMSnoop {}{M.~Misiak { et~al.}, ``Updated next-to-next-to-leading-order
  {QCD} predictions for the weak radiative {\PB}-meson decays'',} \textit{
  Phys. Rev. Lett.} \textbf{ 114} (2015) 221801,
  \href{http://dx.doi.org/10.1103/PhysRevLett.114.221801}{\doi{10.1103/PhysRevLett.114.221801}},
  \href{http://www.arXiv.org/abs/1503.01789}{\texttt{arXiv:1503.01789}}.

\bibitem{Czakon:2015exa}
M.~Czakon\hrefCMSnoop {}{ { et~al.}, ``The ${(Q_7, Q_{1,2})}$ contribution to
  {$\PAB\to\HepParticle{X}{s}{}\PGg$} at {$\mathcal{O}(\alpS^2)$}'',} \textit{
  JHEP} \textbf{ 04} (2015) 168,
  \href{http://dx.doi.org/10.1007/JHEP04(2015)168}{\doi{10.1007/JHEP04(2015)168}},
  \href{http://www.arXiv.org/abs/1503.01791}{\texttt{arXiv:1503.01791}}.

\bibitem{CMS:2011acs}
\hrefCMSnoop {}{{CMS Collaboration}, ``Measurement of the \ttbar production
  cross section and the top quark mass in the dilepton channel in ${\Pp\Pp}$
  collisions at $\sqrt{s}={7\TeV}$'',} \textit{ JHEP} \textbf{ 07} (2011) 049,
  \href{http://dx.doi.org/10.1007/JHEP07(2011)049}{\doi{10.1007/JHEP07(2011)049}},
  \href{http://www.arXiv.org/abs/1105.5661}{\texttt{arXiv:1105.5661}}.

\bibitem{CMS:2012sas}
\hrefCMSnoop {}{{CMS Collaboration}, ``Measurement of the top-quark mass in
  \ttbar events with lepton+jets final states in ${\Pp\Pp}$ collisions at
  $\sqrt{s}={7\TeV}$'',} \textit{ JHEP} \textbf{ 12} (2012) 105,
  \href{http://dx.doi.org/10.1007/JHEP12(2012)105}{\doi{10.1007/JHEP12(2012)105}},
  \href{http://www.arXiv.org/abs/1209.2319}{\texttt{arXiv:1209.2319}}.

\bibitem{CMS:2012tdr}
\hrefCMSnoop {}{{CMS Collaboration}, ``Measurement of the top-quark mass in
  \ttbar events with dilepton final states in ${\Pp\Pp}$ collisions at
  $\sqrt{s}={7\TeV}$'',} \textit{ Eur. Phys. J. C} \textbf{ 72} (2012) 2202,
  \href{http://dx.doi.org/10.1140/epjc/s10052-012-2202-z}{\doi{10.1140/epjc/s10052-012-2202-z}},
  \href{http://www.arXiv.org/abs/1209.2393}{\texttt{arXiv:1209.2393}}.

\bibitem{CMS:2013wbt}
\hrefCMSnoop {}{{CMS Collaboration}, ``Measurement of masses in the \ttbar
  system by kinematic endpoints in ${\Pp\Pp}$ collisions at
  $\sqrt{s}={7\TeV}$'',} \textit{ Eur. Phys. J. C} \textbf{ 73} (2013) 2494,
  \href{http://dx.doi.org/10.1140/epjc/s10052-013-2494-7}{\doi{10.1140/epjc/s10052-013-2494-7}},
  \href{http://www.arXiv.org/abs/1304.5783}{\texttt{arXiv:1304.5783}}.

\bibitem{CMS:2013lqq}
\hrefCMSnoop {}{{CMS Collaboration}, ``Measurement of the top-quark mass in
  all-jets \ttbar events in ${\Pp\Pp}$ collisions at $\sqrt{s}={7\TeV}$'',}
  \textit{ Eur. Phys. J. C} \textbf{ 74} (2014) 2758,
  \href{http://dx.doi.org/10.1140/epjc/s10052-014-2758-x}{\doi{10.1140/epjc/s10052-014-2758-x}},
  \href{http://www.arXiv.org/abs/1307.4617}{\texttt{arXiv:1307.4617}}.

\bibitem{CMS:2014rml}
\hrefCMSnoop {}{{CMS Collaboration}, ``Determination of the top-quark pole mass
  and strong coupling constant from the \ttbar production in ${\Pp\Pp}$
  collisions at $\sqrt{s}={7\TeV}$'',} \textit{ Phys. Lett. B} \textbf{ 728}
  (2014) 496,
  \href{http://dx.doi.org/10.1016/j.physletb.2013.12.009}{\doi{10.1016/j.physletb.2013.12.009}},
  \href{http://www.arXiv.org/abs/1307.1907}{\texttt{arXiv:1307.1907}}.
  [Corrigendum: \DOI{10.1016/j.physletb.2014.08.040}].

\bibitem{CMS:2015lbj}
\hrefCMSnoop {}{{CMS Collaboration}, ``Measurement of the top quark mass using
  proton-proton data at $\sqrt{s}=7$ and {8\TeV}'',} \textit{ Phys. Rev. D}
  \textbf{ 93} (2016) 072004,
  \href{http://dx.doi.org/10.1103/PhysRevD.93.072004}{\doi{10.1103/PhysRevD.93.072004}},
  \href{http://www.arXiv.org/abs/1509.04044}{\texttt{arXiv:1509.04044}}.

\bibitem{CMS:2016yys}
\hrefCMSnoop {}{{CMS Collaboration}, ``Measurement of the \ttbar production
  cross section in the ${\Pe\PGm}$ channel in proton-proton collisions at
  $\sqrt{s}=7$ and {8\TeV}'',} \textit{ JHEP} \textbf{ 08} (2016) 029,
  \href{http://dx.doi.org/10.1007/JHEP08(2016)029}{\doi{10.1007/JHEP08(2016)029}},
  \href{http://www.arXiv.org/abs/1603.02303}{\texttt{arXiv:1603.02303}}.

\bibitem{CMS:2016iru}
\hrefCMSnoop {}{{CMS Collaboration}, ``Measurement of the top quark mass using
  charged particles in ${\Pp\Pp}$ collisions at $\sqrt{s}={8\TeV}$'',} \textit{
  Phys. Rev. D} \textbf{ 93} (2016) 092006,
  \href{http://dx.doi.org/10.1103/PhysRevD.93.092006}{\doi{10.1103/PhysRevD.93.092006}},
  \href{http://www.arXiv.org/abs/1603.06536}{\texttt{arXiv:1603.06536}}.

\bibitem{CMS:2016ixg}
\hrefCMSnoop {}{{CMS Collaboration}, ``Measurement of the mass of the top quark
  in decays with a {\PJGy} meson in ${\Pp\Pp}$ collisions at {8\TeV}'',}
  \textit{ JHEP} \textbf{ 12} (2016) 123,
  \href{http://dx.doi.org/10.1007/JHEP12(2016)123}{\doi{10.1007/JHEP12(2016)123}},
  \href{http://www.arXiv.org/abs/1608.03560}{\texttt{arXiv:1608.03560}}.

\bibitem{CMS:2017xrt}
\hrefCMSnoop {}{{CMS Collaboration}, ``Measurement of the \ttbar production
  cross section using events with one lepton and at least one jet in ${\Pp\Pp}$
  collisions at $\sqrt{s}={13\TeV}$'',} \textit{ JHEP} \textbf{ 09} (2017) 051,
  \href{http://dx.doi.org/10.1007/JHEP09(2017)051}{\doi{10.1007/JHEP09(2017)051}},
  \href{http://www.arXiv.org/abs/1701.06228}{\texttt{arXiv:1701.06228}}.

\bibitem{CMS:2017mpr}
\hrefCMSnoop {}{{CMS Collaboration}, ``Measurement of the top quark mass using
  single top quark events in proton-proton collisions at $\sqrt{s}={8\TeV}$'',}
  \textit{ Eur. Phys. J. C} \textbf{ 77} (2017) 354,
  \href{http://dx.doi.org/10.1140/epjc/s10052-017-4912-8}{\doi{10.1140/epjc/s10052-017-4912-8}},
  \href{http://www.arXiv.org/abs/1703.02530}{\texttt{arXiv:1703.02530}}.

\bibitem{CMS:2017pcy}
\hrefCMSnoop {}{{CMS Collaboration}, ``Measurement of the jet mass in highly
  boosted \ttbar events from ${\Pp\Pp}$ collisions at $\sqrt{s}=$ {8\TeV}'',}
  \textit{ Eur. Phys. J. C} \textbf{ 77} (2017) 467,
  \href{http://dx.doi.org/10.1140/epjc/s10052-017-5030-3}{\doi{10.1140/epjc/s10052-017-5030-3}},
  \href{http://www.arXiv.org/abs/1703.06330}{\texttt{arXiv:1703.06330}}.

\bibitem{CMS:2017znf}
\hrefCMSnoop {}{{CMS Collaboration}, ``Measurement of the top quark mass in the
  dileptonic \ttbar decay channel using the mass observables ${M}_{\PQb\Pell}$,
  ${M}_{{\mathrm{T}2}}$, and $m_{\PQb\Pell\PGn}$ in ${\Pp\Pp}$ collisions at
  $\sqrt{s}={8\TeV}$'',} \textit{ Phys. Rev. D} \textbf{ 96} (2017) 032002,
  \href{http://dx.doi.org/10.1103/PhysRevD.96.032002}{\doi{10.1103/PhysRevD.96.032002}},
  \href{http://www.arXiv.org/abs/1704.06142}{\texttt{arXiv:1704.06142}}.

\bibitem{CMS:2018quc}
\hrefCMSnoop {}{{CMS Collaboration}, ``Measurement of the top quark mass with
  lepton+jets final states using ${\Pp\Pp}$ collisions at
  $\sqrt{s}={13\TeV}$'',} \textit{ Eur. Phys. J. C} \textbf{ 78} (2018) 891,
  \href{http://dx.doi.org/10.1140/epjc/s10052-018-6332-9}{\doi{10.1140/epjc/s10052-018-6332-9}},
  \href{http://www.arXiv.org/abs/1805.01428}{\texttt{arXiv:1805.01428}}.

\bibitem{CMS:2018tye}
\hrefCMSnoop {}{{CMS Collaboration}, ``Measurement of the top quark mass in the
  all-jets final state at $\sqrt{s}={13\TeV}$ and combination with the
  lepton+jets channel'',} \textit{ Eur. Phys. J. C} \textbf{ 79} (2019) 313,
  \href{http://dx.doi.org/10.1140/epjc/s10052-019-6788-2}{\doi{10.1140/epjc/s10052-019-6788-2}},
  \href{http://www.arXiv.org/abs/1812.10534}{\texttt{arXiv:1812.10534}}.

\bibitem{CMS:2018fks}
\hrefCMSnoop {}{{CMS Collaboration}, ``Measurement of the \ttbar production
  cross section, the top quark mass, and the strong coupling constant using
  dilepton events in ${\Pp\Pp}$ collisions at $\sqrt{s}={13\TeV}$'',} \textit{
  Eur. Phys. J. C} \textbf{ 79} (2019) 368,
  \href{http://dx.doi.org/10.1140/epjc/s10052-019-6863-8}{\doi{10.1140/epjc/s10052-019-6863-8}},
  \href{http://www.arXiv.org/abs/1812.10505}{\texttt{arXiv:1812.10505}}.

\bibitem{CMS:2019esx}
\hrefCMSnoop {}{{CMS Collaboration}, ``Measurement of \ttbar normalised
  multi-differential cross sections in ${\Pp\Pp}$ collisions at
  $\sqrt{s}={13\TeV}$, and simultaneous determination of the strong coupling
  strength, top quark pole mass, and parton distribution functions'',} \textit{
  Eur. Phys. J. C} \textbf{ 80} (2020) 658,
  \href{http://dx.doi.org/10.1140/epjc/s10052-020-7917-7}{\doi{10.1140/epjc/s10052-020-7917-7}},
  \href{http://www.arXiv.org/abs/1904.05237}{\texttt{arXiv:1904.05237}}.

\bibitem{CMS:2019jul}
\hrefCMSnoop {}{{CMS Collaboration}, ``Running of the top quark mass from
  proton-proton collisions at $\sqrt{s}={13\TeV}$'',} \textit{ Phys. Lett. B}
  \textbf{ 803} (2020) 135263,
  \href{http://dx.doi.org/10.1016/j.physletb.2020.135263}{\doi{10.1016/j.physletb.2020.135263}},
  \href{http://www.arXiv.org/abs/1909.09193}{\texttt{arXiv:1909.09193}}.

\bibitem{CMS:2019fak}
\hrefCMSnoop {}{{CMS Collaboration}, ``Measurement of the jet mass distribution
  and top quark mass in hadronic decays of boosted top quarks in ${\Pp\Pp}$
  collisions at $\sqrt{s}={13\TeV}$'',} \textit{ Phys. Rev. Lett.} \textbf{
  124} (2020) 202001,
  \href{http://dx.doi.org/10.1103/PhysRevLett.124.202001}{\doi{10.1103/PhysRevLett.124.202001}},
  \href{http://www.arXiv.org/abs/1911.03800}{\texttt{arXiv:1911.03800}}.

\bibitem{CMS:2021jnp}
\hrefCMSnoop {}{{CMS Collaboration}, ``Measurement of the top quark mass using
  events with a single reconstructed top quark in ${\Pp\Pp}$ collisions at
  $\sqrt{s}={13\TeV}$'',} \textit{ JHEP} \textbf{ 12} (2021) 161,
  \href{http://dx.doi.org/10.1007/JHEP12(2021)161}{\doi{10.1007/JHEP12(2021)161}},
  \href{http://www.arXiv.org/abs/2108.10407}{\texttt{arXiv:2108.10407}}.

\bibitem{ATLAS:2022aof}
\hrefCMSnoop {}{{ATLAS and CMS Collaborations}, ``Combination of inclusive
  top-quark pair production cross-section measurements using {ATLAS} and {CMS}
  data at $\sqrt{s}=7$ and {8\TeV}'',} \textit{ JHEP} \textbf{ 07} (2023) 213,
  \href{http://dx.doi.org/10.1007/JHEP07(2023)213}{\doi{10.1007/JHEP07(2023)213}},
  \href{http://www.arXiv.org/abs/2205.13830}{\texttt{arXiv:2205.13830}}.

\bibitem{CMS:2022emx}
\hrefCMSnoop {}{{CMS Collaboration}, ``Measurement of the top quark pole mass
  using {\ttbar}+jet events in the dilepton final state in proton-proton
  collisions at $\sqrt{s}={13\TeV}$'',} \textit{ JHEP} \textbf{ 07} (2023) 077,
  \href{http://dx.doi.org/10.1007/JHEP07(2023)077}{\doi{10.1007/JHEP07(2023)077}},
  \href{http://www.arXiv.org/abs/2207.02270}{\texttt{arXiv:2207.02270}}.

\bibitem{CMS:2022kqg}
\hrefCMSnoop {}{{CMS Collaboration}, ``Measurement of the differential \ttbar
  production cross section as a function of the jet mass and extraction of the
  top quark mass in hadronic decays of boosted top quarks'',} \textit{ Eur.
  Phys. J. C} \textbf{ 83} (2023) 560,
  \href{http://dx.doi.org/10.1140/epjc/s10052-023-11587-8}{\doi{10.1140/epjc/s10052-023-11587-8}},
  \href{http://www.arXiv.org/abs/2211.01456}{\texttt{arXiv:2211.01456}}.

\bibitem{CMS:2023ebf}
\hrefCMSnoop {}{{CMS Collaboration}, ``Measurement of the top quark mass using
  a profile likelihood approach with the lepton+jets final states in
  proton-proton collisions at $\sqrt{s}={13\TeV}$'',} \textit{ Eur. Phys. J. C}
  \textbf{ 83} (2023) 963,
  \href{http://dx.doi.org/10.1140/epjc/s10052-023-12050-4}{\doi{10.1140/epjc/s10052-023-12050-4}},
  \href{http://www.arXiv.org/abs/2302.01967}{\texttt{arXiv:2302.01967}}.

\bibitem{CMS:2023wnd}
\hrefCMSnoop {}{{ATLAS and CMS Collaborations}, ``Combination of measurements
  of the top quark mass from data collected by the {ATLAS} and {CMS}
  experiments at $\sqrt{s}=7$ and {8\TeV}'',} \textit{ Phys. Rev. Lett.}
  \textbf{ 132} (2024) 261902,
  \href{http://dx.doi.org/10.1103/PhysRevLett.132.261902}{\doi{10.1103/PhysRevLett.132.261902}},
  \href{http://www.arXiv.org/abs/2402.08713}{\texttt{arXiv:2402.08713}}.

\bibitem{ATLAS:2021urs}
\href {https://cds.cern.ch/record/2777332}{{ATLAS Collaboration}, ``Towards a
  precise interpretation for the top quark mass parameter in {ATLAS} {Monte
  Carlo} samples'',} ATLAS PUB Note ATL-PHYS-PUB-2021-034, 2021.

\bibitem{CMS:2006myw}
\href {https://cds.cern.ch/record/922757}{{CMS Collaboration}, ``{CMS} physics
  technical design report, volume {I}: Detector performance and software'',}
  CMS Technical Proposal CERN-LHCC-2006-001, CMS-TDR-8.1, 2006.

\bibitem{Nason:1987xz}
\hrefCMSnoop {}{P.~Nason, S.~Dawson, and R.~K. Ellis, ``The total cross section
  for the production of heavy quarks in hadronic collisions'',} \textit{ Nucl.
  Phys. B} \textbf{ 303} (1988) 607,
  \href{http://dx.doi.org/10.1016/0550-3213(88)90422-1}{\doi{10.1016/0550-3213(88)90422-1}}.

\bibitem{Beenakker:1988bq}
\hrefCMSnoop {}{W.~Beenakker, H.~Kuijf, W.~L. van Neerven, and J.~Smith,
  ``{QCD} corrections to heavy-quark production in ${\Pp\PAp}$ collisions'',}
  \textit{ Phys. Rev. D} \textbf{ 40} (1989) 54,
  \href{http://dx.doi.org/10.1103/PhysRevD.40.54}{\doi{10.1103/PhysRevD.40.54}}.

\bibitem{Bernreuther:2004jv}
\hrefCMSnoop {}{W.~Bernreuther, A.~Brandenburg, Z.~G. Si, and P.~Uwer, ``Top
  quark pair production and decay at hadron colliders'',} \textit{ Nucl. Phys.
  B} \textbf{ 690} (2004) 81,
  \href{http://dx.doi.org/10.1016/j.nuclphysb.2004.04.019}{\doi{10.1016/j.nuclphysb.2004.04.019}},
  \href{http://www.arXiv.org/abs/hep-ph/0403035}{\texttt{arXiv:hep-ph/0403035}}.

\bibitem{Czakon:2011xx}
\hrefCMSnoop {}{M.~Czakon and A.~Mitov, ``\textsc{top++}: a program for the
  calculation of the top-pair cross-section at hadron colliders'',} \textit{
  Comput. Phys. Commun.} \textbf{ 185} (2014) 2930,
  \href{http://dx.doi.org/10.1016/j.cpc.2014.06.021}{\doi{10.1016/j.cpc.2014.06.021}},
  \href{http://www.arXiv.org/abs/1112.5675}{\texttt{arXiv:1112.5675}}.

\bibitem{Kidonakis:2014pja}
\hrefCMSnoop {}{N.~Kidonakis, ``{NNNLO} soft-gluon corrections for the
  top-quark \pt and rapidity distributions'',} \textit{ Phys. Rev. D} \textbf{
  91} (2015) 031501,
  \href{http://dx.doi.org/10.1103/PhysRevD.91.031501}{\doi{10.1103/PhysRevD.91.031501}},
  \href{http://www.arXiv.org/abs/1411.2633}{\texttt{arXiv:1411.2633}}.

\bibitem{Czakon:2015owf}
\hrefCMSnoop {}{M.~Czakon, D.~Heymes, and A.~Mitov, ``High-precision
  differential predictions for top-quark pairs at the {LHC}'',} \textit{ Phys.
  Rev. Lett.} \textbf{ 116} (2016) 082003,
  \href{http://dx.doi.org/10.1103/PhysRevLett.116.082003}{\doi{10.1103/PhysRevLett.116.082003}},
  \href{http://www.arXiv.org/abs/1511.00549}{\texttt{arXiv:1511.00549}}.

\bibitem{Czakon:2018nun}
M.~Czakon\hrefCMSnoop {}{ { et~al.}, ``Resummation for (boosted) top-quark pair
  production at {NNLO+NNLL'} in {QCD}'',} \textit{ JHEP} \textbf{ 05} (2018)
  149,
  \href{http://dx.doi.org/10.1007/JHEP05(2018)149}{\doi{10.1007/JHEP05(2018)149}},
  \href{http://www.arXiv.org/abs/1803.07623}{\texttt{arXiv:1803.07623}}.

\bibitem{Catani:2019hip}
S.~Catani\hrefCMSnoop {}{ { et~al.}, ``Top-quark pair production at the {LHC}:
  fully differential {QCD} predictions at {NNLO}'',} \textit{ JHEP} \textbf{
  07} (2019) 100,
  \href{http://dx.doi.org/10.1007/JHEP07(2019)100}{\doi{10.1007/JHEP07(2019)100}},
  \href{http://www.arXiv.org/abs/1906.06535}{\texttt{arXiv:1906.06535}}.

\bibitem{Kidonakis:2019yji}
\hrefCMSnoop {}{N.~Kidonakis, ``Top-quark double-differential distributions at
  approximate {N\textsuperscript{3}LO}'',} \textit{ Phys. Rev. D} \textbf{ 101}
  (2020) 074006,
  \href{http://dx.doi.org/10.1103/PhysRevD.101.074006}{\doi{10.1103/PhysRevD.101.074006}},
  \href{http://www.arXiv.org/abs/1912.10362}{\texttt{arXiv:1912.10362}}.

\bibitem{Czakon:2017wor}
M.~Czakon\hrefCMSnoop {}{ { et~al.}, ``Top-pair production at the {LHC} through
  {NNLO} {QCD} and {NLO} {EW}'',} \textit{ JHEP} \textbf{ 10} (2017) 186,
  \href{http://dx.doi.org/10.1007/JHEP10(2017)186}{\doi{10.1007/JHEP10(2017)186}},
  \href{http://www.arXiv.org/abs/1705.04105}{\texttt{arXiv:1705.04105}}.

\bibitem{Czakon:2020qbd}
\hrefCMSnoop {}{M.~Czakon, A.~Mitov, and R.~Poncelet, ``{NNLO} {QCD}
  corrections to leptonic observables in top-quark pair production and
  decay'',} \textit{ JHEP} \textbf{ 05} (2021) 212,
  \href{http://dx.doi.org/10.1007/JHEP05(2021)212}{\doi{10.1007/JHEP05(2021)212}},
  \href{http://www.arXiv.org/abs/2008.11133}{\texttt{arXiv:2008.11133}}.

\bibitem{CMS:2023qyl}
\hrefCMSnoop {}{{CMS Collaboration}, ``First measurement of the top quark pair
  production cross section in proton-proton collisions at
  $\sqrt{s}={13.6\TeV}$'',} \textit{ JHEP} \textbf{ 08} (2023) 204,
  \href{http://dx.doi.org/10.1007/JHEP08(2023)204}{\doi{10.1007/JHEP08(2023)204}},
  \href{http://www.arXiv.org/abs/2303.10680}{\texttt{arXiv:2303.10680}}.

\bibitem{Cortese:1991fw}
\hrefCMSnoop {}{S.~Cortese and R.~Petronzio, ``The single top production
  channel at {Tevatron} energies'',} \textit{ Phys. Lett. B} \textbf{ 253}
  (1991) 494,
  \href{http://dx.doi.org/10.1016/0370-2693(91)91758-N}{\doi{10.1016/0370-2693(91)91758-N}}.

\bibitem{PhysRevD.34.155}
\hrefCMSnoop {}{S.~S.~D. Willenbrock and D.~A. Dicus, ``Production of heavy
  quarks from {\PW}-gluon fusion'',} \textit{ Phys. Rev. D} \textbf{ 34} (1986)
  155,
  \href{http://dx.doi.org/10.1103/PhysRevD.34.155}{\doi{10.1103/PhysRevD.34.155}}.

\bibitem{Campbell:2020fhf}
\hrefCMSnoop {}{J.~Campbell, T.~Neumann, and Z.~Sullivan, ``Single-top-quark
  production in the $t$-channel at {NNLO}'',} \textit{ JHEP} \textbf{ 02}
  (2021) 040,
  \href{http://dx.doi.org/10.1007/JHEP02(2021)040}{\doi{10.1007/JHEP02(2021)040}},
  \href{http://www.arXiv.org/abs/2012.01574}{\texttt{arXiv:2012.01574}}.

\bibitem{Brucherseifer:2014ama}
\hrefCMSnoop {}{M.~Brucherseifer, F.~Caola, and K.~Melnikov, ``On the {NNLO}
  {QCD} corrections to single-top production at the {LHC}'',} \textit{ Phys.
  Lett. B} \textbf{ 736} (2014) 58,
  \href{http://dx.doi.org/10.1016/j.physletb.2014.06.075}{\doi{10.1016/j.physletb.2014.06.075}},
  \href{http://www.arXiv.org/abs/1404.7116}{\texttt{arXiv:1404.7116}}.

\bibitem{Berger:2017zof}
\hrefCMSnoop {}{E.~L. Berger, J.~Gao, and H.~X. Zhu, ``Differential
  distributions for $t$-channel single top-quark production and decay at
  next-to-next-to-leading order in {QCD}'',} \textit{ JHEP} \textbf{ 11} (2017)
  158,
  \href{http://dx.doi.org/10.1007/JHEP11(2017)158}{\doi{10.1007/JHEP11(2017)158}},
  \href{http://www.arXiv.org/abs/1708.09405}{\texttt{arXiv:1708.09405}}.

\bibitem{Kidonakis:2010tc}
\hrefCMSnoop {}{N.~Kidonakis, ``Next-to-next-to-leading logarithm resummation
  for $s$-channel single top quark production'',} \textit{ Phys. Rev. D}
  \textbf{ 81} (2010) 054028,
  \href{http://dx.doi.org/10.1103/PhysRevD.81.054028}{\doi{10.1103/PhysRevD.81.054028}},
  \href{http://www.arXiv.org/abs/1001.5034}{\texttt{arXiv:1001.5034}}.

\bibitem{Frixione:2008yi}
S.~Frixione\hrefCMSnoop {}{ { et~al.}, ``Single-top hadroproduction in
  association with a {\PW} boson'',} \textit{ JHEP} \textbf{ 07} (2008) 029,
  \href{http://dx.doi.org/10.1088/1126-6708/2008/07/029}{\doi{10.1088/1126-6708/2008/07/029}},
  \href{http://www.arXiv.org/abs/0805.3067}{\texttt{arXiv:0805.3067}}.

\bibitem{White:2009yt}
\hrefCMSnoop {}{C.~D. White, S.~Frixione, E.~Laenen, and F.~Maltoni,
  ``Isolating ${\PW\PQt}$ production at the {LHC}'',} \textit{ JHEP} \textbf{
  11} (2009) 074,
  \href{http://dx.doi.org/10.1088/1126-6708/2009/11/074}{\doi{10.1088/1126-6708/2009/11/074}},
  \href{http://www.arXiv.org/abs/0908.0631}{\texttt{arXiv:0908.0631}}.

\bibitem{Bevilacqua:2010qb}
G.~Bevilacqua\hrefCMSnoop {}{ { et~al.}, ``Complete off-shell effects in top
  quark pair hadroproduction with leptonic decay at next-to-leading order'',}
  \textit{ JHEP} \textbf{ 02} (2011) 083,
  \href{http://dx.doi.org/10.1007/JHEP02(2011)083}{\doi{10.1007/JHEP02(2011)083}},
  \href{http://www.arXiv.org/abs/1012.4230}{\texttt{arXiv:1012.4230}}.

\bibitem{Denner:2010jp}
\hrefCMSnoop {}{A.~Denner, S.~Dittmaier, S.~Kallweit, and S.~Pozzorini,
  ``Next-to-leading-order {QCD} corrections to ${\PWp\PWm\bbbar}$ production at
  hadron colliders'',} \textit{ Phys. Rev. Lett.} \textbf{ 106} (2011) 052001,
  \href{http://dx.doi.org/10.1103/PhysRevLett.106.052001}{\doi{10.1103/PhysRevLett.106.052001}},
  \href{http://www.arXiv.org/abs/1012.3975}{\texttt{arXiv:1012.3975}}.

\bibitem{Denner:2012yc}
\hrefCMSnoop {}{A.~Denner, S.~Dittmaier, S.~Kallweit, and S.~Pozzorini, ``{NLO}
  {QCD} corrections to off-shell top-antitop production with leptonic decays at
  hadron colliders'',} \textit{ JHEP} \textbf{ 10} (2012) 110,
  \href{http://dx.doi.org/10.1007/JHEP10(2012)110}{\doi{10.1007/JHEP10(2012)110}},
  \href{http://www.arXiv.org/abs/1207.5018}{\texttt{arXiv:1207.5018}}.

\bibitem{Kidonakis:2010ux}
\hrefCMSnoop {}{N.~Kidonakis, ``Two-loop soft anomalous dimensions for single
  top quark associated production with a {\PWm} or {\PSHm}'',} \textit{ Phys.
  Rev. D} \textbf{ 82} (2010) 054018,
  \href{http://dx.doi.org/10.1103/PhysRevD.82.054018}{\doi{10.1103/PhysRevD.82.054018}},
  \href{http://www.arXiv.org/abs/1005.4451}{\texttt{arXiv:1005.4451}}.

\bibitem{Czarnecki:1998qc}
\hrefCMSnoop {}{A.~Czarnecki and K.~Melnikov, ``Two-loop {QCD} corrections to
  top quark width'',} \textit{ Nucl. Phys. B} \textbf{ 544} (1999) 520,
  \href{http://dx.doi.org/10.1016/S0550-3213(98)00844-X}{\doi{10.1016/S0550-3213(98)00844-X}},
  \href{http://www.arXiv.org/abs/hep-ph/9806244}{\texttt{arXiv:hep-ph/9806244}}.

\bibitem{Chetyrkin:1999ju}
\hrefCMSnoop {}{K.~G. Chetyrkin, R.~Harlander, T.~Seidensticker, and
  M.~Steinhauser, ``Second order {QCD} corrections to
  ${\Gamma(\PQt\to\PW\PQb)}$'',} \textit{ Phys. Rev. D} \textbf{ 60} (1999)
  114015,
  \href{http://dx.doi.org/10.1103/PhysRevD.60.114015}{\doi{10.1103/PhysRevD.60.114015}},
  \href{http://www.arXiv.org/abs/hep-ph/9906273}{\texttt{arXiv:hep-ph/9906273}}.

\bibitem{Blokland:2004ye}
\hrefCMSnoop {}{I.~Blokland, A.~Czarnecki, M.~{\'S}lusarczyk, and F.~Tkachov,
  ``Heavy-to-light-quark decays with two-loop accuracy'',} \textit{ Phys. Rev.
  Lett.} \textbf{ 93} (2004) 062001,
  \href{http://dx.doi.org/10.1103/PhysRevLett.93.062001}{\doi{10.1103/PhysRevLett.93.062001}},
  \href{http://www.arXiv.org/abs/hep-ph/0403221}{\texttt{arXiv:hep-ph/0403221}}.

\bibitem{Blokland:2005vq}
\hrefCMSnoop {}{I.~Blokland, A.~Czarnecki, M.~{\'S}lusarczyk, and F.~Tkachov,
  ``Next-to-next-to-leading order calculations for heavy-to-light decays'',}
  \textit{ Phys. Rev. D} \textbf{ 71} (2005) 054004,
  \href{http://dx.doi.org/10.1103/PhysRevD.71.054004}{\doi{10.1103/PhysRevD.71.054004}},
  \href{http://www.arXiv.org/abs/hep-ph/0503039}{\texttt{arXiv:hep-ph/0503039}}.
  [Erratum: \DOI{10.1103/PhysRevD.79.019901}].

\bibitem{Chen:2022wit}
\hrefCMSnoop {}{L.-B. Chen, H.~T. Li, J.~Wang, and Y.~Wang, ``Analytic result
  for the top-quark width at next-to-next-to-leading order in {QCD}'',}
  \textit{ Phys. Rev. D} \textbf{ 108} (2023) 054003,
  \href{http://dx.doi.org/10.1103/PhysRevD.108.054003}{\doi{10.1103/PhysRevD.108.054003}},
  \href{http://www.arXiv.org/abs/2212.06341}{\texttt{arXiv:2212.06341}}.

\bibitem{Bigi:1986jk}
I.~Bigi\hrefCMSnoop {}{ { et~al.}, ``Production and decay properties of
  ultra-heavy quarks'',} \textit{ Phys. Lett. B} \textbf{ 181} (1986) 157,
  \href{http://dx.doi.org/10.1016/0370-2693(86)91275-X}{\doi{10.1016/0370-2693(86)91275-X}}.

\bibitem{Chatrchyan:2008zzk}
\hrefCMSnoop {}{{CMS Collaboration}, ``The {CMS} experiment at the {CERN}
  {LHC}'',} \textit{ JINST} \textbf{ 3} (2008) S08004,
  \href{http://dx.doi.org/10.1088/1748-0221/3/08/S08004}{\doi{10.1088/1748-0221/3/08/S08004}}.

\bibitem{CMS:2017yfk}
\hrefCMSnoop {}{{CMS Collaboration}, ``Particle-flow reconstruction and global
  event description with the {CMS} detector'',} \textit{ JINST} \textbf{ 12}
  (2017) P10003,
  \href{http://dx.doi.org/10.1088/1748-0221/12/10/P10003}{\doi{10.1088/1748-0221/12/10/P10003}},
  \href{http://www.arXiv.org/abs/1706.04965}{\texttt{arXiv:1706.04965}}.

\bibitem{CMS-TDR-15-02}
\hrefCMSnoop {}{{CMS Collaboration}, ``Technical proposal for the {Phase-II}
  upgrade of the {Compact Muon Solenoid}'',} CMS Technical Proposal
  CERN-LHCC-2015-010, CMS-TDR-15-02, 2015.
\newblock
  \href{http://dx.doi.org/10.17181/CERN.VU8I.D59J}{\doi{10.17181/CERN.VU8I.D59J}}.

\bibitem{CMS-PAS-JME-14-001}
\href {https://cds.cern.ch/record/1751454}{{CMS Collaboration}, ``Pileup
  removal algorithms'',} CMS Physics Analysis Summary CMS-PAS-JME-14-001, 2014.

\bibitem{CMS:2020ebo}
\hrefCMSnoop {}{{CMS Collaboration}, ``Pileup mitigation at {CMS} in {13\TeV}
  data'',} \textit{ JINST} \textbf{ 15} (2020) P09018,
  \href{http://dx.doi.org/10.1088/1748-0221/15/09/P09018}{\doi{10.1088/1748-0221/15/09/P09018}},
  \href{http://www.arXiv.org/abs/2003.00503}{\texttt{arXiv:2003.00503}}.

\bibitem{Bertolini:2014bba}
\hrefCMSnoop {}{D.~Bertolini, P.~Harris, M.~Low, and N.~Tran, ``Pileup per
  particle identification'',} \textit{ JHEP} \textbf{ 10} (2014) 059,
  \href{http://dx.doi.org/10.1007/JHEP10(2014)059}{\doi{10.1007/JHEP10(2014)059}},
  \href{http://www.arXiv.org/abs/1407.6013}{\texttt{arXiv:1407.6013}}.

\bibitem{Cacciari:2008gp}
\hrefCMSnoop {}{M.~Cacciari, G.~P. Salam, and G.~Soyez, ``The anti-\kt jet
  clustering algorithm'',} \textit{ JHEP} \textbf{ 04} (2008) 063,
  \href{http://dx.doi.org/10.1088/1126-6708/2008/04/063}{\doi{10.1088/1126-6708/2008/04/063}},
  \href{http://www.arXiv.org/abs/0802.1189}{\texttt{arXiv:0802.1189}}.

\bibitem{Cacciari:2011ma}
\hrefCMSnoop {}{M.~Cacciari, G.~P. Salam, and G.~Soyez, ``{\FASTJET} user
  manual'',} \textit{ Eur. Phys. J. C} \textbf{ 72} (2012) 1896,
  \href{http://dx.doi.org/10.1140/epjc/s10052-012-1896-2}{\doi{10.1140/epjc/s10052-012-1896-2}},
  \href{http://www.arXiv.org/abs/1111.6097}{\texttt{arXiv:1111.6097}}.

\bibitem{CMS:2019ctu}
\hrefCMSnoop {}{{CMS Collaboration}, ``Performance of missing transverse
  momentum reconstruction in proton-proton collisions at $\sqrt{s}={13\TeV}$
  using the {CMS} detector'',} \textit{ JINST} \textbf{ 14} (2019) P07004,
  \href{http://dx.doi.org/10.1088/1748-0221/14/07/P07004}{\doi{10.1088/1748-0221/14/07/P07004}},
  \href{http://www.arXiv.org/abs/1903.06078}{\texttt{arXiv:1903.06078}}.

\bibitem{CMS:2016lmd}
\hrefCMSnoop {}{{CMS Collaboration}, ``Jet energy scale and resolution in the
  {CMS} experiment in ${\Pp\Pp}$ collisions at {8\TeV}'',} \textit{ JINST}
  \textbf{ 12} (2017) P02014,
  \href{http://dx.doi.org/10.1088/1748-0221/12/02/P02014}{\doi{10.1088/1748-0221/12/02/P02014}},
  \href{http://www.arXiv.org/abs/1607.03663}{\texttt{arXiv:1607.03663}}.

\bibitem{CMS:2017wtu}
\hrefCMSnoop {}{{CMS Collaboration}, ``Identification of heavy-flavour jets
  with the {CMS} detector in ${\Pp\Pp}$ collisions at {13\TeV}'',} \textit{
  JINST} \textbf{ 13} (2018) P05011,
  \href{http://dx.doi.org/10.1088/1748-0221/13/05/P05011}{\doi{10.1088/1748-0221/13/05/P05011}},
  \href{http://www.arXiv.org/abs/1712.07158}{\texttt{arXiv:1712.07158}}.

\bibitem{Abbott:1998dc}
\hrefCMSnoop {}{{\DZERO} Collaboration, ``Direct measurement of the top quark
  mass at {\DZERO}'',} \textit{ Phys. Rev. D} \textbf{ 58} (1998) 052001,
  \href{http://dx.doi.org/10.1103/PhysRevD.58.052001}{\doi{10.1103/PhysRevD.58.052001}},
  \href{http://www.arXiv.org/abs/hep-ex/9801025}{\texttt{arXiv:hep-ex/9801025}}.

\bibitem{CMS-NOTE-2006-023}
J.~D'Hondt\href {https://cds.cern.ch/record/926540}{ { et~al.}, ``Fitting of
  event topologies with external kinematic constraints in {CMS}'',} CMS Note
  CMS-NOTE-2006-023, 2006.

\bibitem{Sonnenschein:2006ud}
\hrefCMSnoop {}{L.~Sonnenschein, ``Analytical solution of \ttbar dilepton
  equations'',} \textit{ Phys. Rev. D} \textbf{ 73} (2006) 054015,
  \href{http://dx.doi.org/10.1103/PhysRevD.73.054015}{\doi{10.1103/PhysRevD.73.054015}},
  \href{http://www.arXiv.org/abs/hep-ph/0603011}{\texttt{arXiv:hep-ph/0603011}}.
  [Erratum: \DOI{10.1103/PhysRevD.78.079902}].

\bibitem{Frixione:2007nw}
\hrefCMSnoop {}{S.~Frixione, G.~Ridolfi, and P.~Nason, ``A positive-weight
  next-to-leading-order {Monte Carlo} for heavy flavour hadroproduction'',}
  \textit{ JHEP} \textbf{ 09} (2007) 126,
  \href{http://dx.doi.org/10.1088/1126-6708/2007/09/126}{\doi{10.1088/1126-6708/2007/09/126}},
  \href{http://www.arXiv.org/abs/0707.3088}{\texttt{arXiv:0707.3088}}.

\bibitem{Nason:2004rx}
\hrefCMSnoop {}{P.~Nason, ``A new method for combining {NLO} {QCD} with shower
  {Monte Carlo} algorithms'',} \textit{ JHEP} \textbf{ 11} (2004) 040,
  \href{http://dx.doi.org/10.1088/1126-6708/2004/11/040}{\doi{10.1088/1126-6708/2004/11/040}},
  \href{http://www.arXiv.org/abs/hep-ph/0409146}{\texttt{arXiv:hep-ph/0409146}}.

\bibitem{Alioli:2010xd}
\hrefCMSnoop {}{S.~Alioli, P.~Nason, C.~Oleari, and E.~Re, ``A general
  framework for implementing {NLO} calculations in shower {Monte Carlo}
  programs: the {\POWHEG} \textsc{box}'',} \textit{ JHEP} \textbf{ 06} (2010)
  043,
  \href{http://dx.doi.org/10.1007/JHEP06(2010)043}{\doi{10.1007/JHEP06(2010)043}},
  \href{http://www.arXiv.org/abs/1002.2581}{\texttt{arXiv:1002.2581}}.

\bibitem{Frixione:2007vw}
\hrefCMSnoop {}{S.~Frixione, P.~Nason, and C.~Oleari, ``Matching {NLO} {QCD}
  computations with parton shower simulations: the {\POWHEG} method'',}
  \textit{ JHEP} \textbf{ 11} (2007) 070,
  \href{http://dx.doi.org/10.1088/1126-6708/2007/11/070}{\doi{10.1088/1126-6708/2007/11/070}},
  \href{http://www.arXiv.org/abs/0709.2092}{\texttt{arXiv:0709.2092}}.

\bibitem{Sjostrand:2014zea}
T.~Sj{\"o}strand\hrefCMSnoop {}{ { et~al.}, ``An introduction to
  {\PYTHIA8.2}'',} \textit{ Comput. Phys. Commun.} \textbf{ 191} (2015) 159,
  \href{http://dx.doi.org/10.1016/j.cpc.2015.01.024}{\doi{10.1016/j.cpc.2015.01.024}},
  \href{http://www.arXiv.org/abs/1410.3012}{\texttt{arXiv:1410.3012}}.

\bibitem{Alwall:2011uj}
J.~Alwall\hrefCMSnoop {}{ { et~al.}, ``${\MADGRAPH5}$: going beyond'',}
  \textit{ JHEP} \textbf{ 06} (2011) 128,
  \href{http://dx.doi.org/10.1007/JHEP06(2011)128}{\doi{10.1007/JHEP06(2011)128}},
  \href{http://www.arXiv.org/abs/1106.0522}{\texttt{arXiv:1106.0522}}.

\bibitem{Gribov:1972ri}
\hrefCMSnoop {}{V.~N. Gribov and L.~N. Lipatov, ``Deep inelastic ${\Pe\Pp}$
  scattering in perturbation theory'',} \textit{ Sov. J. Nucl. Phys.} \textbf{
  15} (1972) 438.

\bibitem{Gribov:1972rt}
\hrefCMSnoop {}{V.~N. Gribov and L.~N. Lipatov, ``{\EE} pair annihilation and
  deep inelastic ${\Pe\Pp}$ scattering in perturbation theory'',} \textit{ Sov.
  J. Nucl. Phys.} \textbf{ 15} (1972) 675.

\bibitem{Lipatov:1974qm}
\hrefCMSnoop {}{L.~N. Lipatov, ``The parton model and perturbation theory'',}
  \textit{ Yad. Fiz.} \textbf{ 20} (1974) 181.

\bibitem{Dokshitzer:1977sg}
\hrefCMSnoop {}{Y.~L. Dokshitzer, ``Calculation of the structure functions for
  deep inelastic scattering and {\EE} annihilation by perturbation theory in
  quantum chromodynamics'',} \textit{ Sov. Phys. JETP} \textbf{ 46} (1977) 641.

\bibitem{Altarelli:1977zs}
\hrefCMSnoop {}{G.~Altarelli and G.~Parisi, ``Asymptotic freedom in parton
  language'',} \textit{ Nucl. Phys. B} \textbf{ 126} (1977) 298,
  \href{http://dx.doi.org/10.1016/0550-3213(77)90384-4}{\doi{10.1016/0550-3213(77)90384-4}}.

\bibitem{Agostinelli:2002hh}
\hrefCMSnoop {}{{GEANT4} Collaboration, ``{\GEANTfour}---a simulation
  toolkit'',} \textit{ Nucl. Instrum. Meth. A} \textbf{ 506} (2003) 250,
  \href{http://dx.doi.org/10.1016/S0168-9002(03)01368-8}{\doi{10.1016/S0168-9002(03)01368-8}}.

\bibitem{ParticleDataGroup:2007}
\hrefCMSnoop {}{{Particle Data Group}, W.~M. Yao { et~al.}, ``Review of
  particle physics'',} \textit{ J. Phys. G.} \textbf{ 33} (2006) 1,
  \href{http://dx.doi.org/10.1088/0954-3899/33/1/001}{\doi{10.1088/0954-3899/33/1/001}}.

\bibitem{Artoisenet:2012st}
\hrefCMSnoop {}{P.~Artoisenet, R.~Frederix, O.~Mattelaer, and R.~Rietkerk,
  ``Automatic spin-entangled decays of heavy resonances in {Monte Carlo}
  simulations'',} \textit{ JHEP} \textbf{ 03} (2013) 015,
  \href{http://dx.doi.org/10.1007/JHEP03(2013)015}{\doi{10.1007/JHEP03(2013)015}},
  \href{http://www.arXiv.org/abs/1212.3460}{\texttt{arXiv:1212.3460}}.

\bibitem{Sjostrand:2006za}
\hrefCMSnoop {}{T.~Sj{\"o}strand, S.~Mrenna, and P.~Z. Skands, ``{\PYTHIA6.4}
  physics and manual'',} \textit{ JHEP} \textbf{ 05} (2006) 026,
  \href{http://dx.doi.org/10.1088/1126-6708/2006/05/026}{\doi{10.1088/1126-6708/2006/05/026}},
  \href{http://www.arXiv.org/abs/hep-ph/0603175}{\texttt{arXiv:hep-ph/0603175}}.

\bibitem{CMS:2011qzf}
\hrefCMSnoop {}{{CMS Collaboration}, ``Measurement of the underlying event
  activity at the {LHC} with $\sqrt{s}={7\TeV}$ and comparison with
  $\sqrt{s}={0.9\TeV}$'',} \textit{ JHEP} \textbf{ 09} (2011) 109,
  \href{http://dx.doi.org/10.1007/JHEP09(2011)109}{\doi{10.1007/JHEP09(2011)109}},
  \href{http://www.arXiv.org/abs/1107.0330}{\texttt{arXiv:1107.0330}}.

\bibitem{CMS:2015wcf}
\hrefCMSnoop {}{{CMS Collaboration}, ``Event generator tunes obtained from
  underlying event and multiparton scattering measurements'',} \textit{ Eur.
  Phys. J. C} \textbf{ 76} (2016) 155,
  \href{http://dx.doi.org/10.1140/epjc/s10052-016-3988-x}{\doi{10.1140/epjc/s10052-016-3988-x}},
  \href{http://www.arXiv.org/abs/1512.00815}{\texttt{arXiv:1512.00815}}.

\bibitem{Buckley:2009bj}
A.~Buckley\hrefCMSnoop {}{ { et~al.}, ``Systematic event generator tuning for
  the {LHC}'',} \textit{ Eur. Phys. J. C} \textbf{ 65} (2010) 331,
  \href{http://dx.doi.org/10.1140/epjc/s10052-009-1196-7}{\doi{10.1140/epjc/s10052-009-1196-7}},
  \href{http://www.arXiv.org/abs/0907.2973}{\texttt{arXiv:0907.2973}}.

\bibitem{CMS:2016kle}
\href {https://cds.cern.ch/record/2235192}{{CMS Collaboration},
  ``Investigations of the impact of the parton shower tuning in {\PYTHIA8} in
  the modelling of \ttbar at $\sqrt{s}=8$ and {13\TeV}'',} CMS Physics Analysis
  Summary CMS-PAS-TOP-16-021, 2016.

\bibitem{CMS:2019csb}
\hrefCMSnoop {}{{CMS Collaboration}, ``Extraction and validation of a new set
  of {CMS} {\PYTHIA8} tunes from underlying-event measurements'',} \textit{
  Eur. Phys. J. C} \textbf{ 80} (2020) 4,
  \href{http://dx.doi.org/10.1140/epjc/s10052-019-7499-4}{\doi{10.1140/epjc/s10052-019-7499-4}},
  \href{http://www.arXiv.org/abs/1903.12179}{\texttt{arXiv:1903.12179}}.

\bibitem{Giele:2002hx}
\hrefCMSnoop {}{W.~Giele { et~al.}, ``The {QCD}/{SM} working group: Summary
  report'',} in \textit{ {Proc. 2nd Les Houches Workshop on Physics at TeV
  Colliders (PhysTeV 2001): Les Houches, France, May 21--June 01, 2001}},
  p.~275.
\newblock 2002.
\newblock
  \href{http://www.arXiv.org/abs/hep-ph/0204316}{\texttt{arXiv:hep-ph/0204316}}.

\bibitem{Whalley:2005nh}
\hrefCMSnoop {}{M.~R. Whalley, D.~Bourilkov, and R.~C. Group, ``The {Les
  Houches} accord {PDFs} ({LHAPDF}) and {LHAGLUE}'',} in \textit{ {Proc. HERA
  and the LHC: A Workshop on the Implications of HERA for LHC Physics: Geneva,
  Switzerland, October 11--13, 2004; and Hamburg, Germany, January 17--21,
  2005}}, p.~575.
\newblock 2005.
\newblock
  \href{http://www.arXiv.org/abs/hep-ph/0508110}{\texttt{arXiv:hep-ph/0508110}}.
\newblock
  \href{http://dx.doi.org/10.5170/CERN-2005-014.575}{\doi{10.5170/CERN-2005-014.575}}.

\bibitem{Bourilkov:2006cj}
\hrefCMSnoop {}{D.~Bourilkov, R.~C. Group, and M.~R. Whalley, ``{LHAPDF}: {PDF}
  use from the {Tevatron} to the {LHC}'',} in \textit{ {Proc. 4th TeV4LHC
  Workshop: Batavia IL, United States, October 20--22, 2005}}.
\newblock 2006.
\newblock
  \href{http://www.arXiv.org/abs/hep-ph/0605240}{\texttt{arXiv:hep-ph/0605240}}.

\bibitem{Lai:2010vv}
H.-L. Lai\hrefCMSnoop {}{ { et~al.}, ``New parton distributions for collider
  physics'',} \textit{ Phys. Rev. D} \textbf{ 82} (2010) 074024,
  \href{http://dx.doi.org/10.1103/PhysRevD.82.074024}{\doi{10.1103/PhysRevD.82.074024}},
  \href{http://www.arXiv.org/abs/1007.2241}{\texttt{arXiv:1007.2241}}.

\bibitem{Martin:2009iq}
\hrefCMSnoop {}{A.~D. Martin, W.~J. Stirling, R.~S. Thorne, and G.~Watt,
  ``Parton distributions for the {LHC}'',} \textit{ Eur. Phys. J. C} \textbf{
  63} (2009) 189,
  \href{http://dx.doi.org/10.1140/epjc/s10052-009-1072-5}{\doi{10.1140/epjc/s10052-009-1072-5}},
  \href{http://www.arXiv.org/abs/0901.0002}{\texttt{arXiv:0901.0002}}.

\bibitem{Ball:2012cx}
\hrefCMSnoop {}{{NNPDF} Collaboration, ``Parton distributions with {LHC}
  data'',} \textit{ Nucl. Phys. B} \textbf{ 867} (2013) 244,
  \href{http://dx.doi.org/10.1016/j.nuclphysb.2012.10.003}{\doi{10.1016/j.nuclphysb.2012.10.003}},
  \href{http://www.arXiv.org/abs/1207.1303}{\texttt{arXiv:1207.1303}}.

\bibitem{NNPDF:2014otw}
\hrefCMSnoop {}{{NNPDF} Collaboration, ``Parton distributions for the {LHC} run
  {II}'',} \textit{ JHEP} \textbf{ 04} (2015) 040,
  \href{http://dx.doi.org/10.1007/JHEP04(2015)040}{\doi{10.1007/JHEP04(2015)040}},
  \href{http://www.arXiv.org/abs/1410.8849}{\texttt{arXiv:1410.8849}}.

\bibitem{Dulat:2015mca}
S.~Dulat\hrefCMSnoop {}{ { et~al.}, ``New parton distribution functions from a
  global analysis of quantum chromodynamics'',} \textit{ Phys. Rev. D} \textbf{
  93} (2016) 033006,
  \href{http://dx.doi.org/10.1103/PhysRevD.93.033006}{\doi{10.1103/PhysRevD.93.033006}},
  \href{http://www.arXiv.org/abs/1506.07443}{\texttt{arXiv:1506.07443}}.

\bibitem{Harland-Lang:2014zoa}
\hrefCMSnoop {}{L.~A. Harland-Lang, A.~D. Martin, P.~Motylinski, and R.~S.
  Thorne, ``Parton distributions in the {LHC} era: {MMHT} 2014 {PDFs}'',}
  \textit{ Eur. Phys. J. C} \textbf{ 75} (2015) 204,
  \href{http://dx.doi.org/10.1140/epjc/s10052-015-3397-6}{\doi{10.1140/epjc/s10052-015-3397-6}},
  \href{http://www.arXiv.org/abs/1412.3989}{\texttt{arXiv:1412.3989}}.

\bibitem{Cacciari:2003fi}
M.~Cacciari\hrefCMSnoop {}{ { et~al.}, ``The \ttbar cross-section at 1.8 and
  {1.96\TeV}: a study of the systematics due to parton densities and scale
  dependence'',} \textit{ JHEP} \textbf{ 04} (2004) 068,
  \href{http://dx.doi.org/10.1088/1126-6708/2004/04/068}{\doi{10.1088/1126-6708/2004/04/068}},
  \href{http://www.arXiv.org/abs/hep-ph/0303085}{\texttt{arXiv:hep-ph/0303085}}.

\bibitem{Alwall:2007fs}
J.~Alwall\hrefCMSnoop {}{ { et~al.}, ``Comparative study of various algorithms
  for the merging of parton showers and matrix elements in hadronic
  collisions'',} \textit{ Eur. Phys. J. C} \textbf{ 53} (2008) 473,
  \href{http://dx.doi.org/10.1140/epjc/s10052-007-0490-5}{\doi{10.1140/epjc/s10052-007-0490-5}},
  \href{http://www.arXiv.org/abs/0706.2569}{\texttt{arXiv:0706.2569}}.

\bibitem{CMS:2015ilk}
\hrefCMSnoop {}{{CMS Collaboration}, ``Measurement of \ttbar production with
  additional jet activity, including {\PQb} quark jets, in the dilepton decay
  channel using ${\Pp\Pp}$ collisions at $\sqrt{s}={8\TeV}$'',} \textit{ Eur.
  Phys. J. C} \textbf{ 76} (2016) 379,
  \href{http://dx.doi.org/10.1140/epjc/s10052-016-4105-x}{\doi{10.1140/epjc/s10052-016-4105-x}},
  \href{http://www.arXiv.org/abs/1510.03072}{\texttt{arXiv:1510.03072}}.

\bibitem{Mrenna:2016sih}
\hrefCMSnoop {}{S.~Mrenna and P.~Skands, ``Automated parton-shower variations
  in {\PYTHIA8}'',} \textit{ Phys. Rev. D} \textbf{ 94} (2016) 074005,
  \href{http://dx.doi.org/10.1103/PhysRevD.94.074005}{\doi{10.1103/PhysRevD.94.074005}},
  \href{http://www.arXiv.org/abs/1605.08352}{\texttt{arXiv:1605.08352}}.

\bibitem{CMS:2018tdx}
\hrefCMSnoop {}{{CMS Collaboration}, ``Measurements of differential cross
  sections of top quark pair production as a function of kinematic event
  variables in proton-proton collisions at $\sqrt{s}={13\TeV}$'',} \textit{
  JHEP} \textbf{ 06} (2018) 002,
  \href{http://dx.doi.org/10.1007/JHEP06(2018)002}{\doi{10.1007/JHEP06(2018)002}},
  \href{http://www.arXiv.org/abs/1803.03991}{\texttt{arXiv:1803.03991}}.

\bibitem{CMS:2018ypj}
\hrefCMSnoop {}{{CMS Collaboration}, ``Measurement of jet substructure
  observables in \ttbar events from proton-proton collisions at
  $\sqrt{s}={13\TeV}$'',} \textit{ Phys. Rev. D} \textbf{ 98} (2018) 092014,
  \href{http://dx.doi.org/10.1103/PhysRevD.98.092014}{\doi{10.1103/PhysRevD.98.092014}},
  \href{http://www.arXiv.org/abs/1808.07340}{\texttt{arXiv:1808.07340}}.

\bibitem{Frederix:2012ps}
\hrefCMSnoop {}{R.~Frederix and S.~Frixione, ``Merging meets matching in
  {\MCATNLO}'',} \textit{ JHEP} \textbf{ 12} (2012) 061,
  \href{http://dx.doi.org/10.1007/JHEP12(2012)061}{\doi{10.1007/JHEP12(2012)061}},
  \href{http://www.arXiv.org/abs/1209.6215}{\texttt{arXiv:1209.6215}}.

\bibitem{Frixione:2023hwz}
\hrefCMSnoop {}{S.~Frixione, S.~Amoroso, and S.~Mrenna, ``Matrix element
  corrections in the {\PYTHIA8} parton shower in the context of matched
  simulations at next-to-leading order'',} \textit{ Eur. Phys. J. C} \textbf{
  83} (2023) 970,
  \href{http://dx.doi.org/10.1140/epjc/s10052-023-12154-x}{\doi{10.1140/epjc/s10052-023-12154-x}},
  \href{http://www.arXiv.org/abs/2308.06389}{\texttt{arXiv:2308.06389}}.

\bibitem{Jezo:2016ujg}
T.~Je{\v{z}}o\hrefCMSnoop {}{ { et~al.}, ``An {NLO+PS} generator for \ttbar and
  ${\PW\PQt}$ production and decay including non-resonant and interference
  effects'',} \textit{ Eur. Phys. J. C} \textbf{ 76} (2016) 691,
  \href{http://dx.doi.org/10.1140/epjc/s10052-016-4538-2}{\doi{10.1140/epjc/s10052-016-4538-2}},
  \href{http://www.arXiv.org/abs/1607.04538}{\texttt{arXiv:1607.04538}}.

\bibitem{Jezo:2023rht}
\hrefCMSnoop {}{T.~Je{\v{z}}o, J.~M. Lindert, and S.~Pozzorini,
  ``Resonance-aware {NLOPS} matching for off-shell $\ttbar+{\PQt\PW}$
  production with semileptonic decays'',} \textit{ JHEP} \textbf{ 10} (2023)
  008,
  \href{http://dx.doi.org/10.1007/JHEP10(2023)008}{\doi{10.1007/JHEP10(2023)008}},
  \href{http://www.arXiv.org/abs/2307.15653}{\texttt{arXiv:2307.15653}}.

\bibitem{CMS:2018htd}
\hrefCMSnoop {}{{CMS Collaboration}, ``Measurement of differential cross
  sections for the production of top quark pairs and of additional jets in
  lepton+jets events from ${\Pp\Pp}$ collisions at $\sqrt{s}={13\TeV}$'',}
  \textit{ Phys. Rev. D} \textbf{ 97} (2018) 112003,
  \href{http://dx.doi.org/10.1103/PhysRevD.97.112003}{\doi{10.1103/PhysRevD.97.112003}},
  \href{http://www.arXiv.org/abs/1803.08856}{\texttt{arXiv:1803.08856}}.

\bibitem{Bahr:2008pv}
M.~B{\"a}hr\hrefCMSnoop {}{ { et~al.}, ``{\HERWIGpp} physics and manual'',}
  \textit{ Eur. Phys. J. C} \textbf{ 58} (2008) 639,
  \href{http://dx.doi.org/10.1140/epjc/s10052-008-0798-9}{\doi{10.1140/epjc/s10052-008-0798-9}},
  \href{http://www.arXiv.org/abs/0803.0883}{\texttt{arXiv:0803.0883}}.

\bibitem{Gieseke:2012ft}
\hrefCMSnoop {}{S.~Gieseke, C.~R{\"o}hr, and A.~Siodmok, ``Colour reconnections
  in {\HERWIGpp}'',} \textit{ Eur. Phys. J. C} \textbf{ 72} (2012) 2225,
  \href{http://dx.doi.org/10.1140/epjc/s10052-012-2225-5}{\doi{10.1140/epjc/s10052-012-2225-5}},
  \href{http://www.arXiv.org/abs/1206.0041}{\texttt{arXiv:1206.0041}}.

\bibitem{Bellm:2015jjp}
\hrefCMSnoop {}{J.~Bellm { et~al.}, ``{\HERWIG7.0}/{\HERWIGpp3.0} release
  note'',} \textit{ Eur. Phys. J. C} \textbf{ 76} (2016) 196,
  \href{http://dx.doi.org/10.1140/epjc/s10052-016-4018-8}{\doi{10.1140/epjc/s10052-016-4018-8}},
  \href{http://www.arXiv.org/abs/1512.01178}{\texttt{arXiv:1512.01178}}.

\bibitem{CMS:2016oae}
\hrefCMSnoop {}{{CMS Collaboration}, ``Measurement of differential cross
  sections for top quark pair production using the lepton+jets final state in
  proton-proton collisions at {13\TeV}'',} \textit{ Phys. Rev. D} \textbf{ 95}
  (2017) 092001,
  \href{http://dx.doi.org/10.1103/PhysRevD.95.092001}{\doi{10.1103/PhysRevD.95.092001}},
  \href{http://www.arXiv.org/abs/1610.04191}{\texttt{arXiv:1610.04191}}.

\bibitem{Skands:2010ak}
\hrefCMSnoop {}{P.~Z. Skands, ``Tuning {Monte Carlo} generators: The {Perugia}
  tunes'',} \textit{ Phys. Rev. D} \textbf{ 82} (2010) 074018,
  \href{http://dx.doi.org/10.1103/PhysRevD.82.074018}{\doi{10.1103/PhysRevD.82.074018}},
  \href{http://www.arXiv.org/abs/1005.3457}{\texttt{arXiv:1005.3457}}.

\bibitem{NNPDF:2017mvq}
\hrefCMSnoop {}{{NNPDF} Collaboration, ``Parton distributions from
  high-precision collider data'',} \textit{ Eur. Phys. J. C} \textbf{ 77}
  (2017) 663,
  \href{http://dx.doi.org/10.1140/epjc/s10052-017-5199-5}{\doi{10.1140/epjc/s10052-017-5199-5}},
  \href{http://www.arXiv.org/abs/1706.00428}{\texttt{arXiv:1706.00428}}.

\bibitem{CMS:2015zrm}
\hrefCMSnoop {}{{CMS Collaboration}, ``Pseudorapidity distribution of charged
  hadrons in proton-proton collisions at $\sqrt{s}={13\TeV}$'',} \textit{ Phys.
  Lett. B} \textbf{ 751} (2015) 143,
  \href{http://dx.doi.org/10.1016/j.physletb.2015.10.004}{\doi{10.1016/j.physletb.2015.10.004}},
  \href{http://www.arXiv.org/abs/1507.05915}{\texttt{arXiv:1507.05915}}.

\bibitem{CMS:2015zev}
\href {https://cds.cern.ch/record/2104473}{{CMS Collaboration}, ``Underlying
  event measurements with leading particles and jets in ${\Pp\Pp}$ collisions
  at $\sqrt{s}={13\TeV}$'',} CMS Physics Analysis Summary CMS-PAS-FSQ-15-007,
  2015.

\bibitem{CMS:2018mdd}
\hrefCMSnoop {}{{CMS Collaboration}, ``Study of the underlying event in top
  quark pair production in ${\Pp\Pp}$ collisions at {13\TeV}'',} \textit{ Eur.
  Phys. J. C} \textbf{ 79} (2019) 123,
  \href{http://dx.doi.org/10.1140/epjc/s10052-019-6620-z}{\doi{10.1140/epjc/s10052-019-6620-z}},
  \href{http://www.arXiv.org/abs/1807.02810}{\texttt{arXiv:1807.02810}}.

\bibitem{Seymour:2013qka}
\hrefCMSnoop {}{M.~H. Seymour and A.~Si{\'o}dmok, ``Constraining {MPI} models
  using $\sigma_{\text{eff}}$ and recent {Tevatron} and {LHC} underlying event
  data'',} \textit{ JHEP} \textbf{ 10} (2013) 113,
  \href{http://dx.doi.org/10.1007/JHEP10(2013)113}{\doi{10.1007/JHEP10(2013)113}},
  \href{http://www.arXiv.org/abs/1307.5015}{\texttt{arXiv:1307.5015}}.

\bibitem{Pumplin:2002vw}
J.~Pumplin\hrefCMSnoop {}{ { et~al.}, ``New generation of parton distributions
  with uncertainties from global {QCD} analysis'',} \textit{ JHEP} \textbf{ 07}
  (2002) 012,
  \href{http://dx.doi.org/10.1088/1126-6708/2002/07/012}{\doi{10.1088/1126-6708/2002/07/012}},
  \href{http://www.arXiv.org/abs/hep-ph/0201195}{\texttt{arXiv:hep-ph/0201195}}.

\bibitem{Gleisberg:2008ta}
T.~Gleisberg\hrefCMSnoop {}{ { et~al.}, ``Event generation with
  {\SHERPA1.1}'',} \textit{ JHEP} \textbf{ 02} (2009) 007,
  \href{http://dx.doi.org/10.1088/1126-6708/2009/02/007}{\doi{10.1088/1126-6708/2009/02/007}},
  \href{http://www.arXiv.org/abs/0811.4622}{\texttt{arXiv:0811.4622}}.

\bibitem{Cascioli:2011va}
\hrefCMSnoop {}{F.~Cascioli, P.~Maierh{\"o}fer, and S.~Pozzorini, ``Scattering
  amplitudes with open loops'',} \textit{ Phys. Rev. Lett.} \textbf{ 108}
  (2012) 111601,
  \href{http://dx.doi.org/10.1103/PhysRevLett.108.111601}{\doi{10.1103/PhysRevLett.108.111601}},
  \href{http://www.arXiv.org/abs/1111.5206}{\texttt{arXiv:1111.5206}}.

\bibitem{Schumann:2007mg}
\hrefCMSnoop {}{S.~Schumann and F.~Krauss, ``A parton shower algorithm based on
  {Catani}--{Seymour} dipole factorisation'',} \textit{ JHEP} \textbf{ 03}
  (2008) 038,
  \href{http://dx.doi.org/10.1088/1126-6708/2008/03/038}{\doi{10.1088/1126-6708/2008/03/038}},
  \href{http://www.arXiv.org/abs/0709.1027}{\texttt{arXiv:0709.1027}}.

\bibitem{CMS:2022awf}
\hrefCMSnoop {}{{CMS Collaboration}, ``{CMS} {\PYTHIA8} colour reconnection
  tunes based on underlying-event data'',} \textit{ Eur. Phys. J. C} \textbf{
  83} (2023) 587,
  \href{http://dx.doi.org/10.1140/epjc/s10052-023-11630-8}{\doi{10.1140/epjc/s10052-023-11630-8}},
  \href{http://www.arXiv.org/abs/2205.02905}{\texttt{arXiv:2205.02905}}.

\bibitem{Argyropoulos:2014zoa}
\hrefCMSnoop {}{S.~Argyropoulos and T.~Sj{\"o}strand, ``Effects of color
  reconnection on \ttbar final states at the {LHC}'',} \textit{ JHEP} \textbf{
  11} (2014) 043,
  \href{http://dx.doi.org/10.1007/JHEP11(2014)043}{\doi{10.1007/JHEP11(2014)043}},
  \href{http://www.arXiv.org/abs/1407.6653}{\texttt{arXiv:1407.6653}}.

\bibitem{ATLAS:2015ytt}
\hrefCMSnoop {}{{ATLAS Collaboration}, ``Measurement of colour flow with the
  jet pull angle in \ttbar events using the {ATLAS} detector at
  $\sqrt{s}={8\TeV}$'',} \textit{ Phys. Lett. B} \textbf{ 750} (2015) 475,
  \href{http://dx.doi.org/10.1016/j.physletb.2015.09.051}{\doi{10.1016/j.physletb.2015.09.051}},
  \href{http://www.arXiv.org/abs/1506.05629}{\texttt{arXiv:1506.05629}}.

\bibitem{Bowler:1981sb}
\hrefCMSnoop {}{M.~G. Bowler, ``{\EE} production of heavy quarks in the string
  model'',} \textit{ Z. Phys. C} \textbf{ 11} (1981) 169,
  \href{http://dx.doi.org/10.1007/BF01574001}{\doi{10.1007/BF01574001}}.

\bibitem{Heister:2001jg}
\hrefCMSnoop {}{{ALEPH} Collaboration, ``Study of the fragmentation of {\PQb}
  quarks into {\PB} mesons at the {\PZ} peak'',} \textit{ Phys. Lett. B}
  \textbf{ 512} (2001) 30,
  \href{http://dx.doi.org/10.1016/S0370-2693(01)00690-6}{\doi{10.1016/S0370-2693(01)00690-6}},
  \href{http://www.arXiv.org/abs/hep-ex/0106051}{\texttt{arXiv:hep-ex/0106051}}.

\bibitem{DELPHI:2011aa}
\hrefCMSnoop {}{{DELPHI} Collaboration, ``A study of the {\PQb}-quark
  fragmentation function with the {DELPHI} detector at \mbox{LEP I} and an
  averaged distribution obtained at the {\PZ} pole'',} \textit{ Eur. Phys. J.
  C} \textbf{ 71} (2011) 1557,
  \href{http://dx.doi.org/10.1140/epjc/s10052-011-1557-x}{\doi{10.1140/epjc/s10052-011-1557-x}},
  \href{http://www.arXiv.org/abs/1102.4748}{\texttt{arXiv:1102.4748}}.

\bibitem{OPAL:2002plk}
\hrefCMSnoop {}{{OPAL} Collaboration, ``Inclusive analysis of the {\PQb} quark
  fragmentation function in {\PZ} decays at {LEP}'',} \textit{ Eur. Phys. J. C}
  \textbf{ 29} (2003) 463,
  \href{http://dx.doi.org/10.1140/epjc/s2003-01229-x}{\doi{10.1140/epjc/s2003-01229-x}},
  \href{http://www.arXiv.org/abs/hep-ex/0210031}{\texttt{arXiv:hep-ex/0210031}}.

\bibitem{SLD:2002poq}
\hrefCMSnoop {}{{SLD} Collaboration, ``Measurement of the {\PQb}-quark
  fragmentation function in {\PZz} decays'',} \textit{ Phys. Rev. D} \textbf{
  65} (2002) 092006,
  \href{http://dx.doi.org/10.1103/PhysRevD.65.092006}{\doi{10.1103/PhysRevD.65.092006}},
  \href{http://www.arXiv.org/abs/hep-ex/0202031}{\texttt{arXiv:hep-ex/0202031}}.
  [Erratum: \DOI{10.1103/PhysRevD.66.079905}].

\bibitem{Peterson:1982ak}
\hrefCMSnoop {}{C.~Peterson, D.~Schlatter, I.~Schmitt, and P.~M. Zerwas,
  ``Scaling violations in inclusive {\EE} annihilation spectra'',} \textit{
  Phys. Rev. D} \textbf{ 27} (1983) 105,
  \href{http://dx.doi.org/10.1103/PhysRevD.27.105}{\doi{10.1103/PhysRevD.27.105}}.

\bibitem{Khachatryan:2014gga}
\hrefCMSnoop {}{{CMS Collaboration}, ``Performance of the {CMS} missing
  transverse momentum reconstruction in ${\Pp\Pp}$ data at
  $\sqrt{s}={8\TeV}$'',} \textit{ JINST} \textbf{ 10} (2015) P02006,
  \href{http://dx.doi.org/10.1088/1748-0221/10/02/P02006}{\doi{10.1088/1748-0221/10/02/P02006}},
  \href{http://www.arXiv.org/abs/1411.0511}{\texttt{arXiv:1411.0511}}.

\bibitem{CMS:2020uim}
\hrefCMSnoop {}{{CMS Collaboration}, ``Electron and photon reconstruction and
  identification with the {CMS} experiment at the {CERN} {LHC}'',} \textit{
  JINST} \textbf{ 16} (2021) P05014,
  \href{http://dx.doi.org/10.1088/1748-0221/16/05/P05014}{\doi{10.1088/1748-0221/16/05/P05014}},
  \href{http://www.arXiv.org/abs/2012.06888}{\texttt{arXiv:2012.06888}}.

\bibitem{CMS:2018rym}
\hrefCMSnoop {}{{CMS Collaboration}, ``Performance of the {CMS} muon detector
  and muon reconstruction with proton-proton collisions at
  $\sqrt{s}={13\TeV}$'',} \textit{ JINST} \textbf{ 13} (2018) P06015,
  \href{http://dx.doi.org/10.1088/1748-0221/13/06/P06015}{\doi{10.1088/1748-0221/13/06/P06015}},
  \href{http://www.arXiv.org/abs/1804.04528}{\texttt{arXiv:1804.04528}}.

\bibitem{CMS:2016ngn}
\hrefCMSnoop {}{{CMS Collaboration}, ``The {CMS} trigger system'',} \textit{
  JINST} \textbf{ 12} (2017) P01020,
  \href{http://dx.doi.org/10.1088/1748-0221/12/01/P01020}{\doi{10.1088/1748-0221/12/01/P01020}},
  \href{http://www.arXiv.org/abs/1609.02366}{\texttt{arXiv:1609.02366}}.

\bibitem{CMS-PAS-SMP-12-008}
\href {https://cds.cern.ch/record/1434360}{{CMS Collaboration}, ``Absolute
  calibration of the luminosity measurement at {CMS}: Winter 2012 update'',}
  CMS Physics Analysis Summary CMS-PAS-SMP-12-008, 2012.

\bibitem{CMS-PAS-LUM-13-001}
\href {https://cds.cern.ch/record/1598864}{{CMS Collaboration}, ``{CMS}
  luminosity based on pixel cluster counting---{Summer} 2013 update'',} CMS
  Physics Analysis Summary CMS-PAS-LUM-13-001, 2013.

\bibitem{CMS:2021xjt}
\hrefCMSnoop {}{{CMS Collaboration}, ``Precision luminosity measurement in
  proton-proton collisions at $\sqrt{s}={13\TeV}$ in 2015 and 2016 at {CMS}'',}
  \textit{ Eur. Phys. J. C} \textbf{ 81} (2021) 800,
  \href{http://dx.doi.org/10.1140/epjc/s10052-021-09538-2}{\doi{10.1140/epjc/s10052-021-09538-2}},
  \href{http://www.arXiv.org/abs/2104.01927}{\texttt{arXiv:2104.01927}}.

\bibitem{CMS-PAS-LUM-17-004}
\href {https://cds.cern.ch/record/2621960}{{CMS Collaboration}, ``{CMS}
  luminosity measurement for the 2017 data-taking period at
  $\sqrt{s}={13\TeV}$'',} CMS Physics Analysis Summary CMS-PAS-LUM-17-004,
  2018.

\bibitem{CMS-PAS-LUM-18-002}
\href {https://cds.cern.ch/record/2676164}{{CMS Collaboration}, ``{CMS}
  luminosity measurement for the 2018 data-taking period at
  $\sqrt{s}={13\TeV}$'',} CMS Physics Analysis Summary CMS-PAS-LUM-18-002,
  2019.

\bibitem{Schmitt:2012kp}
\hrefCMSnoop {}{S.~Schmitt, ``\textsc{TUnfold}, an algorithm for correcting
  migration effects in high energy physics'',} \textit{ JINST} \textbf{ 7}
  (2012) T10003,
  \href{http://dx.doi.org/10.1088/1748-0221/7/10/T10003}{\doi{10.1088/1748-0221/7/10/T10003}},
  \href{http://www.arXiv.org/abs/1205.6201}{\texttt{arXiv:1205.6201}}.

\bibitem{Schmitt:2016orm}
\hrefCMSnoop {}{S.~Schmitt, ``Data unfolding methods in high energy physics'',}
  in \textit{ {Proc. 12th Conference on Quark Confinement and the Hadron
  Spectrum (Confinement XII): Thessaloniki, Greece, August 28--September 04,
  2016}}.
\newblock 2017.
\newblock
  \href{http://www.arXiv.org/abs/1611.01927}{\texttt{arXiv:1611.01927}}.
\newblock [EPJ Web Conf. 137 (2017) 11008].
  \href{http://dx.doi.org/10.1051/epjconf/201713711008}{\doi{10.1051/epjconf/201713711008}}.

\bibitem{DAGOSTINI1995487}
\hrefCMSnoop {}{G.~D'Agostini, ``A multidimensional unfolding method based on
  {Bayes}' theorem'',} \textit{ Nucl. Instrum. Meth. A} \textbf{ 362} (1995)
  487,
  \href{http://dx.doi.org/10.1016/0168-9002(95)00274-X}{\doi{10.1016/0168-9002(95)00274-X}}.

\bibitem{dagostini2010improved}
\hrefCMSnoop {}{G.~D'Agostini, ``Improved iterative {Bayesian} unfolding'',} in
  \textit{ {Proc. Alliance Workshop on Unfolding and Data Correction: Hamburg,
  Germany, May 27--28, 2010}}.
\newblock 2010.
\newblock \href{http://www.arXiv.org/abs/1010.0632}{\texttt{arXiv:1010.0632}}.

\bibitem{Hocker:1995kb}
\hrefCMSnoop {}{A.~H{\"o}cker and V.~Kartvelishvili, ``{SVD} approach to data
  unfolding'',} \textit{ Nucl. Instrum. Meth. A} \textbf{ 372} (1996) 469,
  \href{http://dx.doi.org/10.1016/0168-9002(95)01478-0}{\doi{10.1016/0168-9002(95)01478-0}},
  \href{http://www.arXiv.org/abs/hep-ph/9509307}{\texttt{arXiv:hep-ph/9509307}}.

\bibitem{TIKHONOV}
\href {https://www.mathnet.ru/eng/dan/v151/i3/p501}{A.~N. Tikhonov, ``On the
  solution of ill-posed problems and the method of regularization'',} \textit{
  Dokl. Akad. Nauk SSSR} \textbf{ 151} (1963) 3.

\bibitem{Brenner:2019lmf}
L.~Brenner\hrefCMSnoop {}{ { et~al.}, ``Comparison of unfolding methods using
  \textsc{RooFitUnfold}'',} \textit{ Int. J. Mod. Phys. A} \textbf{ 35} (2020)
  2050145,
  \href{http://dx.doi.org/10.1142/S0217751X20501456}{\doi{10.1142/S0217751X20501456}},
  \href{http://www.arXiv.org/abs/1910.14654}{\texttt{arXiv:1910.14654}}.

\bibitem{Stanley:2021xce}
\hrefCMSnoop {}{M.~Stanley, P.~Patil, and M.~Kuusela, ``Uncertainty
  quantification for wide-bin unfolding: one-at-a-time strict bounds and
  prior-optimized confidence intervals'',} \textit{ JINST} \textbf{ 17} (2022)
  P10013,
  \href{http://dx.doi.org/10.1088/1748-0221/17/10/P10013}{\doi{10.1088/1748-0221/17/10/P10013}},
  \href{http://www.arXiv.org/abs/2111.01091}{\texttt{arXiv:2111.01091}}.

\bibitem{Collaboration:2267573}
\href {https://cds.cern.ch/record/2267573}{{CMS Collaboration}, ``Object
  definitions for top quark analyses at the particle level'',} CMS Note
  CMS-NOTE-2017-004, 2017.

\bibitem{Buckley:2010ar}
A.~Buckley\hrefCMSnoop {}{ { et~al.}, ``\textsc{rivet} user manual'',} \textit{
  Comput. Phys. Commun.} \textbf{ 184} (2013) 2803,
  \href{http://dx.doi.org/10.1016/j.cpc.2013.05.021}{\doi{10.1016/j.cpc.2013.05.021}},
  \href{http://www.arXiv.org/abs/1003.0694}{\texttt{arXiv:1003.0694}}.

\bibitem{Bierlich:2019rhm}
C.~Bierlich\hrefCMSnoop {}{ { et~al.}, ``Robust independent validation of
  experiment and theory: \textsc{rivet} version 3'',} \textit{ SciPost Phys.}
  \textbf{ 8} (2020) 026,
  \href{http://dx.doi.org/10.21468/SciPostPhys.8.2.026}{\doi{10.21468/SciPostPhys.8.2.026}},
  \href{http://www.arXiv.org/abs/1912.05451}{\texttt{arXiv:1912.05451}}.

\bibitem{Herren:2017osy}
\hrefCMSnoop {}{F.~Herren and M.~Steinhauser, ``Version 3 of \textsc{RunDec}
  and \textsc{CRunDec}'',} \textit{ Comput. Phys. Commun.} \textbf{ 224} (2018)
  333,
  \href{http://dx.doi.org/10.1016/j.cpc.2017.11.014}{\doi{10.1016/j.cpc.2017.11.014}},
  \href{http://www.arXiv.org/abs/1703.03751}{\texttt{arXiv:1703.03751}}.

\bibitem{Hoang:2021fhn}
\hrefCMSnoop {}{A.~H. Hoang, C.~Lepenik, and V.~Mateu, ``\textsc{REvolver}:
  Automated running and matching of couplings and masses in {QCD}'',} \textit{
  Comput. Phys. Commun.} \textbf{ 270} (2022) 108145,
  \href{http://dx.doi.org/10.1016/j.cpc.2021.108145}{\doi{10.1016/j.cpc.2021.108145}},
  \href{http://www.arXiv.org/abs/2102.01085}{\texttt{arXiv:2102.01085}}.

\bibitem{Mazzitelli:2022scc}
\hrefCMSnoop {}{J.~Mazzitelli, ``{NNLO} study of top-quark mass renormalization
  scheme uncertainties in {Higgs} boson production'',} \textit{ JHEP} \textbf{
  09} (2022) 065,
  \href{http://dx.doi.org/10.1007/JHEP09(2022)065}{\doi{10.1007/JHEP09(2022)065}},
  \href{http://www.arXiv.org/abs/2206.14667}{\texttt{arXiv:2206.14667}}.

\bibitem{Hoang:2017suc}
A.~H. Hoang\hrefCMSnoop {}{ { et~al.}, ``The {MSR} mass and the
  $\mathcal{O}({\Lambda}_{{\mathrm{QCD}}})$ renormalon sum rule'',} \textit{
  JHEP} \textbf{ 04} (2018) 003,
  \href{http://dx.doi.org/10.1007/JHEP04(2018)003}{\doi{10.1007/JHEP04(2018)003}},
  \href{http://www.arXiv.org/abs/1704.01580}{\texttt{arXiv:1704.01580}}.

\bibitem{Tarrach:1980up}
\hrefCMSnoop {}{R.~Tarrach, ``The pole mass in perturbative {QCD}'',} \textit{
  Nucl. Phys. B} \textbf{ 183} (1981) 384,
  \href{http://dx.doi.org/10.1016/0550-3213(81)90140-1}{\doi{10.1016/0550-3213(81)90140-1}}.

\bibitem{Kronfeld:1998di}
\hrefCMSnoop {}{A.~S. Kronfeld, ``Perturbative pole mass in {QCD}'',} \textit{
  Phys. Rev. D} \textbf{ 58} (1998) 051501,
  \href{http://dx.doi.org/10.1103/PhysRevD.58.051501}{\doi{10.1103/PhysRevD.58.051501}},
  \href{http://www.arXiv.org/abs/hep-ph/9805215}{\texttt{arXiv:hep-ph/9805215}}.

\bibitem{Beneke:2016cbu}
\hrefCMSnoop {}{M.~Beneke, P.~Marquard, P.~Nason, and M.~Steinhauser, ``On the
  ultimate uncertainty of the top quark pole mass'',} \textit{ Phys. Lett. B}
  \textbf{ 775} (2017) 63,
  \href{http://dx.doi.org/10.1016/j.physletb.2017.10.054}{\doi{10.1016/j.physletb.2017.10.054}},
  \href{http://www.arXiv.org/abs/1605.03609}{\texttt{arXiv:1605.03609}}.

\bibitem{Hoang:2017btd}
\hrefCMSnoop {}{A.~H. Hoang, C.~Lepenik, and M.~Preisser, ``On the light
  massive flavor dependence of the large order asymptotic behavior and the
  ambiguity of the pole mass'',} \textit{ JHEP} \textbf{ 09} (2017) 099,
  \href{http://dx.doi.org/10.1007/JHEP09(2017)099}{\doi{10.1007/JHEP09(2017)099}},
  \href{http://www.arXiv.org/abs/1706.08526}{\texttt{arXiv:1706.08526}}.

\bibitem{Hoang:2017kmk}
\hrefCMSnoop {}{A.~H. Hoang, S.~Mantry, A.~Pathak, and I.~W. Stewart,
  ``Extracting a short distance top mass with light grooming'',} \textit{ Phys.
  Rev. D} \textbf{ 100} (2019) 074021,
  \href{http://dx.doi.org/10.1103/PhysRevD.100.074021}{\doi{10.1103/PhysRevD.100.074021}},
  \href{http://www.arXiv.org/abs/1708.02586}{\texttt{arXiv:1708.02586}}.

\bibitem{Bachu:2020nqn}
B.~Bachu\hrefCMSnoop {}{ { et~al.}, ``Boosted top quarks in the peak region
  with {N\textsuperscript{3}LL} resummation'',} \textit{ Phys. Rev. D} \textbf{
  104} (2021) 014026,
  \href{http://dx.doi.org/10.1103/PhysRevD.104.014026}{\doi{10.1103/PhysRevD.104.014026}},
  \href{http://www.arXiv.org/abs/2012.12304}{\texttt{arXiv:2012.12304}}.

\bibitem{Abdallah:2008xh}
\hrefCMSnoop {}{{DELPHI} Collaboration, ``Measurement of the mass and width of
  the {\PW} boson in {\EE} collisions at $\sqrt{s}=161$--{209\GeV}'',} \textit{
  Eur. Phys. J. C} \textbf{ 55} (2008) 1,
  \href{http://dx.doi.org/10.1140/epjc/s10052-008-0585-7}{\doi{10.1140/epjc/s10052-008-0585-7}},
  \href{http://www.arXiv.org/abs/0803.2534}{\texttt{arXiv:0803.2534}}.

\bibitem{Barlow:1993dm}
\hrefCMSnoop {}{R.~Barlow and C.~Beeston, ``Fitting using finite {Monte Carlo}
  samples'',} \textit{ Comput. Phys. Commun.} \textbf{ 77} (1993) 219,
  \href{http://dx.doi.org/10.1016/0010-4655(93)90005-W}{\doi{10.1016/0010-4655(93)90005-W}}.

\bibitem{Conway:2011in}
\hrefCMSnoop {}{J.~S. Conway, ``Incorporating nuisance parameters in
  likelihoods for multisource spectra'',} in \textit{ {Proc. 2011 Workshop on
  Statistical Issues Related to Discovery Claims in Search Experiments and
  Unfolding (PHYSTAT 2011): Geneva, Switzerland, January 17--20, 2011}}.
\newblock 2011.
\newblock \href{http://www.arXiv.org/abs/1103.0354}{\texttt{arXiv:1103.0354}}.
\newblock
  \href{http://dx.doi.org/10.5170/CERN-2011-006.115}{\doi{10.5170/CERN-2011-006.115}}.

\bibitem{Aaboud:2016igd}
\hrefCMSnoop {}{{ATLAS Collaboration}, ``Measurement of the top quark mass in
  the $\ttbar\to{}$dilepton channel from $\sqrt{s}={8\TeV}$'',} \textit{ Phys.
  Lett. B} \textbf{ 761} (2016) 350,
  \href{http://dx.doi.org/10.1016/j.physletb.2016.08.042}{\doi{10.1016/j.physletb.2016.08.042}},
  \href{http://www.arXiv.org/abs/1606.02179}{\texttt{arXiv:1606.02179}}.

\bibitem{Aad:2015nba}
\hrefCMSnoop {}{{ATLAS Collaboration}, ``Measurement of the top quark mass in
  the $\ttbar\to{}$lepton+jets and $\ttbar\to{}$dilepton channels using
  $\sqrt{s}={7\TeV}$ {ATLAS} data'',} \textit{ Eur. Phys. J. C} \textbf{ 75}
  (2015) 330,
  \href{http://dx.doi.org/10.1140/epjc/s10052-015-3544-0}{\doi{10.1140/epjc/s10052-015-3544-0}},
  \href{http://www.arXiv.org/abs/1503.05427}{\texttt{arXiv:1503.05427}}.

\bibitem{Aaboud:2018zbu}
\hrefCMSnoop {}{{ATLAS Collaboration}, ``Measurement of the top quark mass in
  the $\ttbar\to{}$lepton+jets channel from $\sqrt{s}={8\TeV}$ {ATLAS} data and
  combination with previous results'',} \textit{ Eur. Phys. J. C} \textbf{ 79}
  (2019) 290,
  \href{http://dx.doi.org/10.1140/epjc/s10052-019-6757-9}{\doi{10.1140/epjc/s10052-019-6757-9}},
  \href{http://www.arXiv.org/abs/1810.01772}{\texttt{arXiv:1810.01772}}.

\bibitem{Alioli:2009je}
\hrefCMSnoop {}{S.~Alioli, P.~Nason, C.~Oleari, and E.~Re, ``{NLO} single-top
  production matched with shower in {\POWHEG}: $s$- and $t$-channel
  contributions'',} \textit{ JHEP} \textbf{ 09} (2009) 111,
  \href{http://dx.doi.org/10.1088/1126-6708/2009/09/111}{\doi{10.1088/1126-6708/2009/09/111}},
  \href{http://www.arXiv.org/abs/0907.4076}{\texttt{arXiv:0907.4076}}.
  [Erratum: \DOI{10.1007/JHEP02(2010)011}].

\bibitem{Frederix:2012dh}
\hrefCMSnoop {}{R.~Frederix, E.~Re, and P.~Torrielli, ``Single-top $t$-channel
  hadroproduction in the four-flavors scheme with {\POWHEG} and {a\MCATNLO}'',}
  \textit{ JHEP} \textbf{ 09} (2012) 130,
  \href{http://dx.doi.org/10.1007/JHEP09(2012)130}{\doi{10.1007/JHEP09(2012)130}},
  \href{http://www.arXiv.org/abs/1207.5391}{\texttt{arXiv:1207.5391}}.

\bibitem{ATLAS:2017rso}
\hrefCMSnoop {}{{ATLAS Collaboration}, ``Fiducial, total and differential
  cross-section measurements of $t$-channel single top-quark production in
  ${\Pp\Pp}$ collisions at {8\TeV} using data collected by the {ATLAS}
  detector'',} \textit{ Eur. Phys. J. C} \textbf{ 77} (2017) 531,
  \href{http://dx.doi.org/10.1140/epjc/s10052-017-5061-9}{\doi{10.1140/epjc/s10052-017-5061-9}},
  \href{http://www.arXiv.org/abs/1702.02859}{\texttt{arXiv:1702.02859}}.

\bibitem{CMS:2019jjp}
\hrefCMSnoop {}{{CMS Collaboration}, ``Measurement of differential cross
  sections and charge ratios for $t$-channel single top quark production in
  proton-proton collisions at $\sqrt{s}={13\TeV}$'',} \textit{ Eur. Phys. J. C}
  \textbf{ 80} (2020) 370,
  \href{http://dx.doi.org/10.1140/epjc/s10052-020-7858-1}{\doi{10.1140/epjc/s10052-020-7858-1}},
  \href{http://www.arXiv.org/abs/1907.08330}{\texttt{arXiv:1907.08330}}.

\bibitem{Aliev:2010zk}
M.~Aliev\hrefCMSnoop {}{ { et~al.}, ``\textsc{hathor}: Hadronic top and heavy
  quarks cross section calculator'',} \textit{ Comput. Phys. Commun.} \textbf{
  182} (2011) 1034,
  \href{http://dx.doi.org/10.1016/j.cpc.2010.12.040}{\doi{10.1016/j.cpc.2010.12.040}},
  \href{http://www.arXiv.org/abs/1007.1327}{\texttt{arXiv:1007.1327}}.

\bibitem{Kant:2014oha}
P.~Kant\hrefCMSnoop {}{ { et~al.}, ``\textsc{hathor} for single top-quark
  production: Updated predictions and uncertainty estimates for single
  top-quark production in hadronic collisions'',} \textit{ Comput. Phys.
  Commun.} \textbf{ 191} (2015) 74,
  \href{http://dx.doi.org/10.1016/j.cpc.2015.02.001}{\doi{10.1016/j.cpc.2015.02.001}},
  \href{http://www.arXiv.org/abs/1406.4403}{\texttt{arXiv:1406.4403}}.

\bibitem{CrystalBallRef}
\href {https://www.slac.stanford.edu/cgi-bin/getdoc/slac-r-236.pdf}{M.~J.
  Oreglia, ``A study of the reactions ${\PGy^\prime\to\PGg\PGg\PGy}$''}.
\newblock PhD thesis, Stanford University, 1980.
\newblock {SLAC-R-236}.

\bibitem{Ikeda:1999aq}
\hrefCMSnoop {}{{Belle} Collaboration, ``A detailed test of the {CsI(T\Pell)}
  calorimeter for {BELLE} with photon beams of energy between {20\MeV} and
  {5.4\GeV}'',} \textit{ Nucl. Instrum. Meth. A} \textbf{ 441} (2000) 401,
  \href{http://dx.doi.org/10.1016/S0168-9002(99)00992-4}{\doi{10.1016/S0168-9002(99)00992-4}}.

\bibitem{Chatrchyan:2011ds}
\hrefCMSnoop {}{{CMS Collaboration}, ``Determination of jet energy calibration
  and transverse momentum resolution in {CMS}'',} \textit{ JINST} \textbf{ 6}
  (2011) P11002,
  \href{http://dx.doi.org/10.1088/1748-0221/6/11/P11002}{\doi{10.1088/1748-0221/6/11/P11002}},
  \href{http://www.arXiv.org/abs/1107.4277}{\texttt{arXiv:1107.4277}}.

\bibitem{Greenberg:2002uu}
\hrefCMSnoop {}{O.~W. Greenberg, ``${CPT}$ violation implies violation of
  {Lorentz} invariance'',} \textit{ Phys. Rev. Lett.} \textbf{ 89} (2002)
  231602,
  \href{http://dx.doi.org/10.1103/PhysRevLett.89.231602}{\doi{10.1103/PhysRevLett.89.231602}},
  \href{http://www.arXiv.org/abs/hep-ph/0201258}{\texttt{arXiv:hep-ph/0201258}}.

\bibitem{CMS:2012hzy}
\hrefCMSnoop {}{{CMS Collaboration}, ``Measurement of the mass difference
  between top and antitop quarks'',} \textit{ JHEP} \textbf{ 06} (2012) 109,
  \href{http://dx.doi.org/10.1007/JHEP06(2012)109}{\doi{10.1007/JHEP06(2012)109}},
  \href{http://www.arXiv.org/abs/1204.2807}{\texttt{arXiv:1204.2807}}.

\bibitem{CMS:2016les}
\hrefCMSnoop {}{{CMS Collaboration}, ``Measurement of the mass difference
  between top quark and antiquark in ${\Pp\Pp}$ collisions at
  $\sqrt{s}={8\TeV}$'',} \textit{ Phys. Lett. B} \textbf{ 770} (2017) 50,
  \href{http://dx.doi.org/10.1016/j.physletb.2017.04.028}{\doi{10.1016/j.physletb.2017.04.028}},
  \href{http://www.arXiv.org/abs/1610.09551}{\texttt{arXiv:1610.09551}}.

\bibitem{BASE:2022yvh}
\hrefCMSnoop {}{{BASE} Collaboration, ``A 16-parts-per-trillion measurement of
  the antiproton-to-proton charge-mass ratio'',} \textit{ Nature} \textbf{ 601}
  (2022) 53,
  \href{http://dx.doi.org/10.1038/s41586-021-04203-w}{\doi{10.1038/s41586-021-04203-w}}.

\bibitem{Cheng:2022omt}
\hrefCMSnoop {}{T.~Cheng, M.~Lindner, and M.~Sen, ``Implications of a
  matter-antimatter mass asymmetry in {Penning}-trap experiments'',} \textit{
  Phys. Lett. B} \textbf{ 844} (2023) 138068,
  \href{http://dx.doi.org/10.1016/j.physletb.2023.138068}{\doi{10.1016/j.physletb.2023.138068}},
  \href{http://www.arXiv.org/abs/2210.10819}{\texttt{arXiv:2210.10819}}.

\bibitem{D0:2011hwd}
\hrefCMSnoop {}{{\DZERO} Collaboration, ``Determination of the pole and
  {$\overline{\mathrm{MS}}$} masses of the top quark from the \ttbar cross
  section'',} \textit{ Phys. Lett. B} \textbf{ 703} (2011) 422,
  \href{http://dx.doi.org/10.1016/j.physletb.2011.08.015}{\doi{10.1016/j.physletb.2011.08.015}},
  \href{http://www.arXiv.org/abs/1104.2887}{\texttt{arXiv:1104.2887}}.

\bibitem{CMS:2021yzl}
\hrefCMSnoop {}{{CMS Collaboration}, ``Measurement and {QCD} analysis of
  double-differential inclusive jet cross sections in proton-proton collisions
  at $\sqrt{s}={13\TeV}$'',} \textit{ JHEP} \textbf{ 02} (2022) 142,
  \href{http://dx.doi.org/10.1007/JHEP02(2022)142}{\doi{10.1007/JHEP02(2022)142}},
  \href{http://www.arXiv.org/abs/2111.10431}{\texttt{arXiv:2111.10431}}.
  [Addendum: \DOI{10.1007/JHEP12(2022)035}].

\bibitem{Biswas:2010sa}
\hrefCMSnoop {}{S.~Biswas, K.~Melnikov, and M.~Schulze, ``Next-to-leading order
  {QCD} effects and the top quark mass measurements at the {LHC}'',} \textit{
  JHEP} \textbf{ 08} (2010) 048,
  \href{http://dx.doi.org/10.1007/JHEP08(2010)048}{\doi{10.1007/JHEP08(2010)048}},
  \href{http://www.arXiv.org/abs/1006.0910}{\texttt{arXiv:1006.0910}}.

\bibitem{CMS:2012exf}
\hrefCMSnoop {}{{CMS Collaboration}, ``Measurement of the \ttbar production
  cross section in the dilepton channel in ${\Pp\Pp}$ collisions at
  $\sqrt{s}={7\TeV}$'',} \textit{ JHEP} \textbf{ 11} (2012) 067,
  \href{http://dx.doi.org/10.1007/JHEP11(2012)067}{\doi{10.1007/JHEP11(2012)067}},
  \href{http://www.arXiv.org/abs/1208.2671}{\texttt{arXiv:1208.2671}}.

\bibitem{Melnikov:2009dn}
\hrefCMSnoop {}{K.~Melnikov and M.~Schulze, ``{NLO} {QCD} corrections to top
  quark pair production and decay at hadron colliders'',} \textit{ JHEP}
  \textbf{ 08} (2009) 049,
  \href{http://dx.doi.org/10.1088/1126-6708/2009/08/049}{\doi{10.1088/1126-6708/2009/08/049}},
  \href{http://www.arXiv.org/abs/0907.3090}{\texttt{arXiv:0907.3090}}.

\bibitem{Bernreuther:2010ny}
\hrefCMSnoop {}{W.~Bernreuther and Z.-G. Si, ``Distributions and correlations
  for top quark pair production and decay at the {Tevatron} and {LHC}'',}
  \textit{ Nucl. Phys. B} \textbf{ 837} (2010) 90,
  \href{http://dx.doi.org/10.1016/j.nuclphysb.2010.05.001}{\doi{10.1016/j.nuclphysb.2010.05.001}},
  \href{http://www.arXiv.org/abs/1003.3926}{\texttt{arXiv:1003.3926}}.

\bibitem{Moch:2014tta}
\hrefCMSnoop {}{S.~Moch { et~al.}, ``High precision fundamental constants at
  the {\TeVns} scale'',} 2014.
  \href{http://www.arXiv.org/abs/1405.4781}{\texttt{arXiv:1405.4781}}.

\bibitem{Heinrich:2013qaa}
G.~Heinrich\hrefCMSnoop {}{ { et~al.}, ``{NLO} {QCD} corrections to
  ${\PWp\PWm\bbbar}$ production with leptonic decays in the light of top quark
  mass and asymmetry measurements'',} \textit{ JHEP} \textbf{ 06} (2014) 158,
  \href{http://dx.doi.org/10.1007/JHEP06(2014)158}{\doi{10.1007/JHEP06(2014)158}},
  \href{http://www.arXiv.org/abs/1312.6659}{\texttt{arXiv:1312.6659}}.

\bibitem{Kieseler:2015jzh}
\hrefCMSnoop {}{J.~Kieseler, K.~Lipka, and S.-O. Moch, ``Calibration of the
  top-quark {Monte Carlo} mass'',} \textit{ Phys. Rev. Lett.} \textbf{ 116}
  (2016) 162001,
  \href{http://dx.doi.org/10.1103/PhysRevLett.116.162001}{\doi{10.1103/PhysRevLett.116.162001}},
  \href{http://www.arXiv.org/abs/1511.00841}{\texttt{arXiv:1511.00841}}.

\bibitem{Dowling:2013baa}
\hrefCMSnoop {}{M.~Dowling and S.-O. Moch, ``Differential distributions for
  top-quark hadro-production with a running mass'',} \textit{ Eur. Phys. J. C}
  \textbf{ 74} (2014) 3167,
  \href{http://dx.doi.org/10.1140/epjc/s10052-014-3167-x}{\doi{10.1140/epjc/s10052-014-3167-x}},
  \href{http://www.arXiv.org/abs/1305.6422}{\texttt{arXiv:1305.6422}}.

\bibitem{Alekhin:2014irh}
\hrefCMSnoop {}{S.~Alekhin { et~al.}, ``{HERAFitter}: Open source {QCD} fit
  project'',} \textit{ Eur. Phys. J. C} \textbf{ 75} (2015) 304,
  \href{http://dx.doi.org/10.1140/epjc/s10052-015-3480-z}{\doi{10.1140/epjc/s10052-015-3480-z}},
  \href{http://www.arXiv.org/abs/1410.4412}{\texttt{arXiv:1410.4412}}.

\bibitem{Baikov:2014qja}
\hrefCMSnoop {}{P.~A. Baikov, K.~G. Chetyrkin, and J.~H. K{\"u}hn, ``Quark mass
  and field anomalous dimensions to $\mathcal{O}({\alpS}^5)$'',} \textit{ JHEP}
  \textbf{ 10} (2014) 076,
  \href{http://dx.doi.org/10.1007/JHEP10(2014)076}{\doi{10.1007/JHEP10(2014)076}},
  \href{http://www.arXiv.org/abs/1402.6611}{\texttt{arXiv:1402.6611}}.

\bibitem{Luthe:2016xec}
\hrefCMSnoop {}{T.~Luthe, A.~Maier, P.~Marquard, and Y.~Schr{\"o}der,
  ``Five-loop quark mass and field anomalous dimensions for a general gauge
  group'',} \textit{ JHEP} \textbf{ 01} (2017) 081,
  \href{http://dx.doi.org/10.1007/JHEP01(2017)081}{\doi{10.1007/JHEP01(2017)081}},
  \href{http://www.arXiv.org/abs/1612.05512}{\texttt{arXiv:1612.05512}}.

\bibitem{Mihaila:2013wma}
\hrefCMSnoop {}{L.~Mihaila, ``Precision calculations in supersymmetric
  theories'',} \textit{ Adv. High Energy Phys.} \textbf{ 2013} (2013) 607807,
  \href{http://dx.doi.org/10.1155/2013/607807}{\doi{10.1155/2013/607807}},
  \href{http://www.arXiv.org/abs/1310.6178}{\texttt{arXiv:1310.6178}}.

\bibitem{Christensen:2005hm}
\hrefCMSnoop {}{N.~D. Christensen and R.~Shrock, ``Implications of dynamical
  generation of standard-model fermion masses'',} \textit{ Phys. Rev. Lett.}
  \textbf{ 94} (2005) 241801,
  \href{http://dx.doi.org/10.1103/PhysRevLett.94.241801}{\doi{10.1103/PhysRevLett.94.241801}},
  \href{http://www.arXiv.org/abs/hep-ph/0501294}{\texttt{arXiv:hep-ph/0501294}}.

\bibitem{Catani:2020tko}
S.~Catani\hrefCMSnoop {}{ { et~al.}, ``Top-quark pair hadroproduction at
  {NNLO}: differential predictions with the {$\overline{\mathrm{MS}}$} mass'',}
  \textit{ JHEP} \textbf{ 08} (2020) 027,
  \href{http://dx.doi.org/10.1007/JHEP08(2020)027}{\doi{10.1007/JHEP08(2020)027}},
  \href{http://www.arXiv.org/abs/2005.00557}{\texttt{arXiv:2005.00557}}.

\bibitem{Defranchis:2022nqb}
\hrefCMSnoop {}{M.~M. Defranchis, J.~Kieseler, K.~Lipka, and J.~Mazzitelli,
  ``Running of the top quark mass at {NNLO} in {QCD}'',} \textit{ JHEP}
  \textbf{ 04} (2024) 125,
  \href{http://dx.doi.org/10.1007/JHEP04(2024)125}{\doi{10.1007/JHEP04(2024)125}},
  \href{http://www.arXiv.org/abs/2208.11399}{\texttt{arXiv:2208.11399}}.

\bibitem{CMS:2017zpm}
\hrefCMSnoop {}{{CMS Collaboration}, ``Measurement of the inclusive \ttbar
  cross section in ${\Pp\Pp}$ collisions at $\sqrt{s}={5.02\TeV}$ using final
  states with at least one charged lepton'',} \textit{ JHEP} \textbf{ 03}
  (2018) 115,
  \href{http://dx.doi.org/10.1007/JHEP03(2018)115}{\doi{10.1007/JHEP03(2018)115}},
  \href{http://www.arXiv.org/abs/1711.03143}{\texttt{arXiv:1711.03143}}.

\bibitem{H1:2015ubc}
\hrefCMSnoop {}{{H1 and ZEUS Collaborations}, ``Combination of measurements of
  inclusive deep inelastic ${\Pepm\Pp}$ scattering cross sections and {QCD}
  analysis of {HERA} data'',} \textit{ Eur. Phys. J. C} \textbf{ 75} (2015)
  580,
  \href{http://dx.doi.org/10.1140/epjc/s10052-015-3710-4}{\doi{10.1140/epjc/s10052-015-3710-4}},
  \href{http://www.arXiv.org/abs/1506.06042}{\texttt{arXiv:1506.06042}}.

\bibitem{CMS:2016qqr}
\hrefCMSnoop {}{{CMS Collaboration}, ``Measurement of the differential cross
  section and charge asymmetry for inclusive ${\Pp\Pp\to\PWpm+\PX}$ production
  at $\sqrt{s}={8\TeV}$'',} \textit{ Eur. Phys. J. C} \textbf{ 76} (2016) 469,
  \href{http://dx.doi.org/10.1140/epjc/s10052-016-4293-4}{\doi{10.1140/epjc/s10052-016-4293-4}},
  \href{http://www.arXiv.org/abs/1603.01803}{\texttt{arXiv:1603.01803}}.

\bibitem{Guzzi:2014wia}
\hrefCMSnoop {}{M.~Guzzi, K.~Lipka, and S.-O. Moch, ``Top-quark pair production
  at hadron colliders: differential cross section and phenomenological
  applications with \textsc{DiffTop}'',} \textit{ JHEP} \textbf{ 01} (2015)
  082,
  \href{http://dx.doi.org/10.1007/JHEP01(2015)082}{\doi{10.1007/JHEP01(2015)082}},
  \href{http://www.arXiv.org/abs/1406.0386}{\texttt{arXiv:1406.0386}}.

\bibitem{CMS:2017iqf}
\hrefCMSnoop {}{{CMS Collaboration}, ``Measurement of double-differential cross
  sections for top quark pair production in ${\Pp\Pp}$ collisions at
  $\sqrt{s}={8\TeV}$ and impact on parton distribution functions'',} \textit{
  Eur. Phys. J. C} \textbf{ 77} (2017) 459,
  \href{http://dx.doi.org/10.1140/epjc/s10052-017-4984-5}{\doi{10.1140/epjc/s10052-017-4984-5}},
  \href{http://www.arXiv.org/abs/1703.01630}{\texttt{arXiv:1703.01630}}.

\bibitem{Alioli:2013mxa}
S.~Alioli\hrefCMSnoop {}{ { et~al.}, ``A new observable to measure the
  top-quark mass at hadron colliders'',} \textit{ Eur. Phys. J. C} \textbf{ 73}
  (2013) 2438,
  \href{http://dx.doi.org/10.1140/epjc/s10052-013-2438-2}{\doi{10.1140/epjc/s10052-013-2438-2}},
  \href{http://www.arXiv.org/abs/1303.6415}{\texttt{arXiv:1303.6415}}.

\bibitem{Mangano:1991jk}
\hrefCMSnoop {}{M.~L. Mangano, P.~Nason, and G.~Ridolfi, ``Heavy-quark
  correlations in hadron collisions at next-to-leading order'',} \textit{ Nucl.
  Phys. B} \textbf{ 373} (1992) 295,
  \href{http://dx.doi.org/10.1016/0550-3213(92)90435-E}{\doi{10.1016/0550-3213(92)90435-E}}.

\bibitem{Dittmaier:2007wz}
\hrefCMSnoop {}{S.~Dittmaier, P.~Uwer, and S.~Weinzierl, ``{NLO} {QCD}
  corrections to $\ttbar+{}$jet production at hadron colliders'',} \textit{
  Phys. Rev. Lett.} \textbf{ 98} (2007) 262002,
  \href{http://dx.doi.org/10.1103/PhysRevLett.98.262002}{\doi{10.1103/PhysRevLett.98.262002}},
  \href{http://www.arXiv.org/abs/hep-ph/0703120}{\texttt{arXiv:hep-ph/0703120}}.

\bibitem{Bevilacqua:2010ve}
\hrefCMSnoop {}{G.~Bevilacqua, M.~Czakon, C.~G. Papadopoulos, and M.~Worek,
  ``Dominant {QCD} backgrounds in {Higgs} boson analyses at the {LHC}: A study
  of ${\Pp\Pp}\to\ttbar+2$ jets at next-to-leading order'',} \textit{ Phys.
  Rev. Lett.} \textbf{ 104} (2010) 162002,
  \href{http://dx.doi.org/10.1103/PhysRevLett.104.162002}{\doi{10.1103/PhysRevLett.104.162002}},
  \href{http://www.arXiv.org/abs/1002.4009}{\texttt{arXiv:1002.4009}}.

\bibitem{Bevilacqua:2011aa}
\hrefCMSnoop {}{G.~Bevilacqua, M.~Czakon, C.~G. Papadopoulos, and M.~Worek,
  ``Hadronic top-quark pair production in association with two jets at
  next-to-leading order {QCD}'',} \textit{ Phys. Rev. D} \textbf{ 84} (2011)
  114017,
  \href{http://dx.doi.org/10.1103/PhysRevD.84.114017}{\doi{10.1103/PhysRevD.84.114017}},
  \href{http://www.arXiv.org/abs/1108.2851}{\texttt{arXiv:1108.2851}}.

\bibitem{ParticleDataGroup:2018ovx}
\hrefCMSnoop {}{{Particle Data Group}, M.~Tanabashi { et~al.}, ``Review of
  particle physics'',} \textit{ Phys. Rev. D} \textbf{ 98} (2018) 030001,
  \href{http://dx.doi.org/10.1103/PhysRevD.98.030001}{\doi{10.1103/PhysRevD.98.030001}}.

\bibitem{Kiyo:2008bv}
Y.~Kiyo\hrefCMSnoop {}{ { et~al.}, ``Top-quark pair production near threshold
  at {LHC}'',} \textit{ Eur. Phys. J. C} \textbf{ 60} (2009) 375,
  \href{http://dx.doi.org/10.1140/epjc/s10052-009-0892-7}{\doi{10.1140/epjc/s10052-009-0892-7}},
  \href{http://www.arXiv.org/abs/0812.0919}{\texttt{arXiv:0812.0919}}.

\bibitem{Piclum:2018ndt}
\hrefCMSnoop {}{J.~Piclum and C.~Schwinn, ``Soft-gluon and {Coulomb}
  corrections to hadronic top-quark pair production beyond {NNLO}'',} \textit{
  JHEP} \textbf{ 03} (2018) 164,
  \href{http://dx.doi.org/10.1007/JHEP03(2018)164}{\doi{10.1007/JHEP03(2018)164}},
  \href{http://www.arXiv.org/abs/1801.05788}{\texttt{arXiv:1801.05788}}.

\bibitem{Makela:2023xnt}
\hrefCMSnoop {}{T.~M{\"a}kel{\"a}, A.~Hoang, K.~Lipka, and S.-O. Moch,
  ``Investigation of the scale dependence in the {MSR} and
  {$\overline{\mathrm{MS}}$} top quark mass schemes for the \ttbar invariant
  mass differential cross section using {LHC} data'',} \textit{ JHEP} \textbf{
  09} (2023) 037,
  \href{http://dx.doi.org/10.1007/JHEP09(2023)037}{\doi{10.1007/JHEP09(2023)037}},
  \href{http://www.arXiv.org/abs/2301.03546}{\texttt{arXiv:2301.03546}}.

\bibitem{bib:ttjPowheg}
\hrefCMSnoop {}{S.~Alioli, S.-O. Moch, and P.~Uwer, ``Hadronic top-quark
  pair-production with one jet and parton showering'',} \textit{ JHEP} \textbf{
  01} (2012) 137,
  \href{http://dx.doi.org/10.1007/JHEP01(2012)137}{\doi{10.1007/JHEP01(2012)137}},
  \href{http://www.arXiv.org/abs/1110.5251}{\texttt{arXiv:1110.5251}}.

\bibitem{bib:ABMP16}
\hrefCMSnoop {}{S.~Alekhin, J.~Bl{\"u}mlein, and S.~Moch, ``{NLO} {PDFs} from
  the {ABMP16} fit'',} \textit{ Eur. Phys. J. C} \textbf{ 78} (2018) 477,
  \href{http://dx.doi.org/10.1140/epjc/s10052-018-5947-1}{\doi{10.1140/epjc/s10052-018-5947-1}},
  \href{http://www.arXiv.org/abs/1803.07537}{\texttt{arXiv:1803.07537}}.

\bibitem{bib:CT18}
T.-J. Hou\hrefCMSnoop {}{ { et~al.}, ``Progress in the {CTEQ-TEA} {NNLO} global
  {QCD} analysis'',} 2019.
  \href{http://www.arXiv.org/abs/1908.11394}{\texttt{arXiv:1908.11394}}.

\bibitem{bib:ttjPheno}
S.~Alioli\hrefCMSnoop {}{ { et~al.}, ``Phenomenology of
  $\ttbar{\HepParticle{j}{}{}}+{\PX}$ production at the {LHC}'',} \textit{
  JHEP} \textbf{ 05} (2022) 146,
  \href{http://dx.doi.org/10.1007/JHEP05(2022)146}{\doi{10.1007/JHEP05(2022)146}},
  \href{http://www.arXiv.org/abs/2202.07975}{\texttt{arXiv:2202.07975}}.

\bibitem{Babaev:2023fim}
A.~Babaev\hrefCMSnoop {}{ { et~al.}, ``Impact of beam-beam effects on absolute
  luminosity calibrations at the {CERN} {Large Hadron Collider}'',} \textit{
  Eur. Phys. J. C} \textbf{ 84} (2024) 17,
  \href{http://dx.doi.org/10.1140/epjc/s10052-023-12192-5}{\doi{10.1140/epjc/s10052-023-12192-5}},
  \href{http://www.arXiv.org/abs/2306.10394}{\texttt{arXiv:2306.10394}}.

\bibitem{Hoang:2000yr}
A.~H. Hoang\hrefCMSnoop {}{ { et~al.}, ``Top-antitop pair production close to
  threshold: Synopsis of recent {NNLO} results'',} in \textit{ {Proc. 4th
  Workshop of the 2nd ECFA/DESY Study on Physics and Detectors for a Linear
  Electron-Positron Collider: Oxford, UK, March 20--23, 1999}}.
\newblock 2000.
\newblock
  \href{http://www.arXiv.org/abs/hep-ph/0001286}{\texttt{arXiv:hep-ph/0001286}}.
\newblock [Eur. Phys. J. direct 2 (2000) 1].
  \href{http://dx.doi.org/10.1007/s1010500c0003}{\doi{10.1007/s1010500c0003}}.

\bibitem{Hagiwara:2008df}
\hrefCMSnoop {}{K.~Hagiwara, Y.~Sumino, and H.~Yokoya, ``Bound-state effects on
  top quark production at hadron colliders'',} \textit{ Phys. Lett. B} \textbf{
  666} (2008) 71,
  \href{http://dx.doi.org/10.1016/j.physletb.2008.07.006}{\doi{10.1016/j.physletb.2008.07.006}},
  \href{http://www.arXiv.org/abs/0804.1014}{\texttt{arXiv:0804.1014}}.

\bibitem{Ju:2020otc}
W.-L. Ju\hrefCMSnoop {}{ { et~al.}, ``Top quark pair production near threshold:
  single/double distributions and mass determination'',} \textit{ JHEP}
  \textbf{ 06} (2020) 158,
  \href{http://dx.doi.org/10.1007/JHEP06(2020)158}{\doi{10.1007/JHEP06(2020)158}},
  \href{http://www.arXiv.org/abs/2004.03088}{\texttt{arXiv:2004.03088}}.

\bibitem{Britzger:2012bs}
{fastNLO} Collaboration, \hrefCMSnoop {}{D.~Britzger, K.~Rabbertz, F.~Stober,
  and M.~Wobisch, ``New features in version 2 of the {fastNLO} project'',} in
  \textit{ {Proc. 20th International Workshop on Deep-Inelastic Scattering and
  Related Subjects (DIS2012): Bonn, Germany, March 26--30, 2012}}.
\newblock 2012.
\newblock \href{http://www.arXiv.org/abs/1208.3641}{\texttt{arXiv:1208.3641}}.
\newblock
  \href{http://dx.doi.org/10.3204/DESY-PROC-2012-02/165}{\doi{10.3204/DESY-PROC-2012-02/165}}.

\bibitem{Carli:2010rw}
T.~Carli\hrefCMSnoop {}{ { et~al.}, ``A posteriori inclusion of parton density
  functions in {NLO} {QCD} final-state calculations at hadron colliders: the
  {APPLGRID} project'',} \textit{ Eur. Phys. J. C} \textbf{ 66} (2010) 503,
  \href{http://dx.doi.org/10.1140/epjc/s10052-010-1255-0}{\doi{10.1140/epjc/s10052-010-1255-0}},
  \href{http://www.arXiv.org/abs/0911.2985}{\texttt{arXiv:0911.2985}}.

\bibitem{Britzger:2022lbf}
D.~Britzger\hrefCMSnoop {}{ { et~al.}, ``{NNLO} interpolation grids for jet
  production at the {LHC}'',} \textit{ Eur. Phys. J. C} \textbf{ 82} (2022)
  930,
  \href{http://dx.doi.org/10.1140/epjc/s10052-022-10880-2}{\doi{10.1140/epjc/s10052-022-10880-2}},
  \href{http://www.arXiv.org/abs/2207.13735}{\texttt{arXiv:2207.13735}}.

\bibitem{Kataev:2022dua}
\hrefCMSnoop {}{A.~L. Kataev and V.~S. Molokoedov, ``Notes on interplay between
  the {QCD} and {EW} perturbative corrections to the pole-running-to-top-quark
  mass ratio'',} \textit{ JETP Lett.} \textbf{ 115} (2022) 704,
  \href{http://dx.doi.org/10.1134/S0021364022600902}{\doi{10.1134/S0021364022600902}},
  \href{http://www.arXiv.org/abs/2201.12073}{\texttt{arXiv:2201.12073}}.

\bibitem{Dittmaier:2022maf}
\hrefCMSnoop {}{S.~Dittmaier and H.~Rzehak, ``Electroweak renormalization based
  on gauge-invariant vacuum expectation values of non-linear {Higgs}
  representations. {Part I}. {Standard} model'',} \textit{ JHEP} \textbf{ 05}
  (2022) 125,
  \href{http://dx.doi.org/10.1007/JHEP05(2022)125}{\doi{10.1007/JHEP05(2022)125}},
  \href{http://www.arXiv.org/abs/2203.07236}{\texttt{arXiv:2203.07236}}.

\bibitem{Fleming:2007qr}
\hrefCMSnoop {}{S.~Fleming, A.~H. Hoang, S.~Mantry, and I.~W. Stewart, ``Jets
  from massive unstable particles: Top-mass determination'',} \textit{ Phys.
  Rev. D} \textbf{ 77} (2008) 074010,
  \href{http://dx.doi.org/10.1103/PhysRevD.77.074010}{\doi{10.1103/PhysRevD.77.074010}},
  \href{http://www.arXiv.org/abs/hep-ph/0703207}{\texttt{arXiv:hep-ph/0703207}}.

\bibitem{Fleming:2007xt}
\hrefCMSnoop {}{S.~Fleming, A.~H. Hoang, S.~Mantry, and I.~W. Stewart, ``Top
  jets in the peak region: Factorization analysis with next-to-leading-log
  resummation'',} \textit{ Phys. Rev. D} \textbf{ 77} (2008) 114003,
  \href{http://dx.doi.org/10.1103/PhysRevD.77.114003}{\doi{10.1103/PhysRevD.77.114003}},
  \href{http://www.arXiv.org/abs/0711.2079}{\texttt{arXiv:0711.2079}}.

\bibitem{Hoang:2019ceu}
\hrefCMSnoop {}{A.~H. Hoang, S.~Mantry, A.~Pathak, and I.~W. Stewart,
  ``Nonperturbative corrections to soft drop jet mass'',} \textit{ JHEP}
  \textbf{ 12} (2019) 002,
  \href{http://dx.doi.org/10.1007/JHEP12(2019)002}{\doi{10.1007/JHEP12(2019)002}},
  \href{http://www.arXiv.org/abs/1906.11843}{\texttt{arXiv:1906.11843}}.

\bibitem{Butenschoen:2016lpz}
M.~Butenschoen\hrefCMSnoop {}{ { et~al.}, ``Top quark mass calibration for
  {Monte Carlo} event generators'',} \textit{ Phys. Rev. Lett.} \textbf{ 117}
  (2016) 232001,
  \href{http://dx.doi.org/10.1103/PhysRevLett.117.232001}{\doi{10.1103/PhysRevLett.117.232001}},
  \href{http://www.arXiv.org/abs/1608.01318}{\texttt{arXiv:1608.01318}}.

\bibitem{Hoang:2018zrp}
\hrefCMSnoop {}{A.~H. Hoang, S.~Pl{\"a}tzer, and D.~Samitz, ``On the cutoff
  dependence of the quark mass parameter in angular ordered parton showers'',}
  \textit{ JHEP} \textbf{ 10} (2018) 200,
  \href{http://dx.doi.org/10.1007/JHEP10(2018)200}{\doi{10.1007/JHEP10(2018)200}},
  \href{http://www.arXiv.org/abs/1807.06617}{\texttt{arXiv:1807.06617}}.

\bibitem{Salam:2010nqg}
\hrefCMSnoop {}{G.~P. Salam, ``Towards jetography'',} \textit{ Eur. Phys. J. C}
  \textbf{ 67} (2010) 637,
  \href{http://dx.doi.org/10.1140/epjc/s10052-010-1314-6}{\doi{10.1140/epjc/s10052-010-1314-6}},
  \href{http://www.arXiv.org/abs/0906.1833}{\texttt{arXiv:0906.1833}}.

\bibitem{Dasgupta:2013ihk}
\hrefCMSnoop {}{M.~Dasgupta, A.~Fregoso, S.~Marzani, and G.~P. Salam, ``Towards
  an understanding of jet substructure'',} \textit{ JHEP} \textbf{ 09} (2013)
  029,
  \href{http://dx.doi.org/10.1007/JHEP09(2013)029}{\doi{10.1007/JHEP09(2013)029}},
  \href{http://www.arXiv.org/abs/1307.0007}{\texttt{arXiv:1307.0007}}.

\bibitem{Kogler:2021kkw}
R.~Kogler, ``Advances in jet substructure at the {LHC}: Algorithms,
  measurements and searches for new physical phenomena''.
\newblock Springer Cham, 2021.
\newblock ISBN 978-3-030-72858-8.
  \href{http://dx.doi.org/10.1007/978-3-030-72858-8}{\doi{10.1007/978-3-030-72858-8}}.

\bibitem{Larkoski:2017jix}
\hrefCMSnoop {}{A.~J. Larkoski, I.~Moult, and B.~Nachman, ``Jet substructure at
  the {Large Hadron Collider}: A review of recent advances in theory and
  machine learning'',} \textit{ Phys. Rept.} \textbf{ 841} (2020) 1,
  \href{http://dx.doi.org/10.1016/j.physrep.2019.11.001}{\doi{10.1016/j.physrep.2019.11.001}},
  \href{http://www.arXiv.org/abs/1709.04464}{\texttt{arXiv:1709.04464}}.

\bibitem{Asquith:2018igt}
\hrefCMSnoop {}{R.~Kogler { et~al.}, ``Jet substructure at the {Large Hadron
  Collider}'',} \textit{ Rev. Mod. Phys.} \textbf{ 91} (2019) 045003,
  \href{http://dx.doi.org/10.1103/RevModPhys.91.045003}{\doi{10.1103/RevModPhys.91.045003}},
  \href{http://www.arXiv.org/abs/1803.06991}{\texttt{arXiv:1803.06991}}.

\bibitem{CMS:2018vzn}
\hrefCMSnoop {}{{CMS Collaboration}, ``Measurements of the differential jet
  cross section as a function of the jet mass in dijet events from
  proton-proton collisions at $\sqrt{s}={13\TeV}$'',} \textit{ JHEP} \textbf{
  11} (2018) 113,
  \href{http://dx.doi.org/10.1007/JHEP11(2018)113}{\doi{10.1007/JHEP11(2018)113}},
  \href{http://www.arXiv.org/abs/1807.05974}{\texttt{arXiv:1807.05974}}.

\bibitem{Bauer:2000ew}
\hrefCMSnoop {}{C.~W. Bauer, S.~Fleming, and M.~E. Luke, ``Summing {Sudakov}
  logarithms in ${\PB\to\PX_{\PQs}\PGg}$ in effective field theory'',} \textit{
  Phys. Rev. D} \textbf{ 63} (2000) 014006,
  \href{http://dx.doi.org/10.1103/PhysRevD.63.014006}{\doi{10.1103/PhysRevD.63.014006}},
  \href{http://www.arXiv.org/abs/hep-ph/0005275}{\texttt{arXiv:hep-ph/0005275}}.

\bibitem{Bauer:2000yr}
\hrefCMSnoop {}{C.~W. Bauer, S.~Fleming, D.~Pirjol, and I.~W. Stewart, ``An
  effective field theory for collinear and soft gluons: Heavy to light
  decays'',} \textit{ Phys. Rev. D} \textbf{ 63} (2001) 114020,
  \href{http://dx.doi.org/10.1103/PhysRevD.63.114020}{\doi{10.1103/PhysRevD.63.114020}},
  \href{http://www.arXiv.org/abs/hep-ph/0011336}{\texttt{arXiv:hep-ph/0011336}}.

\bibitem{Bauer:2001ct}
\hrefCMSnoop {}{C.~W. Bauer and I.~W. Stewart, ``Invariant operators in
  collinear effective theory'',} \textit{ Phys. Lett. B} \textbf{ 516} (2001)
  134,
  \href{http://dx.doi.org/10.1016/S0370-2693(01)00902-9}{\doi{10.1016/S0370-2693(01)00902-9}},
  \href{http://www.arXiv.org/abs/hep-ph/0107001}{\texttt{arXiv:hep-ph/0107001}}.

\bibitem{Bauer:2001yt}
\hrefCMSnoop {}{C.~W. Bauer, D.~Pirjol, and I.~W. Stewart, ``Soft-collinear
  factorization in effective field theory'',} \textit{ Phys. Rev. D} \textbf{
  65} (2002) 054022,
  \href{http://dx.doi.org/10.1103/PhysRevD.65.054022}{\doi{10.1103/PhysRevD.65.054022}},
  \href{http://www.arXiv.org/abs/hep-ph/0109045}{\texttt{arXiv:hep-ph/0109045}}.

\bibitem{Bauer:2002nz}
C.~W. Bauer\hrefCMSnoop {}{ { et~al.}, ``Hard scattering factorization from
  effective field theory'',} \textit{ Phys. Rev. D} \textbf{ 66} (2002) 014017,
  \href{http://dx.doi.org/10.1103/PhysRevD.66.014017}{\doi{10.1103/PhysRevD.66.014017}},
  \href{http://www.arXiv.org/abs/hep-ph/0202088}{\texttt{arXiv:hep-ph/0202088}}.

\bibitem{Larkoski:2014wba}
\hrefCMSnoop {}{A.~J. Larkoski, S.~Marzani, G.~Soyez, and J.~Thaler, ``Soft
  drop'',} \textit{ JHEP} \textbf{ 05} (2014) 146,
  \href{http://dx.doi.org/10.1007/JHEP05(2014)146}{\doi{10.1007/JHEP05(2014)146}},
  \href{http://www.arXiv.org/abs/1402.2657}{\texttt{arXiv:1402.2657}}.

\bibitem{Dehnadi:2023msm}
\hrefCMSnoop {}{B.~Dehnadi, A.~H. Hoang, O.~L. Jin, and V.~Mateu, ``Top quark
  mass calibration for {Monte Carlo} event generators---an update'',} \textit{
  JHEP} \textbf{ 12} (2023) 065,
  \href{http://dx.doi.org/10.1007/JHEP12(2023)065}{\doi{10.1007/JHEP12(2023)065}},
  \href{http://www.arXiv.org/abs/2309.00547}{\texttt{arXiv:2309.00547}}.

\bibitem{Stewart:2015waa}
I.~W. Stewart\hrefCMSnoop {}{ { et~al.}, ``{XCone}: ${N}$-jettiness as an
  exclusive cone jet algorithm'',} \textit{ JHEP} \textbf{ 11} (2015) 072,
  \href{http://dx.doi.org/10.1007/JHEP11(2015)072}{\doi{10.1007/JHEP11(2015)072}},
  \href{http://www.arXiv.org/abs/1508.01516}{\texttt{arXiv:1508.01516}}.

\bibitem{Dokshitzer:1997in}
\hrefCMSnoop {}{Y.~L. Dokshitzer, G.~D. Leder, S.~Moretti, and B.~R. Webber,
  ``Better jet clustering algorithms'',} \textit{ JHEP} \textbf{ 08} (1997)
  001,
  \href{http://dx.doi.org/10.1088/1126-6708/1997/08/001}{\doi{10.1088/1126-6708/1997/08/001}},
  \href{http://www.arXiv.org/abs/hep-ph/9707323}{\texttt{arXiv:hep-ph/9707323}}.

\bibitem{Wobisch:1998wt}
\hrefCMSnoop {}{M.~Wobisch and T.~Wengler, ``Hadronization corrections to jet
  cross-sections in deep inelastic scattering'',} in \textit{ {Proc. Workshop
  on Monte Carlo Generators for HERA Physics: Hamburg, Germany, April 27--30,
  1998}}, p.~270.
\newblock 1998.
\newblock
  \href{http://www.arXiv.org/abs/hep-ph/9907280}{\texttt{arXiv:hep-ph/9907280}}.

\bibitem{Thaler:2015xaa}
\hrefCMSnoop {}{J.~Thaler and T.~F. Wilkason, ``Resolving boosted jets with
  {XCone}'',} \textit{ JHEP} \textbf{ 12} (2015) 051,
  \href{http://dx.doi.org/10.1007/JHEP12(2015)051}{\doi{10.1007/JHEP12(2015)051}},
  \href{http://www.arXiv.org/abs/1508.01518}{\texttt{arXiv:1508.01518}}.

\bibitem{Stewart:2010tn}
\hrefCMSnoop {}{I.~W. Stewart, F.~J. Tackmann, and W.~J. Waalewijn, ``${N}$
  jettiness: An inclusive event shape to veto jets'',} \textit{ Phys. Rev.
  Lett.} \textbf{ 105} (2010) 092002,
  \href{http://dx.doi.org/10.1103/PhysRevLett.105.092002}{\doi{10.1103/PhysRevLett.105.092002}},
  \href{http://www.arXiv.org/abs/1004.2489}{\texttt{arXiv:1004.2489}}.

\bibitem{Thaler:2010tr}
\hrefCMSnoop {}{J.~Thaler and K.~Van~Tilburg, ``Identifying boosted objects
  with ${N}$-subjettiness'',} \textit{ JHEP} \textbf{ 03} (2011) 015,
  \href{http://dx.doi.org/10.1007/JHEP03(2011)015}{\doi{10.1007/JHEP03(2011)015}},
  \href{http://www.arXiv.org/abs/1011.2268}{\texttt{arXiv:1011.2268}}.

\bibitem{Thaler:2011gf}
\hrefCMSnoop {}{J.~Thaler and K.~Van~Tilburg, ``Maximizing boosted top
  identification by minimizing ${N}$-subjettiness'',} \textit{ JHEP} \textbf{
  02} (2012) 093,
  \href{http://dx.doi.org/10.1007/JHEP02(2012)093}{\doi{10.1007/JHEP02(2012)093}},
  \href{http://www.arXiv.org/abs/1108.2701}{\texttt{arXiv:1108.2701}}.

\bibitem{Apollinari:2015bam}
\hrefCMSnoop {}{G.~Apollinari { et~al.}, ``{High-Luminosity Large Hadron
  Collider (HL-LHC)}: Preliminary design report'',} CERN Technical Proposal
  CERN-2015-005, 2015.
\newblock
  \href{http://dx.doi.org/10.5170/CERN-2015-005}{\doi{10.5170/CERN-2015-005}}.

\bibitem{CMS:2017lum}
\hrefCMSnoop {}{{CMS Collaboration}, ``The {Phase-2} upgrade of the {CMS}
  tracker'',} CMS Technical Proposal CERN-LHCC-2017-009, CMS-TDR-014, 2017.
\newblock
  \href{http://dx.doi.org/10.17181/CERN.QZ28.FLHW}{\doi{10.17181/CERN.QZ28.FLHW}}.

\bibitem{Hebbeker:2017bix}
\href {https://cds.cern.ch/record/2283189}{{CMS Collaboration}, ``The {Phase-2}
  upgrade of the {CMS} muon detectors'',} CMS Technical Proposal
  CERN-LHCC-2017-012, CMS-TDR-016, 2017.

\bibitem{Butler:2019rpu}
\href {https://cds.cern.ch/record/2667167}{{CMS Collaboration}, ``A {MIP}
  timing detector for the {CMS} {Phase-2} upgrade'',} CMS Technical Proposal
  CERN-LHCC-2019-003, CMS-TDR-020, 2019.

\bibitem{CMS:2017jpq}
\href {https://cds.cern.ch/record/2293646}{{CMS Collaboration}, ``The {Phase-2}
  upgrade of the {CMS} endcap calorimeter'',} CMS Technical Proposal
  CERN-LHCC-2017-023, CMS-TDR-019, 2017.

\bibitem{CMS:barrelTDR_no_inspire}
\href {https://cds.cern.ch/record/2283187}{{CMS Collaboration}, ``The {Phase-2}
  upgrade of the {CMS} barrel calorimeters'',} CMS Technical Proposal
  CERN-LHCC-2017-011, CMS-TDR-015, 2017.

\bibitem{CMS:2017gvo}
\href {https://cds.cern.ch/record/2262606}{{CMS Collaboration}, ``{ECFA} 2016:
  Prospects for selected standard model measurements with the {CMS} experiment
  at the {High-Luminosity LHC}'',} CMS Physics Analysis Summary
  CMS-PAS-FTR-16-006, 2017.

\bibitem{CMS:EDQ_note_no_inspire}
\href {https://cds.cern.ch/record/2650976}{{CMS Collaboration}, ``Expected
  performance of the physics objects with the upgraded {CMS} detector at the
  {HL-LHC}'',} CMS Note CMS-NOTE-2018-006, 2018.

\bibitem{CMS:BRILTDR_no_inspire}
\href {https://cds.cern.ch/record/2759074}{{CMS Collaboration}, ``The {Phase-2}
  upgrade of the {CMS} beam radiation, instrumentation, and luminosity
  detectors'',} CMS Technical Proposal CERN-LHCC-2021-008, CMS-TDR-023, 2021.

\bibitem{Dainese:2019rgk}
\hrefCMSnoop {}{A.~Dainese { et~al.}, ``Physics at the {HL-LHC}, and
  perspectives at the {HE-LHC}'',} CERN Report CERN-2019-007, 2019.
\newblock
  \href{http://dx.doi.org/10.23731/CYRM-2019-007}{\doi{10.23731/CYRM-2019-007}}.

\bibitem{Paukkunen:2014zia}
\hrefCMSnoop {}{H.~Paukkunen and P.~Zurita, ``{PDF} reweighting in the
  {Hessian} matrix approach'',} \textit{ JHEP} \textbf{ 12} (2014) 100,
  \href{http://dx.doi.org/10.1007/JHEP12(2014)100}{\doi{10.1007/JHEP12(2014)100}},
  \href{http://www.arXiv.org/abs/1402.6623}{\texttt{arXiv:1402.6623}}.

\end{thebibliography}\endgroup

\clearpage
\appendix

\section{Glossary of acronyms}
\label{sec:glossary}

\begin{longtable}[l]{ll}
AMWT & Analytical matrix weighting technique \\
BBR & Beam-beam remnants \\
BDT & Boosted decision tree \\
BSM & Beyond the standard model \\
CA & Cambridge--Aachen \\
CHS & Charged hadron subtraction \\
CKM & Cabibbo--Kobayashi--Maskawa \\
\CL & Confidence level \\
CMS & Compact Muon Solenoid \\
\CP & Charge conjugation parity \\
CR & Colour reconnection \\
DIS & Deep inelastic scattering \\
ECAL & Electromagnetic calorimeter \\
EFT & Effective field theory \\
ERD & Early resonance decay \\
EW & Electroweak \\
FKR & Full kinematic reconstruction \\
FSR & Final-state radiation \\
GIM & Glashow--Iliopoulos--Maiani \\
HCAL & Hadronic calorimeter \\
HL-LHC & High-Luminosity Large Hadron Collider \\
ISR & Initial-state radiation \\
JER & Jet energy resolution \\
JES & Jet energy scale \\
JMS & Jet mass scale \\
\JSF & Jet scale factor \\
KINb & Kinematic method using \PQb tagging \\
LHC & Large Hadron Collider \\
LKR & Loose kinematic reconstruction \\
LO & Leading order \\
MB & Minimum bias \\
MC & Monte Carlo \\
ME & Matrix element \\
MPI & Multiple-parton interactions \\
MPV & Most probable value \\
\msbar & Modified minimal subtraction \\
MSR & Low-scale short-distance mass scheme derived from the \msbar mass \\
NLO & Next-to-leading order \\
NN & Neural network \\
NNLL & Next-to-next-to-leading logarithm \\
NNLO & Next-to-next-to-leading order \\
PDF & Parton distribution function \\
PF & Particle flow \\
PS & Parton shower \\
PU & Pileup \\
PUPPI & Pileup-per-particle identification \\
QCD & Quantum chromodynamics \\
RGE & Renormalisation group equation \\
RMS & Root mean square \\
SCET & Soft-collinear effective theory \\
SM & Standard model \\
SMEFT & Standard model effective field theory \\
UE & Underlying event \\
2D & Two-dimensional \\
4FS & Four-flavour number scheme \\
5FS & Five-flavour number scheme \\
\end{longtable}
\cleardoublepage \section{The CMS Collaboration \label{app:collab}}\begin{sloppypar}\hyphenpenalty=5000\widowpenalty=500\clubpenalty=5000
\cmsinstitute{Yerevan Physics Institute, Yerevan, Armenia}
{\tolerance=6000
A.~Hayrapetyan, A.~Tumasyan\cmsAuthorMark{1}\cmsorcid{0009-0000-0684-6742}
\par}
\cmsinstitute{Institut f\"{u}r Hochenergiephysik, Vienna, Austria}
{\tolerance=6000
W.~Adam\cmsorcid{0000-0001-9099-4341}, J.W.~Andrejkovic, T.~Bergauer\cmsorcid{0000-0002-5786-0293}, S.~Chatterjee\cmsorcid{0000-0003-2660-0349}, K.~Damanakis\cmsorcid{0000-0001-5389-2872}, M.~Dragicevic\cmsorcid{0000-0003-1967-6783}, A.~Hoang\cmsAuthorMark{2}\cmsorcid{0000-0002-8424-9334}, P.S.~Hussain\cmsorcid{0000-0002-4825-5278}, M.~Jeitler\cmsAuthorMark{3}\cmsorcid{0000-0002-5141-9560}, N.~Krammer\cmsorcid{0000-0002-0548-0985}, A.~Li\cmsorcid{0000-0002-4547-116X}, D.~Liko\cmsorcid{0000-0002-3380-473X}, I.~Mikulec\cmsorcid{0000-0003-0385-2746}, J.~Schieck\cmsAuthorMark{3}\cmsorcid{0000-0002-1058-8093}, R.~Sch\"{o}fbeck\cmsorcid{0000-0002-2332-8784}, D.~Schwarz\cmsorcid{0000-0002-3821-7331}, M.~Sonawane\cmsorcid{0000-0003-0510-7010}, S.~Templ\cmsorcid{0000-0003-3137-5692}, W.~Waltenberger\cmsorcid{0000-0002-6215-7228}, C.-E.~Wulz\cmsAuthorMark{3}\cmsorcid{0000-0001-9226-5812}
% ...
\par}
\vskip\cmsinstskip
\dag:~Deceased\\
$^{1}$Also at Yerevan State University, Yerevan, Armenia\\
$^{2}$Also at University of Vienna, Vienna, Austria\\
$^{3}$Also at TU Wien, Vienna, Austria\\
% ...
\end{sloppypar}
\end{document}